\documentclass[LaM,noexaminfo]{sapthesis}
\usepackage[english]{babel}
\usepackage{hyperref}
\usepackage{amsmath} 
\usepackage{amsfonts}
\usepackage{amssymb}
\usepackage{subcaption}
\usepackage{multirow}
\newtheorem{thm}{Theorem}[section]
\usepackage{graphicx}
\usepackage{enumitem}
\usepackage{float}
\usepackage[sort&compress,square,sort,comma,numbers]{natbib}

\title{Early-universe cosmology in \\ Einstein-scalar-Gauss-Bonnet gravity} 
\author{Laura Sberna} 
\IDnumber{1453550}
\submitdate{2015-2016} 
\advisor{Paolo Pani}
\course{Fisica} 
\courseorganizer{Dipartimento di Fisica}
\copyyear{2017} 
\examdate{26/09/2016}
\authoremail{laura92sb@gmail.com} 
\examiner{...}

\begin{document}
\maketitle
\begin{abstract}
Theoretical arguments and cosmological observations suggest that Einstein's theory of general relativity needs to be modified at high energies. One of the best motivated higher-curvature extensions of general relativity is Einstein-scalar-Gauss-Bonnet gravity, in which a scalar field is coupled to quadratic curvature invariants. This theory is inspired by an effective string-theory model and its predictions dramatically differ from Einstein's theory in high-curvature regions --~such as the interior of black holes and the early universe~-- where it aims at resolving curvature singularities.\\
In this work we derive cosmological solutions in Einstein-scalar-Gauss-Bonnet gravity for quadratic and for exponential coupling functions, and for any spatial curvature. We discuss already known solutions and find new nonsingular, inflationary, and bouncing solutions. We study the linear stability of these solutions and the absence of ghosts, finding that all the aforementioned solutions are unstable against tensor perturbations. We then introduce a simple, quadratic potential for the scalar field. In some cases the presence of a mass term cures the tensor instability. The proposed model is therefore a viable and attractive candidate to inflation, in which the scalar field is naturally provided by the gravitational sector.

\end{abstract}

\tableofcontents
\newpage
\chapter*{Relevant Physical Constants and Conventions}
\begin{align*}\label{key}
&\text{Newton constant}  \quad &G = 6.674 \times 10^{-11} \ N \cdot m^2 \cdot kg^{-2}\\
&\text{Speed of light} \quad &c=2.99792458 \times 10^8 \ m \cdot s^{-1} \\
&\text{Reduced Planck constant} \quad &6.58211928(15) \times 10^{16} \ eV \cdot s \\
&\text{Planck mass} \quad &1.220910 \times 10^{19} \ GeV /c^2 \\
&\text{Planck time} \quad &5.39116(13) \times 10^{-44} \ s
\end{align*}
Throughout this work we use units such that:
\begin{equation}\label{key}
\kappa= \frac{1}{M_{Pl}^2} =8\pi G\equiv 1.
\end{equation}
We call them reduced Planck units, as $ M_{Pl} $ is the reduced Planck mass.
We use mostly positive metric signature $ (-, +, +, +) $ and Misner, Thorn and Wheeler's \cite{Gravitation} sign convention for the Riemann tensor.
Commas denote partial derivatives, e.g. $ ()_{,\mu} = \frac{\partial }{\partial x^{\mu}}$ , while semicolons $ ()_{;\mu}$and nablas $ \nabla_{\mu} $ denote covariant derivatives.

\chapter*{Introduction}
The beauty of Einstein's theory lies in its simplicity. Despite, or thanks to, this simplicity, General Relativity has been remarkably successful as a classical theory of gravitational interactions from scales of millimetres through to kiloparsecs. A major prediction of this hundred-year-old theory, the existence of gravitational waves, has been recently confirmed by the detections of the LIGO experiment \cite{Ligo1,Ligo2}. On the whole, for General Relativity, internal consistency and theoretical elegance stand side by side with continuing experimental confirmation (see \cite{Clifford,Yunes,Testing} for a review of the status of experimental tests). It takes a closer look to unveil imperfections in this seemingly flawless theory.\\
From the theoretical point of view, one may quote, for instance, the non renormalizability of the theory, as stated in \cite{Sagnotti} (though an ultimate proof of non renormalizability is still missing to this date). This feature makes General Relativity a purely classical theory, which cannot undergo the quantum limit without modification. Despite the effort made by the scientific community to find this modification or completion, which goes under the name of Quantum Gravity, none of the proposed theories is entitled to lay claim to the crown.\\
Another theoretical issue of the General theory of Relativity is the rather generic prediction of singularities, stated by the Hawking-Penrose singularity theorems \cite{HawkPen}. A good physical principle seems to be that the prediction of a singularity by a theory indicates the breaking down of the theory itself. In fact, at the singularity, the theoretical model fails in providing a description of reality.\\
On the observational side, mysterious dark energy and dark matter are presently required to interpret astrophysical and cosmological measurements. The introduction of such unknown physical entities poses several conundrums and conceptual issues.\\
To be more precise, not one of the imperfections we enumerated can be considered as a true failure: no experimental data capable of disproving Einstein's theory has been collected so far. Taken individually, each of these issues may be resolved, more or less satisfactorily, within the standard framework. Looking at the bigger picture, though, single issues add up to make a clearly audible cracking sound: a step forward, something new, seems to be required to improve our description of the Universe at all scales.\\
Provided they pass known tests and they are consistent with General Relativity, modified theories of gravity \cite{Clifton} could shed light on some or all of the mentioned issues. This is the reason why the present work has been devoted to the study of an alternative theory of gravity: scalar-Gauss-Bonnet theory \cite{AntoModuliCorr,ARTSingfree}. The theory is studied in relation to a stage and physical setting where General Relativity still lacks experimental tests, showing theoretical fragility at the same time: early-time cosmology and the initial singularity.\\
The study is partly motivated by the link between scalar-Gauss-Bonnet theory and a Quantum Gravity candidate: string theory. String theory is today one of the best proposals for both a quantum theory of gravity and a unified theory of all known interactions. However, string theory is not at all primed or at least hinted at by experimental data: there is no direct experimental evidence for it. Therefore, each and every attempt to provide a testable prediction to the theory, like studying the implications of Gauss-Bonnet gravity, deserves to be carefully carried out. \\
Gauss-Bonnet gravity, whether we consider it as a complete, classical theory with its own right to be, or a low energy prediction of string theory, is an alternative to Einstein's gravity. The theory involves the Gauss-Bonnet invariant and introduces a scalar field, coupled non-minimally to gravity through a generic coupling function and the Gauss-Bonnet term itself. Some of its features make scalar-Gauss-Bonnet shine among the plethora of modified-gravity models:
\begin{enumerate}
\item it is consistent with known tests and at the same time,
\item it gives the opportunity to put to test the quantum gravity proposal of string theory;
\item it preserves Einstein's theory simplicity and spirit, through a (non minimal) coupling between matter, i.e. the scalar field, and curvature;
\item it introduces a minimal number of extra degrees of freedom (one, a scalar field) needed to construct new physics in the gravitational sector.
\end{enumerate}
Property number 4 refers to the fact that, under some basic assumptions, Einstein's is the one and only theory describing an interacting, Lorentz-invariant, massless helicity-2 particle (Lovelock theorem, \cite{Lovelock}): gravity as we know it. Adding a field is one way to circumvent the Lovelock theorem, and the choice of a scalar field is very common in modified-gravity.\\
When, to this list of good properties, one adds the existence, in GB gravity, of singularity-free solutions, one really get the feeling that the theory is a promising candidate. Such solutions, besides solving one of the long-standing problems of our theory of gravity, could allow us to describe the early Universe up to very high energy scales. They even open up the possibility that the gravitational sector, in our Universe, remained completely classical at early times. However, all non singular solutions found in GB gravity are plagued with some pathology, i.e. ghosts or instabilities. Unhealthy behaviour of singularity-free solutions is shared by many other modified-gravity theories, such as Eddington inspired Born-Infeld gravity \cite{Ferreira1,Ferreira2} (see \cite{Rubakov} for a review of non singular theories and their pathologies). 
Notwithstanding, chances are still that non singular solutions without pathologies can be obtained in GB gravity. Our work aims to better understand both singularity-free solutions and their instabilities, as well as the other equally important aspects of GB cosmology at early times. Our three key themes are:
\begin{itemize}
\item the existence of non singular cosmological solutions; 
\item the stability of cosmological solutions;
\item  the structure and phenomenology of cosmological solutions with an inflationary stage.
\end{itemize}
With the intent of providing a comprehensive background to our study, in Chapter 1 we briefly review the physical principles of cosmology in general, and of scalar field cosmology, perturbation theory and the inflationary paradigm in particular. For the same purpose, Chapter 2 is devoted to defining singularities as mathematical and physical objects. The latter Chapter also gives some hints on how singularities can be avoided in General Relativity and modified theories of gravity, introducing the reader to the core question of our work. The theory under study, Gauss-Bonnet gravity, is presented in some detail in Chapter 3, with a particular focus on its early time properties. In this Chapter the reader may find the modified Friedmann equations, and the definition of the quantities which determine the solutions' stability. Known and new cosmological solutions are numerically found and described in Chapter 4 and Chapter 5 for the two choices of coupling functions, quadratic and exponential respectively. The stability of the solutions is addressed in these Chapters as well. Furthermore, in Chapter 4 we discuss the recently proposed pure Gauss-Bonnet approximation \cite{KantiEarlyTime,Comment} and compare it to our findings. A further development is made in Chapter 6, where a potential is added to theory in its simplest form. The potential allows us to study inflationary solutions, along with their stability and phenomenological implications. Finally, in the Conclusion, we discuss and interpret our main results. Not only we give our interpretation of our original results, and also suggest how the understanding of Gauss-Bonnet early-time cosmology could be expanded.

\chapter{Standard Cosmology}
The aim of this Chapter is to introduce the first principles of standard and inflationary cosmology. First, we briefly recall the principles of General Relativity, assuming the reader to be familiar with it. Then, after a concise review of Friedmann-Robertson-Walker (FRW) metric, we describe cosmological evolution driven by a perfect fluid and by a simple free scalar field. We briefly introduce cosmological perturbation theory, discuss the short-comings of the conventional Big Bang theory and their resolution due to the inflationary paradigm. We will naturally focus on elements relevant for our main theme, i.e. early universe cosmology. This material is part of any cosmology textbook: our references are \cite{Weinberg,Kolb,TASI,Durrer,LLInflation}.

\section{Einstein field equations}
In General Relativity (GR), the dynamics of the metric tensor is determined by Einstein equations:
\begin{equation}\label{EE}
G_{\mu\nu}=\kappa T_{\mu\nu}+\Lambda g_{\mu\nu},
\end{equation}
where $\kappa=8\pi G$ and $G = 6.674 \times 10^{−11} N \ m^2/kg^2$ is Newton's constant. $G_{\mu\nu}$ is the usual Einstein tensor:
\begin{equation}
G_{\mu\nu}\equiv R_{\mu\nu}-\dfrac{1}{2}g_{\mu\nu}R,
\end{equation}
with $R_{\mu\nu}$ and $R$ defined in terms of the metric:
\begin{equation}
\begin{split}
&R\equiv g^{\mu\nu}R_{\mu\nu}, \quad R_{\mu\nu}=\Gamma^{\alpha}_{\mu\nu,\alpha}-
\Gamma^{\alpha}_{\mu\alpha,\nu}+\Gamma^{\alpha}_{\beta\alpha}\Gamma^{\beta}_{\mu\nu}
-\Gamma^{\alpha}_{\beta\nu}\Gamma^{\beta}_{\mu\alpha},\\
&\Gamma^{\alpha}_{\mu\nu}=\dfrac{1}{2}g^{\alpha\lambda}(g_{\mu\lambda,\nu}+g_{\lambda\nu,\mu}-g_{\mu\nu,\lambda}).
\end{split}
\end{equation}
For the sake of completeness we have considered a cosmological constant $\Lambda$, which is necessary in the Concordance Cosmological Model to account for present observed accelerated expansion (see for example PLANCK latest results \cite{PLANCK}). From now on we will switch to reduced Planck units.\\
Eq.s \eqref{EE} are obtained, through the variational principle $\frac{\delta S}{\delta g_{\mu\nu}}=0$, from Einstein-Hilbert action:
\begin{equation}\label{EHaction}
S_{E-H}=\int \! \mathrm{d}^4x \ \sqrt{-g}(\dfrac{1}{2\kappa}R +\Lambda) \ + S_{matter}.
\end{equation}

\section{Cosmological Principle and FRW metric}
Cosmology describes the structure and evolution of the universe on the largest scales. Here we can apply the \emph{Cosmological Principle}, i.e. the hypothesis that the Universe is spatially homogeneous (same at every point) and isotropic (same in every direction). It is then possible to choose coordinates $\lbrace t,r,\varphi \theta\rbrace$ for which the metric takes the form:
\begin{equation}\label{FRW}
ds^2=-dt^2+a(t)^2\left[ \dfrac{dr^2}{1-kr^2}+r^2d\Omega^2 \right], \quad \quad d\Omega^2=d\theta^2 + \sin^2 (\theta) d \varphi^2.
\end{equation}
Here the \emph{scale factor} $a(t)$, an unknown function, characterizes the relative size of space-like hyper-surfaces $\Sigma$ at different times. The evolution of FRW Universe is completely described by the one function $a(t)$ of the cosmic time. The curvature parameter $k$ is a constant, which for a suitable choice of units for $r$ can be chosen to have value $+1$, for positively curved $\Sigma$, $-1$, for negatively curved $\Sigma$, or $0$, for flat $\Sigma$. This metric ansatz uses comoving coordinates: free falling bodies, such as galaxies without forces acting on them, keep fixed coordinates $r,\theta,\varphi$. The corresponding physical distance is generally time dependent and given by $R(t)=a(t)r$.\\
By a coordinate transformation the metric \eqref{FRW} may be written as:
\begin{equation}
ds^2=-dt^2+a(t)^2\left[d\chi^2+ \Phi_k(\chi)^2 d\Omega^2  \right],
\end{equation}
where 
\begin{equation}\label{FRWchi}
r^2=\Phi\equiv \begin{cases}
\sinh^2 \chi \quad k=-1\\
\chi^2 \quad k=0\\
\sin^2 \chi \quad k=+1\\
\end{cases}
\end{equation}
The rate of expansion of the Universe is given by the \emph{Hubble parameter}:
\begin{equation}
H(t)=\dfrac{\dot{a}}{a}.
\end{equation}
which has unit of inverse time and is positive for an expanding universe (and
negative for a collapsing one). It sets the fundamental scale of the FRW space-time: the characteristic time-scale of the homogeneous universe is the Hubble time, $t_H \sim H^{-1}$, and the characteristic length-scale is the Hubble length, $d_H \sim H$ (in our units). The Hubble time sets the scale for the age of the universe. The Hubble length sets the size of the observable universe and is also called \emph{Hubble horizon}, as it provides an estimate of the distance light can travel while the Universe expands appreciably. \\
For later use, we also define the comoving horizon as the comoving distance travelled by light from initial time:
\begin{equation}\label{comhor}
d(t)= a(t) \int_0^t \; dt' \dfrac{1}{a(t')}= a(t) \int_0^a \; d \ln a \; \dfrac{1}{a H}.
\end{equation} 
Since we will later use it, we introduce here a different coordinate system, defining the \emph{conformal time}:
\begin{equation}\label{conft}
d \tau = \dfrac{dt}{a(t)}, \quad \Rightarrow \quad \tau = \int \; \dfrac{dt}{a(t)}.
\end{equation}
In the new reference frame, the FRW line element reads:
\begin{equation}
ds^2= a(\tau) \left[ d\tau^2 + d\chi^2+ \Phi_k(\chi)^2 d\Omega^2 \right].
\end{equation}

\section{Cosmology with a perfect fluid}
The source of a homogeneous and isotropic Universe, the energy-momentum tensor $T_{\mu\nu}$ appearing in Eq.s \eqref{EE}, must satisfy the Cosmological principle, too. To be consistent with the symmetries of the metric, the stress-energy tensor must be diagonal and spatial components must be equal. This result can be elegantly written as:
\begin{equation}
T_{\mu\nu}=(\rho(t)+p(t))u^{\mu}u^{\nu}+p(t)g^{\mu\nu},
\end{equation}
where $u^{\mu}$ is a time-like four-velocity. This means that, under the homogeneity and isotropy assumptions, the energy-momentum tensor of the Universe takes the same form as for a perfect fluid. However, also an imperfect fluid with bulk viscosity would satisfy the symmetry requirements. In a frame comoving with the fluid we may choose $u^{\mu}=(1,0,0,0)$ and obtain
\begin{equation}
T^{\mu}_{\nu}= \left( \begin{matrix}
-\rho &0&0&0\\
0&p&0&0\\
0&0&p&0\\
0&0&0&p\\
\end{matrix} \right),
\end{equation} 
The Einstein Equations take the form of two coupled, non-linear ordinary differential equations, also called the \emph{Friedmann Equations}:
\begin{equation}\label{F1}
H^2\equiv \left( \dfrac{\dot{a}}{a} \right)=\dfrac{1}{3}\rho +\dfrac{k}{a^2},
\end{equation}
and
\begin{equation}\label{F2}
H^2+\dot{H}\equiv  \dfrac{\ddot{a}}{a} =-\dfrac{1}{6}(\rho +3p).
\end{equation}
First of all, we notice that if the Universe is filled with ordinary matter, i.e. matter satisfying the strong energy condition:
$\rho+3p>0$ of Theorem \ref{th:HP}, Eqn. \ref{F2}  implies $\ddot{a}<0$. This, for an expanding universe (i.e. $\dot{a}>0$), indicates the existence of a singularity in the finite past: $a(t_0)=0$. This is the anticipated break-down of GR in the early Universe, motivating this thesis.\\
Note that Friedmann equation \eqref{F1} can be written as
\begin{equation}
\Omega -1=\dfrac{k}{H^2 a^2},
\end{equation}
having defined the \emph{critical density} $\rho_c=3{H_0}^2$ and the ratio:
\begin{equation}
\Omega = \dfrac{\rho}{\rho_c}.
\end{equation}
Then there is a correspondence between the sign of $k$, the spatial curvature, and the sign of $\Omega-1$:
\begin{equation}
\begin{split}
k=+1 \quad &\Rightarrow \quad \Omega >1 \quad \textrm{closed},\\
k=0 \quad &\Rightarrow \quad \Omega =1 \quad \textrm{flat},\\
k=-1 \quad &\Rightarrow \quad \Omega <1 \quad \textrm{open}.\\
\end{split}
\end{equation}
We define the equation of state parameter $w$, and assume it is independent of time:
\begin{equation}\label{eos}
w=\dfrac{p}{\rho}.
\end{equation}
Finally, the fluid must satisfy a continuity equation, given by a suitable combination of Einstein equations or as a component of the \lq\lq conservation" of the energy-momentum tensor $T^{\mu\nu}_{;\nu}=0$:
\begin{equation}\label{conteq}
\dot{\rho}+3H(\rho+p)=0.
\end{equation}
Friedmann equations \eqref{F1} \eqref{F2}, the equation of state \eqref{eos}, and the continuity equation \eqref{conteq} completely determine the cosmological dynamics.\\
The continuity equation can be re-written and integrated:
\begin{equation}
\dfrac{d \ln \rho}{d \ln a}=-3(1+w), \quad \Rightarrow \quad \rho \propto a^{-3(1+w)}.
\end{equation}
Together with the Friedmann equation \eqref{F1} this leads to the time evolution of the scale factor:
\begin{equation}
a(t) \propto \begin{cases}
t^{\dfrac{2}{3(1+w)}}   &w\neq -1,\\
e^{H t}   &w=-1.\\
\end{cases}
\end{equation}
In Table \ref{table:FRW} we explicitly see the evolution of the Universe for different matter contents and a flat background ($k=0$).
\begin{table}[H]\label{tab:cosmology}
\centering
\begin{tabular}{|c|c|c|c|c|}
\hline
\textbf{Content}&w&$\rho(a)$&$a(t)$&$t_i$\\ \hline
MD&0&$a^{-3}$&$t^{2/3}$&$0$\\ \hline
RD&1/3&$a^{-4}$&$t^{1/2}$&$0$\\ \hline
$\Lambda$&-1&$const$&$e^{Ht}$&$-\infty$\\ \hline
\end{tabular}
\vspace{2mm}
\caption{FRW solutions for a flat universe dominated by radiation (RD), matter (MD) or a cosmological constant ($\Lambda$).}
\label{table:FRW}
\end{table}
Again we stress that ordinary matter leads to a cosmological initial singularity, i.e. $a \rightarrow 0$.\\

\section{Cosmology with a scalar field}\label{FriedScal}
We write Friedmann equations in presence of a free, massless scalar field minimally coupled to gravity (see action in Section \ref{actioneq} with $f(\phi)=0$). We neglect the cosmological constant term:
\begin{equation}\label{ActScal}
S=S_{E-H} \; + \; \int \! \mathrm{d}^4x \; \sqrt{-g} (-\dfrac{1}{2}\partial_{\mu}\phi\partial^{\mu}\phi - V(\phi)).
\end{equation}
The scalar field equation is:
\begin{equation}\label{EqScal}
\dfrac{1}{\sqrt{-g}}\partial_{\mu}\left(\sqrt{-g}\partial^{\mu}\phi\right)=
V'(\phi),
\end{equation}
A free, massless scalar field has $V=0$.The dynamics of a FRW metric with generic curvature $k$, together with that of the scalar field, is give by the system:
\begin{equation}\label{eqsSc}
\begin{cases}
\ddot{\phi}+3H\phi=0,\\
\dfrac{3k}{a^2}+3H^2-\dfrac{\dot{\phi}^2}{2}=0,\\
\dfrac{k}{a^2}+3H^2+2\dot{H}+\dfrac{\dot{\phi}^2}{2}=0.\\
\end{cases}
\end{equation}
\subsection{Constant scalar field: vacuum cosmology}
We first assume that the scalar field takes a constant value, $\phi(t)= const$, $\dot{\phi}=0$. This configuration is equivalent to consider an empty Universe, as Eq.s \eqref{eqsSc} reduce to the vacuum Freidmann equations:
\begin{equation}
\begin{cases}
(3k+\dot{a}^2)/a^2=0,\\
(k+\dot{a}^2+2a\ddot{a})/a^2=0.\\
\end{cases}
\end{equation}
From the first equation we see that $\dot{a}^2=-3k$. Depending on the value of the spatial curvature we find different solutions:
\begin{itemize}
\item[-] $k=0$: with $\dot{a}=0$ both equations are satisfied, so $a(t)=a_0$, and we have a static universe with arbitrary radius;
\item[-] $k=1$: the only possible solution is a static universe with infinite radius, so that $\dfrac{1}{a^2}=0$; 
\item[-] $k=-1$: in this case $\dot{a}=\pm \sqrt{3}$, so 
\begin{equation}
a(t)=\pm \sqrt{3}t+a_0, \quad \quad H(t)=\dfrac{\sqrt{3}}{\sqrt{3}t\mp a_0}.
\end{equation}or simply $a\sim t$. This is the most interesting case: we have a linearly expanding universe, and the expansion rate is larger than in radiation or matter dominated eras of standard cosmology. Clearly, a singularity is present at some initial time.
\end{itemize}
\subsection{Dynamical scalar field on flat FRW}
\begin{figure}[!ht]
 \centering
 \begin{subfigure}{.46\textwidth}
 \centering
    \includegraphics[width=1\textwidth]{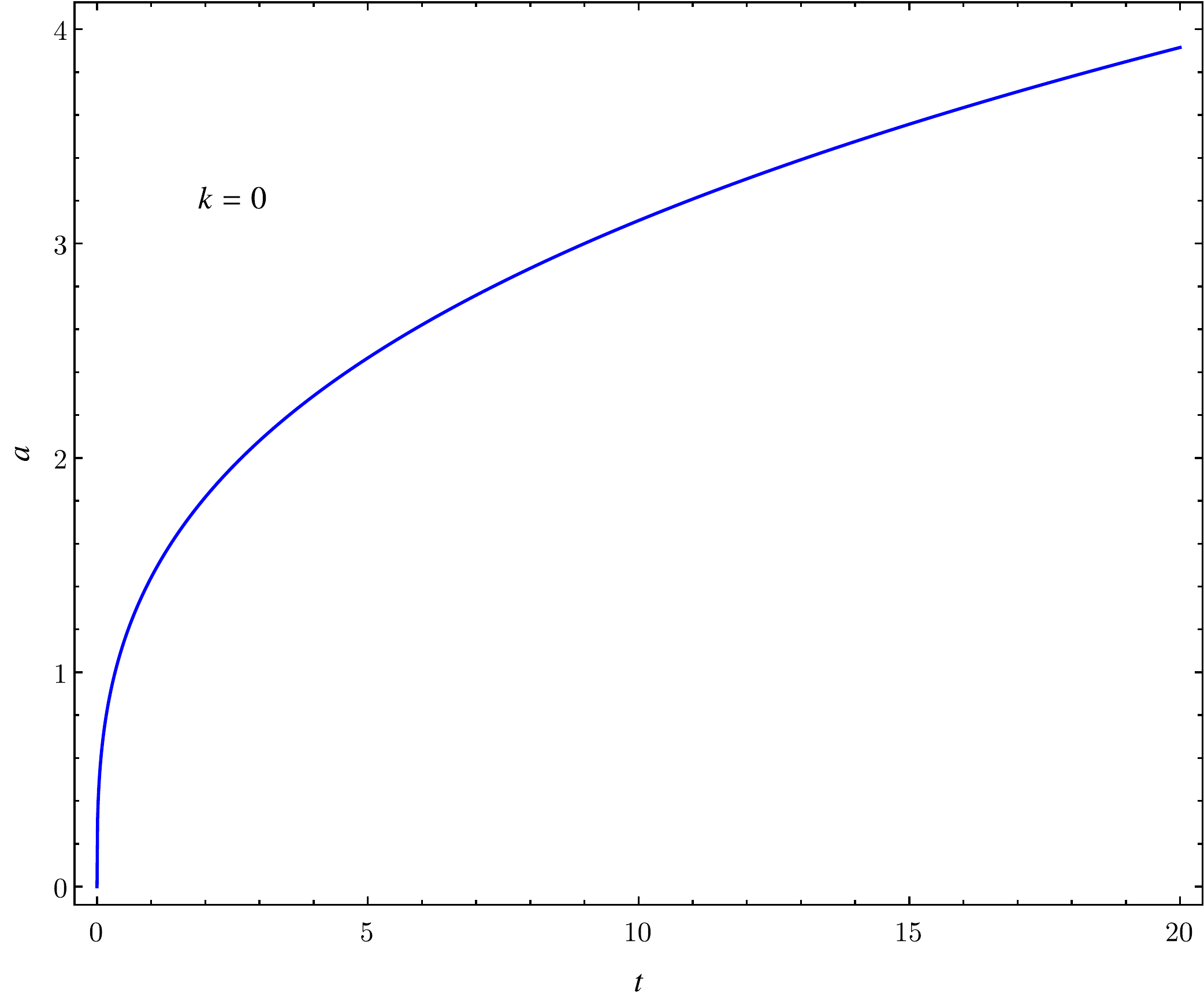}
  \caption{Scale factor as a function of time.}\label{fig:a0}
\end{subfigure}
\begin{subfigure}{.46\textwidth}
\centering
    \includegraphics[width=1\textwidth]{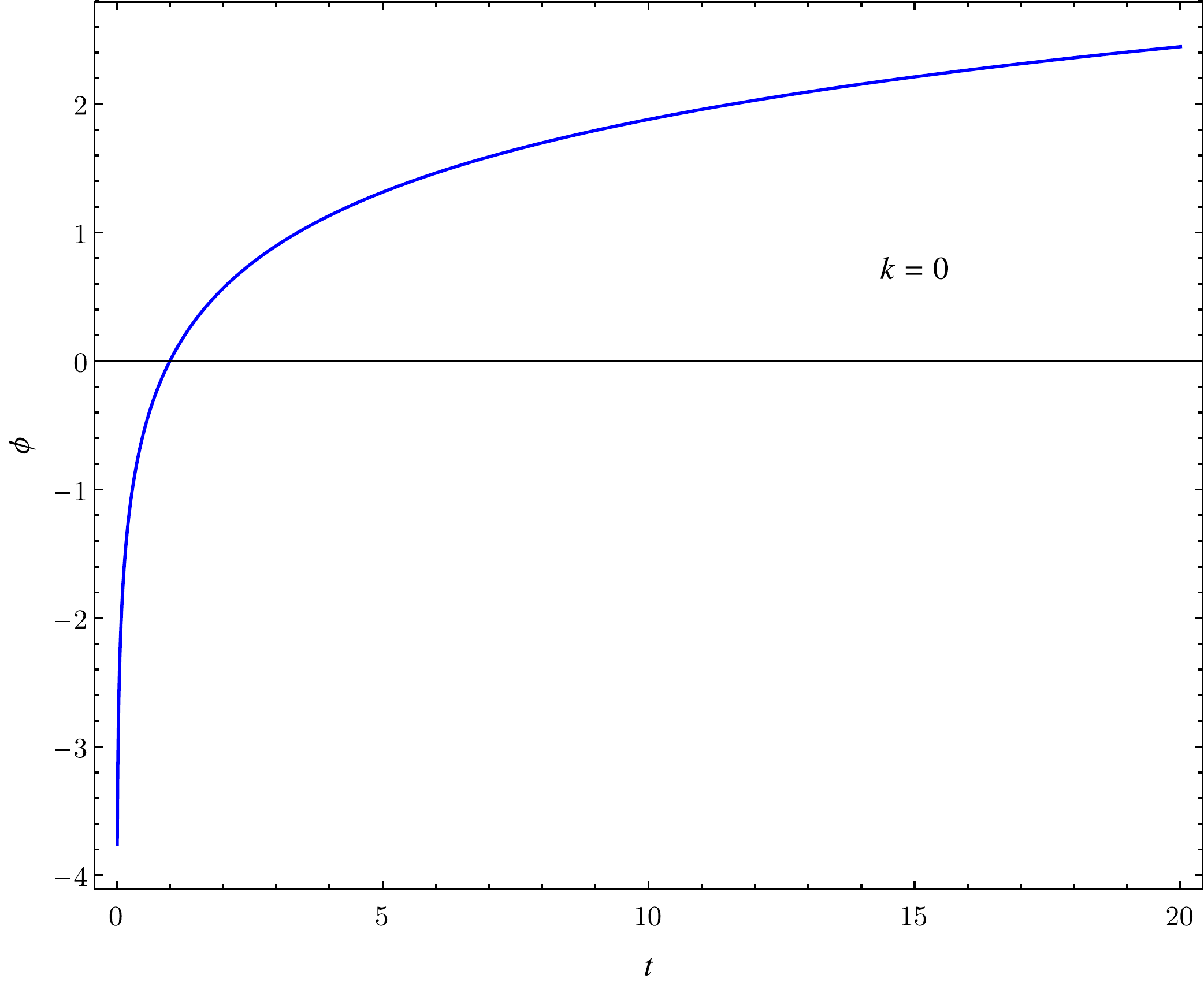}
    \caption{Scalar field as a function of time.}\label{fig:phi0}
    \end{subfigure}\\[1ex]
\begin{subfigure}{\linewidth}
\centering
\includegraphics[width=.46\textwidth]{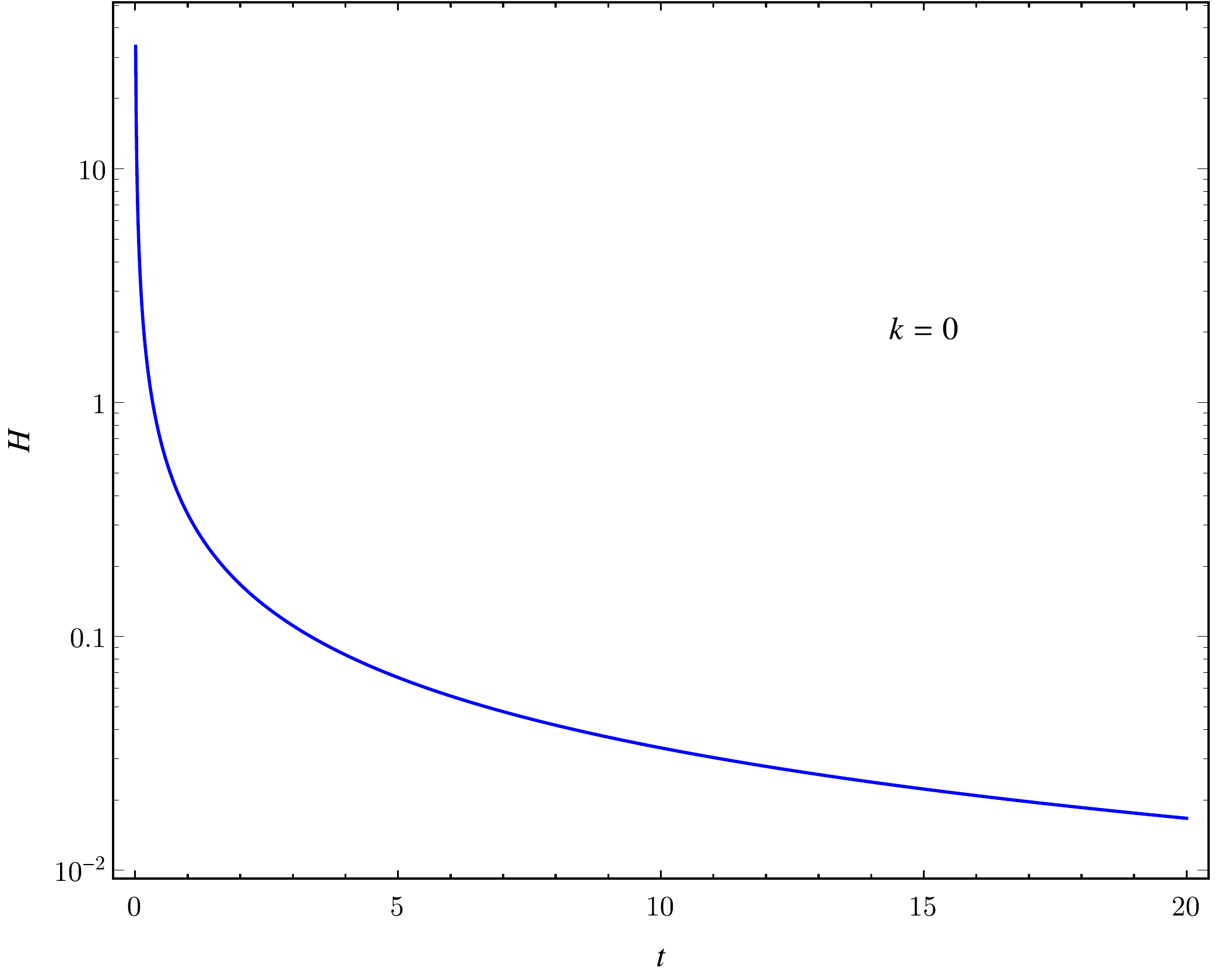}
\caption{Hubble parameter as a function of time.}\label{fig:H0}
\end{subfigure}
   \caption{GR solution for a flat universe with an evolving scalar field. We choose $a_0=0$, $c=1$, $\phi_0=0$.}
\end{figure}

Next we let the scalar field evolve with time ($\dot{\phi}\neq 0$) and again separately study the three cases. \\
We start with a flat background, $k=0$. In this case one can find the analytic solution:
\begin{equation}
a(t)=\pm 3^{1/3}(c \ t+a_0)^{1/3}; \quad  \phi(t)=\phi_0 \pm \sqrt{\dfrac{2}{3}}\ln(c \ t+a_0); \quad H(t)= \dfrac{c}{3(c \ t+a_0)};
\end{equation}
where $c$, $a_0$, $\phi_0$ are integration constants. \\
This solution corresponds to an expanding universe with a singularity at a finite cosmological time, and is displayed in Figures \ref{fig:a0}, \ref{fig:phi0}, \ref{fig:H0}. Both the scale factor and the scalar field exhibit an initial singularity. The expansion rate is smaller than the one in the two epochs of standard cosmology. This can be explained by the presence of the scalar field as a source in Einstein equations: we can deduce that the slowing power of the scalar field is more effective than that of a perfect fluid. \\ 
\subsection{Dynamical scalar field on curved FRW}
\begin{figure}[!ht]
  \centering
      \includegraphics[width=0.5\textwidth]{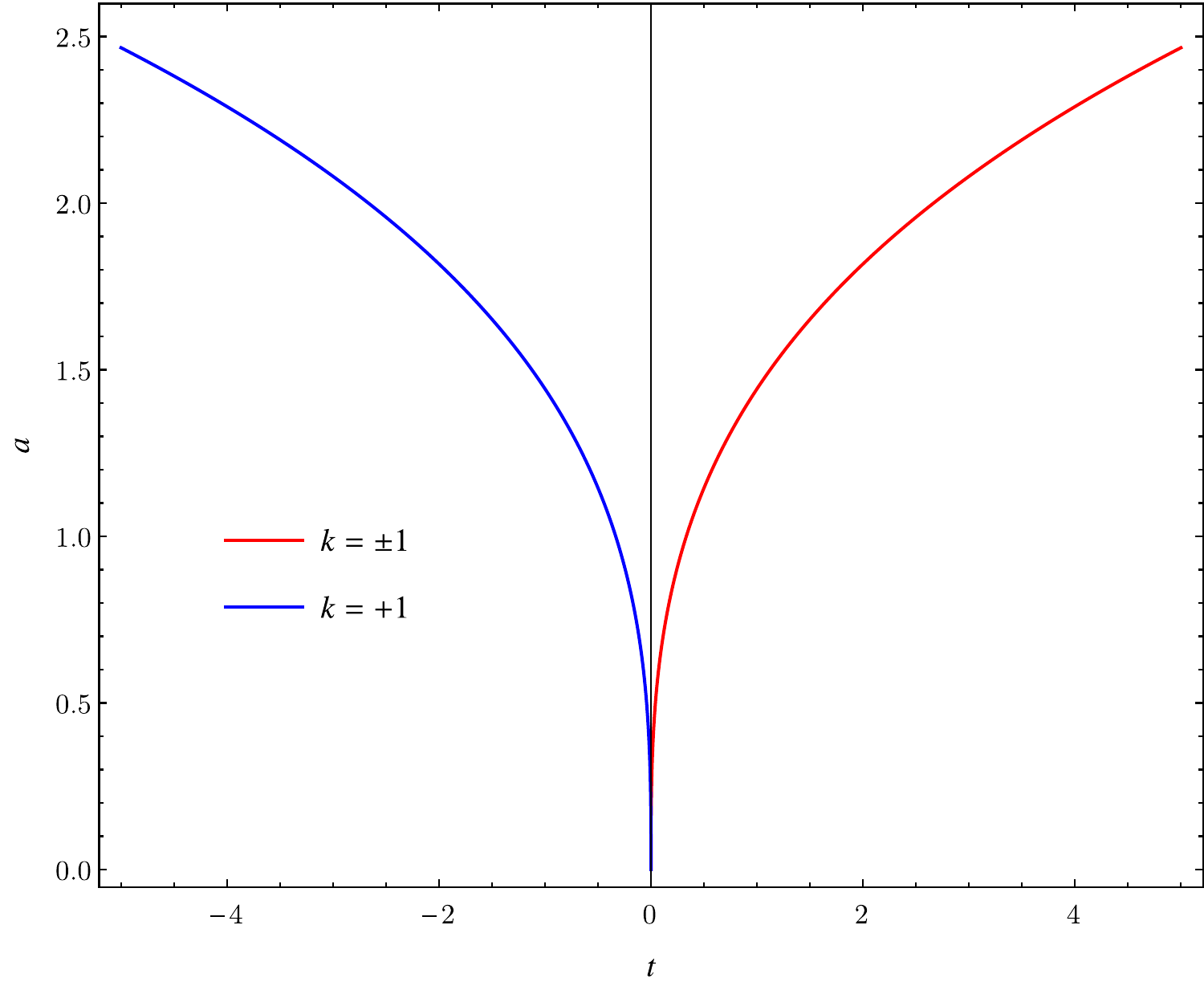}
       \caption{Asymptotic behaviour of the scale factor near the singularity for $k=\pm$, in GR, with an evolving scalar field. We choose $c=1$, $c'=0$ so that singularity occurs at $t=0$.}\label{GRk=+-1a}
\end{figure}

Here we move on to the curved case, $k\neq0$. The equations cannot be solved analytically, but we can study the asymptotic behaviour of the solution near the singularity. Following Kanti \textit{et al.} in \cite{KantiSing}, we introduce the new variables:
\begin{equation}\label{xydef}
x=\dot{\phi}, \quad y=\dfrac{1}{a^2}>0.
\end{equation}
The second equation in \eqref{eqsSc} can be solved for $x$, giving:
\begin{equation}\label{eqx}
x=s\sqrt{6\left(ky+H^2\right)},
\end{equation}
where $s=\pm 1$. The remaining equations become:
\begin{equation}
\begin{split}\label{eqHy}
\dot{H}&=\dfrac{dH}{d\phi}x=-3H^2-2ky,\\
\dot{y}&=\dfrac{dy}{d\phi}x=-2yH.
\end{split}
\end{equation}
The system above may be reduced to a single equation:
\begin{equation}
\dot{H}=\dfrac{dH}{dy}\dot{y}=\dfrac{dH}{dy}\left(-2yH\right)=-3H^2-2ky \quad \Rightarrow \quad 2ydH-3Hdy=\dfrac{2ky}{H}dy.
\end{equation}
If we multiply both sides of the last expression by $\dfrac{H}{y^4}$, we get, integrating once:
\begin{equation}\label{Hy}
d\left(\dfrac{H^2}{y^3}\right)=-k \ d\left(\dfrac{1}{y^2}\right) \quad \Rightarrow \quad H^2+ky=c_1 y^3,
\end{equation}
with $c_1$ being a positive integration constant, because we have integrated positive quantities. We obtain $H$ (choosing $H>0$) from the above expression and substitute it in the differential equation for $y$ \eqref{eqHy}:
\begin{equation}
\dot{y}=\dfrac{dy}{d\phi}s\sqrt{6\left(c_1 y^3\right)}=-2y\sqrt{y \left(c_1 y^2-k \right)}.
\end{equation}
We can solve the equation obtained and find $y$ as a function of the scalar field:
\begin{equation}\label{soly}
y(\phi )= \frac{k  c_1 e^{s\sqrt{\frac{2}{3}} s \ \phi}+{c_2}^2 e^{-\sqrt{\frac{2}{3}} s \ \phi }}{2 c_1c_2}.
\end{equation}
From this solution it is evident that there is an invariance under the interchange of the signs of $s$ and $\phi$. Then we can keep the sign of $s=1$ fixed while allowing $\phi$ to take both signs.\\
We focus now on the singularity region. A cosmological singularity occurs when $a(t)\rightarrow0$, which means $y\rightarrow+\infty$. From expression \eqref{soly} we conclude that:
\begin{itemize}
\item[-]for $k=1$, $y$ goes to infinity when $\phi\rightarrow\pm\infty$;
\item[-]for $k=-1$, the singularity arises only when $\phi\rightarrow-\infty$, because of the condition $y>0$ \eqref{xydef}.  
\end{itemize} 
Near the singularity we can then evaluate the approximate behaviour:
\begin{equation}
\begin{split}
y&\simeq\dfrac{c_2}{2c_1}e^{-\sqrt{\frac{2}{3}} \phi} \quad \textrm{for}\quad \phi\rightarrow -\infty \quad\textrm{and}\quad k=\pm1;\\
y&\simeq\dfrac{k}{2c_2}e^{\sqrt{\frac{2}{3}} \phi} \quad \textrm{for}\quad \phi\rightarrow \infty \quad\textrm{and}\quad k=+1.\\
\end{split}
\end{equation}
From expression \eqref{Hy} we could also derive the corresponding behaviour for $H$. Finally we use the differential equation for the scalar field \eqref{eqx} to deduce its dependence on time near the singularity, put it into the expression of $y$ and obtain the behaviour of the scale factor in the same region:
\begin{equation}
\begin{split}
a(t)&\simeq(c'+3\sqrt{c}t)^{1/3} \quad \textrm{for} \quad k=\pm1;\\
a(t)&\simeq(c'-3\sqrt{c}t)^{1/3}\quad \textrm{for} \quad k=+1.
\end{split}
\end{equation} 
The expressions above describe the asymptotic behaviour of a Universe with a singularity at finite time. Both branches are displayed in Figure \ref{GRk=+-1a}. The closed Universe ($k=1$) posses two branches of singular solutions, while for the open Universe ($k=-1$) there is only one branch, as in the flat case ($k=0$).

\section{Cosmological perturbation theory}
In this Section we treat the general relativistic theory of linear perturbations. The use of linear perturbation theory is justified by the smallness of fluctuations we want to apply our results to, such as the inhomogeneities of the CMB (at the $\sim 10^{-5}$ level) \cite{PLANCK}. \\
In order to perform a perturbative expansion, all the quantities $X(t,\vec{x})$ involved (the metric, matter fields, etc.) are split into a homogeneous background and a spatially dependent perturbation:
\begin{equation}
X(t,\vec{x})=\bar{X}(t)+ \delta X(t,\vec{x}), \quad X= \phi, \ g_{\mu\nu}, \ \rho, \textrm{ etc.},
\end{equation}
where barred quantities are always the unperturbed ones. Assuming $\vert \delta X (t,\vec{x}) \vert \ll \vert X(t,\vec{x}) \vert$, we can expand Einstein equations at linear order and obtain:
\begin{equation}
\delta G_{\mu\nu}= \delta T{\mu\nu},
\end{equation} 
the first order perturbation equations (in our Planckian units). In the study of perturbations we will always refer to a spatially \emph{flat} FRW background.

\subsection{Gauge choice}
The first fundamental problem we face in cosmological perturbation theory is the choice of a gauge. The theory is diffeomorphism invariant, but the split between background and perturbations is not unique and depends on the choice of coordinates, i.e. the \emph{gauge choice}. While in FRW space-time it was easy to identify the comoving frame as the privileged one, because the Universe looked homogeneous and isotropic in it, an inhomogeneous space-time has often not a preferred coordinate choice. The gauge freedom can lead to ambiguous results: on the one hand, a suitable coordinate choice can introduce \textit{fictitious}, non physical perturbations even in FRW space-time; on the other hand, carefully chosen coordinates can remove a \textit{real} perturbation from an inhomogeneous Universe. \\
In order to solve the ambiguity between real and fake perturbations we will introduce gauge-invariant quantities, i.e. quantities that do not depend on the coordinate choice. Gauge-invariant perturbations, then, will be real and physical. Since the perturbation equations are covariant, it is always possible to express them in terms of gauge invariant variables \cite{Bardeen}. An alternative solution to the gauge problem is to fix the gauge, but we will not treat gauge-fixing here.\\
We consider infinitesimal transformations which leave the background metric invariant, i.e. which deviate only in first order from identity. The infinitesimal diffeomorphism can be represented as (being the diffeormophism group a Lie group):
\begin{equation}
x^{\mu'}\rightarrow x^{\mu}+\epsilon \epsilon^{\mu}.
\end{equation}
The group generators are Lie derivatives in the direction defined by the vector $h^{\mu}$:
\begin{equation}
L_\epsilon=\epsilon^{\mu} \dfrac{\partial}{\partial x^{\mu}}.
\end{equation}
The generic quantity $X$ transforms as:
\begin{equation}\label{TransfX}
X\rightarrow \alpha(X) \quad \textrm{with} \quad \alpha=\mathbb{I}+\epsilon L_{\epsilon},
\end{equation}
and is said to be gauge-invariant if $\alpha(X) =X$.

\subsection{SVT decompositions}
Thanks to the great deal of symmetry possessed by the spatially flat, homogeneous and isotropic background, a decomposition of the metric and the stress-energy perturbations
into independent scalar (S), vector (V) and tensor (T) components is allowed. In order to define this decomposition, we go to the Fourier space:
\begin{equation}
X_{\vec{k}}(t)=\int \; d^3 x e^{i \vec{k} \cdot \vec{x}} X(t,\vec{x}).
\end{equation}
The linearly perturbed equations are translation invariant, thus each Fourier mode is independent and can be studied separately. Scalar, vector and tensor components are distinguished by their helicity. A component has helicity $e$ if, under a rotation of an angle $\psi$, transforms as:
\begin{equation}
X_{\vec{k}} \rightarrow e^{ie \psi} \ X_{\vec{k}}.
\end{equation}
Scalar, vector and tensor perturbations have helicity $0$, $\pm 1$ and $\pm 2$, respectively. In real space (as opposed to Fourier space), SVT components are defined by their transformations on spatial hypersurfaces.\\
The advantage of this decompositions is that, at the linear order, each type of perturbation evolve independently and can be studied separately.

\subsection{Perturbed metric and matter}
We define the perturbations of the scalar field and the metric field around FRW background through the following expressions:
\begin{equation}
\phi(t,\vec{x})=\bar{\phi}(t)+\delta \phi(t,\vec{x}) , \quad g_{\mu\nu}(t,\vec{x})= \bar{g}_{\mu\nu}(t)+ \delta g_{\mu\nu}(t,\vec{x}).
\end{equation}
We choose to parametrize metric perturbations as:
\begin{equation}\label{pertmetr}
ds^2=-(1+2\Phi)dt^2+2 a B_i dt dx^i + a^2 [(1-2\Psi)\delta_{ij} + E_{ij}]dx^i dx^j,
\end{equation}
where $\Phi$, $B_i$, $\Psi$, $E_{ij}$ are functions of space and time. We also define the SVT components in real space:
\begin{equation}
\begin{split}
B_i&=\partial _i B +S_i, \quad \textrm{with} \quad \partial ^i S_i=0;\\
E_{ij}&=  2\partial _{(i} \partial _{j)} E + 2\partial (_i F_j) + h_{ij} .
\end{split}
\end{equation}
Is easy to understand how to perform the SVT decomposition of a three-vector: we can split any three-vector into the gradient of a scalar and a divergenceless vector. A similar procedure is applied to the 3-tensor, with $\partial (_i F_j)= 1/2 \left( \partial_i F_j +\partial_j F_i \right) $ ,$F_i$ divergenceless, $h_{ij}$ divergenceless and traceless. The ten degrees of freedom of the metric have thus been divided between $4$ scalars, $2$ vectors and $1$ tensor.\\
Inflation does not produce vector perturbations ($F_i$, $S_i$) and even if they were produced, they would be damped away by the expansion of the Universe. For this reason we choose to ignore vector perturbations and focus on scalar and tensor perturbations, which can be observed in the present Universe as density fluctuations and gravitational waves respectively. \\
Under the gauge transformation defined in \eqref{TransfX}, tensor fluctuations are invariant, while scalar perturbations transform.\\

We also define the small perturbations of the stress-energy tensor:
\begin{equation}
T^{\mu}_{\nu}(t,\vec{x})= \bar{T}^{\mu}_{\nu}(t) + \delta T^{\mu}_{\nu} (t,\vec{x}) .
\end{equation}
The SVT decomposition reads:
\begin{equation}
\begin{split}
T^0_0&=-\left( \bar{\rho} +\delta \rho \right),\\
T^0_i&= \left( \bar{\rho} +\bar{p} \right) a \ v_i,\\
T^i_0&= - \left( \bar{\rho} +\bar{p} \right) \left( v^i-B^i \right) ,\\
T^i_j&=\delta^i_j \left( \bar{p} +\delta p \right) +\Sigma^i_j.
\end{split}
\end{equation}
The perturbed stress-energy tensor acquires a momentum density $(\delta q)_i=\left( \bar{\rho} +\bar{p} \right) \ v_i$ and  an anisotropic stress $\Sigma^i_j$, which is gauge invariant. The three-momentum density can be further decomposed in its scalar and vector part: $(\delta q)_i= \partial_i \delta q + (\delta q)'_i$.

\subsection{Gauge invariant perturbation theory}
We are finally ready to introduce the gauge invariant variables (\textbf{Bardeen variables}). For example, we define the \emph{Bardeen potential} $\Psi_B$:
\begin{equation}
\Psi_B= \Psi+a^2 H \left( \dot{E} -B/a  \right).
\end{equation}
An important gauge-invariant scalar quantity is the \emph{curvature perturbation on uniform-density hypersurfaces}, $\zeta$:
\begin{equation}
-\zeta= \Psi- \dfrac{H}{\dot{\bar{\rho}}}\delta \rho.
\end{equation}
The quantity defined above measures the spatial curvature $R^{(3)}$ of constant-density hypersurfaces. Here we also define adiabatic matter
perturbations, which have the property that the local state of matter at some space-time point $(t, \vec{x})$ of the perturbed universe is the same as in the background universe at some slightly different time $t'(\vec{x})=t+\delta t (\vec{x})$. The perturbed Einstein equations tell us that the variable $\zeta$ remains constant outside the horizon ($k\ll aH$) for adiabatic matter perturbations.\\
Another important gauge-invariant scalar is the \emph{comoving curvature perturbation}, $\mathcal{R}$:
\begin{equation}
\mathcal{R}=\Psi - \dfrac{H}{\bar{\rho} +\bar{p}}\delta q.
\end{equation}
The quantity $\mathcal{R}$ is related to the intrinsic curvature of three-surfaces of constant time, i.e. comoving hypersurfaces. From the perturbation equation (not reported here) one can deduce that $\mathcal{R}$, too, is conserved outside the horizon for adiabatic matter perturbations.\\
Tensor perturbations, i.e. gravitational waves, are described by the two polarization modes $h=h_{+},h_{\times}$ of the already gauge-invariant quantity $h_{ij}$. The first-order Einstein equations for tensor perturbations in the Fourier space ($h=h(t,\vec{k})$) read:
\begin{equation}\label{eqperturbGR}
\ddot{h}+3H\dot{h}+\dfrac{k^2}{a^2} h=0.
\end{equation}
The linearized Einstein equations also relate $\zeta$, $\mathcal{R}$ and $\Psi_B$ as follows:
\begin{equation}\label{zetaRPsi}
-\zeta=\mathcal{R} + \left( \dfrac{k}{aH}\right)^2 \dfrac{2\bar{\rho}}{3(\bar{\rho} +\bar{p})} \Psi_B.
\end{equation}
From the equation above we infer that $\zeta$ and $\mathcal{R}$ are approximately equal $\zeta \approx \mathcal{R} $ on super-horizon scales ($k\ll aH$).

\subsection{Statistics}
The fundamental statistical quantity to characterize perturbations are the power spectral density $P(k)$ and the power spectrum $ \Delta (k)$, which, if the fluctuations are Gaussian, contain \emph{all} the statistical information. For example, the power spectrum of the comoving curvature perturbation is defined as:
\begin{equation}\label{PSR}
\langle \mathcal{R}_{\vec{k}} \mathcal{R}_{\vec{k}'} \rangle = (2\pi)^3 \delta (\vec{k}+\vec{k}') 
P_{\mathcal{R}}(k), \quad \Delta^2_{\mathcal{R}}(k)= \dfrac{k^3}{2 \pi^2} P_{\mathcal{R}}(k).
\end{equation}
We can also define the scale-dependence of the power spectrum, the \emph{scalar spectral index} (or tilt):
\begin{equation}\label{nsdef}
n_s-1= \dfrac{d \ln \Delta ^2 _s(k) }{d \ln k}.
\end{equation}
The power spectrum will be \emph{scale-invariant} if $n_s=1$.
The power spectrum of a single polarization mode of tensor perturbations is:
\begin{equation}
\langle h_{\vec{k}} h_{\vec{k}'} \rangle = (2\pi)^3 \delta (\vec{k}+\vec{k}') 
P_{h}(k), \quad \Delta^2_{h}(k)= \dfrac{k^3}{2 \pi^2} P_{h}(k),
\end{equation}
while we can define the power spectrum of tensor perturbations as the sum of the power spectra for the two polarizations:
\begin{equation}
\Delta _T ^2=2\Delta _h ^2.
\end{equation}
The spectral index is traditionally defined as:
\begin{equation}\label{ntdef}
n_T=\dfrac{d \ln \Delta ^2 _T(k) }{d \ln k}.
\end{equation}
We also define the \emph{tensor-to-scalar} ratio as:
\begin{equation}\label{rdef}
r= \dfrac{ \Delta ^2 _T(k)}{ \Delta ^2 _s(k)}
\end{equation}

\section{Inflation}
\label{sec:inflation}
Inflation is now a well established paradigm of a consistent cosmology. Although as a basic mechanism it has been studied in the context of a wide variety of models, none of them has come to stand out and it is therefore interesting to consider the possibility of achieving an observationally supported inflationary model in low-energy
string effective actions such as the one we are considering in this work. In the present Section we briefly outline the reasons that made this paradigm stand out and its basic principles.

\subsection{Short-comings of Standard Cosmology}
The general purpose of physics is to predict the future evolution of a system given a set of initial conditions. It would seem therefore rather unfair to ask our cosmological model to explain the initial conditions of the Universe. On the other hand, it would be very disappointing if only very special and finely-tuned initial conditions would lead to the universe as we see it.\\
In this section we will explain that many properties of th standard model of cosmology can be traced back to an awkwardly fine-tuned set of initial conditions. We should stress, then, that what we are going to enumerate are not strict inconsistencies of the model. Rather, everything can be solved simply \textit{assuming} peculiar conditions. This point was made specific by Collins and Hawking\cite{ColHaw}, who showed that the set of initial data that evolve to a Universe qualitatively similar to ours is of measure zero\footnote{They show that the set of spatially homogeneous models which approach isotrpy at infinite times is of measure zero in the space of all spatially homogeneous models. This means that the FRW model is unstable to homogeneous and anisotropic perturbations.}. The reason why the inflationary paradigm is so successful is that it solves the initial condition problem dynamically: via inflation, the Universe can grow as we know it out of generic initial conditions.\\

Here we provide a list of the shortcomings of the standard scenario:
\begin{itemize}
\item The homogeneous and isotropic FRW space-time is very good in describing the Universe within our Hubble volume, but is also a very special solution of Einstein equations. On the one hand we have the CMB (Cosmic Microwave Background), proving the smoothness of the Universe to about one part in $\sim 10^{5}$ \cite{PLANCK}. On the other hand, for the Universe to be so smooth at the time when the CMB  was emitted, inhomogeneities had to be much smaller at earlier times. But back then particle horizons were even smaller and the Universe, according to standard evolution, was mostly causally disconnected. There is therefore no hope to find a physical explanation for why causally-separated patches should be so smooth: the high degree of homogeneity has to be assumed as an initial condition. This problem is often referred to as the \textbf{homogeneity} or \textbf{horizon problem}.\\

\begin{figure}[!ht]
  \centering
      \includegraphics[width=0.5\textwidth]{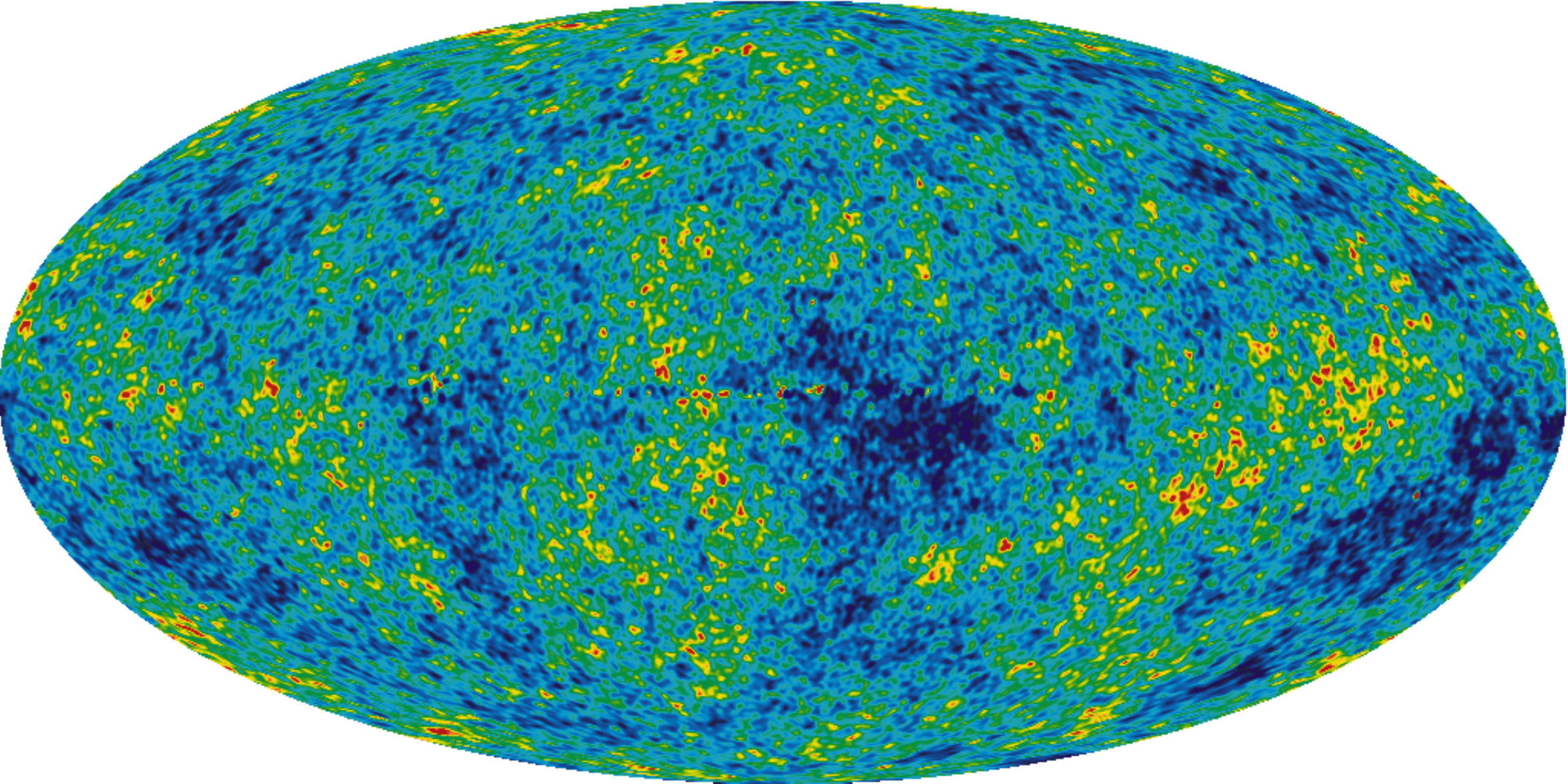}
       \caption{Temperature fluctuations in the CMB radiation, proving that the early universe wasn't perfectly homogeneous. Image from \cite{TASI}.}\label{cmb}
\end{figure}

\item Inhomogeneities are negligible on the large scale, but are present on then smaller scales: around the Universe we can see a plethora of small scale structures. Primordial inhomogeneities need to be assumed to account for the \textbf{structure formation}, from which we can approximately infer their amplitude. But, again, particle horizons in the early epoch preclude the production of perturbations on the scale of interest. Primordial fluctuations need as well to be assumed as initial conditions, rather than being produced by a physical process.\\ 
\item Th observed Universe is consistent with $\Omega_0=1$, or $k=0$. This value, hinting at a perfectly flat Universe, is an unstable fixed point; therefore, in standard Big Bang cosmology without inflation, the near-\textbf{flatness} observed today requires again an extreme fine-tuning of $\Omega$ close to 1 in the early universe.\\
\end{itemize}
To the ones listed above, one could add the \emph{entropy} problem, the \emph{unwanted relics} problem, and the \emph{cosmological constant} problem, which we will not describe in detail. 

\subsection{Basic principles of inflation}
\label{sec:Basic}
The basic idea of inflation is that the Universe in an early epoch underwent an accelerated expansion driven by the dominant contribution of a vacuum energy. Many of the shortcomings discussed in the previous Section are related to the fact that the comoving horizon is smaller in the early epoch, provided that we apply the radiation (or ordinary matter) dominated evolution up to early times. The horizon problem would not show if the Universe evolved differently back then, i.e. if the comoving horizon were bigger, in the early eras, than it is expected according to standard evolution, because physical processes could account for homogeneity as well as flatness and primordial fluctuations. This can be stated, referring to the definition of the comoving horizon \eqref{comhor}, as the requirement:
\begin{equation}
\dfrac{d}{dt}\left( \dfrac{1}{aH} \right)<0.
\end{equation}
It easy to show that the requirement above (the requirement of decreasing horizon) implies accelerated expansion:
\begin{equation}
\dfrac{d}{dt}\left( \dfrac{1}{aH} \right)=-\dfrac{\ddot{a}}{(aH)^2} \quad \Rightarrow \ddot{a}>0.
\end{equation}
A particular, accelerating solution of Friedmann equations is de Sitter space-time (see third line in table \ref{tab:cosmology}), where $H=$const and $\ddot{a}=H^2 a$.\\ 
We can also define the inflationary parameter $\epsilon_H=-\dfrac{\dot{H}}{H^2}$, with which the condition for inflation becomes:
\begin{equation}
\dfrac{\ddot{a}}{a}=H^2+\dot{H}=H^2(1-\epsilon_H)>0 \quad \Rightarrow \quad \epsilon_H<1.
\end{equation}
However, the second Friedmann equation \eqref{F2} indicates that accelerated expansion implies violation of the Strong Energy condition, i.e. $\rho+3p<0$ or $w<-1/3$. Ordinary matter cannot accaunt for inflation: we need to introduce the \emph{inflaton}.

\subsection{The inflaton}
The simplest models of inflation involve a single, canonical scalar field $\phi$, the inflaton, minimally coupled with Einstein's gravity and self-coupled through a potential, $V(\phi)$. Its dynamics is thus governed by the action \eqref{ActScal} and equation \eqref{EqScal}. On FRW background, the energy-momentum tensor of the inflaton assumes the form of a perfect fluid with:
\begin{equation}
\begin{split}
\rho_{\phi}&=\dfrac{1}{2} \dot{\phi}^2 +V,\\
p_{\phi}&=\dfrac{1}{2} \dot{\phi}^2 -V,\\
\end{split}
\end{equation}
and then
\begin{equation}
w_{\phi}=\dfrac{\dfrac{1}{2} \dot{\phi}^2 -V}{\dfrac{1}{2} \dot{\phi}^2 +V}.
\end{equation}
We clearly see that when the dynamics of the scalar field is characterized by potential energy dominating over kinetic energy, the inflaton can account for negative pressure and accelerated expansion.
For the sake of simplicity we focus on the flat case, and we will do so in all the Sections to come. The system of equations governing the dynamics reads, for $k=0$:
\begin{equation}
\begin{cases}\label{eqphik=0}
\ddot{\phi}+3H\dot{\phi}+V'=0,\\
H^2=\dfrac{1}{3}\left( \dfrac{1}{2} \dot{\phi}^2 +V \right),\\
3H^2+2\dot{H}+\dfrac{\dot{\phi}^2}{2}=0.
\end{cases}
\end{equation}
The term $3H\dot{\phi}$ acts as a friction on the evolution of the inflaton, and is much bigger when the potential takes on large values (see the second equation in \eqref{eqphik=0}).

\subsection{Slow-roll inflation}
\label{sec:slowroll}
The inflationary parameter defined in Section \eqref{sec:Basic}, $\epsilon_H$, is also called the slow-roll parameter. A scalar field can account for $\epsilon_H<1$ if:
\begin{equation}\label{src1}
\dot{\phi}^2 \ll V,
\end{equation}
which indicate that the inflaton should evolve slowly (\emph{slowly roll} along the potential). For inflation to solve the initial condition problem we need the accelerated expansion to last sufficiently long. Then, we need to define a second slow-roll parameter:
\begin{equation}
\eta_H =\epsilon_H-\dfrac{1}{2\epsilon_H}\dfrac{d\epsilon_H}{dN},
\end{equation}
$\vert \eta \vert < 1$ ensures that the fractional change of $\epsilon$ per e-fold is small, so that acceleration persist sufficiently. As far as the scalar field dynamics is concerned, accelerated expansion will only be sustained for a sufficiently long period of time if also the second time derivative of $\phi$ is small enough:
\begin{equation}\label{src2}
\vert \ddot{\phi} \vert \ll \vert 3H\dot{\phi} \vert, \; \vert V' \vert.
\end{equation}
We can write the second slow-roll parameter in terms of the scalar field:
\begin{equation}
\eta_H = - \dfrac{\ddot{\phi}}{\dot{\phi}H}.
\end{equation}
One can verify that requiring \eqref{src1}, \eqref{src2} is equivalent to require $\epsilon_H$, $\vert \eta_H \vert<1$. Exact Einstein equations also imply the useful relation:
\begin{equation}\label{dotphieps}
\dot{\phi}^2=-2 H^2 \epsilon_H.
\end{equation} 
Under approximations \eqref{src1}, \eqref{src2} the background evolution is given by: 
\begin{equation}\label{sreqs}
\begin{split}
\begin{cases}
H^2&\approx \dfrac{V}{3},\\
\dot{\phi} &\approx - \dfrac{V'}{3H}.
\end{cases}
\end{split}
\end{equation}

The conditions described so far are called \emph{slow-roll conditions} and may be expressed as conditions on the shape of the inflationary potential:
\begin{equation}\label{potentialparam}
\begin{split}
\epsilon_V&=\dfrac{1}{2} \left( \dfrac{V'}{V}^2 \right),\\
\eta_V&=\dfrac{V''}{V},
\end{split}
\end{equation}
We should distinguish between the \emph{potential slow roll parameters} defined above, and the \emph{Hubble slow roll parameters} $\epsilon_H$, $\eta_H$. Under slow roll approximation they are related by  $\epsilon_H\approx\epsilon_V$ and $\eta_H \approx\eta_V-\epsilon_V$. The scheme which takes the potential as the fundamental quantity is more widely used in literature. The advantage of the Hubble scheme, however, is that it also applies to models where inflation is driven by terms other than the scalar field potential. In Gauss-Bonnet gravity we will use Hubble slow roll parameters to take into account the role of the Gauss-Bonnet invariant as a source of inflation.\\
Inflation will come to an end when its conditions are violated:
\begin{equation}\label{infcd1}
\epsilon_H(\phi_f)\equiv \quad \textrm{or} \quad \epsilon_V\approx 1.
\end{equation}
We can compute the amount of expansion occured between the beginning and the end of inflation, the number of \emph{e-foldings}, $N=\ln \dfrac{a_f}{a_i}$:
\begin{equation}\label{infcd2}
N(\phi_i)=\int_{t_i}^{t_f} \; H \; dt = \int_{\phi_i}^{\phi_f} \; \dfrac{H}{\dot{\phi}} \; d\phi \approx \int_{\phi_i}^{\phi_f} \; \dfrac{V}{V'} \; d\phi,
\end{equation}
where we have used slow roll approximated equations \eqref{sreqs}.\\
Solving the horizon and flatness problems requires that the total number of inflationary e-foldings exceeds about 60,
\begin{equation}
N(\phi_i)\gtrsim 60.
\end{equation}
If we define $\phi_{cmb}$ as the value of the inflaton at the moment when the fluctuations observable in the CMB radiation were produced, then $N(\phi_{cmb})\approx 50-60$.\\

\subsection{Quantum fluctuations of the inflaton}
\label{sec:qflGR}
Even if driven by a uniform scalar field $\phi(t)$, inflation possesses the means to produce density perturbations on cosmologically interesting scales. In this Section we will briefly explain how the quantum fluctuations of a scalar field in de Sitter space-time can produce density perturbations. \\
First we make an important distinction, which we have already referred to in this Chapter. A perturbation of wavenumber $k$ is outiside the horizon if:
\begin{equation}
k \ll aH \quad \textrm{(super-horizon)},
\end{equation} 
and is inside the horizon if:
\begin{equation}
k \gg aH \quad  \textrm{(sub-horizon)}.
\end{equation}
While the wavenumber is fixed, the horizon evolves with time. Cosmologically relevant perturbations are produced inside the horizon, at the beginning of inflation. However, since the horizon becomes smaller with time during the inflationary phase (this is the peculiar property of inflation, the one we required to solve horizon-related problems!), eventually all perturbations exit the horizon. Depending on their scale, perturbations will later re-enter the horizon during matter or radiation dominated eras, when the horizon grows with time. Recall that the gauge-invariant quantities $\mathcal{R}$ and $\zeta$ for adiabatic and super-horizon perturbations remain constant and thus do not depend on the details of after-inflation evolution. Once the perturbation becomes sub-horizon again, we can use the values of $\mathcal{R}$  and $\zeta$ calculated at horizon exit and  determine the fluctuations that eventually result in the CMB anisotropies. \\ 

\begin{figure}[!ht]
  \centering
      \includegraphics[width=0.9\textwidth]{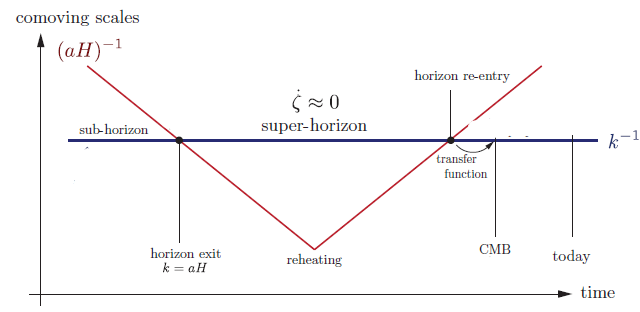}
       \caption{The evolution of the horizon with respect to a fixed comoving wavenumber: horizon exit and re-entering of the mode $k$. Image from \cite{TASI}.}\label{horizoncr}
\end{figure}

In the following discussion we will assume the reader to be acquainted with quantum field theory on Minkowski space-time. We outline the steps one needs to perform to calculate scalar and tensor perturbations as produced by quantum fluctuations during inflation. For the sake of simplicity we first focus on $\mathcal{R}$ alone.\\
\begin{enumerate}[label=\arabic{*})]
\item We first need to expand the action \eqref{ActScal} to second order in perturbations. Working on the action guarantees the correct normalization for the quantization of fluctuations. Before proceeding, we fix the diffeomorphism freedom by selecting a gauge; we are free to do so, since we will be eventually interested in gauge-invariant quantities. We choose a gauge where  the inflaton is unperturbed and all scalar degrees of freedom are parametrized by the metric fluctuation $\mathcal{R}$:
\begin{equation}\label{comovinggauge}
\delta \phi=0, \quad g_{ij} = a^2[(1 + 2\mathcal{R})\delta_{ij} + h_{ij} ], \quad \partial_i h_{ij}=h^i_{i}=0.
\end{equation}
The remaining metric perturbations are related to $\mathcal{R}$ by perturbed Einstein equations, such as \eqref{zetaRPsi}. We are now ready to expand the action \eqref{ActScal} to second order in $\mathcal{R}$:
\begin{equation}
S^{(2)}=\dfrac{1}{2} \int \; d^4 x \; a^3 \dfrac{\dot{\phi}^2}{H^2} \left[ \dot{\mathcal{R}}^2-\dfrac{1}{a^2} (\partial _i\mathcal{R} )^2 \right].
\end{equation}
We work with conformal time $\tau$ defined in Eq. \eqref{conft} and introduce the \emph{Mukhanov-Sasaki} variable:
\begin{equation}
v(\tau, \vec{x})=z  \mathcal{R}, \quad \textrm{with} \quad z^2=a^2\dfrac{\dot{\phi}^2}{H^2}.
\end{equation}
With these coordinates, the action (now called \emph{Mukhanov action}) takes the form:
\begin{equation}\label{actz}
S^{(2)}=\dfrac{1}{2} \int \; d\tau d^3 x \left[ v'^2 + (\partial _i v)^2 + \dfrac{z''}{z} v^2  \right],
\end{equation}
where a prime denotes differentiation with respect to conformal time.\\
\item From the action \eqref{actz} we derive the equation of motion for the mode of wavelength $k$:
\begin{equation}\label{Meq}
v^{''}_k+\left( k^2 -\dfrac{z^{''}}{z} \right) v_k=0.
\end{equation}
We can drop the vector notation since the equation only depend on the magnitude. As always, we were able to treat each Fourier mode separately because of the three-dimensional translation invariance of the background space-time. \emph{Mukhanov equation} \eqref{Meq} can be considered as an harmonic oscillator equation, with a conformal time-dependent effective frequency $\omega^2_k= k^2 m^2_{\textrm{eff}} $, with $m^2_{\textrm{eff}}= -\dfrac{z''}{z}$.\\
\item We quantize the field v, promoting it to a quantum operator:
\begin{equation}
\hat{v}(\tau,\vec{x})= \int \; \dfrac{d^3 k}{(2\pi)^{3}} \left(  v_k(\tau) \hat{a}_{\vec{k}} e^{i\vec{k}\cdot \vec{x}} + v^*_k(\tau) \hat{a}^{\dagger}_{\vec{k}} e^{-i\vec{k}\cdot \vec{x}} \right),
\end{equation}
where the mode functions satisfy:
\begin{equation}
v_k''+ \omega^2_k v_k=0, \quad \omega_k= \sqrt{k^2 +m^2_{\textrm{eff}}(\tau)}.
\end{equation}
Equivalently, we can directly promote the Fourier components to quantum operators:
\begin{equation}
\hat{v}_{\vec{k}}= v_{k}(\tau) \ \hat{a}_{\vec{k}} + v^*_{-k}(\tau) \ \hat{a}^{\dagger}_{-\vec{k}}.
\end{equation}
The operators $\hat{a}_{\vec{k}}$, $\hat{a}^{\dagger}_{-\vec{k}}$ are defined by the canonical commutation relation:
\begin{equation}
[\hat{a}_{\vec{k}},\hat{a}^{\dagger}_{\vec{k}'}]=(2\pi)^3 \delta(\vec{k}-\vec{k}').
\end{equation} 
However, to obtain the commutators above we have to impose the normalization condition:
\begin{equation}\label{cd1M}
\langle v_k(\tau),v_{k'}(\tau) \rangle \equiv i \left( v^*_k v_{k'} + v^*_{k'} v_{k}\right) = 1.
\end{equation}
This provides one of the two boundary conditions needed to solve Eq. \eqref{Meq}. The second boundary condition is obtained specifying the vacuum state, defined as the state annihilated by all the destruction operators:
\begin{equation}
\hat{a}_{\vec{k}} \vert 0 \rangle =0, \quad \forall \vec{k}.
\end{equation}
Eq. \eqref{Meq}, describing a complicated, interacting quantum field, do not guarantee the uniqueness of the vacuum defined above. A unique vacuum (\emph{Bunch-Davies vacuum})\cite{BDV} may be defined in the far past, when all the comoving scales were well inside the Hubble horizon and the effect of gravity was negligible. In the limit $\tau \rightarrow - \infty$, we recover a free field in Minkowski space-time and the mode equation reduces to:
\begin{equation}\label{modelimit}
v_k''+k^2  v_k=0.
\end{equation}
This is the mode equation of a massless, free scalar field, or an harmonic oscillator with time-independent frequency. The vacuum of the harmonic oscillator is unique, and coincides with the minimum energy state. The solution of \eqref{modelimit} is:
\begin{equation}\label{cd2M}
 \lim_{\tau \to -\infty} \; v_k=\dfrac{e^{-i k \tau}}{(2 k)^{1/2}}.
\end{equation} 
We require that the solution of \eqref{Meq} satisfies this limit, obtaining a second boundary condtion. Eq.s \eqref{cd1M} and \eqref{cd2M} provide all of the boundary conditions needed to solve Mukhanov equation.\\
\item Since Eq. \eqref{Meq} depends on background dynamics, it cannot be solved in full generality. However, we are able to solve it in a purely de Sitter space-time ($\epsilon_H=0$, $H=$const), where $a \sim 1/ \tau$ and thus $z''/z= 2/ \tau^2 $. The mode equation in this special case reads:
\begin{equation}\label{DSM}
v_k''+ \left( k^2 - \dfrac{2}{\tau^2} \right)  v_k=0.
\end{equation}
One can verify that:
\begin{equation}
v_k=\alpha \dfrac{e^{-ik\tau}}{\sqrt{2k}}\left( 1- \dfrac{i}{k\tau} \right) + \beta \dfrac{e^{+ik\tau}}{\sqrt{2k}}\left( 1+ \dfrac{i}{k\tau} \right) 
\end{equation}
is a solution of Eq. \eqref{DSM}. The presence of two free parameters reveals the non-uniqueness of the mode functions. However, we may fis $\alpha$ and $\beta$ by considering the quantization conditions \eqref{cd1M}, \eqref{cd2M}. This leads to the unique \emph{Bunch-Davies mode functions}:
\begin{equation}
v_k= \dfrac{e^{-ik\tau}}{\sqrt{2k}}\left( 1- \dfrac{i}{k\tau} \right)
\end{equation}
We have now all the elements needed to quantum correlation functions, i.e. quantum fluctuations.
\end{enumerate}
We have so far focused on scalar perturbations. We briefly summarize the corresponding calculation for tensor perturbations, following the same steps:
\begin{enumerate}[label=\arabic{*})]
\item From second-order perturbed Einstein-Hilbert Action we obtain the first-order perturbation equation for tensor modes \eqref{eqperturbGR}.\\
\item We define the canonically normalized field $v_k= \dfrac{a}{2}h_k$ for each polarization, and switch to conformal time. The corresponding second-order action is:
\begin{equation}
S^{(2)}=\sum_{s=\times,+} \dfrac{1}{2} \int \; d \tau d^3 k 
\left[ (v^{s'}_{\vec{k}})^2- \left( k^2 -\dfrac{a''}{a} \right) (v^s_{\vec{k}})^2 \right],
\end{equation}
where a prime denotes differentiation with respect to conformal time.
\item We quantize the field $v$ and solve the first-order equation exactly in de Sitter space.\\ 
\end{enumerate}

\subsection{Inflation-produced power spectrum}
\label{sec:psGR}
\begin{figure}[!ht]
  \centering
      \includegraphics[width=0.95\textwidth]{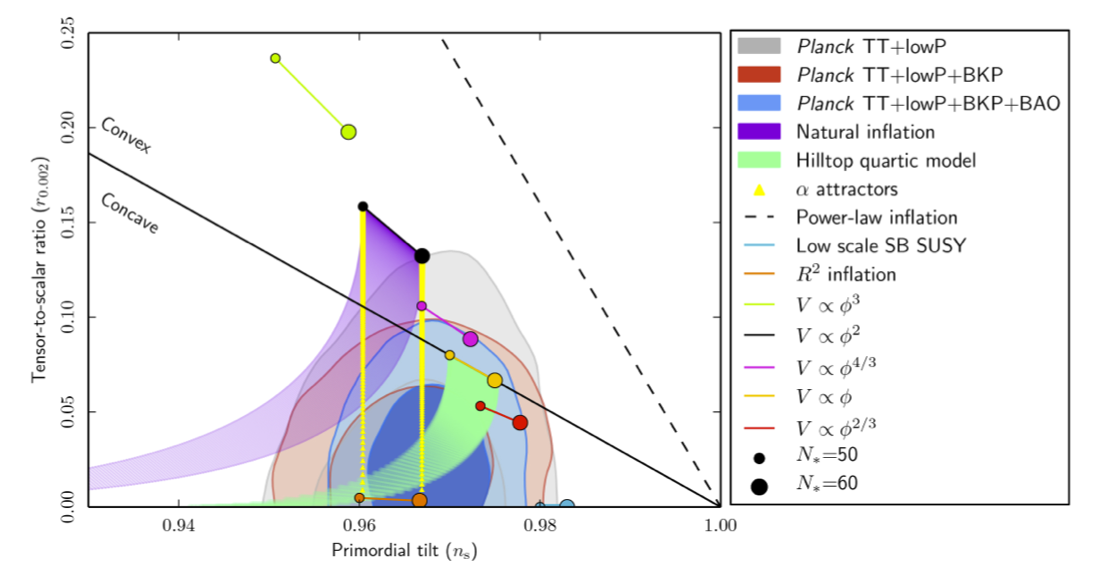}
       \caption{Marginalized joint 68 \% and 95 \%  CL regions for $n_s$ and $r$ 0.002 from Planck in combination with other data sets, compared to the theoretical predictions of selected inflationary models\cite{PLANCKinfl}}.\label{PlanckInfla}
\end{figure}

In this Section we compute the power spectrum of scalar perturbations at horizon crossing produced by quantum fluctuations in de Sitter space-time. We first compute the correlation function of the field $\hat{\Psi}_k= a^{-1} \hat{v}_k$:
\begin{equation}
\begin{split}
\langle \hat{\Psi}_k(\tau),\hat{\Psi}_{k'}(\tau) \rangle =& (2 \pi)^3 \delta (\vec{k}+\vec{k'}) \dfrac{\vert v_k (\tau) \vert ^2}{a^2}=\\
&= (2 \pi)^3 \delta (\vec{k}+\vec{k'}) \dfrac{H^2}{2k^3} (1+k^2 \tau ^2)\simeq\\
&\simeq (2 \pi)^3 \delta (\vec{k}+\vec{k'}) \dfrac{H^2}{2k^3},
\end{split}
\end{equation}  
where the last expression is approached on super-horizon scales, $\vert k \tau \vert \ll 1$. Now we can compute the correlation function of $\mathcal{R}=\Psi H /\dot{\phi}$, focusing on horizon-crossing time $a(\tau_{hc}) H(\tau_{hc})= k$:
\begin{equation}
\langle \hat{\mathcal{R}}_k(\tau),\hat{\mathcal{R}}_{k'}(\tau) \rangle \biggr\rvert_{hc} = (2 \pi)^3 \delta (\vec{k}+\vec{k'}) \dfrac{H^2_{hc}}{2k^3} \dfrac{H^2_{hc}}{\dot{\phi}^2_{hc}}.
\end{equation}
Finally, referring to the definition \eqref{PSR}, we can write down the power spectrum:
\begin{equation}
\Delta^2_{\mathcal{R}}(k)= \dfrac{H^2_{hc}}{(2\pi)^2} \dfrac{H^2_{hc}}{\dot{\phi}^2_{hc}}= \dfrac{H^2}{8\pi ^2} \dfrac{1}{\epsilon_H} \biggr\rvert_{hc} .
\end{equation}
In the last expression we have used \eqref{dotphieps}. This formula applies exactly to de Sitter inflation, but is also approximately correct for quasi-de Sitter space and slow roll inflation.\\
An analogous procedure leads to the power spectrum, evaluated at horizon crossing, of tensor perturbations:
\begin{equation}
\Delta^2_T(k)= 2 \Delta^2_{h}(k)= 2 \dfrac{H^2}{\pi^2} \biggr\rvert_{hc}.
\end{equation}
The power spectra are well described by the parameters $n_s$, $n_T$ and $r$, defined in \eqref{nsdef}, \eqref{ntdef} and \eqref{rdef} respectively, and depend on the time evolution of the Hubble parameter. Let us compute:
\begin{equation}\label{calcns}
n_s-1= \dfrac{d \ln \Delta ^2 _s(k) }{d \ln k} = \dfrac{d \ln \Delta ^2 _s(k) }{d N} \dfrac{d N }{d \ln k}. 
\end{equation}
We can use:
\begin{equation}
\dfrac{d \ln \Delta ^2 _s(k) }{d N}= 2 \dfrac{d \ln H }{d N}- \dfrac{d \ln \epsilon_H }{d N} =-4 \epsilon_H + 2 \eta_H,
\end{equation} 
while second term in \eqref{calcns} becomes, thanks to the horizon-crossing condition,
\begin{equation}
k=aH \quad \Rightarrow \quad \ln k = N + \ln H,
\end{equation}
so that
\begin{equation}
\dfrac{d N}{d \ln k}= \left( \dfrac{d \ln k}{d N} \right) ^{-1}= \left( 1+ \dfrac{d \ln H}{d N} \right)^{-1}= \left( 1 - \epsilon_H \right)^{-1} \simeq 1 + \epsilon_H.
\end{equation}
Eq. \eqref{calcns} reads, at first order in slow-roll parameters:
\begin{equation}
n_s-1= (-4 \epsilon_H + 2\eta_H) (1 + \epsilon_H)\simeq  (-4 \epsilon_H + 2\eta_H ),
\end{equation}
where every term is evaluated at horizon-crossing. Similarly, one may find:
\begin{equation}
n_T \simeq - 2 \epsilon_H.
\end{equation}
Finally, if we substitute potential slow-roll parameters, we get:
\begin{equation}
n_s-1= 2 \eta_V -6 \epsilon_V, \quad n_T= - 2 \epsilon_V, \quad r=16 \epsilon_V=-8n_T .
\end{equation}
The expressions above directly relate observable fluctuations with the shape of the inflationary potential. We also find a consistency relation between $n_T$ and $r$, which are not independent quantities.\\ 
As a useful example we may consider the quadratic potential, $V=m^2 \phi^2$. This model belongs to the class of monomial potential inflationary models, a class called chaotic inflation and due to Linde \cite{Chao}. Since in this case $\epsilon_V=\eta_V= \dfrac{1}{2 N_{cmb}}$, the values predicted by such a potential are:
\begin{equation}\label{nsrNcmb}
n_s= 1- \dfrac{2}{N_{cmb}} \simeq 0.97, \quad r= \dfrac{8}{N_{cmb}}\simeq 0.13,
\end{equation}
if we substitute $N_{cmb}=60$. This prediction, together with that of many other models of inflation, is compared to PLANCK 2015 data \cite{PLANCKinfl} in Figure \ref{PlanckInfla}. 

\subsection{Reheating and criticism on the paradigm}
Inflation would not be a viable physical model if another phenomenon, called \emph{reheating}, did not take place. Qualitatively, reheating is the phase when all known matter was created. Reheating is a necessary completion of the inflationary paradigm: after inflation, all matter/radiation has been diluted by the exponential expansion and the Universe is void with very good approximation. A mechanism is needed to fill the Universe back with matter. We will try to give a broad-brush picture of how this mechanism works.\\
After inflation, the inflaton ens up in the minimum of its potential and oscillates around it, while most of its energy $\rho_{\phi}$ turns into kinetic energy. The spatially coherent oscillations of the inflaton correspond to a condensate of $\phi$ particles. This particles, if the inflaton is suitably coupled, can decay into lighter Standard Model degrees of freedom. The decay products dump the inflaton oscillations and thermalize, resulting in a reheated, radiation dominated Universe.\\

In the last Sections we have shown how inflations succeeds in explaining why the Universe is so flat and uniform, predicting at the same time the properties of matter perturbations. Observational data (see, again, Figure \ref{PlanckInfla}) are today in good agreement with the inflationary paradigm, though favouring Starobinski $R^2$ model \cite{Staro} over monomial potentials. Despite the absence of discrepancies between inflationary predictions and observations, a debate has sprung out about the theoretical soundness of the paradigm. Inflation seem, in fact, to re-propose a fine-tuning problem: its parameters need to be carefully selected as to give rise to the present Universe. Moreover, most inflationary models lead to eternal inflation and to a multiverse, where \lq\lq anything that can happen will happen;
in fact, it will happen an infinite number of times" (Alan Guth). A model of the Universe with infinite different patches can hardly be said to make predictions at all \cite{debate}. Alternatives to inflation have been proposed, and the forthcoming gravitational wave experiments will probably discriminate between the paradigm ant its competitors.

\chapter{Singularities in General Relativity}
In this chapter we define space-time singularities and enounce Hawking and Penrose's singularity theorem. We apply these results to our standard model of the Universe, finding an initial singularity (the Big Bang singularity). Finally, we look for a way to escape the singularity theorems, both in the framework of General Relativity and beyond it.

\section{What is a singularity?}
\begin{figure}[!ht]
  \centering
    \reflectbox{%
      \includegraphics[width=0.5\textwidth]{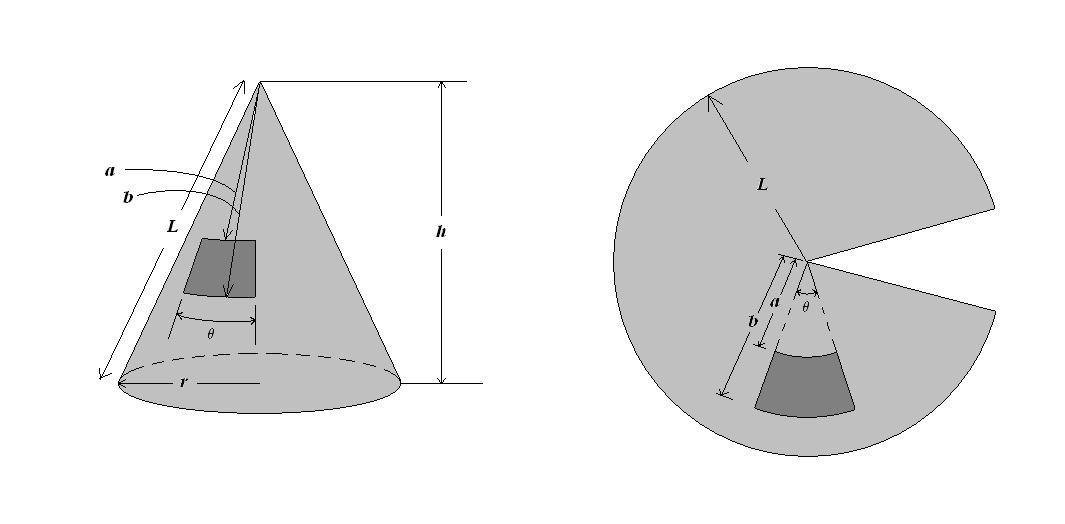}}
       \caption{A cone obtained from a Minkowski space-time (in 2 dimensions), with a conical singularity.}\label{cone}
\end{figure}

It is well known that some solutions of Einstein equations contain points where the geometric description of the space-time breaks down. Such space-times are called singular. Two major examples of singular space-times are the Schwarzschild space-time and a generic Friedmann-Robertson-Walker (FRW) space-time.\\
In the Schwarzschild space-time, it is relatively easy to understand that the black hole center $r=0$ represents a singularity, because the curvature becomes infinite as $r\rightarrow 0$, indicating the presence of unboundedly large tidal forces. Also, every time-like or null geodesic entering the Schwarzschild radius cannot escape hitting the center within a finite proper time (or affine parameter). No light rays or time-like geodesics can be continued past that point because the geometry is singular and the geodesic equation becomes undefined (i.e. it does not predict how a body will move past the center). Thus the center is a place where the geometry of the space-time cannot be described by General Relativity. \\
In an FRW space-time with the scale factor $a(t)\rightarrow 0$ as $t\rightarrow 0$, the curvature is unbounded near $t=0$. Similarly to the Schwarzschild case, geodesics cannot be extended to the past beyond $t=0$, so one cannot describe what happened to an observer before that time. We can safely say, then, that $t=0$ is the \lq\lq beginning of time'' for an FRW space-time, and that the three-surface $t=0$ is singular.\\
However, when we attempt to define a space-time singularity in a more general way, we encounter many difficulties. For example, singularities cannot be considered as a part of the space-time, i.e. we cannot say that a singularity is a point or a region of the space-time under consideration. Space-time is a manifold endowed with a metric, defined everywhere on the manifold. Then singularities such as that of Schwarzschild's or Friedmann's metric cannot be not part of the space-time, because the metric is not defined there. We conclude that we should not address singularities as \lq\lq regions of space-time\rq\rq: they are neither a \lq\lq place" nor a \lq\lq time".\\
Difficulties are encountered even if we try to define singularities on boundaries of space-time, as it can be done only in simple examples. \\
Surely many singularities, known as curvature singularities, may be easily characterized by an \lq\lq explosion" of curvature scalars. Intuitively, we feel that a singularity should involve the blowing up of curvature. But not all singularities share this feature:
\begin{itemize}
\item[-]curvature scalars can become unboundedly large simply when going to space-time infinity (when the space-time is not asymptotically flat), and we don't want to describe this as a singular behaviour;
\item[-]space-time can be singular without the curvature scalars exploding.
\end{itemize}
Let's provide an example of the latter case. Suppose we cut a piece of a Minkowski space-time, and then glue together the remaining edges to form a space-time cone (see Figure \ref{cone}). The summit of the cone is what we call a \emph{conical singularity}, yet curvature scalars don't exhibit explosions approaching that point. In fact, $R^{\mu}_{\nu\rho\sigma}=0$ everywhere, since we are dealing with flat geometry.\\
So, characterizing a singularity by the \lq\lq blowing up" of curvature is not enough, because it doesn't take into account all the possible pathological situations that we wish to comprehend in the definition of a singularity. What seemed a trivial question - what is a space-time singularity? -, is still far from being answered. \cite{Wald,Win} 

\subsection{Schmidt definition}
If singular points need to be excluded, cut out from the manifold, one would expect the space-time to be incomplete in some sense, even though we have included all the points we possibly could. We are referring here to an \emph{inextendible} manifold. The $C^n $ manifold $(\mathcal{M'},g)$ is an extension of the $C^n $ manifold $(\mathcal{M},g)$ if there is an isometric embedding $\mu: \mathcal{M}\rightarrow \mathcal{M'}$. Then a $C^n $ manifold is inextendible if there is no $C^n $ extension where $\mu(\mathcal{M})$ in not equal to $\mathcal{M'}$.\\
Now we are ready to give Schmidt definition of a space-time singularity. Consider, in a space-time manifold M, all possible null and time-like geodesics, and all time-like curves with finite (bounded) acceleration.\\
If
\begin{itemize}
\item[-] one of them ends after a finite proper length, or affine parameter for null geodesics; 
\item[-] the manifold is inextendable, because, for example but not necessarily, of infinite curvature;
\end{itemize}
then the termination point and the neighbouring ones form a space-time singularity. The singularity is also a curvature singularity if any of the geometric scalars is unbounded on the incomplete curve or if, measured in a basis parallely propagated to the curve, components of the Riemann tensor are unbounded. \\
Time-like geodesic incompleteness has an immediate physical meaning: a freely falling observer or particle has an history which does not exist after (or before) a finite lapse of proper time. The inclusion, in Schmidt's definition of a singularity, of null geodesics is also physically significant, as they represent the histories of massless particles. The same can be said for time-like curves with bounded acceleration, which could represent, for example, an observer with some propulsion system, such as a rocket.\\ 
Note that space-like geodesics are not included in the definition above. If we reject the idea of tachionic particles, we see that there is no physical reason to include space-like geodesic in the definition of a singular space-time. Suppose a space-time has only space-like geodesic incompleteness, so that no real observer can reach the singularity. If we cannot physically reach it, is it still a true singularity?\\ \cite{HawkEllis,Gravitation} 
Even if not completely satisfactory, Schmidt's definition is a minimum requirement for a space-time to be considered singular. Last but not least, it provides the means to construct many singularity theorems, to which we devote the following Section.  \\

\section{Singularity theorems} 
\label{HPth}
Many singularity theorems were proposed since 1965 by Penrose, Hawking and Geroch. Singularity theorems often make use of the concept of \emph{trapped surface}. A trapped surface is a 2-dimensional closed surface such that all ingoing (and outgoing) null, surface-orthogonal geodesics converge towards each other. This kind of surfaces signal the presence of a singularity nearby. An example of a trapped surface is every $r=$const surface with $r<2M$ in Schwarzschild geometry: light-rays emitted from such a surface are doomed to fall (converge) on the singularity $r=0$.\\
The last and most satisfactory of the series of theorems mentioned above was Hawking-Penrose singularity theorem \cite{HawkPen}. It states the unavoidability of singularities in General Relativity, under certain conditions which include generic gravitational collapse (realistic, non highly symmetric) and the present Universe.\\
\begin{thm}[Hawking-Penrose, 1970]\label{th:HP}
A space-time $(\mathcal{M},g)$ necessarily contains a singularity,  in the sense that is not time-like and null geodesically complete, if the following conditions hold:
\begin{enumerate}
\item Einstein equations \eqref{EE} (without the cosmological constant);
\item \label{HP2} M contains no time-like closed curves (causality condition);
\item \label{HP3} $\forall \ x^{\mu} \in M$, $\forall \  u^{\alpha}$ time-like, (Strong Energy condition, SEC)
$$R_{\alpha\beta}u^{\alpha}u^{\beta}=\left( T_{\alpha\beta}-\dfrac{1}{2}g_{\alpha\beta}T \right)u^{\alpha}u^{\beta}\geqslant 0 ;$$
\item \label{HP4}$\forall \ u^{\alpha}$ time-like or null,
$$u_{[\alpha}R_{\beta]\gamma\lambda [\epsilon}u_{\rho]}u^{\gamma}u^{\lambda}\neq 0,$$
at some point (generality condition);
\item \label{HP5} M contains either
\begin{itemize}
\item[-]a trapped surface;
\item[-] a point P where the divergence of the convergence of all null geodesics through P changes sign;
\item[-] a compact space-like hyper-surface. 
\end{itemize}
\end{enumerate}
\end{thm} 
Condition \ref{HP2} simply requires the space-time examined to be physical with respect to causality.\\   
If we identify $\rho$, the energy density, and $p^i$, the principal pressures, as the stress-energy tensor's eigenvalues, condition \ref{HP3} requires
\begin{equation}
\rho+\Sigma_i p^i \geq 0.
\end{equation} 
If, for example, we consider a perfect fluid, so that the stress-energy tensor takes the form $T_{\mu\nu}=(\rho(t)+p(t))u^{\mu}u^{\nu}+p(t)g^{\mu\nu}$,
SEC read:
\begin{equation}
\rho+ p \geq 0, \quad \rho+3 p \geq 0.
\end{equation}
Condition \ref{HP4} requires that every time-like or null geodesic enters a region where the curvature is not aligned with the geodesic itself.\\
The theorem does not provide any information on the nature of the singularity nor does it say whether the singularity is in the future or in the past. Moreover, General Relativity is taken as one of the hypothesis, but the theorem does not depend on the full Einstein equations: its result apply also to any modification of GR where gravity is always attractive. This is not the case, however, of Gauss-Bonnet gravity.\\

\subsection{Big Bang singularity}
In this section we show that the conditions of Hawking and Penrose's singularity theorem are satisfied by our present model of the Universe, indicating that there was a singularity, the Big Bang singularity, at the beginning of time. In particular, we prove the existence of trapped surfaces in FRW space-time.\\
The CMB radiation as well as the distribution of galaxies indicate that our Universe is very well described by FRW metric:
\begin{equation}\label{FRWBB}
ds^2=-dt^2+a(t)^2 d\gamma^2,
\end{equation} 
where $d\gamma^2$ is the metric of a three-space of constant curvature $k$. We shall now show that in any FRW space-time containing ordinary matter, with positive energy density, and if $\Lambda=0$, there is a closed trapped surface. We then conclude that our model of the Universe satisfies all the hypothesis of the singularity theorem, including condition \ref{HP5}.\\ 
We can write $d\gamma^2$ as:
\begin{equation}
d\gamma^2=d\chi^2+ \Phi_k(\chi)^2 d\Omega^2,
\end{equation}
see \eqref{FRWchi}, in the previous Chapter.
Consider a two-sphere $\mathcal{S}$ of radius $\chi_0$ lying in the surface $t=t_0$. Consider also the families of past-directed null geodesics orthogonal to $\mathcal{S}$. The geodesics will intersect the surfaces $t=$const in two two-spheres of radii:
\begin{equation}\label{radius}
\chi(t)_{\pm}=\chi_0\pm \int^{t}_{t_0} \dfrac{dt'}{a(t')}.
\end{equation}
The surface area of a two-sphere of radius $\chi$ will be $S(\chi)=4\pi a(t)^2\Phi(\chi)^2$. Then two families of geodesics will be converging in the past if, for both values of $\chi$ given \eqref{radius}:
\begin{equation}\label{focus}
\dfrac{d}{dt}\left(a(t)^2\Phi(\chi)^2\right)\bigg|_{t=t_0,\chi_{\pm}}.
\end{equation}
Equation above, in fact, means that both families of null geodesics departing orthogonally from $\mathcal{S}$ have encountered in the past smaller two-spheres.\\
Using \eqref{radius}, the condition reads:
\begin{equation}
\dfrac{\dot{a}(t_0)}{a(t_0)}>\pm \dfrac{\Phi'(t_0)}{a(t_0)\Phi(t_0)}.
\end{equation}  
But substituting Friedmann equation with $\Lambda=0$, $\dfrac{\dot{a}}{a}=(1/3 \rho +k/a^2)^{1/2}$, we obtain:
\begin{equation}
\left( 1/3 \rho(t_0) a(t_0)^2 +k \right)^{1/2}>\pm \dfrac{\Phi'(t_0)}{\Phi(t_0)}.
\end{equation}
Then the condition for $\mathcal{S}$ to be a trapped source holds if $a(t_0)\chi(t_0)>\sqrt{3/\rho_0}$ when $k=+1,0$, or if $a(t_0)\chi(t_0)>\textrm{min}\lbrace \sqrt{3/\rho_0}, 1/16 \rbrace$ when $k=-1$. To find a trapped surface, we just need, chosen a time $t_0$ such as the present time, to find a big enough radius $\chi_0$.\\
Intuitively we can say that in the Universe, at the present time or at any time, must exist a  sphere sufficiently small to be within the Schwarzschild sphere, i.e. $a(t_0)\chi(t_0)<r_S=2m=1/3 \rho_0 a(t_0)^3\chi(t_0)^3$. It is easy to see that the latter condition is equivalent to the one found with the more rigorous calculation \cite{HawkEllis}.\\

In the derivation above we have supposed the absence of a cosmological constant, $\Lambda=0$. Many models with a sufficiently large cosmological constant show no Big Bang singularity, but its present measured value \cite{PLANCK} is too small to be dynamically relevant in the early Universe (its importance grows with time). One might suspect that the Big Bang singularity is a consequence of the perfect symmetry assumed, and that anisotropic or inhomogeneous models would not show such a feature; however, this seems not to be the case.\\  
In the next Section we will discuss how the problem of singularities has been tackled within General Relativity, and whether a different theory of gravity can solve it.

\section{Escaping the singularity theorems..}
\subsection{..in General Relativity}
Given the singularity theorems, showing that singularities occur under quite general conditions, the foundations of Einstein's theory seem to be in danger. Holding in mind that the predictive power of a theory, and hence the possibility of a physical interpretation, breaks down at a space-time singularity, we should conclude that General Relativity predicts its own failure. \\
The \emph{cosmic censorship} (CC) is a mathematical conjecture proposed in hope to redeem General Relativity, stating that \emph{naked singularities} in the theory only appear in \lq\lq physically unreasonable" models. The idea of cosmic censorship was introduced by Roger Penrose in 1969 \cite{Censor}. Today, his statement is appealing but still open to debate. We now should clarify some points of the conjecture, and distinguish between \emph{weak} and \emph{strong} cosmic censorship (WCC and SCC):
\begin{itemize}
\item[-] in the WCC a singularity is not naked if it is hidden behind an event horizon. The WCC states that all physically realistic process should deposit the singularity behind an event horizon; 
\item[-] the SCC requires the space-time $(\mathcal{M},g)$  to contain a \emph{Cauchy surface}, i.e. a space-like hyper-surface $\Sigma$ (\emph{time slice}) whose total domain of dependence $D(\Sigma) = D^+(\Sigma) \cup D^−(\Sigma)$\footnote{The future domain of dependence $D^+(\Sigma)$  is defined to be the set of all points
$p\in \mathcal{M}$, with the property that every past-directed inextendible time-like curve starting at p intersects $\Sigma$.} is all of $\mathcal{M}$. A space-time containing a Cauchy surface is said to be \emph{globally hyperbolic} and satisfies the SCC.
\end{itemize}
We should also try to explain what a \lq\lq physically unreasonable" model is. A model can be so in different senses: involving unrealistic idealizations, rare features, or being simply and literally physically impossible.\\
The CC hypothesis leans on \lq\lq the faith that General Relativity has some built-in mechanism for preserving modesty by clothing naked singularities" (Earman, \cite{Earman}). However both a clear definition of what a naked singularity is and a proof of even the simpler WCC are still missing. Moreover, \lq\lq nakedness" of the Big Bang singularity (in which we are mostly interested in this work) is generally regarded as unavoidable and is not addressed in the debate around the cosmic censorship \cite{Earman,WaldCC}.  

\subsection{..in modified gravity}
When a singularity is encountered in physical theory, the general conclusion is that the failure does not lie in the physical world, but rather in our theoretical description. Such a view denies that singularities are real features of the space-time, and asserts that they are instead merely artifices of our current (flawed) physical theories. Our concern should turn, then, to the quest for a more fundamental theory, whether quantum or classic, where singularities do not appear.\\  
Many proposed modifications of Einstein's General Relativity can be accounted for defining an effective energy-momentum tensor, so that the field equations read:
\begin{equation}
G_{\mu\nu}=T^{\textrm{eff}}_{\mu\nu}.
\end{equation}
The newly defined $T^{\textrm{eff}}_{\mu\nu}=T_{\mu\nu}+\textrm{...}$ will contain the old energy-momentum tensor and every modification of Einstein equation arising in the proposed theory. The terms introduced by the modified theory can drastically change the properties of the energy-momentum tensor. For example, the metric or the new degrees of freedom introduced in $T^{\textrm{eff}}_{\mu\nu}$ can allow the theory to escape the singularity theorems, breaking the Strong Energy conditions. In the next Chapter we will explicitly see how this works for Gauss-Bonnet gravity.\\

\chapter{Gauss-Bonnet gravity}
\label{chap:GB}
In this Chapter we introduce scalar-Gauss-Bonnet theory. We write down the action and the equations of motion, and summarize the main results obtained in this framework. Due to the breadth of current and past research about this theory, a truly comprehensive review is probably impossible, and certainly beyond the scope of this thesis. Subsequently, we focus on the dynamics of an FRW metric, and discuss some results which will come in handy in the Chapters to follow.

\section{Action and equations of motion}\label{actioneq}
In the present work we study Einstein-scalar-Gauss-Bonnet gravity. The model is described by the action:
\begin{equation}\label{action}
S=\int \! \mathrm{d}^4x \ \sqrt{-g} (\dfrac{1}{2\kappa}R-\dfrac{1}{2}\partial_{\mu}\phi\partial^{\mu}\phi+ \dfrac{f(\phi)}{8} R^2_{GB}) \ + S_{matter},
\end{equation}
where $R_{GB}^2=R_{\mu\nu\rho\sigma}R^{\mu\nu\rho\sigma}-4R_{\mu\nu}R^{\mu\nu}+R^2$ is the Gauss-Bonnet (GB) curvature invariant, $\phi=\phi(x)$ is a scalar field, and $f(\phi)$ is a generic function of the scalar field.\\
Matter fields other that the scalar $ \phi $, e.g. radiation, fluids, dark energy, or other scalars that drive different stages of cosmic evolution, are all contained in the $ S_{matter} $ term. However, they are considered subdominant and, hence, negligible during the very early time stages of the evolution of the Universe. Although they play an important role after this early period, we will never take them into account in the present work.\\
The evolution of the scalar field is described by a scalar field equation, which we can derive from action \eqref{action} using the variational principle. The variation of the action with respect to the scalar field gives:
\begin{equation}
\delta S=\int \! \mathrm{d}^4x \lbrace -\sqrt{-g}g^{\mu\nu}(\partial_{\mu}\delta\phi)( \partial_{\nu}\phi)+\sqrt{-g} \dfrac{f'(\phi)}{8}R^2_{GB} \delta\phi \rbrace.
\end{equation}
After integrating by parts the first term, we obtain the scalar field equation. Vari- ation of the action with respect to the metric gives the modified Einstein equations (here we will not show their derivation step by step). Finally we can present the scalar field and metric equations together \cite{ARTSingfree}:
\begin{equation}\label{eqs!}
\begin{cases}

\dfrac{1}{\sqrt{-g}}\partial_{\mu}\left(\sqrt{-g}\partial^{\mu}\phi\right)=
-\dfrac{f'(\phi)}{8}R^2_{GB},\\

G_{\mu\nu}=\partial_{\mu}\phi{\partial}_{\nu}\phi
-\dfrac{1}{2}g_{\mu\nu}\left({\partial} _{\rho}\phi\right)^2-K_{\mu\nu},

\end{cases}
\end{equation}
where we have defined:
\begin{equation}\label{Kmunu}
K_{\mu\nu}=\left(g_{\mu\rho}g_{\nu\lambda}+g_{\mu\lambda}g_{\nu\rho}\right)
\eta^{\kappa\lambda\alpha\beta}\nabla_{\gamma}\left[\tilde{R}^{\rho\gamma}_{\alpha\beta}
\ \partial_{\kappa}f(\phi)/8\right],
\end{equation}
with
\begin{equation*}
\qquad \eta^{\mu\nu\rho\sigma}=\epsilon^{\mu\nu\rho\sigma}\dfrac{1}{\sqrt{-g}}, \quad \epsilon^{0ijk}=- \epsilon_{ijk},
\end{equation*}
\begin{equation*}
\tilde{R}^{\mu\nu}_{\rho\sigma}=\eta^{\mu\nu\alpha\beta}R_{\alpha\beta\rho\sigma}.
\end{equation*}
The coupling $f(\phi)$ between the GB curvature invariant and a scalar field is a universal feature of all four-dimensional effective field theory manifestations of heterotic super string and M-theory \cite{AntoModuliCorr}. This coupling function can take on different forms in effective string theory, whether we consider the scalar field to be a \emph{modulus} field or a \emph{dilaton} field (see, for example \cite{AntoModuliCorr, ARTSingfree}). For the modulus field the coupling is of the form:
\begin{equation}
f(\phi)=\lambda \ln \left[ 2 e^{\phi} \eta^4 (ie^{\phi}) \right],
\end{equation}
where $\eta$ is the Dedekind function:
\begin{equation}
\eta (\tau)=q^{1/12} \prod_{n=1}^\infty \left( 1-q^{2n} \right) \quad q=e^{i\pi\tau}.
\end{equation}
For a dilaton field, on the other hand, the coupling is much simpler:
\begin{equation}
f(\phi)=\lambda e^{\phi}.
\end{equation}
The dilaton is a scalar field always present in the gravitational sector of the string effective action, even to the lowest order. Its coupling with the Gauss-Bonnet invariant arises as a first order term in the expansion in powers of $\lambda \simeq \alpha '=\lambda_s^2 / 2\pi$, together with a term $(\nabla \phi)^4$, to be included in action \eqref{action}. Higher order corrections become more and more important when the curvature grows, and the perturbative expansion fails when the curvature radius becomes of the same order of the fundamental string length. The modulus field or, more frequently, several moduli fields, arise in string theory as a product of compactification from $D$ to 4 dimensions.\\ 
In effective string theory the coupling parameter $\lambda$ for the modulus field is proportional to the trace anomaly and its sign and value may depend on the composition of massless spectrum of the particular string model. When the scalar field is interpreted as a dilaton, instead, the string coupling is always positive \cite{GasperiniStringCosmology}.\\
Throughout this thesis we will bear in mind the existing connection between string theory and Gauss-Bonnet gravity, but we will not restrict ourself to a particular string model. Conversely, we consider a variety of coupling functions, and let the parameter $\lambda$ run freely, even adopting the negative sign for a dilaton-like coupling. \\

\section{Properties of Einstein-scalar-Gauss-Bonnet theory}
In four dimensions, the Gauss-Bonnet scalar $R_{GB}^2$ is a topological invariant. The action:
\begin{equation}
\int \; \mathrm{d}^4x \; \sqrt{-g} \; R_{GB}^2
\end{equation}
can be reduced to a surface term and thus does not contribute to the equations of motion. A non minimal coupling to a scalar field, at least, is needed to have Gauss-Bonnet term contribute to the dynamics.\\
The addition of  a scalar-Gauss-Bonnet term to Einstein-Hilbert action provides the theory with a number of desired features as well as novel solutions: this modified theory of gravity has uncovered many interesting possibilities not realized by the Einstein-Hilbert action, maintaining anyhow many of its desired features.\\ 
First of all, despite being a combination of higher curvature terms, Gauss-Bonnet invariant gives rise to still second order equations: thanks to the particular combination of curvature squared scalars, terms containing more than 2 partial derivatives of the metric cancel out. This makes the theory one of the simplest higher-curvature modifications of Einstein's General Relativity. Second order equations also imply that GB gravity is ghost-free: there are no spurious degrees of freedom \cite{GBGhost}. The structure of the GB invariant also guarantees that the theory is not plagued by Ostrogradski instability, which is associated with all Lagrangians with more than one time derivative when higher derivatives cannot be eliminated by integration. Finally, like the Einstein-Hilbert action, Gauss-Bonnet term gives rise to field equations which are divergence free, even though this property descends from the scalar equation and not from a geometrical identity, as in General Relativity. Note that Gauss-Bonnet is the only quadratic curvature term which has all of these properties.\\
All versions of string theory in 10 dimensions (except
type II) include Gauss-Bonnet term with a field-dependent coupling as the leading order $\alpha '$ correction \cite{StringGB}, where $\alpha '$  is the so-called Regge slope, or inverse of the string tension. Therefore Gauss-Bonnet gravity is a perfect framework to study the low energy limit of string theories. Besides, it is also an interesting modified gravity model \textit{per se}. A coupling between Gauss-Bonnet term and a scalar field is present in the best motivated scalar-tensor
theories which respect most of GR’s symmetries.\\   
The theory under study is indistinguishable from General Relativity at the post-
Newtonian order, having the same PPN (Parameterized Post Newtonian) parameters, $\beta=\gamma\equiv 1$\cite{PNEDGB}. Therefore it can't be ruled out by Solar system tests, as it trivially satisfies the provided constraints. The peculiarities of the theory emerge when taking into account fully nonlinear effects, and the ideal place to look for them is in early universe cosmology, or near compact objects such as black holes.\\
Black hole solutions exist in GB gravity \cite{KantiBH,MaedaBH}, and they have a regular horizon and flat asymptotics. These black holes can
evade the classical \emph{no-scalar-hair} theorem, and be
dressed with classical nontrivial dilaton hair. Stability \cite{Schiappa,KantiLinStab,MaedaLinStab} as well as astrophysical and observational implications \cite{EDGBQPO,PaniCardosoAstroCand,RotEDGBBH,DBHGravWaves} of dilatonic BHs have been carefully studied and could lead to the possibility of ruling out such objects and theories through future experiments.\\
Gauss-Bonnet theory has recently been intensly studied as a candidate explanation for the observed acceleration of the Universe (see, for example \cite{LeiNeuGBcosmologies,NeuCartDarkEnergy}).\\
Nonsingular cosmological solutions have been studied in a variety of string-inspired models containing Gauss-Bonnet term \cite{ARTSingfree,KantiSing,Easther,RizosSing,KantiEarlyTime,KantiGBInflation,Comment,nonsingephi,AltriNonSing,AnisotropicNonSing,ATUNonSing}. We will discuss the properties of these solutions in more detail in Chapter \ref{chap:phi2}. Here we only outline the main results. nonsingular solutions in presence of Gauss-Bonnet invariant were found: 
\begin{itemize}
\item[-] in the case of even monomial coupling ($f=\phi^n$, $n$ even) in FRW background with every spatial curvature \cite{ARTSingfree,KantiSing};
\item[-] in the case of the complete string theory coupling for modulus and dilaton fields \cite{Easther};
\item[-] in the case of approximated modulus coupling with or without an Ekpyrotic-like\footnote{The first version of Ekpyrotic Universe is a cosmological scenario, introduced by J. Khoury, B. Ovrut, Paul Steinhardt and Neil Turok \cite{Turok}, in which the hot Big Bang is produced by the collision of a brane in the bulk space. This collision is generally described in the four dimensional space-time with a negative exponential potential.} potential \cite{nonsingephi};
\item[-] in the case of anisotropic Bianchi I model, with both dilaton and modulus fields \cite{AnisotropicNonSing};
\end{itemize}
Inflationary scenarios have been intensely studied in string cosmology. Conserved cosmological quantities in a very general model containing Gauss-Bonnet term were first calculated by Hwang and Noh in \cite{PerturbInfl}. Within the so-called \emph{Pre Big-Bang scenario}, for example, inflationary solutions were studied and quantum originated tensor and scalar perturbations were calculated \cite{AltriInflPerturb}. This was done also for models with a generic coupling $F(\phi)R$ between the dilaton and the Ricci scalar and a kinetic term $\omega (\phi) (\nabla \phi)^2$, in the Jordan frame \cite{AltriInflPerturb2}.\\
Inflation driven by a Gauss-Bonnet-coupled scalar field was studied also in not so strictly stringy scenarios \cite{LeiNeuGBcosmologies, KohObsCos, GSPerturb, GSSlowRoll, JHGGBInflation, PlanckReheating, CircPolGravWaves}. When dealing with the inflationary properties of our solutions, we will mainly refer to the latter approach.\\
We close our brief introduction to GB gravity by reviewing the observational constraints discussed by Esposito-Farese in \cite{EspositoFarese}. As we have already pointed out, solar-system observations are in agreement with both GR and Einstein-scalar-Gauss-Bonnet gravity. Constraing our model through solar-system tests is only possible when also cosmological data are taken into account. If we use the present accelerated expansion to renconstruct the form of $f(\phi)$, and then compare it with solar-system data, we can succesfully constraint the coupling function. Esposito-Farese thus finds that the theory could be rejected because of a high degree of fine-tuning. However, we should stress that the present dark energy content could be explained with a different mechanism, which could co-exist with the GB modification: in this case the reconstruction of $f(\phi)$ from cosmological obeservations would not be possibile. The model is not ruled out if a modification as simple as the introduction of a potential $V(\phi)$ or a cosmological constant $\Lambda$ is made.  

\section{Early-universe cosmology in Gauss-Bonnet gravity}
\subsection{Field equations}
\label{sec:cosmoeqs}
We assume that the line element has the FLRW form (Eq. \eqref{FRW}) and assume, consistently, a homogeneous time-dependent scalar field:
\begin{equation}\label{metricansatz}
\phi=\phi(t).
\end{equation}
Using this ansatz, the scalar and gravitational field equations \eqref{eqs!} reduce to the following system of three ordinary coupled differential equations:
\begin{equation}\label{eqf}
\begin{cases}
\ddot{\phi}+3H\dot{\phi}-3f' \left(H^2+\dfrac{k}{a^2}\right)\left(H^2+\dot{H}\right)=0,\\
3\left(1+\dot{f}H\right)\left(H^2+\dfrac{k}{a^2}\right)=\dfrac{\dot{\phi}^2}{2},\\
2\left(1+\dot{f}H\right)\left(H^2+\dot{H}\right)+
\left(1+\ddot{f}\right)\left(H^2+\dfrac{k}{a^2}\right)=-\dfrac{\dot{\phi}^2}{2}.\\
\end{cases}
\end{equation}
Here and throughout the work, a prime will denote differentiation with respect to the function's argument, while the dot will always represent differentiation with respect to coordinate time: 
\begin{equation*}
f'=\dfrac{df}{d\phi}, \quad \quad \dot{f}=\dfrac{df}{dt}.
\end{equation*}
The second equation in \eqref{eqf} is not a dynamical equation but a constraint. This means that its time derivative identically vanishes along the solutions of the system \eqref{eqf}:
\begin{equation}
\dfrac{\partial}{\partial t}\left( 3\left(1+\dot{f}H\right)\left(H^2+\dfrac{k}{a^2}\right)-\dfrac{\dot{\phi}^2}{2} \right)=0.
\end{equation}
We have explicitly verified the statement above for the two choices of the coupling functions.

Looking at action \eqref{action}, we notice that the coupling parameter $\lambda$ has dimensions $[\lambda]=1/[t]^2$. Eq.s \eqref{eqf} in flat background ($k=0$) possess a symmetry under the simultaneous rescaling of $\lambda \rightarrow \alpha \lambda$ and $t \rightarrow \sqrt{\alpha} t $. Then we can put $\vert\lambda\vert=1$ and express solutions as functions of the a-dimensional quantity $t/\sqrt{\vert \lambda \vert}$. In short, the value of $\vert \lambda \vert$ only determines the time-scale of the cosmological model. When the background is not flat ($k=\pm 1$), in order to preserve the symmetry we need to further rescale $a \rightarrow \sqrt{\alpha} a$. 

\subsection{Energy conditions}
\label{sec:ec}
In this section we discuss whether the Strong Energy conditions (SEC), $\rho+p>0$, $\rho+3p>0$, can be violated in Gauss-Bonnet gravity. The violation of SEC allows for nonsingular cosmological solutions to arise, since Hawking Penrose singularity theorem (Section \ref{HPth}) does not hold anymore.\\
Assuming a perfect fluid form for the effective energy-momentum tensor of the theory, we define:
\begin{equation}
T^{\textrm{eff}}_{\mu\nu}=G_{\mu\nu}, \quad \rho=T^{\textrm{eff}}_{00}, \quad \quad p \ g_{rr}= T^{\textrm{eff}}_{rr}.
\end{equation}
Using equations \eqref{eqf} to replace $\ddot{\phi}$, $\dot{H}$ and $f'(\phi)$, we can explicitly write the quantities above. The total energy density can be written as:
\begin{equation}
\rho= \dfrac{\dot{\phi}^2}{2}-3\dot{f}H^3 - 3 \dfrac{k}{a^2}\dot{f}H,
\end{equation}
where we see the contribution of the kinetic energy of the scalar field, and two contributions arising from the Gauss-Bonnet term, one of which disappears in flat three-space.\\
We focus on the simplest case, $k=0$, for which the two expressions simplify as follows:
\begin{equation}\label{encond0}
\begin{split}
\rho&+p=2H^2\frac{B+12 H(t)^2 \dot{\phi}^2 \left(\dot{\phi}^2 f''(\phi)+4\right)}{B}\\
\rho&+3p=\dfrac{72 H^4 \dot{\phi}^2}{B}\left( 4+\dot{\phi}^2f''(\phi) \right)
\end{split}
\end{equation}
where $B=-12 H^2 \dot{\phi}^2+36 H^4+5 \dot{\phi}^4$. \\
$B$ is a polynomial in $\dot{\phi}^2$ with no real roots, so it is always positive definite, $B>0$. Then energy conditions are violated if and only if:
\begin{equation}
4+\dot{\phi}^2f''(\phi)<0
\end{equation} 
The condition above reduces, in the case of quadratic and exponential coupling, respectively, to:
\begin{equation}
 4 + 2 \lambda  \dot{\phi}^2  <0 \quad \quad 4+ \lambda \beta^2 \dot{\phi}^2 e^{\beta \phi}<0
\end{equation} 
It is clear from these explicit expressions that a violation of strong energy conditions is only possible if $\lambda<0$.\\
When $k\neq0$ the conditions are a slightly more complicated. For $k=1$ one can perform an analogous analysis and prove that energy conditions are violated only if $\lambda<0$. For $k=-1$, on the other hand, one can show that energy conditions can be violated for both signs of $\lambda$\cite{KantiSing}.\\
The two expressions \eqref{encond0} should be compared with the ones obtained in \cite{KantiSing}, with which they coincide except for the definition of the coupling function, $f(\phi)=-2\delta\xi(\phi)$\footnote{We have used here a different sign for the coupling function: while in Kanti's 1999 paper energy conditions can be violeted only for $\delta>0$, with our notation we need $f<0$, or $\lambda<0$ if we explicit a coupling constant.}. \\

Another way to express the strong energy conditions for a flat space-time is by defining the parameter $\Gamma$, similar to the equation of state parameter $\Gamma =w+1$:
\begin{equation}\label{Gammadef}
p=(\Gamma -1)\rho.
\end{equation}  
Using the background equations \eqref{eqf}, for a flat universe:
\begin{equation}
\Gamma=-\dfrac{2}{3} \dfrac{\dot{H}}{H^2}.
\end{equation}
Strong Energy conditions are thus violated when:
\begin{equation}
\Gamma<0 \quad \text{or} \quad \Gamma<2/3 \quad \Rightarrow \quad \Gamma<2/3.
\end{equation}

\subsection{Tensor perturbations}
\label{sec:GBstability}
Once a solution of Gauss-Bonnet theory is found, we are required to check if it is a viable model for our Universe, i.e. if it is consistent with observational data. In the case of quadratic coupling function, for example, it was recently shown by G. Hikmawan, J. Soda and others \cite{Comment} that inflationary solutions with flat spatial geometry are to be rejected because of unstable tensor perturbations. In the present section we derive, following their  argument, the equations for tensor perturbations in Gauss-Bonnet gravity with an arbitrary coupling $f$. In the next two chapters we will use the results of this section to analyse the properties of our numerical solutions, improving and expanding the results of Ref. \cite{Comment}.\\
We proceed to study stability in the context of first order perturbation theory, following the work of Sakagami, Kawai and Soda, who first calculated scalar, vector and tensor perturbations in scalar-Gauss-Bonnet theory \cite{KSSPerturbations,KSSNovelInstab,KSSInstability}. A gauge invariant perturbation method is used, so that conclusions about stability are not plagued by gauge artifacts. Sakagami, Kawai and Soda show that there are no growing scalar modes, and that vector perturbations decrease as the universe expands. Scalar, vector and tensor perturbations are decoupled, as in GR, so that we can focus our attention to tensor perturbations alone.\\
Tensor perturbations on a flat FRW background are defined by:
\begin{equation}\label{pertmetric}
ds^2=-dt^2+a(t)^2(\delta_{ij}+h_{ij})dx^idx^j,
\end{equation}
with the transverse and traceless condition:
\begin{equation}
h^{ij}_{,j}=h^i_i=0.
\end{equation}
This corresponds to the perturbed metric of Eq. \eqref{pertmetr}, provided we turn off all non-tensor degrees of freedom. Here we are exploiting the freedom to fix a gauge, choosing the \emph{comoving gauge}: inhomogeneities are all attributed to the metric and the scalar field does not carry any perturbation, $\delta \phi=0$. We have already used this gauge to study the fluctuations produced by inflation in General Relativity, see Eq. \eqref{comovinggauge}.\\
By substituting the metric \eqref{pertmetric} in the action of our model, and expanding the action to $O(h^2)$, in order to obtain $O(h)$ equations, we obtain the \lq\lq perturbed" action:
\begin{equation}\label{pertaction}
\begin{split}
S_h & =\dfrac{1}{8} \int d^4x a^3 \lbrace [ \dot{h_{ij}}\dot{h^{ij}}-\dfrac{1}{a^2} h_{ij,k}h^{ij,k}+\\
&-(4\dot{H}+6H^2+\dot{\phi}^2)h_{ij}h^{ij}]
-\ddot{f}\left[\dfrac{1}{a^2}h_{ij,k}h^{ij,k}+2H^2)h_{ij}h^{ij}\right]+\\
&+\dot{f}\left[ -H\dot{h_{ij}}\dot{h^{ij}}+4H(\dot{H}+H^2)h_{ij}h^{ij}\right]\rbrace.
\end{split}
\end{equation}
Using the background equations the above expression can be simplified:
\begin{equation}\label{simpertaction}
S_h=\dfrac{1}{8}\int d^4x a^3 \left[ \alpha\dot{h_{ij}}\dot{h^{ij}} -\dfrac{1}{a^2}\left(1+\ddot{f}\right)h_{ij,k}h^{ij,k}\right].
\end{equation}
We have introduced $\alpha$, which only depends on the background variables:
\begin{equation}\label{alpha}
\alpha=1+H\dot{f}.
\end{equation}
As we can see from the constraint equation (second equation in the system \eqref{eqf}), $\alpha$ has a clear physical meaning: it is proportional to the fraction of the scalar field kinetic energy over the geometrical one. Moreover, as $\alpha$ shows up in front of the kinetic term in the action \eqref{simpertaction}, the model is \emph{ghost-free} only if 
\begin{equation}\label{gfcond}
\textbf{ghost-free condition: }\alpha>0.
\end{equation}
After applying the variational principle with one integration by parts,
and setting $h_{ij}(t,\vec{x})=h_{ij}(t)e^{\vec{k} \cdot \vec{x}}$, we obtain the equation for tensor perturbations:\\
\begin{equation}\label{eqperturb}
\ddot{h}_{ij}+\left(3H+\dfrac{\dot{\alpha}}{\alpha} \right)\dot{h}_{ij}+\dfrac{k^2}{a^2}\dfrac{1+\ddot{f}}{\alpha} h_{ij}=0.
\end{equation}
The equation above can be compared to the one found in General Relativity \eqref{eqperturbGR}.\\
Gravitational waves will grow arbitrarily if the factor in front of $h_{ij}$ is negative and big. First of all, then, we can say that instability, if present, has a time-scale growing when the wavelength becomes smaller. In addition, Sakagami, Kawai and Soda showed that the friction term in equation \eqref{eqperturb}, namely
\begin{equation}\label{frictionterm}
F=3H+\dfrac{\dot{\alpha}}{\alpha},
\end{equation}
When $ F < 0 $, this term cannot be regarded as a friction contribution anymore. $ F $ should not be neglected in the stability analysis, as it can produce or enhance instability. \\ 
To better discuss stability we can define the effective speed of sound:\\
\begin{equation}\label{scond}
c_s^2=\dfrac{1+\ddot{f}}{\alpha}.
\end{equation}
Tensor perturbations are stable only if:
\begin{equation}
\textbf{stability condition: }c_s^2>0.
\end{equation}
An additional condition should be imposed to prevent tensor modes from becoming
superluminal:
\begin{equation}\label{key}
c_s^2 \leqslant 1. 
\end{equation}
Superluminal modes can create problems with causality; however, superluminal propagation can also be admitted under specific conditions. In this thesis superlu- minal modes will be marked as stable and will not be rejected: the complexity of the issue requires further study, which is left to a future extension of the present work.\\
We can join the two conditions \eqref{gfcond}, \eqref{scond}, obtaining the comprehensive requirement:
\begin{equation}
\left( 1+\ddot{f} \right) >0, \quad \textrm{and} \quad \alpha >0.
\end{equation}
Using the background equations \eqref{eqf} (the constraint and Einstein equation) we can put the above condition in a different form. Thanks to the already defined function $\Gamma$ \eqref{Gammadef} we can finally write:
\begin{equation}\label{condition}
\textbf{stability + ghost-free condition: }\alpha \left(3\Gamma-5\right)>0.
\end{equation}
We easily see that, providing it's ghost-free, the solution is stable if:
\begin{equation}
\Gamma>5/3
\end{equation}
Comparing this inequality with the Strong Energy conditions, we see that \emph{no nonsingular solution}, for which $\Gamma<2/3$ is necessary, \emph{can be stable}.\\

If we consider inflationary solutions, rather than nonsingular ones, we need to define, as one of the slow-roll parameters:
\begin{equation}
\epsilon= - \dfrac{\dot{H}}{H^2}= \dfrac{3}{2} \Gamma.
\end{equation}
The slow-roll approximation requires $\epsilon_H < 1$. But the stability condition, expressed as a function of the parameter $\epsilon_H$, becomes:
\begin{equation}
\alpha \left(2\epsilon_H -5\right)>0.
\end{equation}
Then we see, as discussed in \cite{Comment}, that also \emph{no accelerated solution}, for which $\epsilon_H < 1$ is necessary, \emph{can be stable}.\\

Throughout our work we will use the quantities \eqref{scond}, \eqref{alpha}, to test the stability of cosmological solutions. If the outcome of the test is negative, i.e. if tensor perturbations are unstable or contain a ghost, first order perturbation theory should be abandoned and the solution should be rejected. When tensor perturbations are unstable the background geometry breaks down quickly, due to high frequency gravitational waves, unless we fine-tune initial conditions of the perturbations. Indeed, if at some initial time we force the amplitude of the perturbations to be sufficiently small, and the instability switches off after a finite lapse of time, then gravitational waves produced could be too weak to destroy the background geometry. However, this would clearly require a high degree of fine-tuning. Moreover, a condition for the initial amplitude of perturbations can be found only if we assume a cut off in the modes' frequency.\\ 
We should stress that an instability does arise even in the radiation era in Einstein gravity as the scale factor approached the singular value $a\sim t^{1/2} \rightarrow 0$. The solution of the perturbation equation in General Relativity \eqref{eqperturbGR} in the radiation era and on flat background is\cite{Weinberg}:
\begin{equation}
h(t)= Re \; \lbrace \; \frac{c_1 e^{2 i k \sqrt{t}}}{\sqrt{t}}+\frac{i c_2 e^{-2 i k \sqrt{t}}}{2 k \sqrt{t}} \; \rbrace.
\end{equation}
Tensor perturbations clearly explode, breaking down the background geometry, when $t\rightarrow 0$. Yet, the current paradigm does not extend the radiation dominated solution all the way back to the Big Bang, while Gauss-Bonnet solutions are proposed exactly for this purpose.\\

The theoretical results of this Section suggest that Gauss-Bonnet gravity does not work in its simplest realization. We will numerically verify instability of all interesting solutions in Chapters \ref{chap:phi2} and \ref{chap:exp}. We will then attempt a modification in order to improve the stability of the theory, and will present our results in Chapter \ref{chap:V}.\\ 
If, on the other hand, the outcome of the test is positive, as happens when a potential is taken into account (Chapter \ref{chap:V}), we need to check the stability and phenomenological implications of scalar and vector perturbations, as well as that of quantum perturbations, before marking the solution as a viable cosmological model.\\

So far, we have discussed tensor perturbations on a flat background. When the background is spatially curved, the equation of the tensor mode $k$ reads \cite{KSSInstability}:
\begin{equation}
\ddot{h}_{ij}+\left(3H+\dfrac{\dot{\alpha}}{\alpha} \right)\dot{h}_{ij}+\dfrac{k^2+2\kappa}{a^2}\dfrac{1+\ddot{f}}{\alpha} h_{ij}=0,
\end{equation}
where $\kappa$ is the the spatial curvature, $c_s$ and $\alpha$ have the same expressions of the flat case, i.e. \eqref{alpha} and \eqref{scond}. However, in this work we will not discuss the stability of spatially curved solutions.

\chapter{Gauss-Bonnet cosmology: quadratic coupling}
\label{chap:phi2}

In this chapter we study cosmological solutions of our theory when the coupling between the scalar field and the Gauss-Bonnet invariant takes a quadratic form. We describe nonsingular and early-time solutions obtained by Kanti \textit{et al.} in this model. Then we summarize our numerical results, and compare them to the ones previously discussed. We also study the stability of nonsingular solutions. 

\section{Field equations}
We address the case of a simple quadratic coupling: 
\begin{equation}
f(\phi)=\lambda \phi^2.
\end{equation}
Of course, this choice is much simpler than the true coupling function arising in the context of super string effective theory and described in Section \ref{actioneq}, $f(\phi)=\ln \left[ 2 e^{\phi} \eta^4\left( ie^{\phi} \right) \right]$. However, the complete string coupling and the quadratic one share a number of important properties. They are both invariant under the change of sign $\phi\rightarrow -\phi$, have a global minimum at $\phi=0$ and diverge when $\phi \rightarrow \pm \infty$. Thus the quadratic coupling can be considered a good approximation of the string one, and we can expect their cosmological solutions to share most of their features.\\
Once we substitute the explicit form of the coupling, the cosmological equations of the theory become: 
\begin{equation}\label{eqphi2}
\begin{cases}
\ddot{\phi}+3H\dot{\phi}-6\lambda\phi \left(H^2+\dfrac{k}{a^2}\right)\left(H^2+\dot{H}\right)=0\\
3\left(1+2\lambda\phi\dot{\phi}H\right)\left(H^2+\dfrac{k}{a^2}\right)=\dfrac{\dot{\phi}^2}{2}\\
2\left(1+2\lambda\phi\dot{\phi}H\right)\left(H^2+\dot{H}\right)+
\left(1+2\lambda\phi\ddot{\phi}+2\lambda\dot{\phi}^2\right)\left(H^2+\dfrac{k}{a^2}\right)=-\dfrac{\dot{\phi}^2}{2}\\
\end{cases}
\end{equation}
\section{Existence of nonsingular solutions}
\label{nonsingKanti}
nonsingular solutions in the quadratic coupling case were first discovered to exist due to the violation of Strong Energy conditions a numerically found by Antoniadis, Rizos and Tamvakis in \cite{ARTSingfree}. However, the analytical argument proving their existence for a variety of couplings (including $\phi^2$) and for every spatial curvature is due to Kanti, Rizos and Tamvakis and described in detail in \cite{KantiSing}.\\
The argument is based on the search for \emph{singular} cosmological solutions, characterized by a singularity at finite time. Referring to equations \eqref{eqf}, they define the quantities:
\begin{equation}
x=\dot{\phi}, \quad z=H, \quad y=\dfrac{1}{a^2}.
\end{equation}
They show that singular solutions are indeed present, and can only take the form:
\begin{equation}
z^2\sim y \sim \dfrac{1}{f(\phi)-f(\phi_s)} \sim \dfrac{1}{t-t_s}, 
\end{equation}
where $\phi_s$ and $t_s$ are the value of the scalar field and time, respectively, at the singularity. Substituting the definition of $y$ as a function of $a$, we find that the scale factor behaves near the singularity as:
\begin{equation}
a(t)\sim \left( t-t_s\right) \rightarrow 0 \quad \textrm{when} \quad t\rightarrow t_s.
\end{equation}
The only assumption of their completely analytical argument is the monomial form of the coupling function, $f=\phi^n$, $n\geq 1$, while spatial curvature is allowed to be $k=0,\pm 1$. If $n$ is even, singular solutions are confined in a region of the phase space of the theory. This restriction leaves the rest of the phase space to be necessarily covered by nonsingular solutions. This proves the existence of nonsingular solutions in Gauss-Bonnet gravity with a quadratic coupling.\\
Kanti \textit{et al.}'s result, together with Easther and Maeda's \cite{Easther}, also show that nonsingular solutions appear with a finite measure of the required initial conditions.

\section{Early-time solutions}
\label{sec:earlytimesol}
\begin{figure}[!ht]
 \centering
 \begin{subfigure}{.46\textwidth}
 \centering
    \includegraphics[width=1\textwidth]{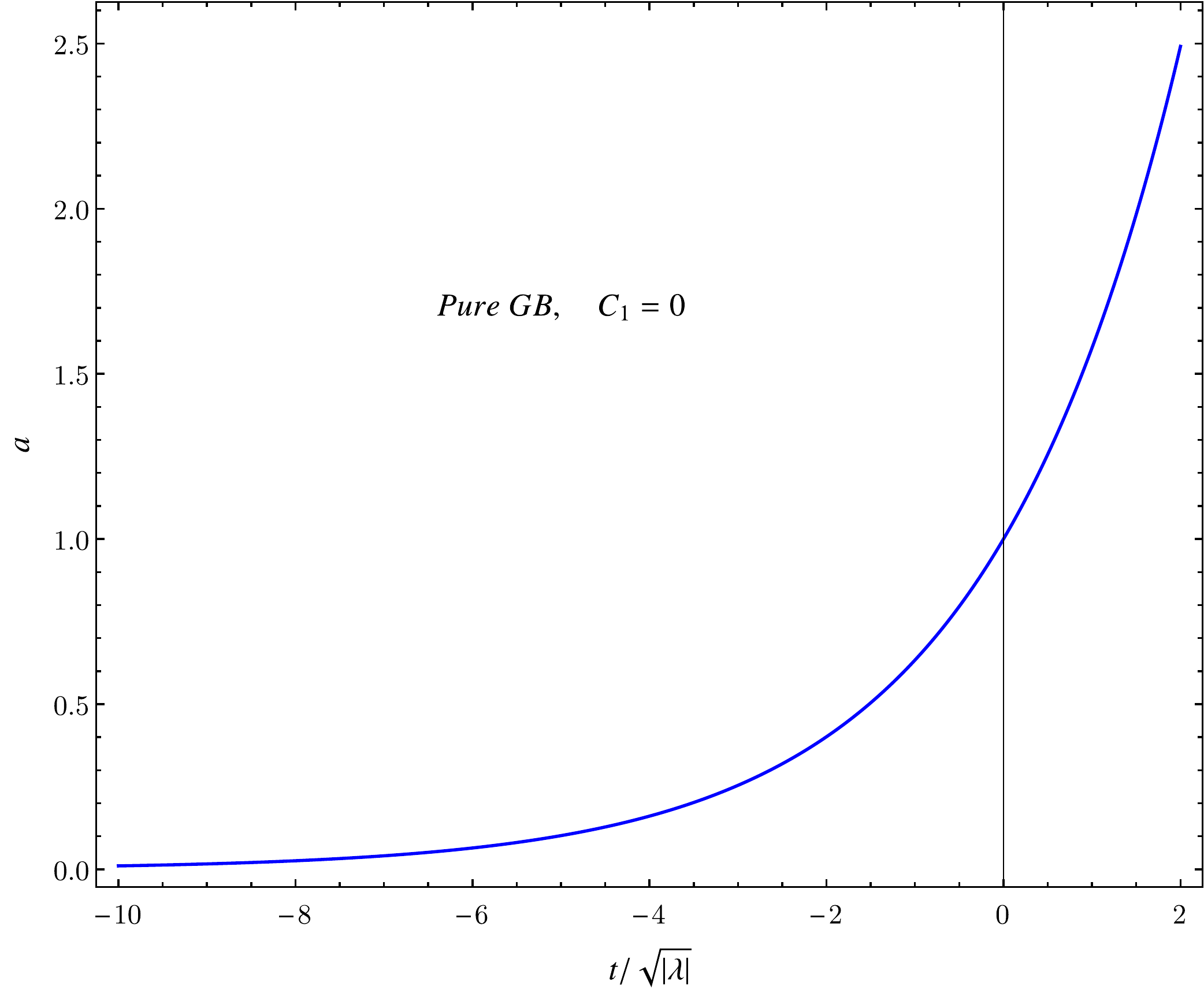}
    \caption{Solution with $C_1=0$ (A, de Sitter) and $a_0=1$.}\label{fig:phi2NORA}
\end{subfigure}
\begin{subfigure}{.46\textwidth}
\centering
    \includegraphics[width=1\textwidth]{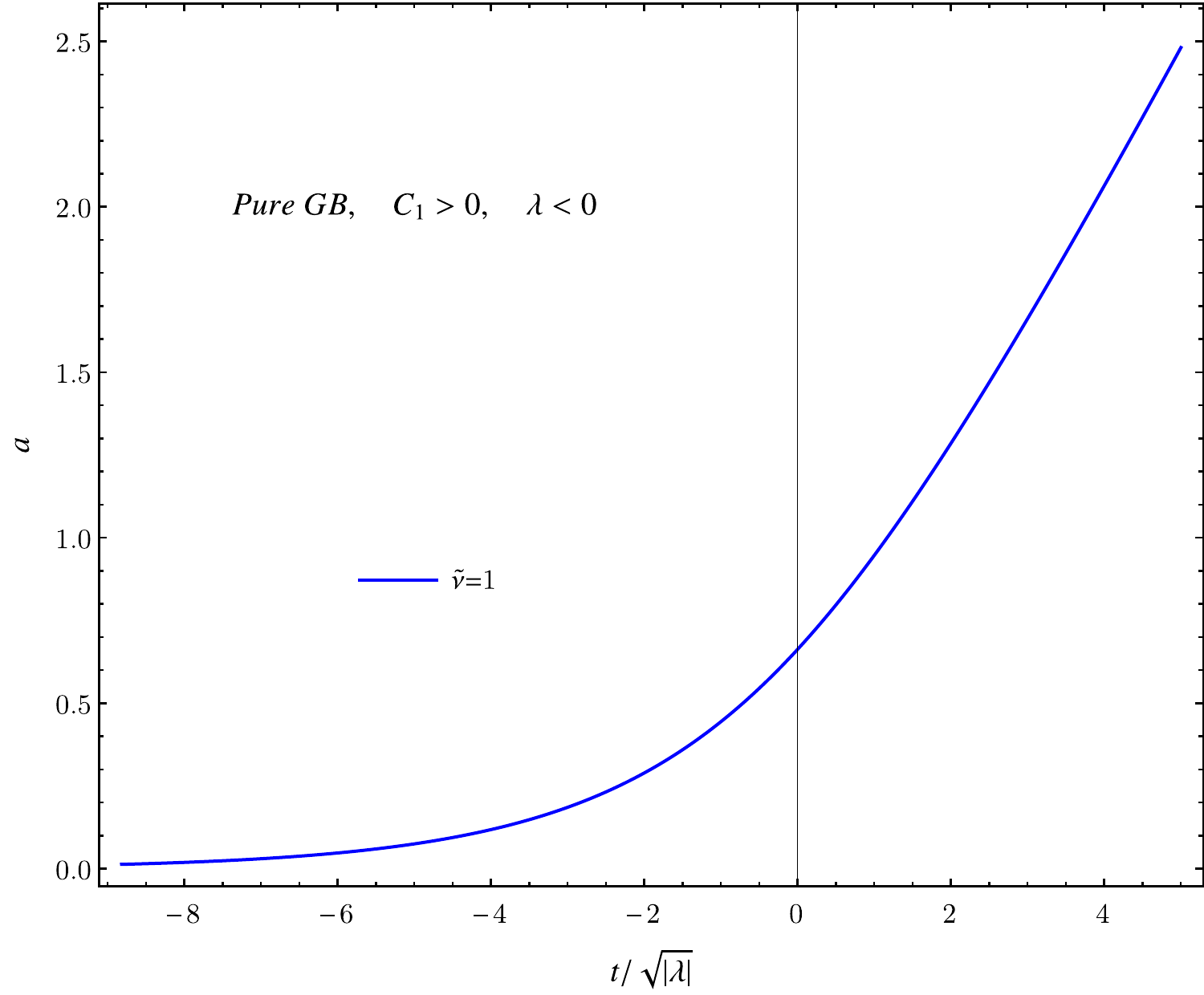}
    \caption{Solution with $C_1>0, \ \lambda <0$ (B) and $\nu=1$.}\label{fig:phi2NORB}
    \end{subfigure}
 \begin{subfigure}{.46\textwidth}
 \centering
    \includegraphics[width=1\textwidth]{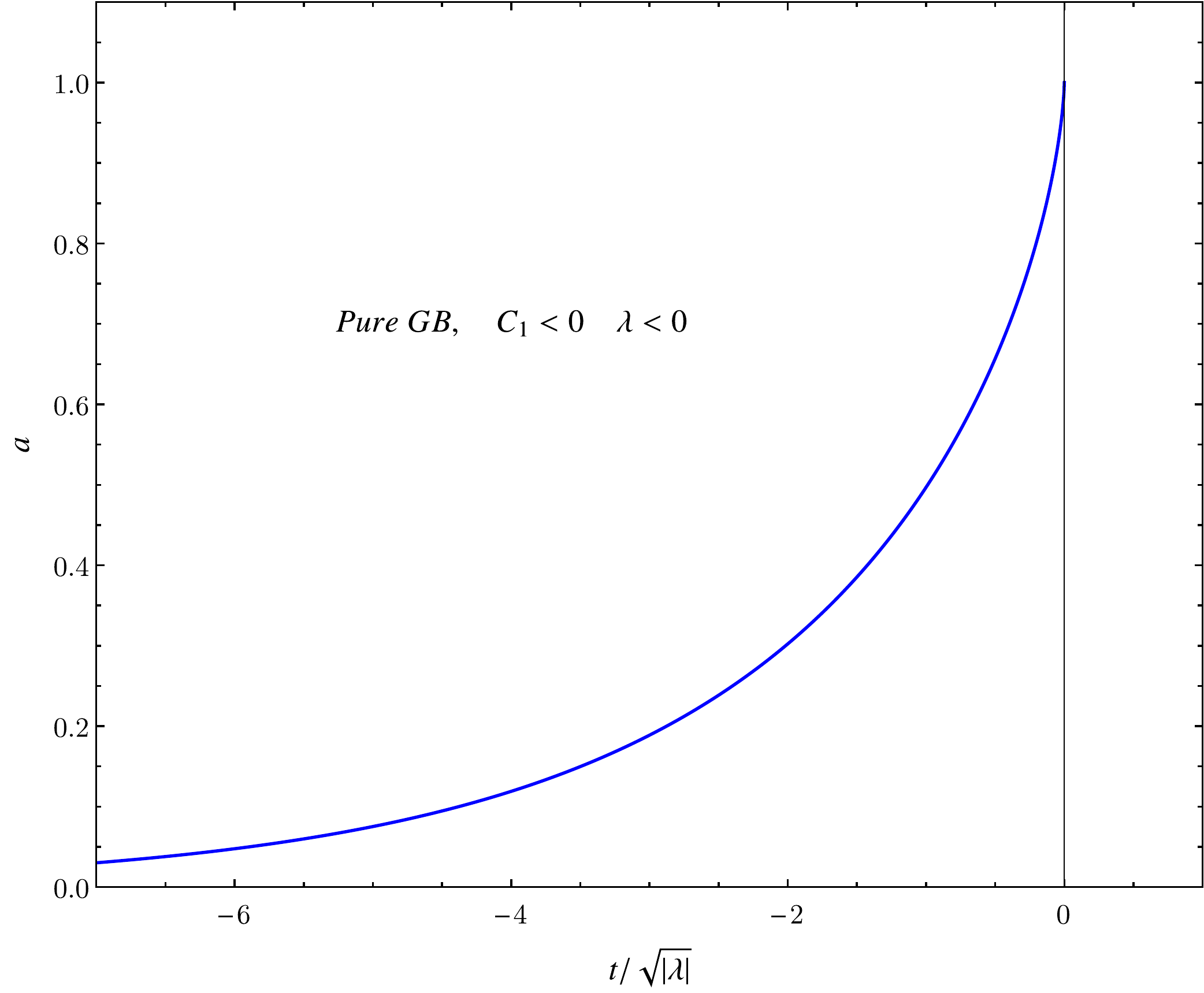}
  \caption{Solution with $C_1<0, \ \lambda <0$ (C) and $\tilde{\nu}=1$.}\label{fig:phi2NORC}
\end{subfigure}
\begin{subfigure}{.46\textwidth}
\centering
    \includegraphics[width=1\textwidth]{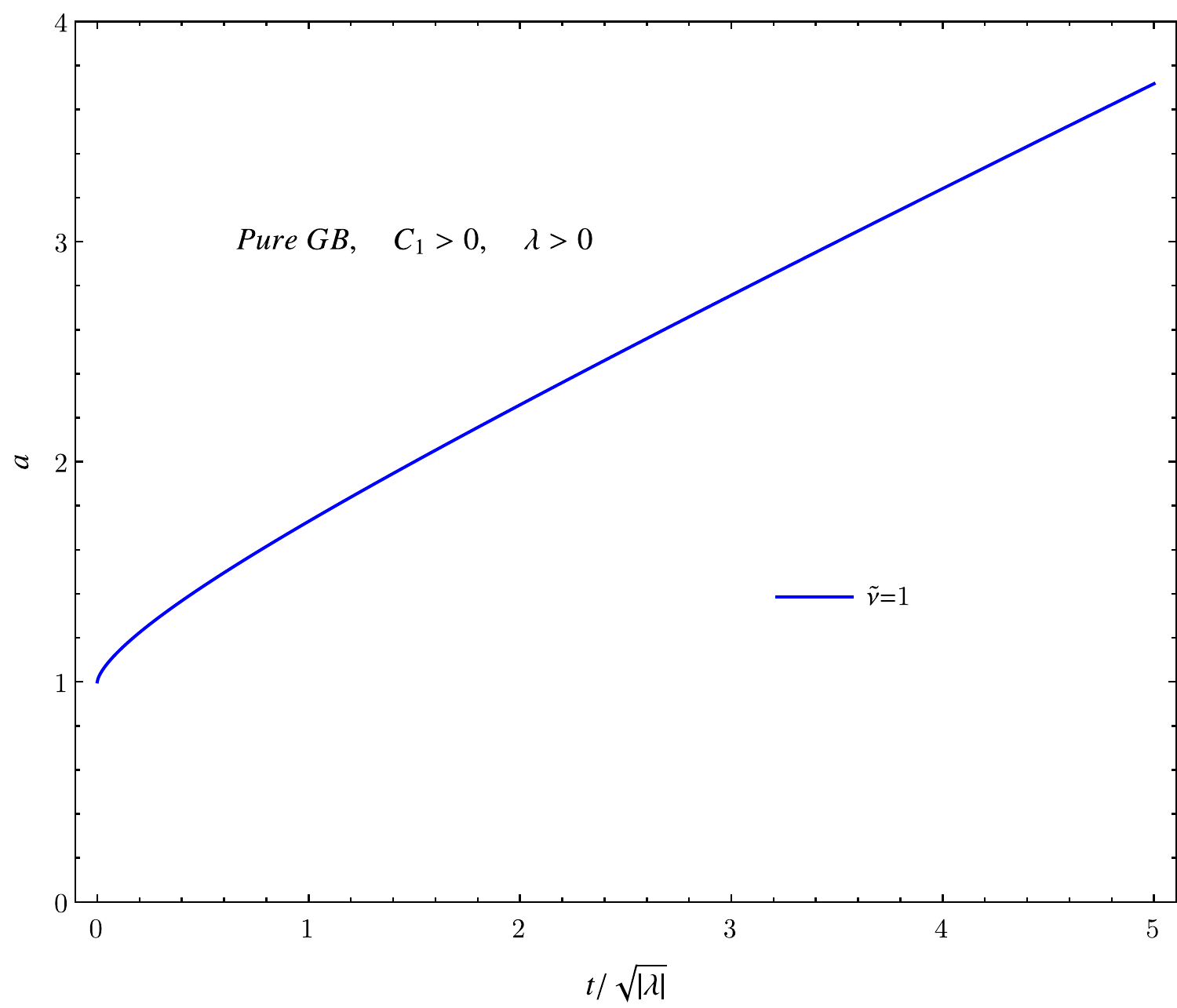}
    \caption{Solution with $C_1>0, \ \lambda >0$ (D) and $\tilde{\nu}=1$.}\label{fig:phi2NORD}
    \end{subfigure}
   \caption{Scale factor as a function of rescaled time, in pure GB theory with quadratic coupling function.}\label{phi2NORCD}
\end{figure}

Recently, Kanti, Gannouji and Dadhich \cite{KantiEarlyTime, KantiGBInflation} proposed a simplifying assumption for early-time Gauss-Bonnet gravity leading to analytical solutions. They argue that the role of the Ricci scalar is negligible for early-time dynamics, where the Gauss-Bonnet term should be dominant, instead. \\
They consider what we may denominate a \emph{pure Gauss-Bonnet} model, to which we can associate the action:
\begin{equation}
S_{pGB}=\int \! \mathrm{d}^4x \ \sqrt{-g} (-\dfrac{1}{2}\partial_{\mu}\phi\partial^{\mu}\phi+ \dfrac{f(\phi)}{8} R^2_{GB}). 
\end{equation}
Pure Gauss-Bonnet theory can also be considered as an approximation of GB gravity. In this approximation, terms arising from the Ricci invariant in the equations of motion are discarded: the unit terms inside brackets in Eq.s \ref{eqphi2} are neglected. This simplification is proved to be correct in the simple case of a linear coupling, $f=\lambda \phi$, and its principle is illustrated with a toy model where the constraint $R_{GB}^2=0$ is imposed. Still they are not able to find an analytical argument supporting the dominance of the Gauss-Bonnet scalar over the Ricci term when the coupling is quadratic. In their opinion, however, the specific form of the coupling function only determines the weight of the Gauss-Bonnet term in the theory, merely defining the point where the dominance takes place. Later, in Section \ref{sec:phi2nonsing}, we will comment on this and argue that this approximation should be treated more cautiously. In the present Section we briefly summarize their results, showing the analytical solutions they find. \\
When $k=0$, i.e. the Universe is flat, Kanti \emph{et al.} obtain the simple equation, in the pure GB model:
\begin{equation}\label{eqPGB}
\begin{cases}
\dfrac{5}{2}\dfrac{1}{\dot{a}^2}=C_1-\dfrac{12\lambda}{a^2}\\
\dfrac{\dot{\phi}}{\phi}=12\lambda \dfrac{\dot{a}^3}{a^3}
\end{cases}
\end{equation}
Four different kinds of cosmological solutions can be derived from the equations above depending on the values of $C_1$ (arising as an integration constant) and the coupling parameter $\lambda$.
\begin{itemize}
\item[A)] For $C_1=0$, solutions only exist when $\lambda<0$, and have the simple form:
\begin{equation}
\begin{split}
a(t)&=a_0 \; e^{\pm H_{dS} t}, \quad H_{dS}=\sqrt{\dfrac{5}{24 \vert \lambda \vert}}\\
\phi(t)&=\phi_0 \; e^{-\dfrac{5}{2}H_{dS}t}
\end{split}
\end{equation}
They thus obtain an inflationary, purely de Sitter solution (Figure \ref{fig:phi2NORA}) without any need for a self-coupling potential $V(\phi)$: the Gauss-Bonnet term is naturally providing a potential, supporting inflation on its own. \\ 
\item[B)] Setting $C_1>0, \ \lambda <0$ and solving equations \eqref{eqPGB} by separation of variables, one obtains the implicit expressions:
\begin{equation}
\begin{split}
&\sqrt{a^2+\nu ^2}+\nu  \log \left(\frac{\sqrt{a^2+\nu ^2}-\nu }{a}\right)=\pm \sqrt{\frac{5}{2 C_1}} (t+t_0) \\
&\phi^2=C_0 \left( \dfrac{2 C_1}{5} \right)^{5/2} \dfrac{\left( a^2 +\nu^2\right)^{5/2}}{a^5}
\end{split}
\end{equation}
where $\nu=12 \vert \lambda \vert /  C_1 $. This represents a nonsingular, expanding ($+$ sign) Universe, see Figure \ref{fig:phi2NORB}. In the limit $a\rightarrow 0$ this solution is smoothly connected to the de Sitter one, so the singularity is only approched for $t\rightarrow -\infty$. Later in time, the scale factor grows linearly (Milne-type phase). The scalar field tends to a constant for $a^2 \gg \nu^2$.\\
\item[C)] When $C_1<0, \ \lambda <0$, one integration of equations \eqref{eqPGB} yields:
\begin{equation}
\begin{split}
&\sqrt{\tilde{\nu} ^2-a^2}+\tilde{\nu}  \log \left(\frac{\tilde{\nu} -\sqrt{\tilde{\nu} ^2-a^2}}{a}\right)= \pm \sqrt{\frac{5}{2 \vert C_1 \vert}} (t+t_0) \\
&\phi^2=C_0 \left( \dfrac{2 \vert C_1 \vert}{5} \right)^{5/2} \dfrac{\left( \tilde{\nu}^2 -a^2 \right)^{5/2}}{a^5}
\end{split}
\end{equation}
where $\tilde{\nu}=12 \vert \lambda \vert / \vert C_1 \vert$. This solution is similar to the previous one, but the scalar factor can only grow up to a maximum value, $a= \tilde{\nu}$ (see Figure \ref{fig:phi2NORC}). 
\item[D)] Setting $C_1>0, \ \lambda >0$, similarly, Kanti \textit{et al.} obtain:
\begin{equation}
\begin{split}
&\sqrt{a^2-\tilde{\nu} ^2}-\tilde{\nu}  \cos ^{-1}\left(\frac{\tilde{\nu} }{a}\right)= \pm \sqrt{\frac{5}{2 C_1}} (t+t_0) \\
&\phi^2=C_0 \left( \dfrac{2  C_1 }{5} \right)^{5/2} \dfrac{\left( a^2-\tilde{\nu}^2 \right)^{5/2}}{a^5}
\end{split}
\end{equation} 
Here the square root requires $a^2 > \tilde{\nu}^2$, so again singularities are not allowed to arise. This solution is shown in Figure \ref{fig:phi2NORD}. 
\end{itemize}
The solutions we have described so far have the property of being completely analytical. They have been proposed as a good approximation of the solutions of the complete theory, which can only be obtained via numerical methods. In the following Section we will study the numerical solutions of complete Gauss-Bonnet gravity, and in Section \ref{sec:phi2nonsing} we will compare our numerical results with the analytical solutions of the approximate model. \\

\clearpage
\section{Numerical analysis}
\label{sec:NumAnphi2}
We proceed to numerically integrate the equations \eqref{eqphi2}. We will use $O(1)$ (in Planck units) initial conditions\footnote{We should stress that, in the present Chapter and the next one, initial conditions are chosen arbitrarily, indicating that interesting solutions exist for generic initial values of the fields. In Chapter \ref{chap:V} we will constraint initial conditions with the requirement of a sufficiently large inflationary phase, thus reducing the number of degrees of freedom.}:
\begin{equation}\label{ic}
\phi(0)=0.5, \quad H(0)=1, \quad a(0)=1.
\end{equation} 
We only integrate first and third equation in \eqref{eqphi2}, together with the obvious one:
\begin{equation}
H=\dfrac{\dot{a}}{a}
\end{equation}
The second equation is a constraint, as we showed in Section \ref{sec:cosmoeqs}, so it can be used to find the initial condition for $\dot{\phi}$. This procedure, however, gives us two possible branches. The constraint can be rewritten as $\dot{\phi}^2-2A\dot{\phi}+B=0$, with:
\begin{equation}
A=6H\phi\left(H^2+\dfrac{k}{a^2}\right), \quad B=-6\left(H^2+\dfrac{k}{a^2}\right).
\end{equation}
The solutions are:
\begin{equation}
\dot{\phi}_{s}=A\pm s\sqrt{A^2-B}, \quad s=\pm 1.
\end{equation}
Here we recognize a symmetry: when both signs of $s$ and $\phi$ change, the solution remains the same, except for a reversed sign of the derivative $\dot{\phi}$. Is easy to check that the entire system of equations is symmetric under this transformation. If we are interested in a solution where the scale factor remains unchanged, but the scalar field changes sign and monotony (as if it was reflected about the $t$ axis), we should change the sign of $\phi(0)$ and $s$ simultaneously.\\
An equivalent procedure to numerically solve the system of equations is to give the initial condition directly on $\dot{\phi}$ and derive the condition on $\phi$ from the constraint and the scalar field equation:
\begin{equation}
\phi=\dfrac{-6ka-6a\dot{a}^2+a^3\dot{\phi}^2}{12\lambda\dot{a}\left(k+\dot{a}^2\right)\dot{\phi}}
\end{equation}
In this case we should choose $\dot{a}(0)\neq0$, $\dot{\phi}(0)\neq0$, $\dot{a}^2(0)\neq -k$. The last condition, of course, if only meaningful when $k=-1$ and reduce to the first one when $k=0$. Here we cannot distinguish between the two branches explicitly, but we can move from one to the other by changing the sign and/or value of $\dot{\phi}$.\\
In our numerical analysis we choose to use the second procedure, giving an initial condition for $\dot{\phi}$ and 
\begin{equation}\label{ic2}
\quad H(0)=1, \quad a(0)=1.
\end{equation} 
We exceptionally use $H(0)=1.5$ when $k=-1$, in order to obtain a nonsingular value for $\phi$. For each solution we will explicitly present the value of $\dot{\phi}$.\\

Numerical integration was performed using Mathematica built-in function NDSolve, with the additional option \lq \lq StiffnessSwitching". This option is very useful when dealing with singular or very steep solutions, such as the ones we encountered. \lq \lq StiffnessSwitching" uses two integration methods: \lq \lq ExplicitModifiedMidpoint" for nonstiff regions of the solution, and \lq \lq LinearlyImplicitEuler" for stiff ones.\\

In the next sections we separately analyse solutions with different spatial curvature ($k=0,+1,-1$) and with different sign of the coupling ($\lambda\gtrless 0$).\\

\subsection{Flat universe (k=0)}
\label{sec:phi2k=0}
\subsubsection*{Positive coupling, $\boldsymbol{\lambda>0}$}
\label{phi2k=0l=1}
The two branches of solutions found with positive coupling are, as expected, singular. Here we describe the first branch, while the second one is reported in the Appendix. The scale factor, shown in Figure \ref{phi2-aphik=0l=1}, starts with a linear growth and then behaves as in GR. The Gauss-Bonnet term prevents the scalar field from being singular at the initial time (Figure \ref{phi2-aphik=0l=1}). The late time behaviour of the scalar field is also GR-like.
\begin{figure}[!ht]
 \centering
 \begin{subfigure}{.47\textwidth}
 \centering
    \includegraphics[width=1\textwidth]{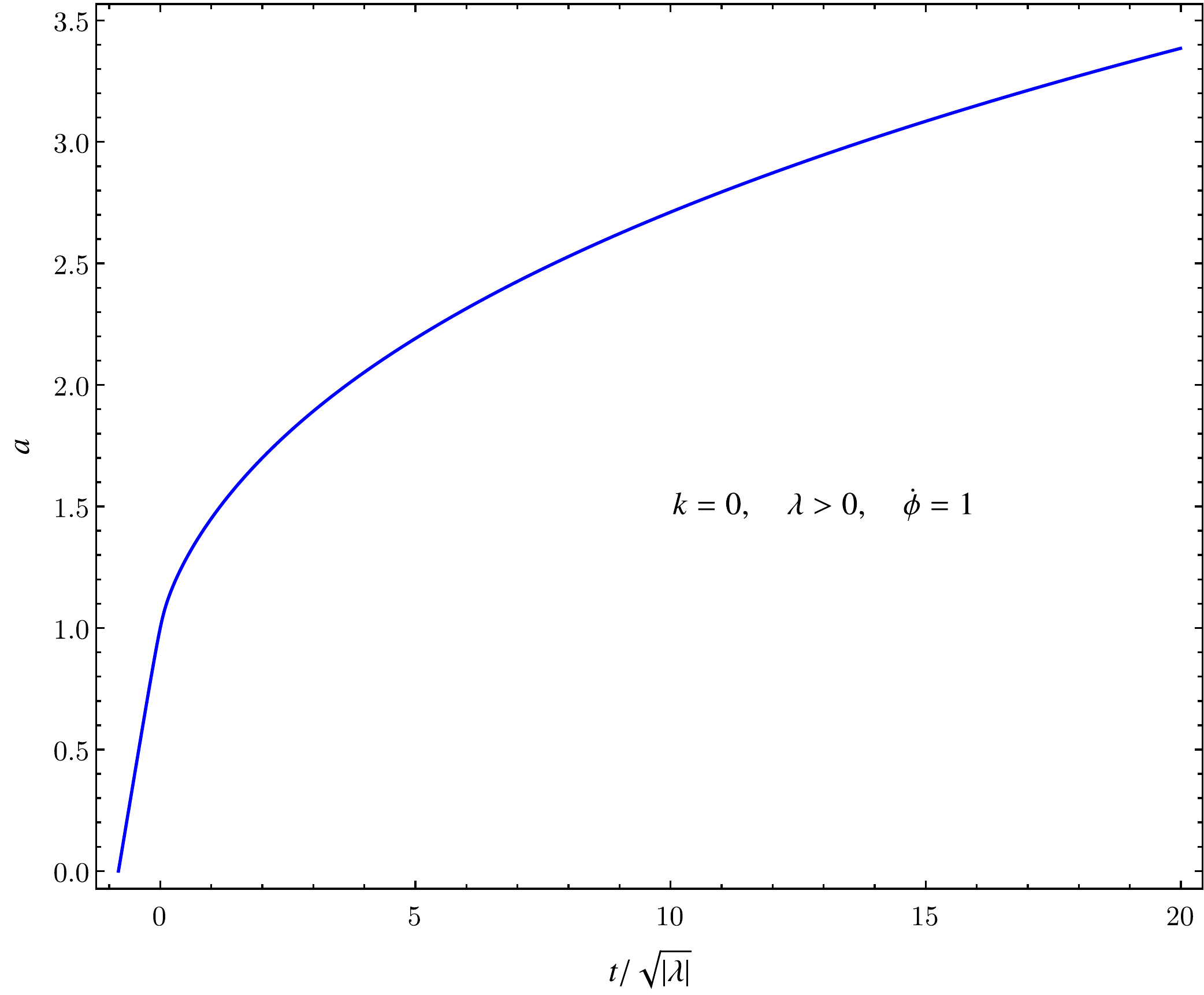}
\end{subfigure}
\begin{subfigure}{.47\textwidth}
\centering
    \includegraphics[width=1\textwidth]{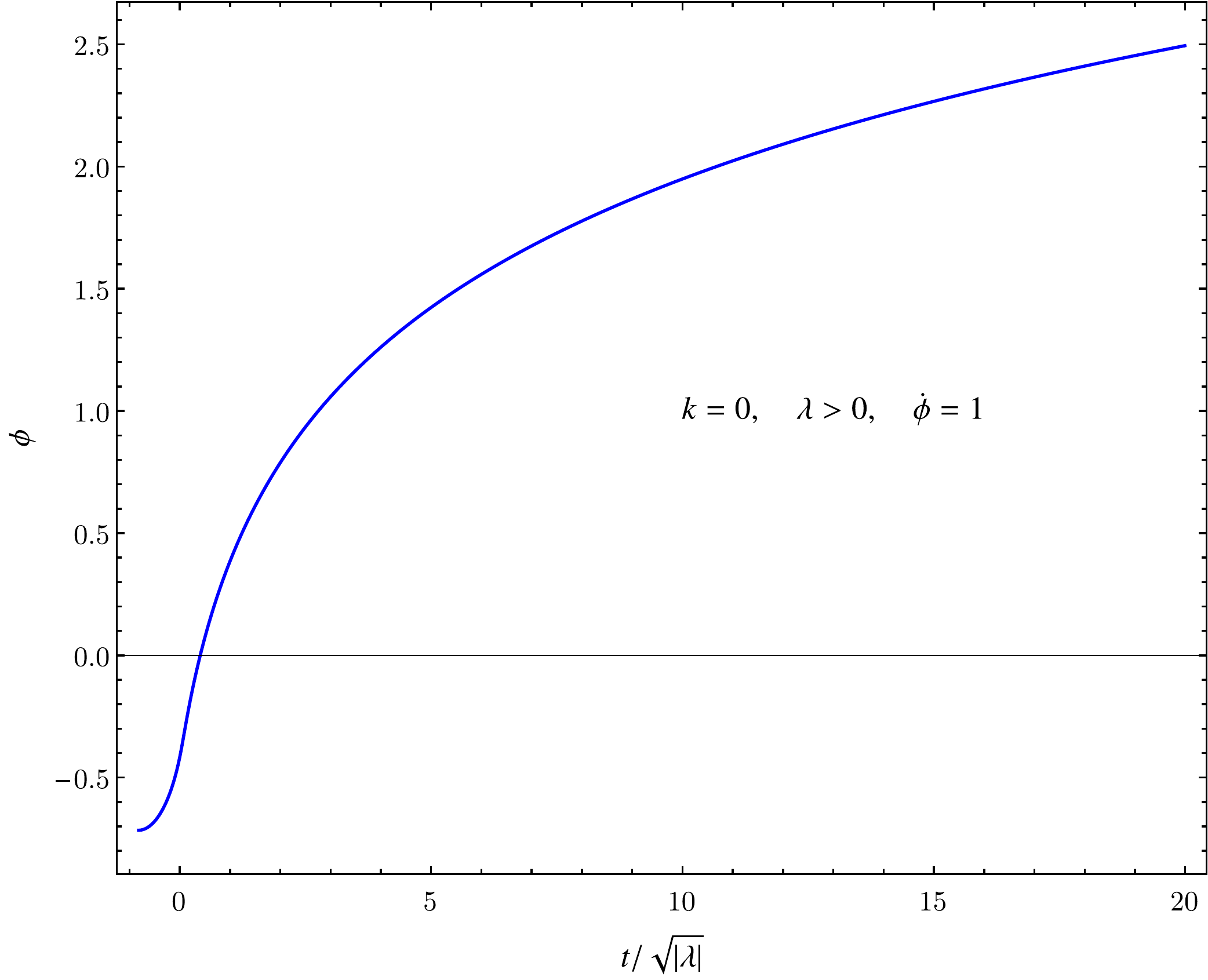}
    \end{subfigure}
    \caption{The scale factor (left) and the scalar field (right) as a function of rescaled time for a singular solution with $k=0$, $\lambda=1$, first branch.}\label{phi2-aphik=0l=1}
\end{figure}

\subsubsection*{Negative coupling, $\boldsymbol{\lambda<0}$}
\begin{figure}[!ht]
  \centering
      \includegraphics[width=0.8\textwidth]{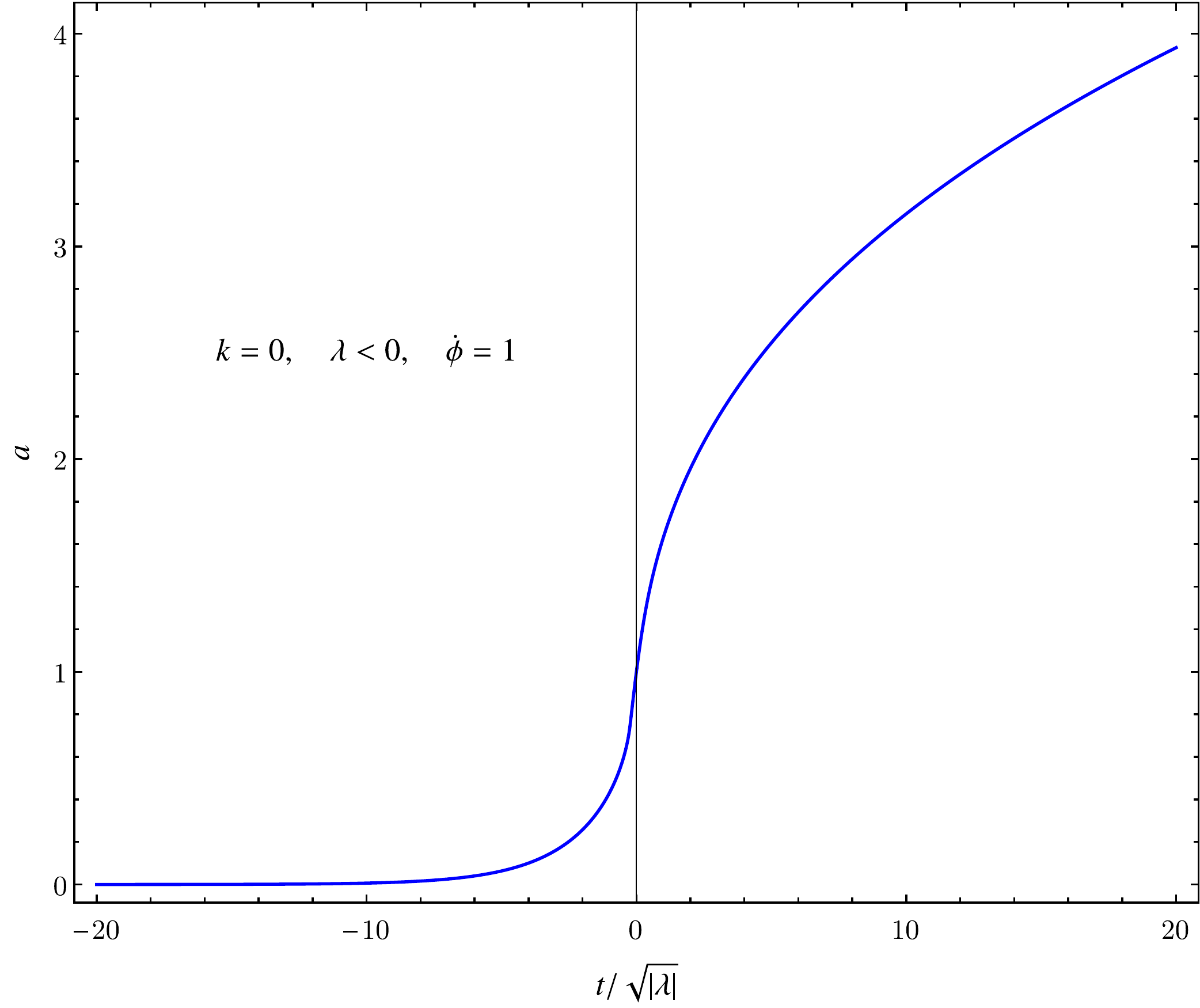}
       \caption{The scale factor as a function of rescaled time for $k=0$, $\lambda=-1$.}\label{phi2-ak=0d=-1}
\end{figure}

\begin{figure}[!ht]
 \centering
 \begin{subfigure}{.47\textwidth}
 \centering
    \includegraphics[width=1\textwidth]{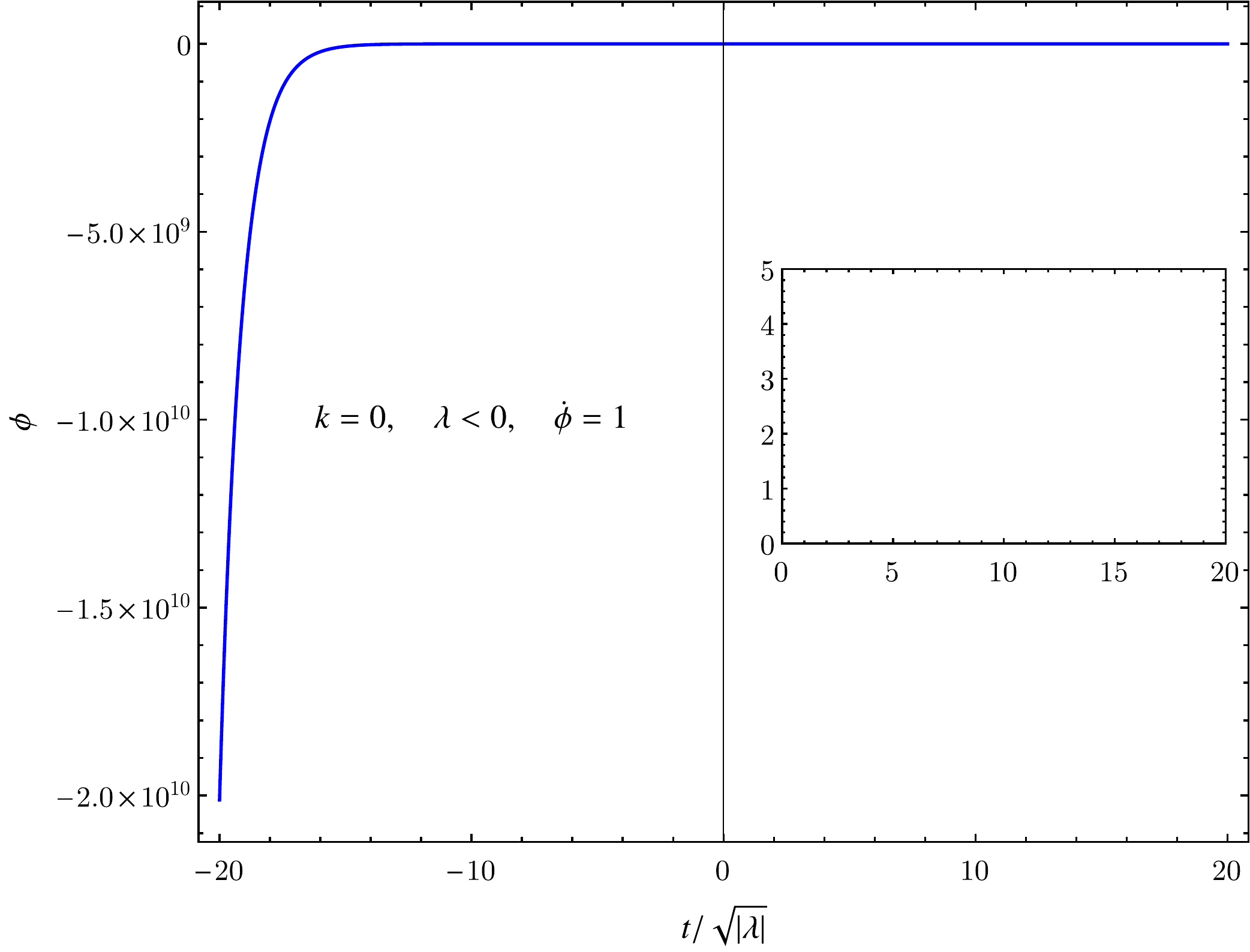}
\end{subfigure}
\begin{subfigure}{.47\textwidth}
\centering
    \includegraphics[width=1\textwidth]{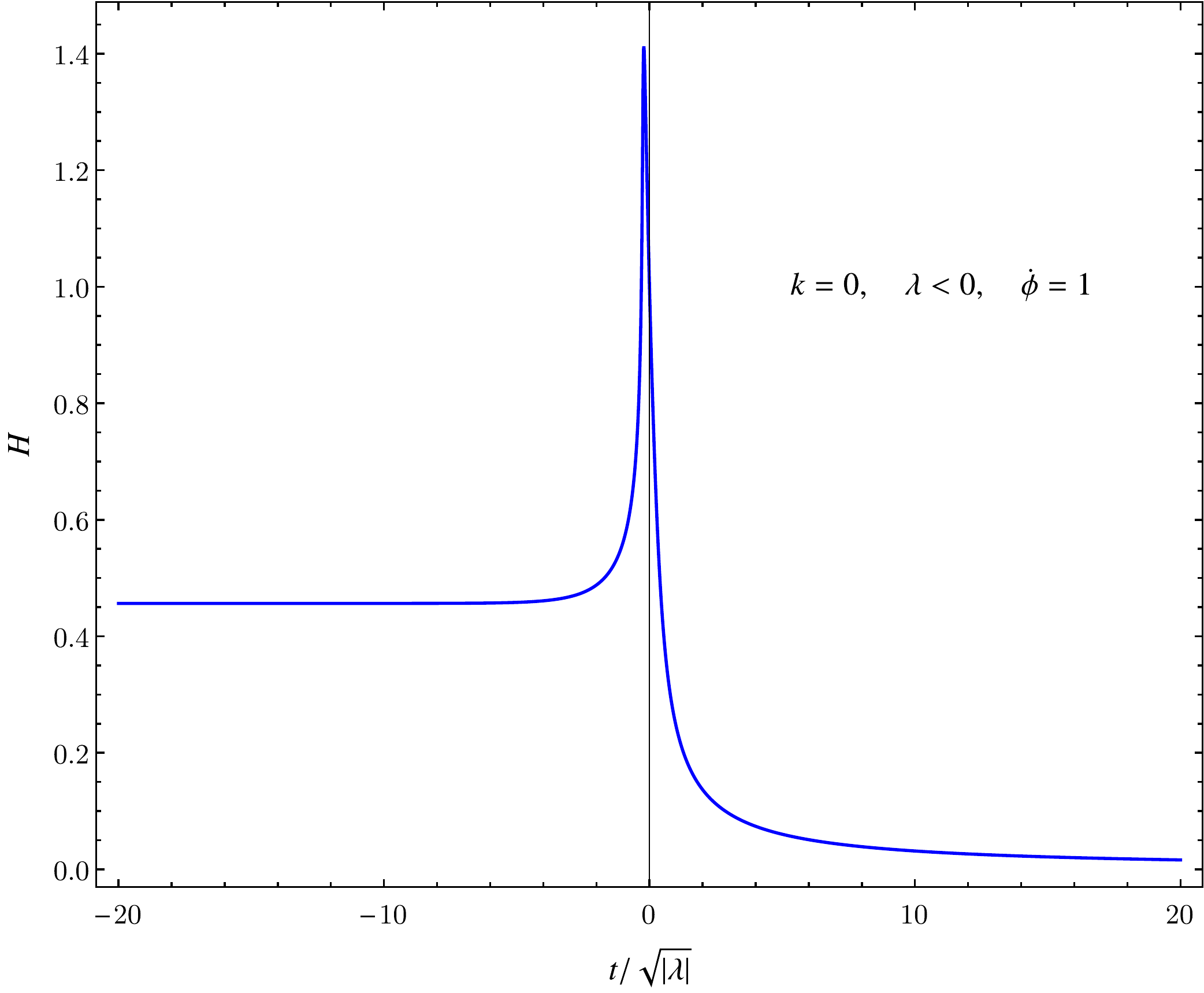}
    \end{subfigure}
    \caption{The scalar field (left) and the Hubble parameter (right) as a function of rescaled time for $k=0$, $\lambda=-1$.}\label{phi2-phiHk=0d=-1bis}
\end{figure}

When we allow the coupling to be negative, we expect to find nonsingular solutions. The negative coupling case, however, posses both singular and nonsingular solutions. We show the former in the Appendix, focusing here on the latter.\\
The scale factor (Figure \ref{phi2-ak=0d=-1}) tends exponentially to zero as time goes to negative infinity, so there is no singularity at finite cosmological time. As the space-time curvature decreases at later times, the GB correction becomes negligible and the Universe expands as in GR with an almost free scalar field. The Hubble parameter, Figure \ref{phi2-phiHk=0d=-1bis}, starts from past infinity with a finite, constant value, reaches a maximum and then tends asymptotically to zero. Clearly the Gauss-Bonnet correction provides a past  de Sitter-Universe. \\
The scalar field, shown in Figure \ref{phi2-phiHk=0d=-1bis}, grows exponentially in the past, so it does not diverge at finite times. In the future, a GR-like solution is recovered.\\
This is clearly an interesting solution, first found by Antoniadis, Rizos and Tamvakis in \cite{ARTSingfree}. Rizos and Tamviakis also proved its existence for a wider variety of coupling functions in \cite{RizosSing}. The issue of its stability will be discussed in Section \ref{sec:phi2nonsing}. In Section \ref{sec:phi2nonsing} we will also compare this numerical solution of the exact equations of the theory with the pure GB approximation.   

\clearpage
\subsection{Positive curvature (k=+1)}
Here we only discuss one of the two branches of solutions, for each sign of the coupling. The second one can be found in the Appendix, whenever its behaviour differs qualitatively from that of the first branch. 

\subsubsection*{Positive coupling, $\boldsymbol{\lambda>0}$}
\begin{figure}[!ht]
 \centering
 \begin{subfigure}{.47\textwidth}
 \centering
    \includegraphics[width=1\textwidth]{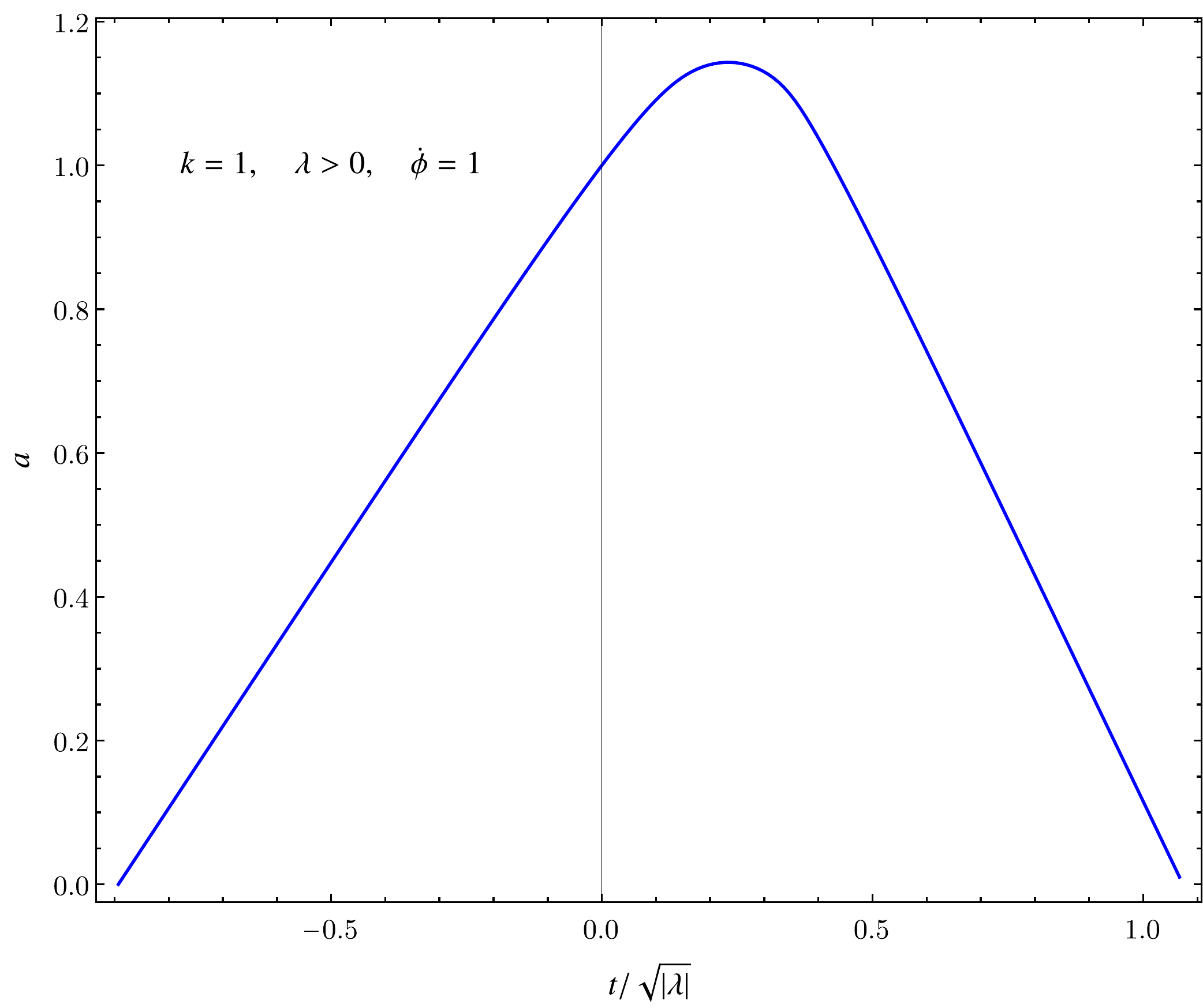}
\end{subfigure}
\begin{subfigure}{.47\textwidth}
\centering
    \includegraphics[width=1\textwidth]{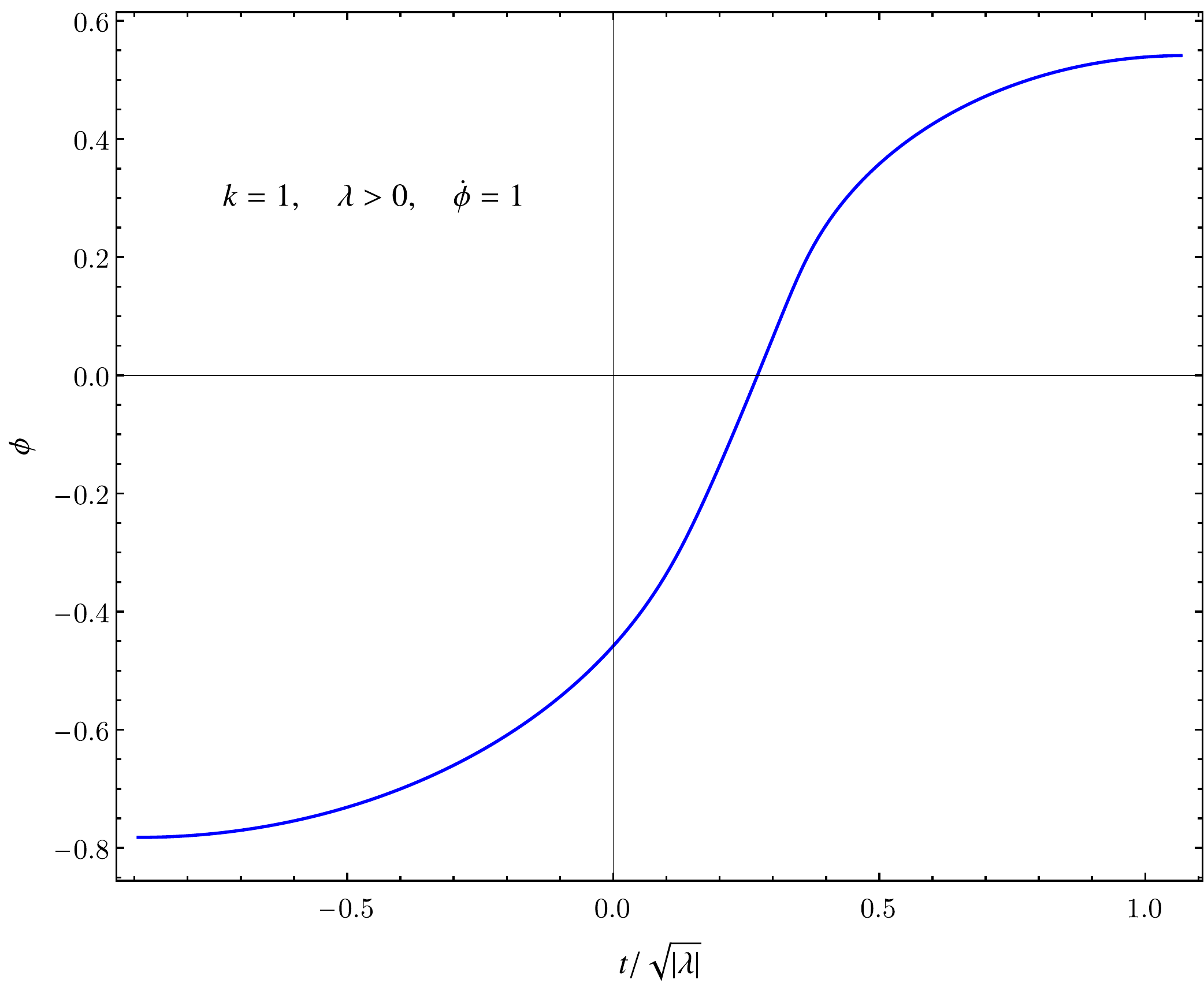}
    \end{subfigure}
        \caption{The scale factor (left) and the scalar field (right) as a function of rescaled time for a singular solution with $k=1$, $\lambda=1$, first branch.}\label{phi2-phiHk=1l=1}
\end{figure}

The behaviour of the scale factor (Figure \ref{phi2-phiHk=1l=1}) shows that the solution has a beginning and an end in time. The scale factor grows linearly from the initial singularity, reaches a maximum ($a_{max}\simeq 1.14$ in the plot) and then decreases linearly to reach the final singularity. The Gauss-Bonnet correction is important in determining the way the scale factor approaches both singularities. We also point out that the scale factor decreases faster than it grows. \\
The scalar field (Figure \ref{phi2-phiHk=1l=1}) approaches both singularities in a regular way, reaching a finite value, and is everywhere else well behaved. \\
The behaviour of the Hubble function, not reported here, is singular at the initial and final time. Both the scale factor and the scalar field behave differently from the GR solution.\\  

\subsubsection*{Negative coupling, $\boldsymbol{\lambda<0}$}
\begin{figure}[!ht]
  \centering
      \includegraphics[width=0.9\textwidth]{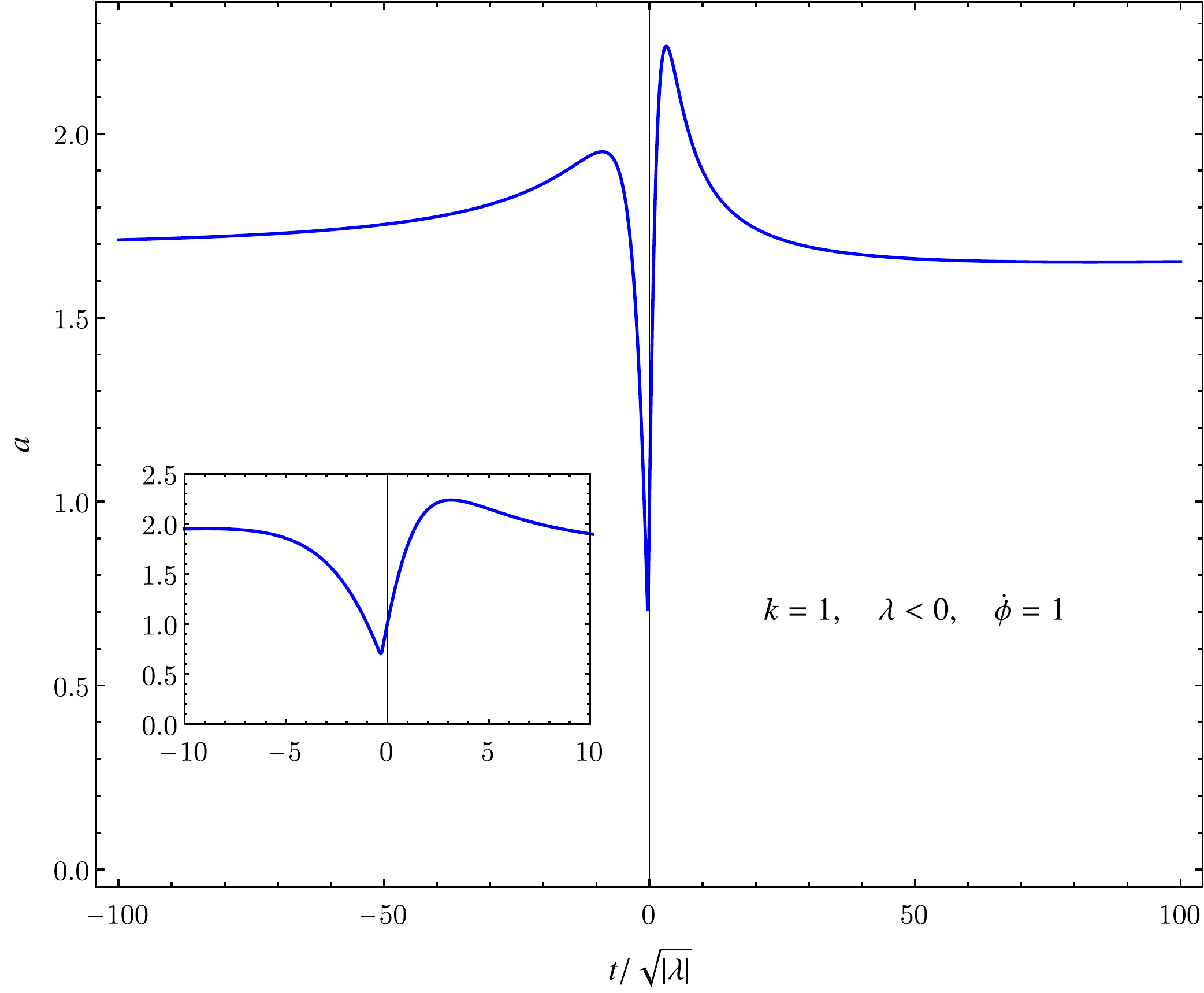}
       \caption{The scale factor as a function of rescaled time for $k=1$, $\lambda=-1$.}\label{phi2-ak=1l=-1}
\end{figure}

\begin{figure}[!ht]
 \centering
 \begin{subfigure}{.47\textwidth}
 \centering
    \includegraphics[width=1\textwidth]{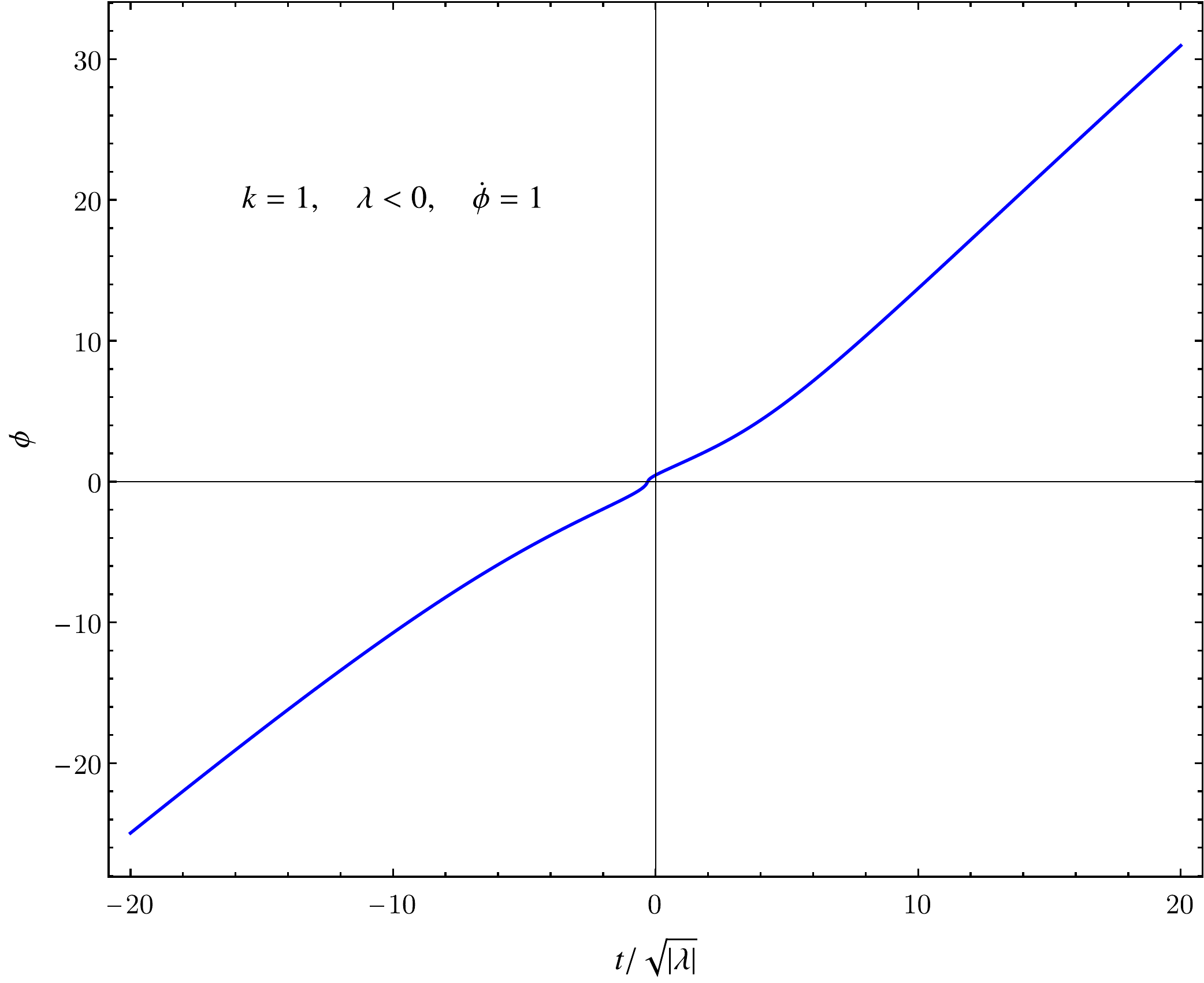}
\end{subfigure}
\begin{subfigure}{.47\textwidth}
\centering
    \includegraphics[width=1\textwidth]{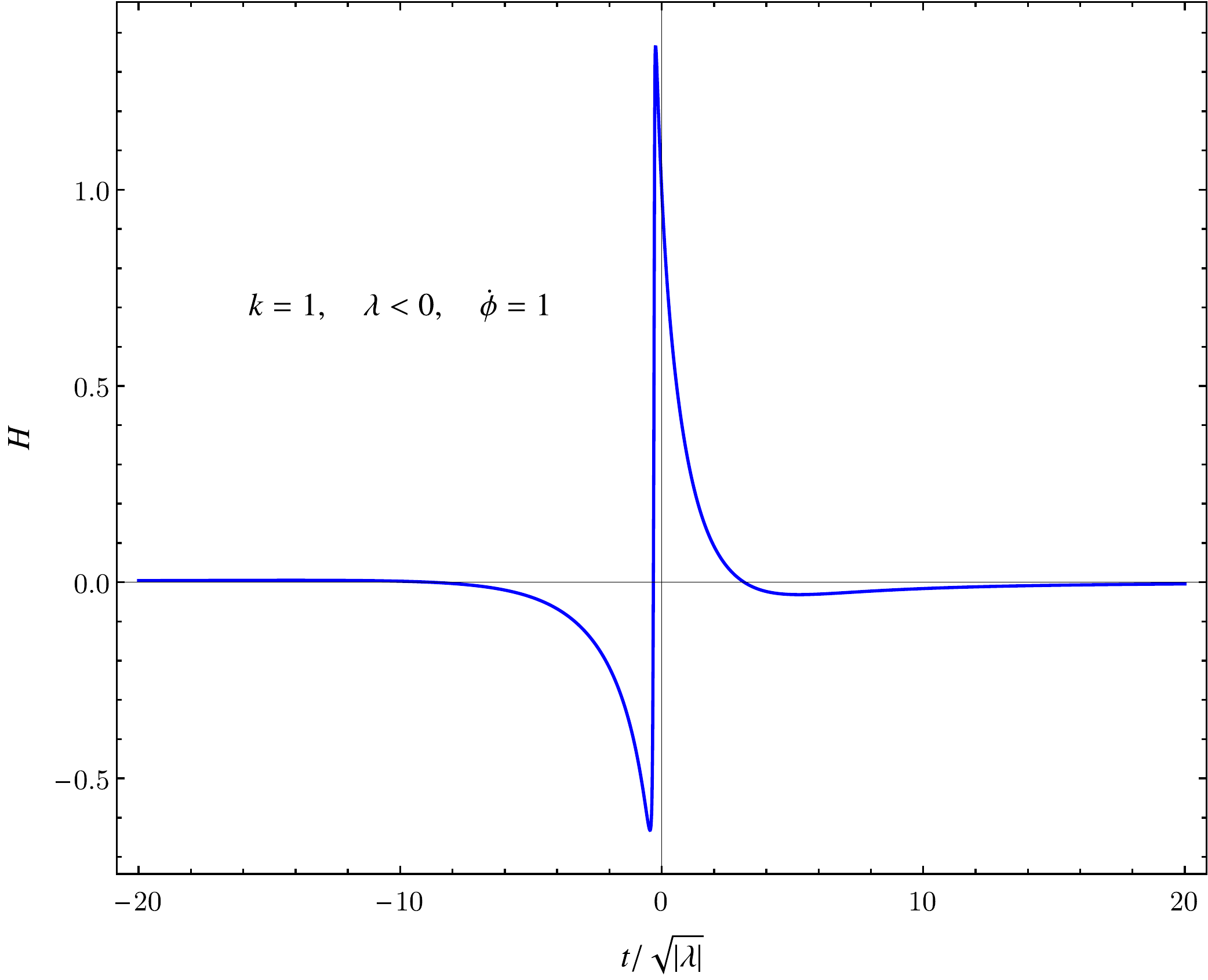}
    \end{subfigure}
    \caption{The scalar field (left) and the Hubble parameter (right) as a function of rescaled time for a nonsingular solutions with $k=1$, $\lambda=-1$.}\label{phi2-phiHk=1l=-1}
\end{figure}

As in the flat space case, by setting $\lambda<0$ we are able to find nonsingular solutions. This family of solutions was already found by Kanti and others\cite{KantiSing}, but using a different choice of variables and giving numerical solutions, as implicit functions of the scalar field $\phi$ rather than time (as illustrated in Section \ref{nonsingKanti}). \\
Comparing the nonsingular solutions found respectively with positive and zero spatial curvature, we see that the qualitative behaviour of the scale factor is rather different. The cosmological solution found here is characterized by a nonsingular \emph{bounce}. As shown in Figure \ref{phi2-ak=1l=-1} the scale factor lingers on a constant value in the past, undergoes a collapse never reaching the singular value and then grows again, stabilizing on a new constant value.\\
The scalar field grows, because of the positive value of $\dot{\phi}(0)$, undergoing a linear growth and remaining finite at finite times. Its behaviour is shown in Figure \ref{phi2-phiHk=1l=-1}\\
In Figure \ref{phi2-phiHk=1l=-1} we also display the Hubble parameter, to stress that no singularity is present in the solution under study. The Hubble parameter is practically zero everywhere except around the bouncing time $t_{bounce}$.  

One can easily guess the asymptotic behaviour of the scale factor and scalar field by looking at Figures \ref{phi2-ak=1l=-1}, \ref{phi2-phiHk=1l=-1}. When taking the limit $t\rightarrow\pm \infty$, under the ansatz
\begin{equation}
a(t)\simeq a_{\infty}, \quad \quad \phi(t)\simeq \phi_\infty +\phi_1 t
\end{equation}
the system of equations \eqref{eqphi2} gives:
\begin{equation}
\begin{cases}
\dfrac{3}{a_{\inf}^2}=\dfrac{\phi_1^2}{2}\\
2+\left(4\lambda+a_{inf}^2\right)\phi_1^2=0\\
\end{cases}
\end{equation}
The solution of the system above is $a_{\infty}=\sqrt{-3\lambda}$, $\phi_1=\sqrt{\dfrac{2}{-\lambda}}$, so that the asympototic behaviour for the closed, nonsingular solution is:
\begin{equation}
a(t)\simeq \sqrt{-3\lambda}, \quad \quad \phi(t)\simeq \phi_\infty +\sqrt{\dfrac{2}{-\lambda}} t.
\end{equation} 
The solution described above is extremely interesting in the context of eternal universe theories, and should be the subject of further studies.

\subsection{Negative curvature (k=-1)}
In the present work, we do not discuss the negative curvature case in any detail. The solutions for this case can be found in \cite{KantiSing}.

\clearpage
\section{Detailed analysis of interesting solutions}
\subsection{Stability of nonsingular solutions}
\label{sec:Inflaphi2}
\begin{figure}[!ht]
  \centering
      \includegraphics[width=0.7\textwidth]{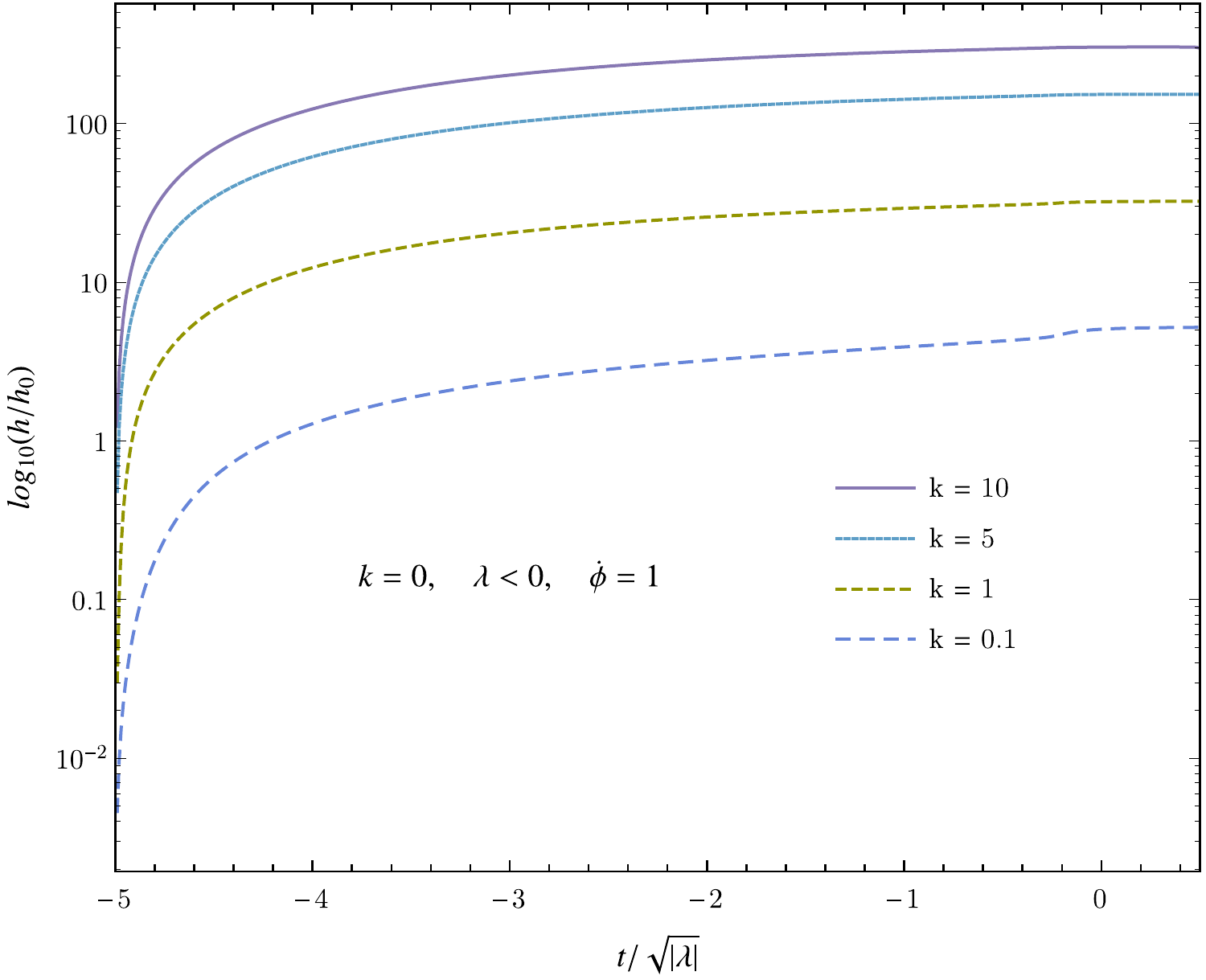}
       \caption{The tensor perturbation as a function of rescaled time for a nonsingular solution with $k=0$, $\lambda=-1$, for different wave numbers.}
       \label{phi2-hk=0l=-1}
\end{figure}

The instability of nonsingular solutions in Gauss-Bonnet gravity with a quadratic coupling and flat background was proved by Kawai, Sakagami and Soda in \cite{KSSPerturbations,KSSNovelInstab,KSSInstability}.\\
Here we present the numerical results of our stability analysis, performed for nonsingular solution arising when $k=0$ and $\lambda<0$. The behaviour of the square sound velocity and the function $\alpha$ at early times is shown in Figure \ref{phi2-csalphak=0d=-1}. We see that, at the time when the solution approaches and avoids the singularity, the squared sound velocity becomes negative and stays negative in the infinite past. Ghosts are not present, as $\alpha>0$ for $\forall t$, but tensor perturbations are unstable and make such an appealing solution unacceptable.\\
An example of the evolution of tensor perturbations is shown in Figure \ref{phi2-hk=0l=-1} for different wave numbers, in logarithmic scale. This evolution is obtained by numerically solving Eq. \eqref{eqperturb}  along the nonsingular background space-time. Instability mainly involves short wavelength modes, as $c_s^2$ possesses a negative lower bound. However, background solution could avoid breaking down only if a natural cut off in the co-moving wave number exists, and suitable initial conditions at an arbitrary time in the past are imposed. \\  

\begin{figure}[!ht]
 \centering
 \begin{subfigure}{.47\textwidth}
 \centering
    \includegraphics[width=1\textwidth]{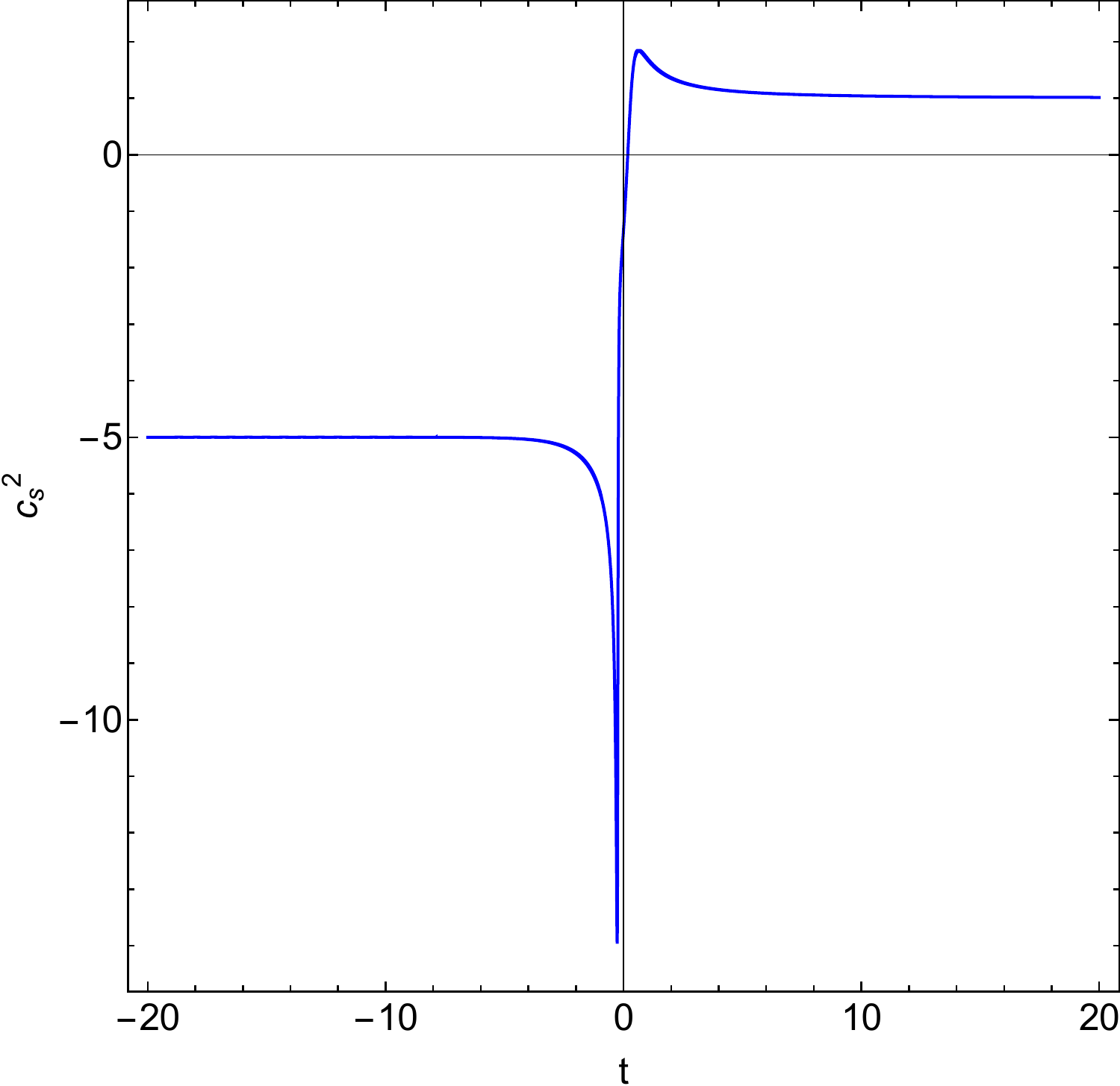}
\end{subfigure}
\begin{subfigure}{.47\textwidth}
\centering
    \includegraphics[width=1\textwidth]{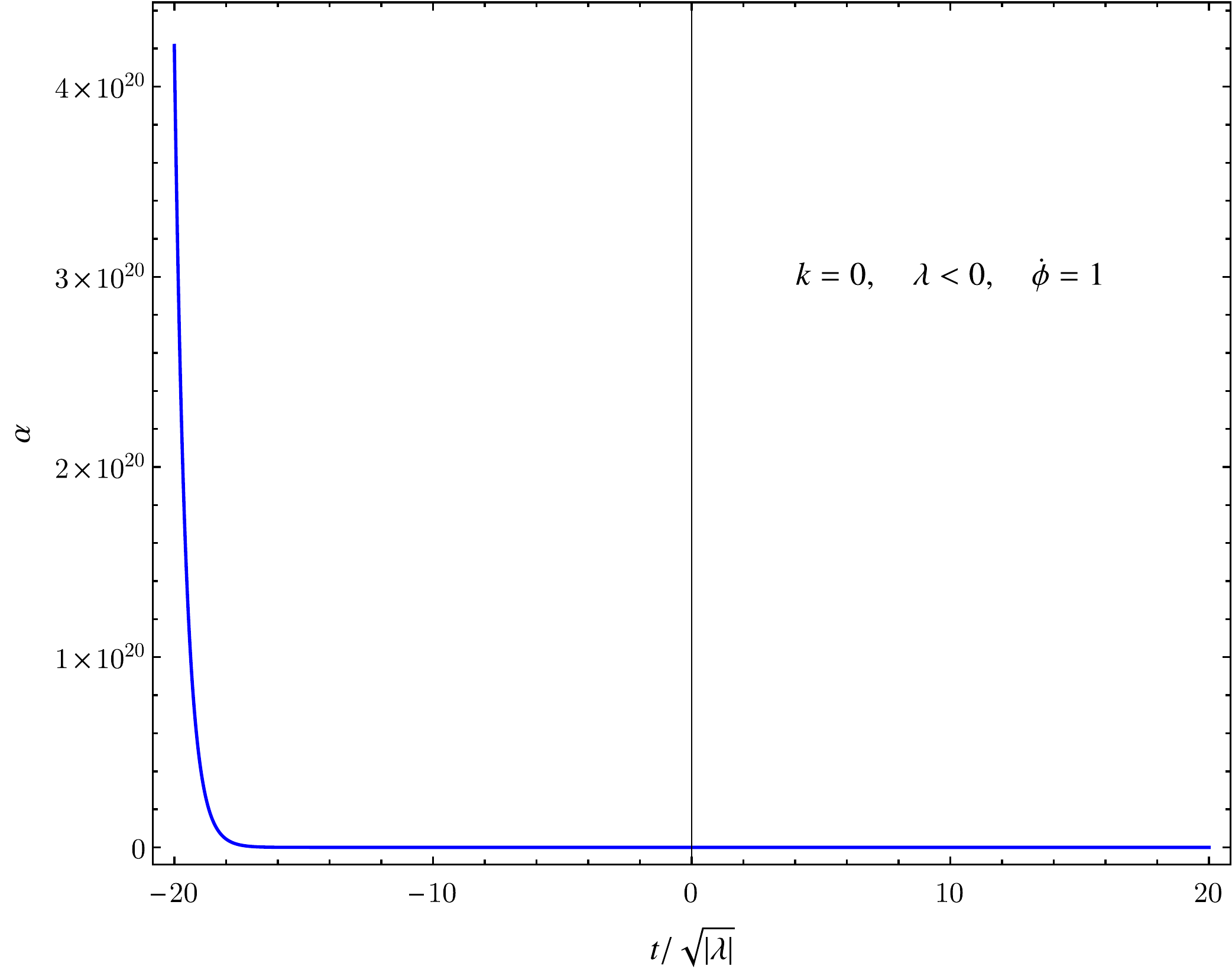}
    \end{subfigure}
    \caption{The squared sound velocity (left) and the ghost factor (right) as a function of rescaled time for the nonsingular solution with $k=0$, $\lambda=-1$.}
\end{figure}\label{phi2-csalphak=0d=-1}

\clearpage
\subsection{Comparison with pure Gauss-Bonnet solutions}
\label{sec:phi2nonsing}
In Section \ref{sec:earlytimesol} we discussed the approximation proposed in Ref.s \cite{KantiEarlyTime, KantiGBInflation}. In the approximated theory, i.e. pure Gauss-Bonnet theory, the terms in Eq.s \eqref{eqphi2} deriving from the scalar curvature $R$ are discarded. Complete and pure Gauss-Bonnet theories are consistent if the following hypothesis are satisfied:
\begin{enumerate}
\item $\dfrac{R}{2}\ll \dfrac{f(\phi)}{8}R^2_{GB} \quad \longrightarrow \quad (\dfrac{f(\phi)}{4}R^2_{GB})/R \gg 1 \quad $  (Condition 1);
\item $H\dot{f}\gg 1 \quad$ (Condition 2);
\item $\ddot{f}\gg 1 \quad $  (Condition 3).
\end{enumerate}

\begin{figure}[!ht]
  \centering
      \includegraphics[width=0.9\textwidth]{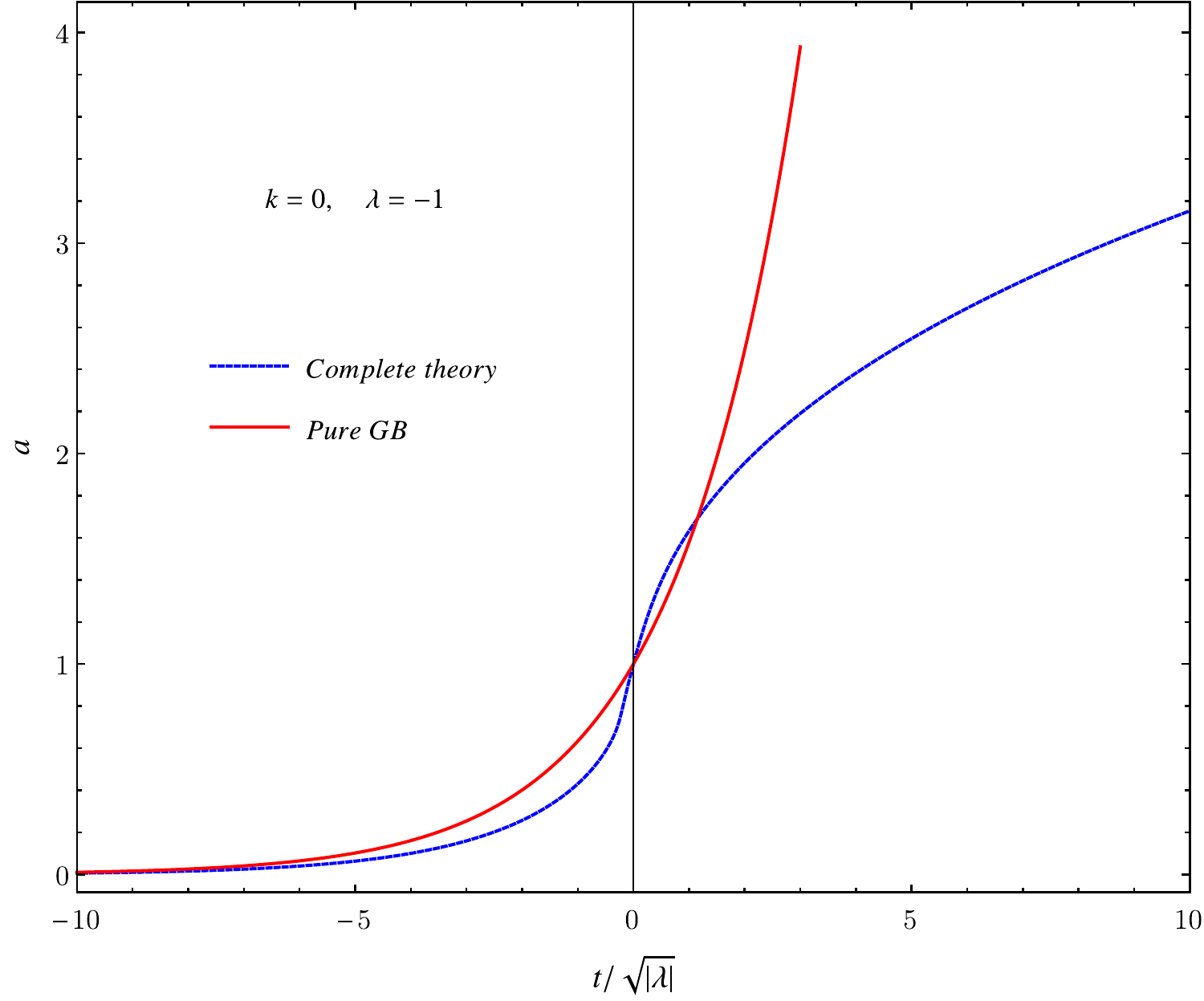}
       \caption{The scale factor as a function of rescaled time for $k=0$, $\lambda=-1$ in pure and complete Gauss-Bonnet theory.}
       \label{phi2CONFR-ak=0l=-1}
\end{figure}

\begin{figure}[!ht]
  \centering
      \includegraphics[width=0.9\textwidth]{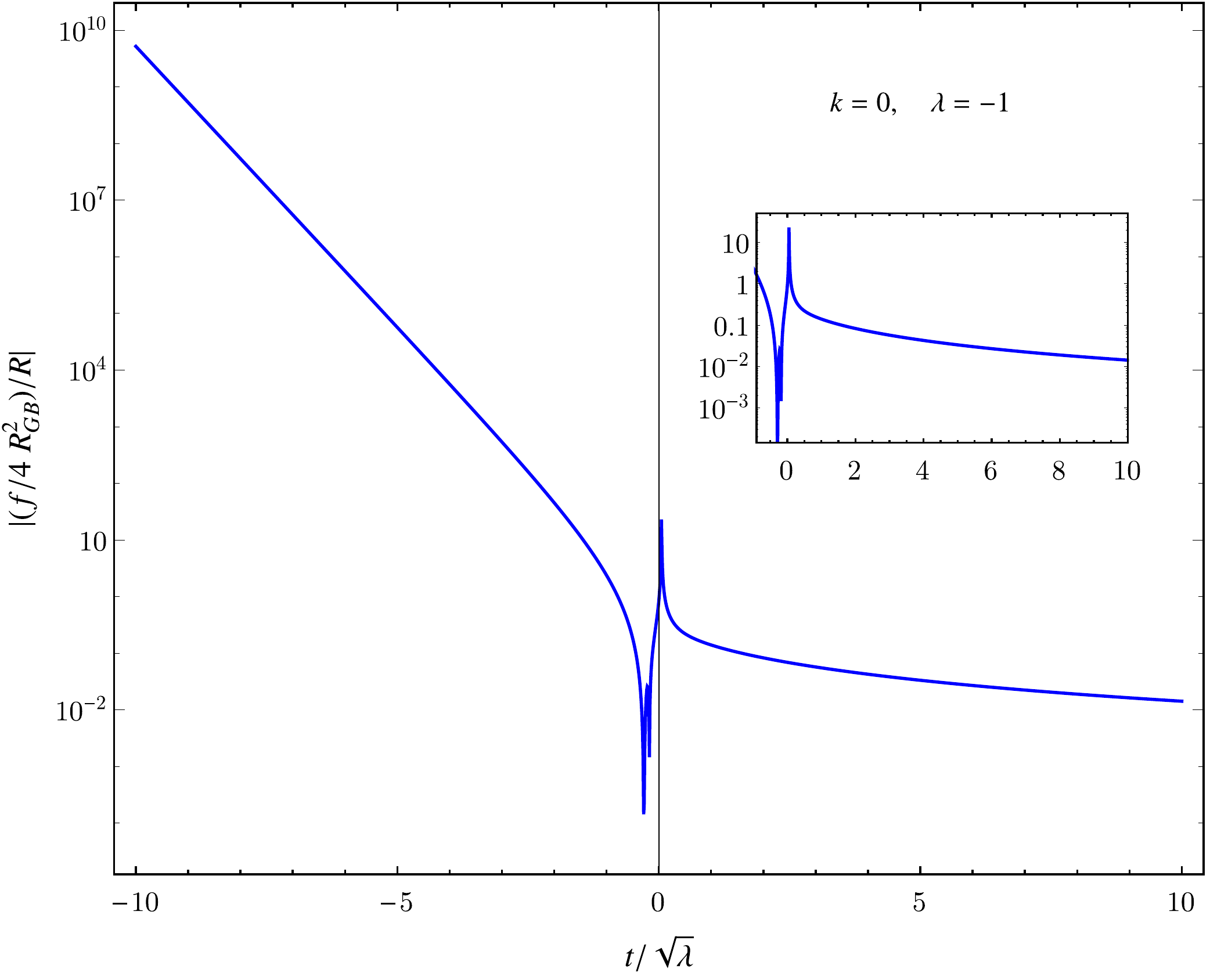}
       \caption{The ratio of Gauss-Bonnet and Ricci terms of the action as a function of rescaled time in the complete theory with $k=0$, $\lambda=-1$.}
       \label{phi2-RGBRk=0l=-1}
\end{figure}

\begin{figure}[!ht]
 \centering
 \begin{subfigure}{.46\textwidth}
 \centering
    \includegraphics[width=1\textwidth]{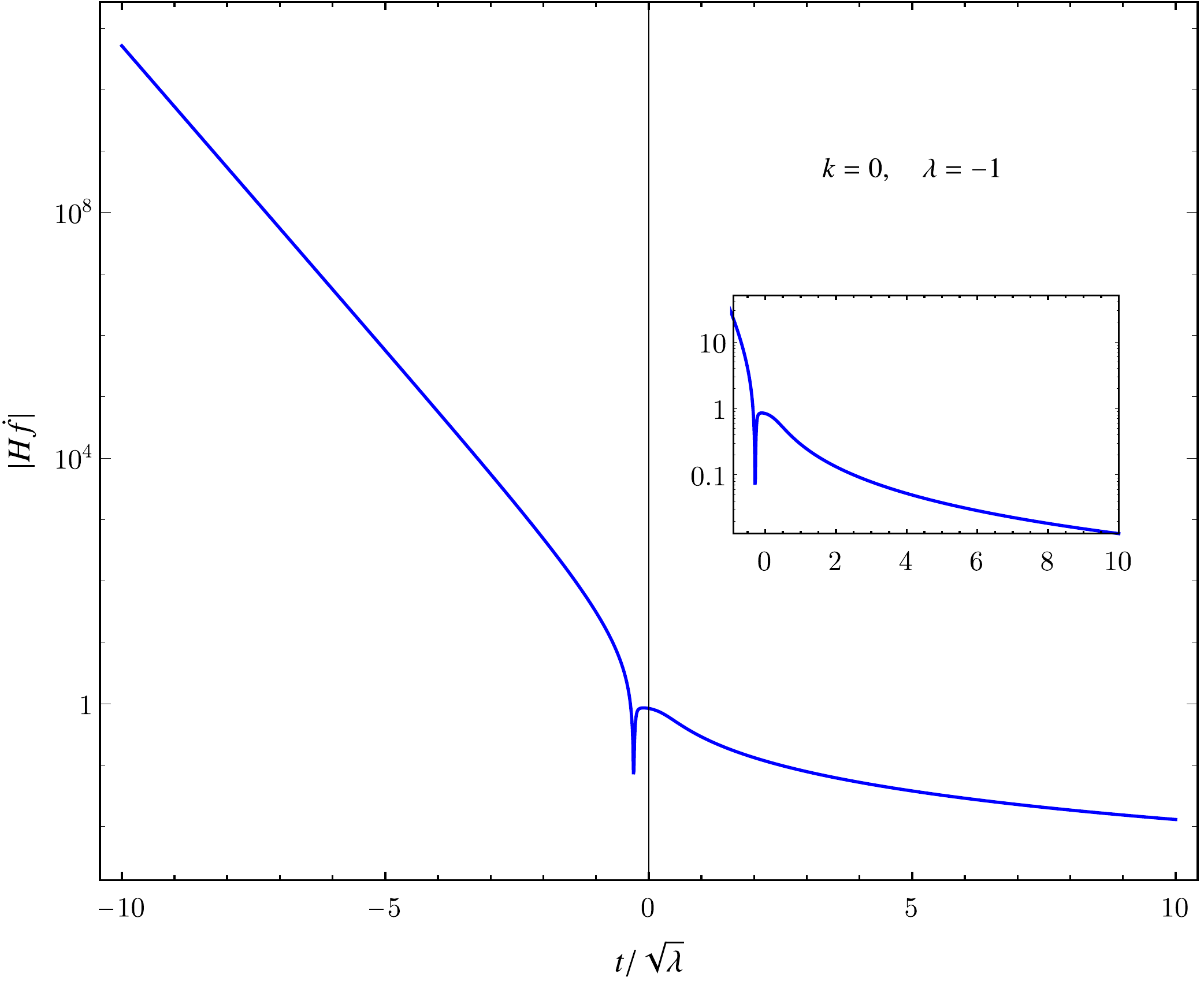}
    \caption{The $H \dot{f}$ term as a function of rescaled time.}\label{phi2-Hf1k=0l=-1}
\end{subfigure}
\begin{subfigure}{.46\textwidth}
\centering
    \includegraphics[width=1\textwidth]{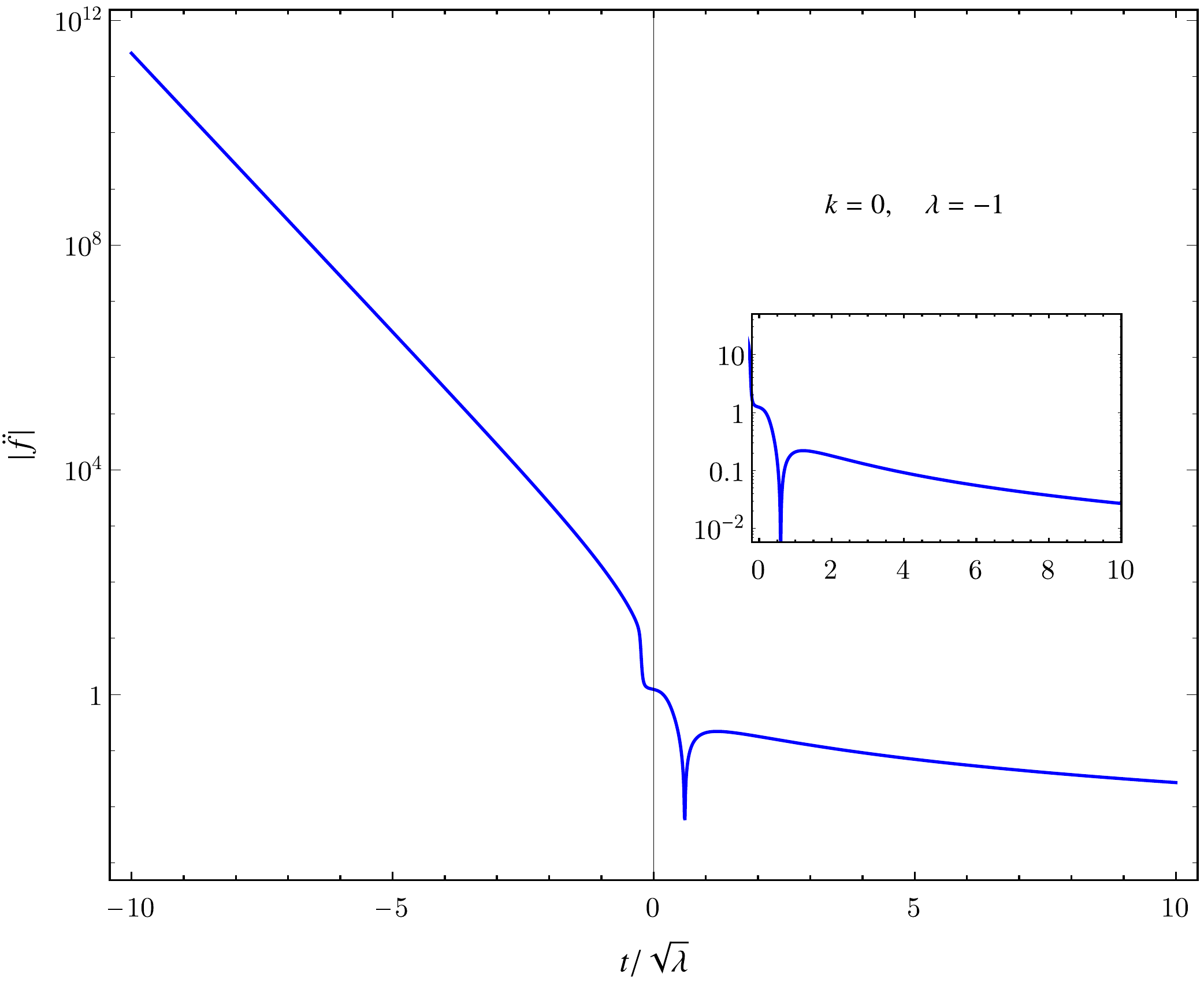}
    \caption{The $\ddot{f}$ term as a function of rescaled time.}\label{phi2-f2k=0l=-1}
    \end{subfigure}
    \caption{Solution of the complete theory with $k=0$, $\lambda=-1$.}
\end{figure}

To check whether these conditions hold, we compare the numerical solutions of the complete theory with the analytical solutions of pure GB. \\
We first consider the non singular solutions arising in flat space-time with a negative coupling, $\lambda=-1$. For pure Gauss-Bonnet theory we have three family of nonsingular solutions (solutions A, B and C in Section \ref{sec:earlytimesol}), all sharing the same early time behaviour. We can thus pick the de Sitter solution (A) as a model. In Figure \ref{phi2CONFR-ak=0l=-1} we show both solutions on the same plot. We see that scale factor in the two solutions has the same asymptotical behaviour as time goes to $- \infty$, begins to differ at early times ($t/ \sqrt{\vert \lambda \vert} \simeq 0$) and then diverges completely at late times. \\
This is what we could have expected looking at the three conditions, calculated along the complete theory's solution and shown in Figure \ref{phi2-RGBRk=0l=-1}, \ref{phi2-Hf1k=0l=-1} and \ref{phi2-f2k=0l=-1}. Condition 1, 2 and 3 are all satisfied at very early times, but are violated at $t \simeq 0$ and later. So, the pure Gauss-Bonnet solution considered here is a good approximation of the complete one only for $t \ll -\sqrt{\vert \lambda \vert}$.\\ 


\begin{figure}[!ht]
  \centering
      \includegraphics[width=0.9\textwidth]{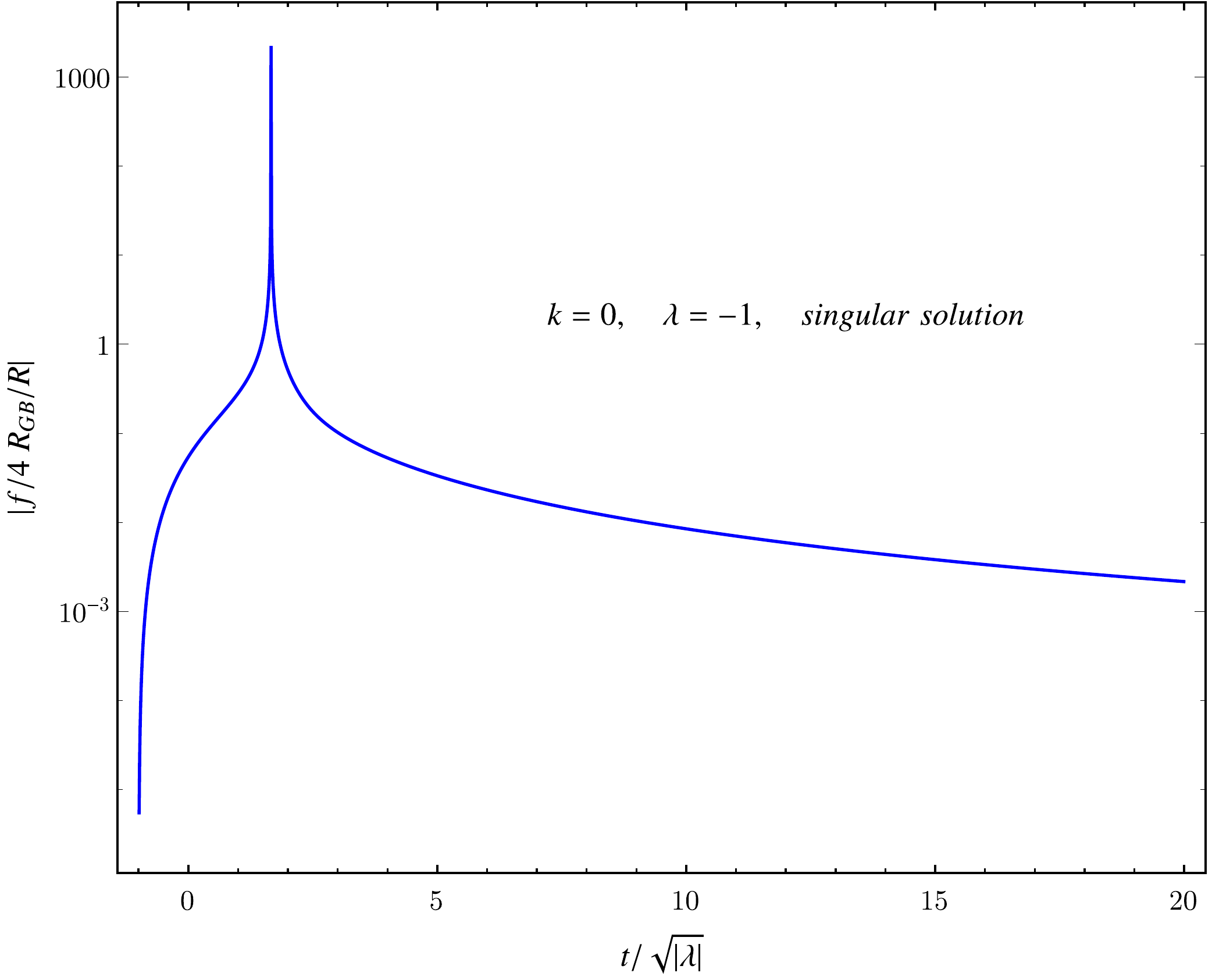}
       \caption{The ratio of Gauss-Bonnet and Ricci terms of the action as a function of rescaled time in the complete theory with $k=0$, $\lambda=-1$, for the singular solution.}\label{phi2-RGBRk=0l=-1bis}
\end{figure}

\begin{figure}[!ht]
 \centering
 \begin{subfigure}{.46\textwidth}
 \centering
    \includegraphics[width=1\textwidth]{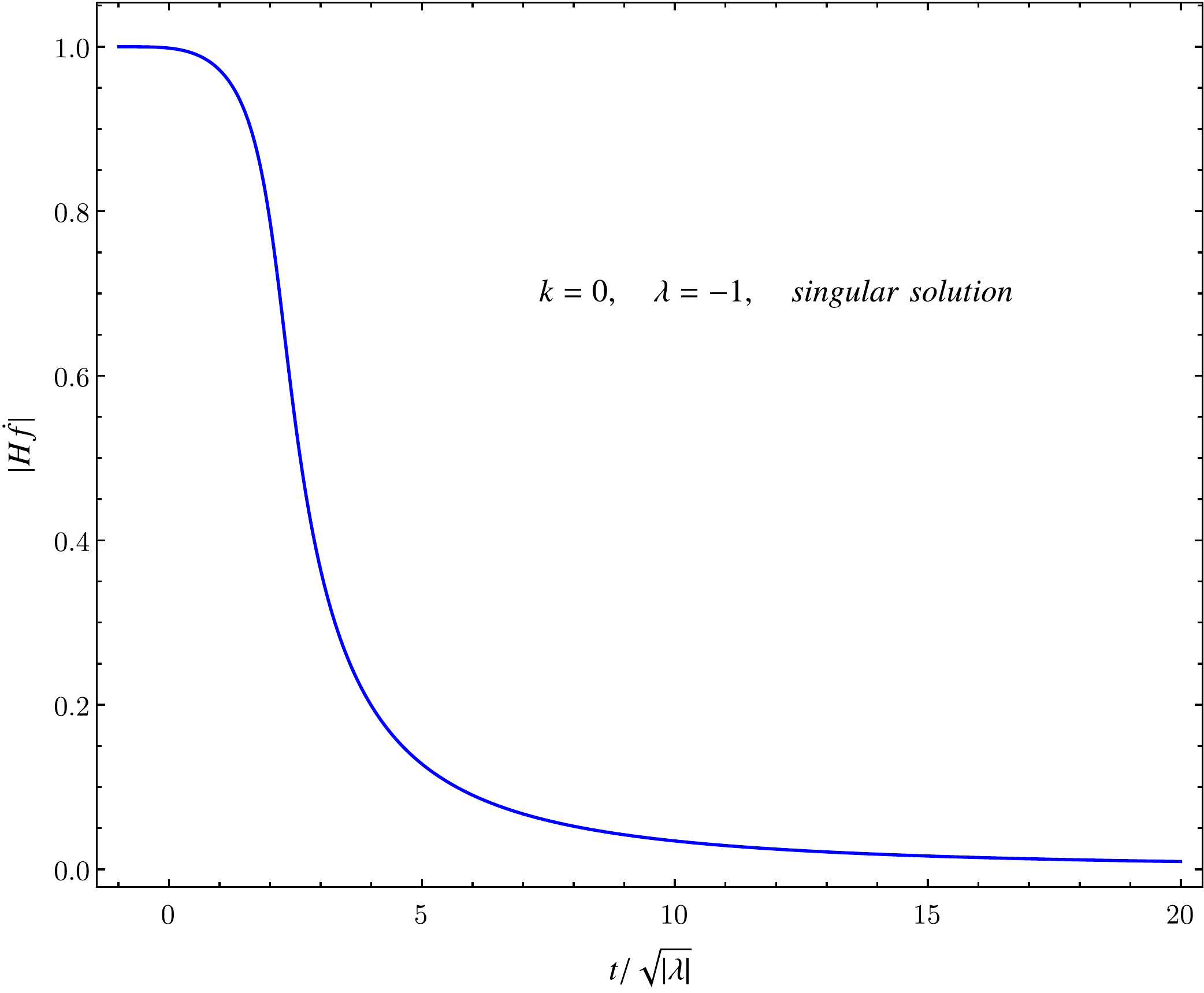}
    \caption{The $H \dot{f}$ term as a function of rescaled time.}\label{phi2-Hf1k=0l=-1bis}
\end{subfigure}
\begin{subfigure}{.46\textwidth}
\centering
    \includegraphics[width=1\textwidth]{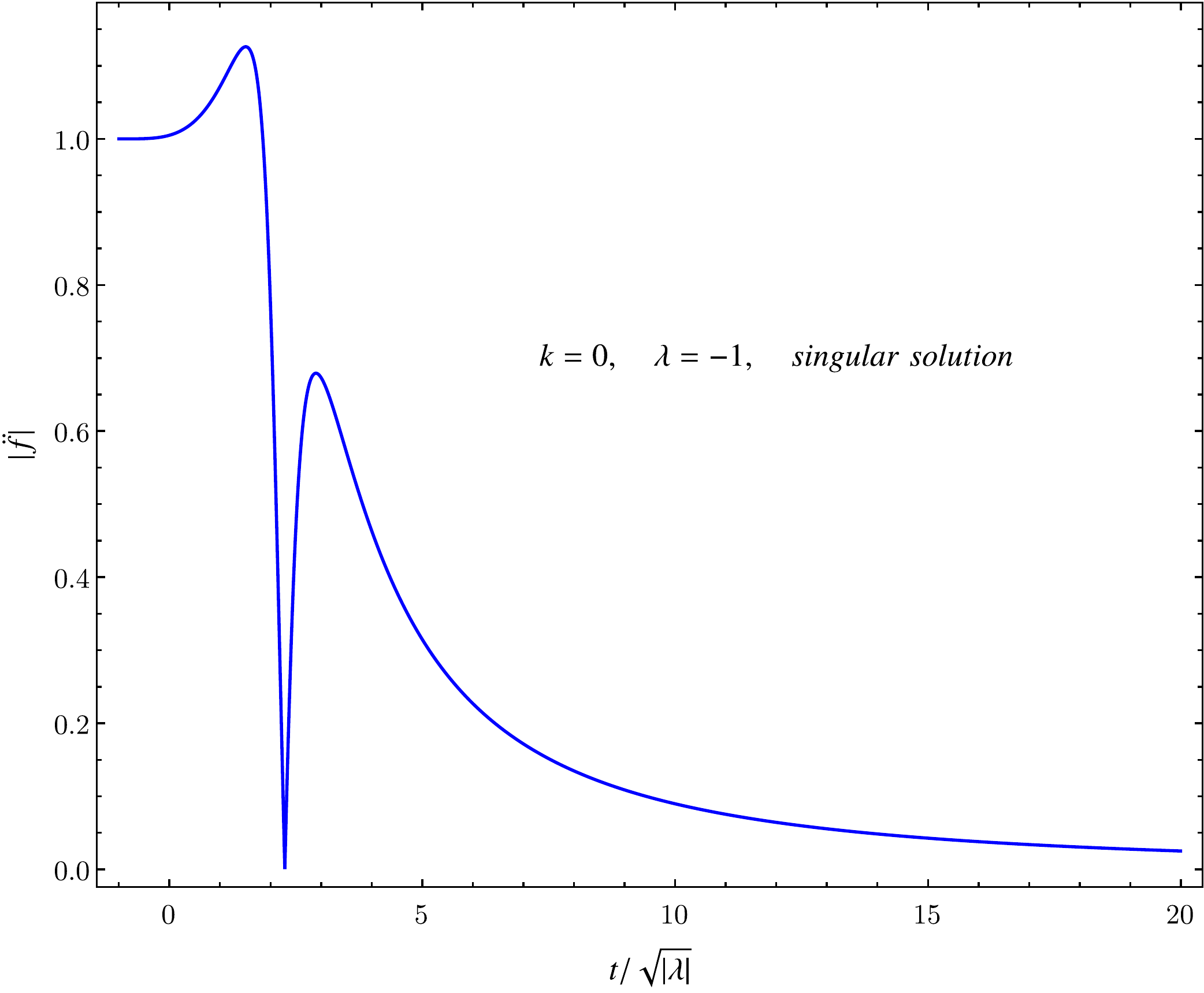}
    \caption{The $\ddot{f}$ term as a function of rescaled time.}\label{phi2-f2k=0l=-1bis}
    \end{subfigure}
    \caption{Singular branch solution of the complete GB theory with $k=0$, $\lambda=-1$.}
\end{figure}

Complete Gauss-Bonnet theory with negative coupling also shows the existence of singular solutions, which are not found in the approximate, early-time theory. We need to check, than, that singular solutions do not satisfy the requirements for pure Gauss-Bonnet theory. Figures \ref{phi2-Hf1k=0l=-1bis}, \ref{phi2-f2k=0l=-1bis} both make clear that Conditions 2 and 3 are not fulfilled near the singularity, as $H\dot{f}$ and $\ddot{f}$ are of order one or smaller. The behaviour of the ratio $fR_{GB}^2/R$ (Figure \ref{phi2-RGBRk=0l=-1bis}) further support our statement: early-time approximation cannot be applied to singular solutions.\\

\begin{figure}[!ht]
  \centering
      \includegraphics[width=0.8\textwidth]{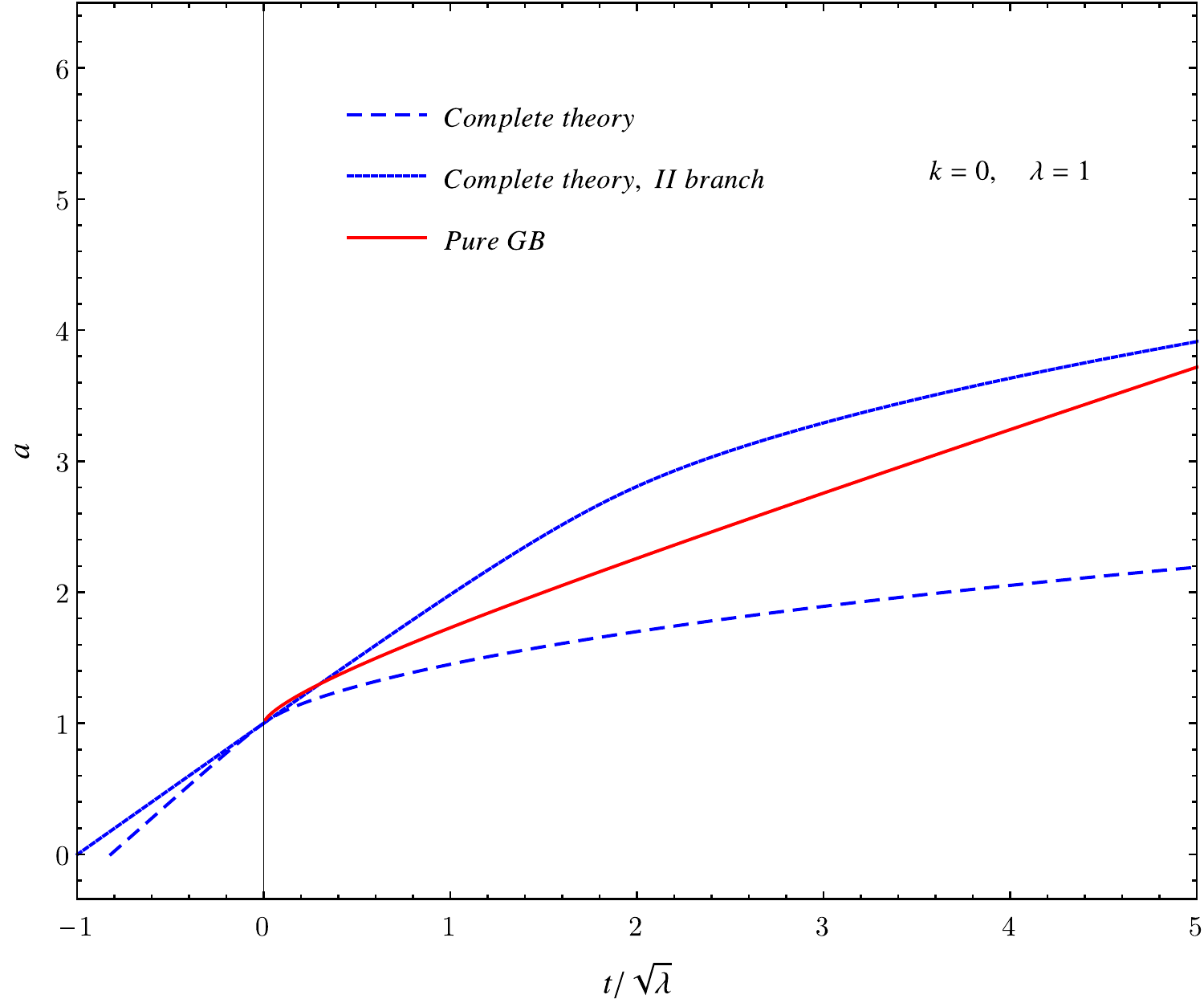}
       \caption{The scale factor as a function of rescaled time for $k=0$, $\lambda=1$ in pure and complete Gauss-Bonnet theory.}
       \label{phi2CONFR-ak=0l=1}
\end{figure}

\begin{figure}[!ht]
  \centering
      \includegraphics[width=0.8\textwidth]{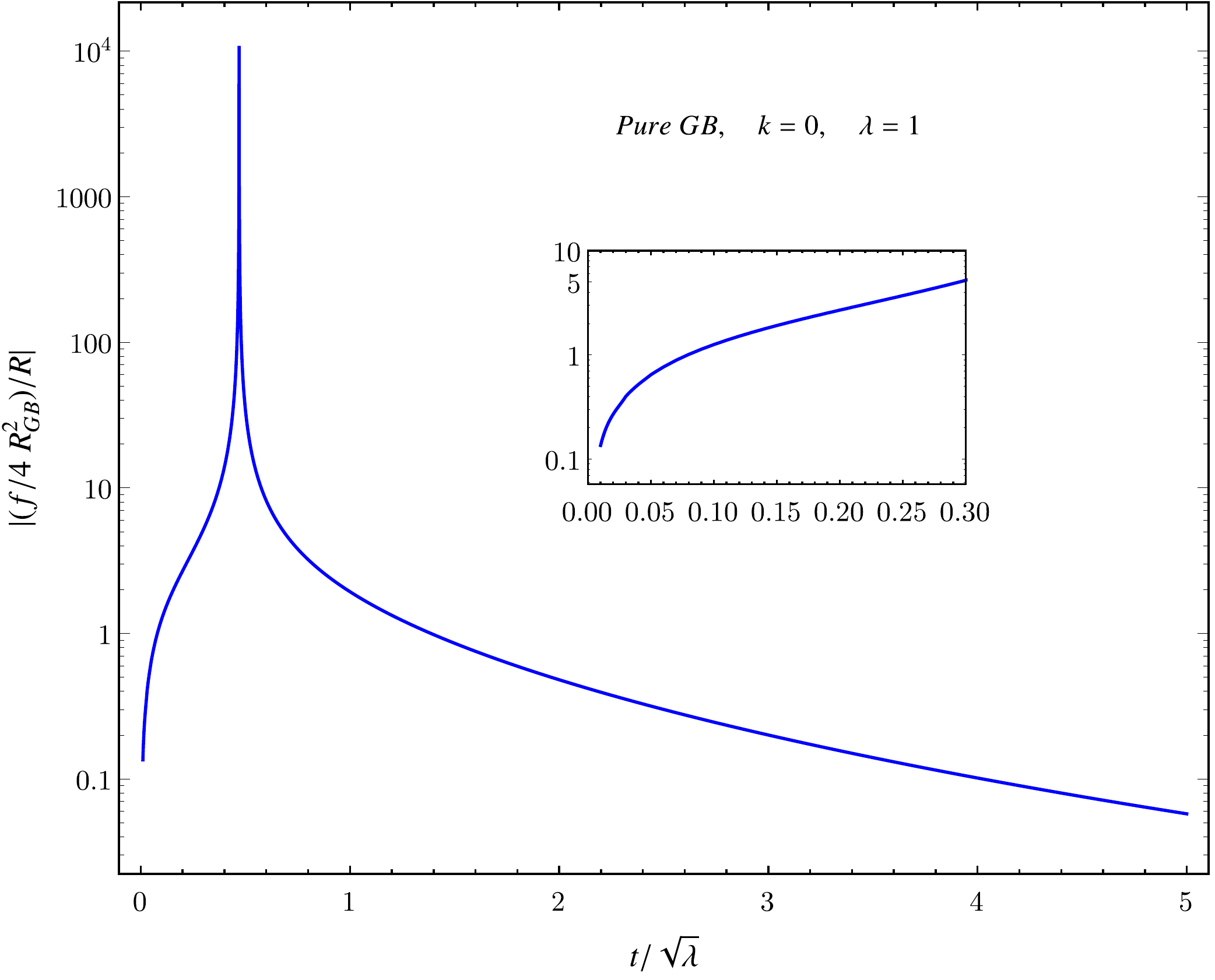}
       \caption{The ratio of Gauss-Bonnet and Ricci terms of the action as a function of rescaled time, along pure Gauss-Bonnet solution with $k=0$, $\lambda=1$.}\label{phi2NOR-RGBRk=0l=-1}
\end{figure}

\begin{figure}[!ht]
 \centering
 \begin{subfigure}{.46\textwidth}
 \centering
    \includegraphics[width=1\textwidth]{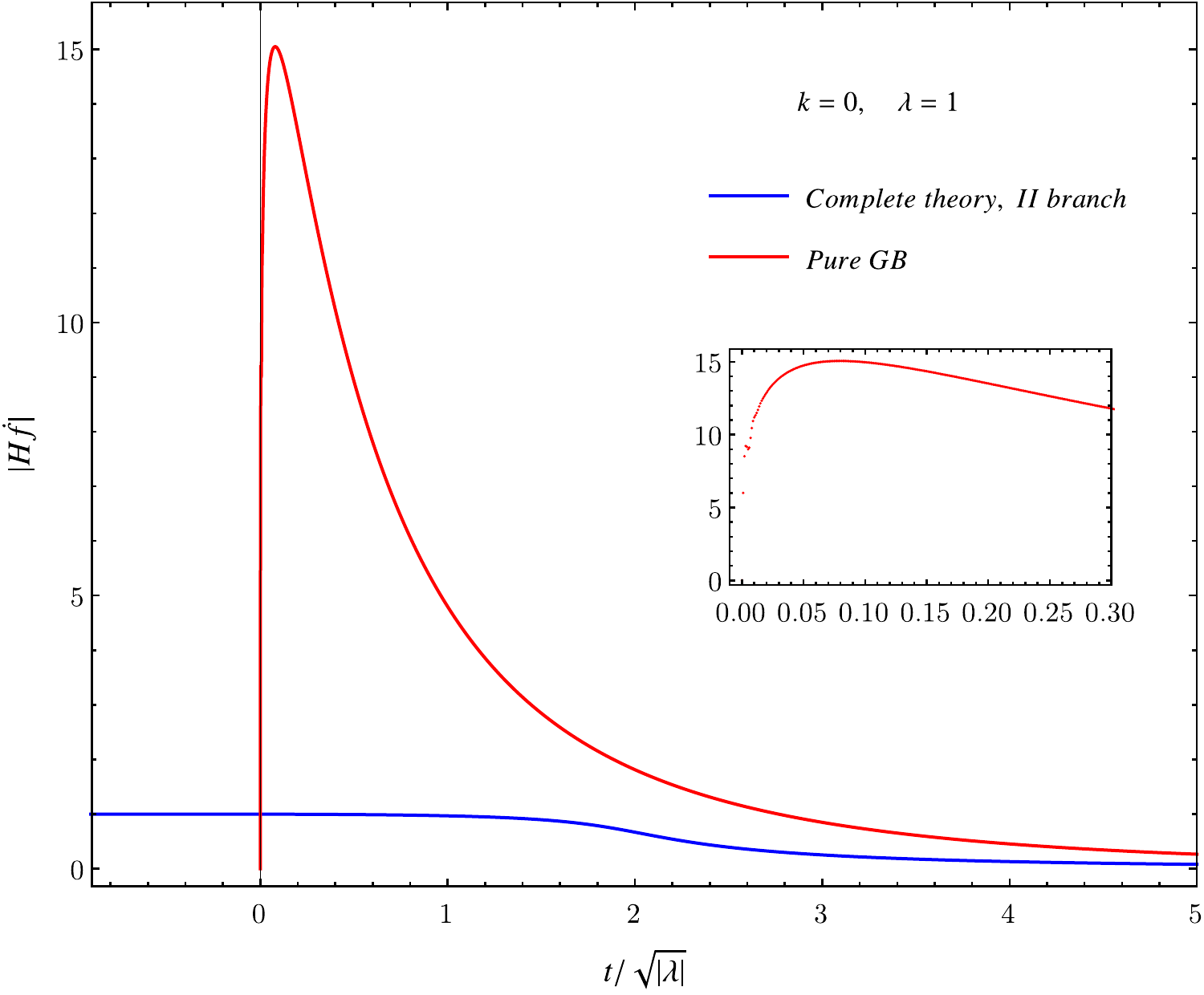}
    \caption{The $H \dot{f}$ term as a function of rescaled time.}\label{phi2NOR-Hf1k=0l=1}
\end{subfigure}
\begin{subfigure}{.46\textwidth}
\centering
    \includegraphics[width=1\textwidth]{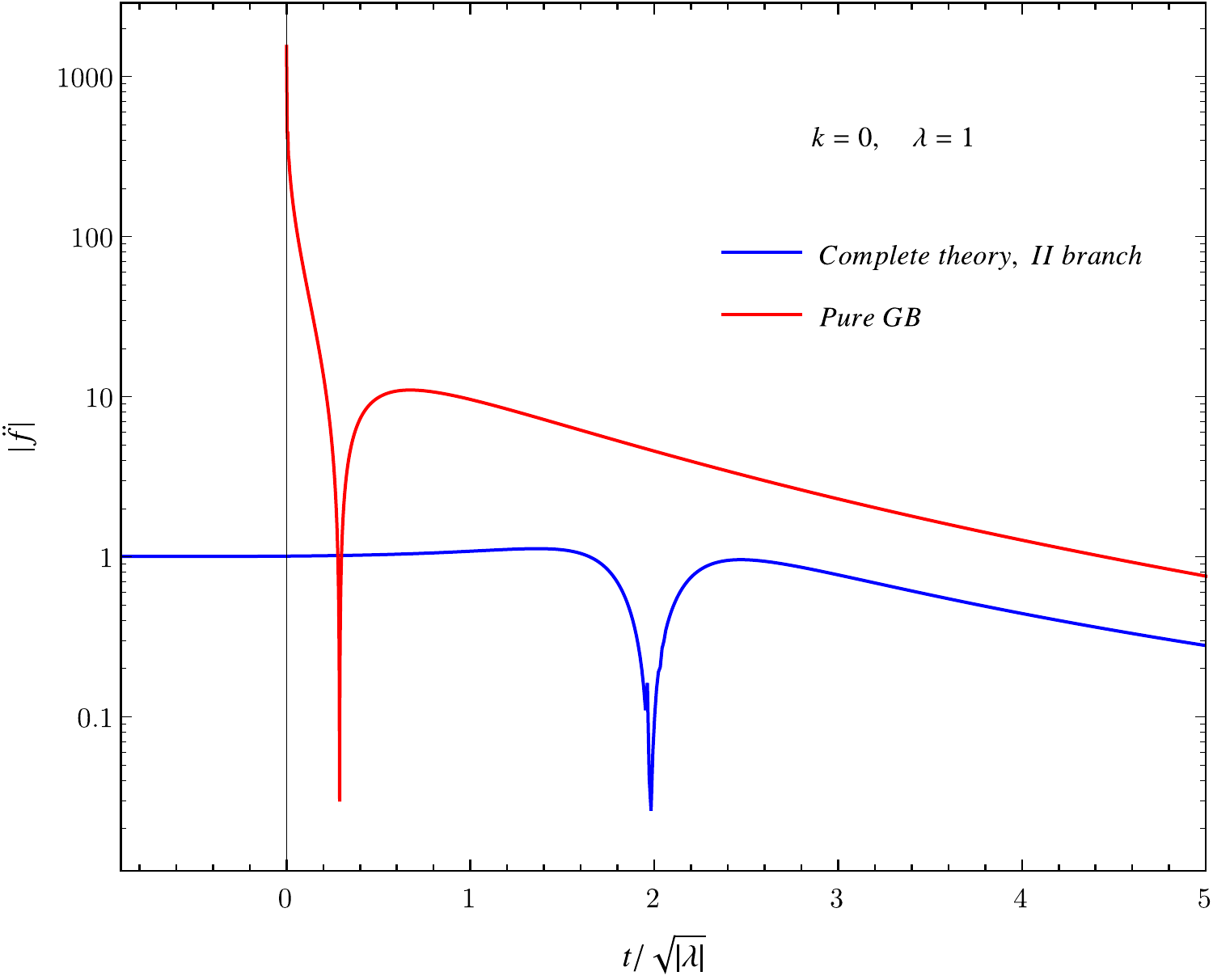}
    \caption{The $\ddot{f}$ term as a function of rescaled time.}\label{phi2NOR-f2k=0l=1}
    \end{subfigure}
    \caption{Pure Gauss-Bonnet solution with $k=0$, $\lambda=1$.}
\end{figure}

We now want to understand why a nonsingular solution is found in pure Gauss-Bonnet theory with $\lambda>0$, despite the fact that, as discussed in Section \ref{nonsingKanti}, only solutions with $\lambda<0$ can be nonsingular in the full theory. The nonsingular solution of pure GB, found in \cite{KantiEarlyTime}, was re-derived in Section \ref{sec:earlytimesol}. nonsingular solutions are forbidden in the complete theory with positive coupling because of the Strong Energy Conditions, an argument that is however not applicable in the pure Gauss-Bonnet case. Even if it doesn't violate any principle, it is rather surprising to find a nonsingular solution which would instead be forbidden in the complete theory. We want then to check if the aforementioned solution satisfies the assumptions of pure Gauss-Bonnet theory.\\
In Figure \ref{phi2CONFR-ak=0l=1} we show the nonsingular solution in pure Gauss-Bonnet gravity together with the two branches of singular solutions found in Section \ref{phi2k=0l=1}, numerically integrating the complete theory. The pure, analytical solution nearly coincides with the numerical second branch at very early times and then becomes quite different. It seems that the complete theory allows to continue the scale factor beyond the minimum value found in the pure approach, up to the singularity. \\
The study of the three conditions support this picture. In Figure \ref{phi2NOR-RGBRk=0l=-1} we show that the Gauss-Bonnet term is of the same order or slightly bigger than the Ricci scalar almost everywhere, except for a small time interval. In particular, the ratio is smaller than one near $t=0$, where the solutions of the complete and the approximated theory differ the most (pure GB solution does not exist for $t<0$). \\
If we look at the second condition for the pure Gauss-Bonnet solution (Figure \ref{phi2NOR-Hf1k=0l=1}), we see that the condition is not fulfilled at the initial time. This plot explains why its behaviour is so different from that of the complete theory's solution, i.e. why it stops at a non-zero value. We also see that for the complete theory's solution $\vert H \dot{f} \vert$ is always order one: if we looked at this plot in the first place, we never would have thought about discarding this term.\\
The third condition, shown in Figure \ref{phi2NOR-f2k=0l=1} for both versions of the theory, is satisfied at early times in pure Gauss-Bonnet, but this is not enough to make the two solutions consistent with each other.\\
We have thus understood why the nonsingular solution is not shared by the two versions of the theory with $\lambda >0$: in this case the Ricci scalar cannot be neglected in the action. This nonsingular solution has its own right to be, if pure GB gravity is considered as the underlying theory, but shouldn't be considered a good approximation of the early time behaviour of the complete GB theory. \\


\chapter{Gauss-Bonnet cosmology: exponential coupling}
\label{chap:exp}

In this chapter we study cosmological solutions of our theory when the coupling between the scalar field and the Gauss-Bonnet invariant takes an exponential form. First, we briefly discuss why nonsingular solutions should arise. Then we summarize our numerical results, and study the stability of both singular and nonsingular solutions. 

\section{Field equations}
We address the case of a dilaton-like scalar field, where the coupling function has an exponential form. We use the most general form
\begin{equation}
f(\phi)=\lambda e^{\beta \phi(t)}.
\end{equation}
This is the so called Einsten-dilaton-Gauss-Bonnet (EDGB) theory, where the cosmological equations take the form: 
\begin{equation}\label{eqephi}
\begin{cases}
\ddot{\phi}+3H\dot{\phi}-3\lambda \beta e^{\beta \phi} \left(H^2+\dfrac{k}{a^2}\right)\left(H^2+\dot{H}\right)=0\\
3\left(1+\lambda \beta e^{\beta \phi} \dot{\phi}H\right)\left(H^2+\dfrac{k}{a^2}\right)=\dfrac{\dot{\phi}^2}{2}\\
2\left(1+\lambda \beta e^{\beta \phi} \dot{\phi}H\right)\left(H^2+\dot{H}\right)+
\left(1+\lambda \beta e^{\beta \phi}(\ddot{\phi}+\beta\dot{\phi}^2) \right)\left(H^2+\dfrac{k}{a^2}\right)=-\dfrac{\dot{\phi}^2}{2}\\
\end{cases}
\end{equation}
However, we stress again that we will partly deviate from the string-inspired EDGB model, allowing the coupling $\lambda$ to take both signs.\\
The set of equations \eqref{eqephi} possess a new symmetry. Under the simultaneous transformations
\begin{equation}
\beta \rightarrow -\beta, \quad t \rightarrow -t, \quad \phi \rightarrow -\phi.
\end{equation}
the equations remain unchanged. This means that we don't need to study the two signs of $\beta$ separately and we can choose $\beta>0$, for instance. We could then obtain solutions with $\beta<0$ by reversing time and the sign of the scalar field. This is not, of course, particularly physically meaningful, as the behaviour of the scale factor remains the same.\\  
By re-scaling the scalar field:
\begin{equation}\label{key}
\phi \rightarrow \frac{1}{\beta} \phi \ ,
\end{equation}
we can predict what happens if the coupling $ \beta $ goes to zero. Eqs. (5.2) in the limit $ \beta \rightarrow 0 $ become:
\begin{align}\label{eqapproxephi}
\begin{cases}
&\ddot{\phi}+3H\dot{\phi}\simeq 0,\\
&\dot{\phi}^2\simeq 0 .
\end{cases}
\end{align}
We conclude that in the small $ \beta $-coupling limit the scalar field decouples from the space-time dynamics and freezes: $ \dot{\phi} = 0 $ is the only solution of Eqs. \eqref{eqapproxephi}.

\section{Nonsingular solutions}
The analytical argument proving the existence of nonsingular solutions for the quadratic coupling cannot be directly extended to the exponential case. The key of the argument is the fact that $f'$ can take both signs, depending on the sign of the scalar field itself. This is true when $f'=n \lambda \phi^{n-1}$ with even $n$, but not when $f'=\beta\lambda e^{\phi}$.  We were not able to construct a different argument analytically proving the existence of nonsingular solutions in the exponential coupling case. However, similarly to the monomial coupling case previously discussed, an exponentially coupled GB term can introduce violations of the Strong Energy conditions (see Section \ref{sec:ec}) and nonsingular solutions are actually found numerically.\\
nonsingular solutions for a dilaton-modulus-Gauss-Bonnet theory where first found, numerically, by Easther and Maeda in \cite{Easther}. The authors focused on a closed universe ($k=1$), flat universe being already discussed within modulus-Gauss-Bonnet theory in \cite{RizosSing,ARTSingfree}. The action they considered includes both the dilaton and the moduli effective string corrections, but the dilaton coupling is only allowed to take positive values because of effective string theory considerations. Thus the dilaton field is not responsible for the nonsingular solutions found by Easther and Maeda. Accordingly, their numerical results show that, when the modulus is turned off, only singular solutions appear. As a result, no direct comparison  with our solutions, to be described in the next Sections,   can be drawn. \\
In \cite{ATUNonSing}, Alexeyev, Toporensky and Ustiansky find nonsingular aymptotics for an exponential coupling of the form $e^{\vert \phi \vert}$. This coupling is chosen because its asymptotic behaviour is similar to that of the complete sting coupling function for a modulus field. Again, no direct comparison  with our results can be drawn.
 
\section{Numerical solutions}
\label{sec:numExp}
We proceed to numerically integrate the equations \eqref{eqephi}. We first analyse the constraint equation to characterize the two branches of solutions. The constraint can be rewritten as follows:
\begin{equation}
\dot{\phi}^2-2A\dot{\phi}+B=0 \quad \textrm{with} \ \ A=3\beta \lambda H e^{\beta \phi}\left(H^2+\dfrac{k}{a^2}\right), \quad B=-6\left(H^2+\dfrac{k}{a^2}\right).
\end{equation}
The roots of the polynomial above are:
\begin{equation}
\dot{\phi}_{s}=A\pm s\sqrt{A^2-B}, \quad s=\pm 1.
\end{equation}
We use the same procedure and initial conditions described in Section \ref{sec:NumAnphi2} for the quadratic coupling. Again, it is necessary to perform a complete analysis of the two branches arising in the system, by tuning the initial condition $\dot{\phi}_0$. In most cases the two branches do not differ much qualitatively, and we will only show one of them. When, however, the branches of solutions show distinct properties, we report the second one in the Appendix.\\    
Similarly to the case of quadratic coupling, we are going to impose the initial condition on $\dot{\phi}$, rather than on $\phi$, and we obtain the latter from the constraint equation as follows:
\begin{equation}\label{logphi}
\phi(t)=\dfrac{1}{\beta}\left[2\pi i \ C+\ln \left(\dfrac{-6ka-6a\dot{a}^2+a^3\dot{\phi}^2}{6\beta\lambda\dot{a}\left(k+\dot{a}^2\right)\dot{\phi}}\right)\right], \quad C \in \mathbb{Z}
\end{equation}
Clearly, we need to set $C=0$ to have a real solution for the scalar field. The expression above also demands the argument of the logarithm to be positive. Fulfilling this condition requires us, in some cases, to adjust the initial conditions.\\
As already pointed out in the opening of this Chapter, we can choose a positive value for $\beta$ and study how the solutions depend on its absolute value.\\

The integration method used has already been described in Section\ref{sec:NumAnphi2}.In the following Sections, we separately analyse solutions with different spatial curvature ($k=0,+1,-1$) and with different sign of the coupling ($\lambda\gtrless 0$).

\subsection{Flat universe (k=0)}
\label{sec:ephik=0}
\subsubsection*{Positive coupling, $\boldsymbol{\lambda>0}$}
\begin{figure}[!ht]
 \centering
 \begin{subfigure}{.47\textwidth}
 \centering
    \includegraphics[width=1\textwidth]{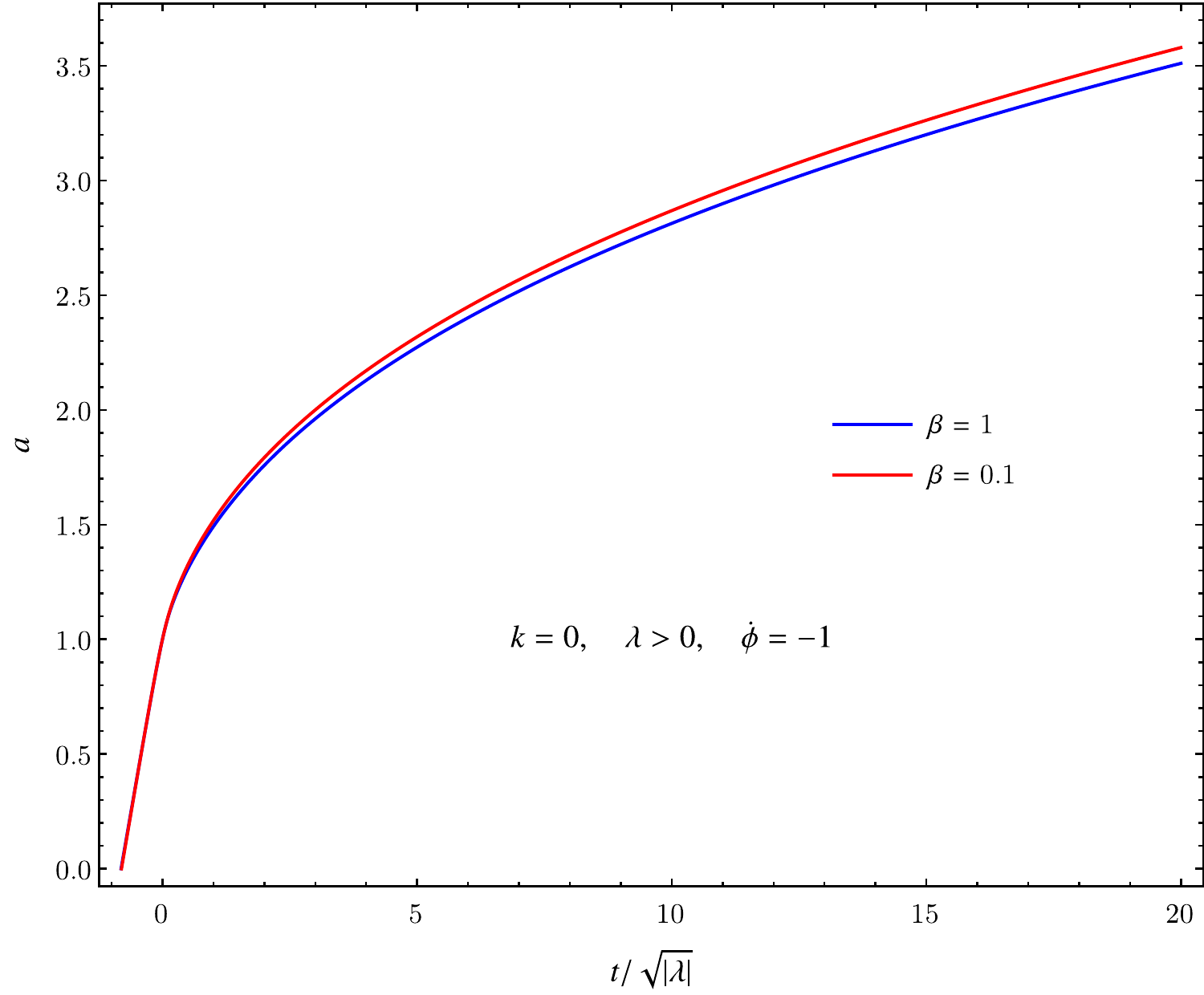}
\end{subfigure}
\begin{subfigure}{.47\textwidth}
\centering
    \includegraphics[width=1\textwidth]{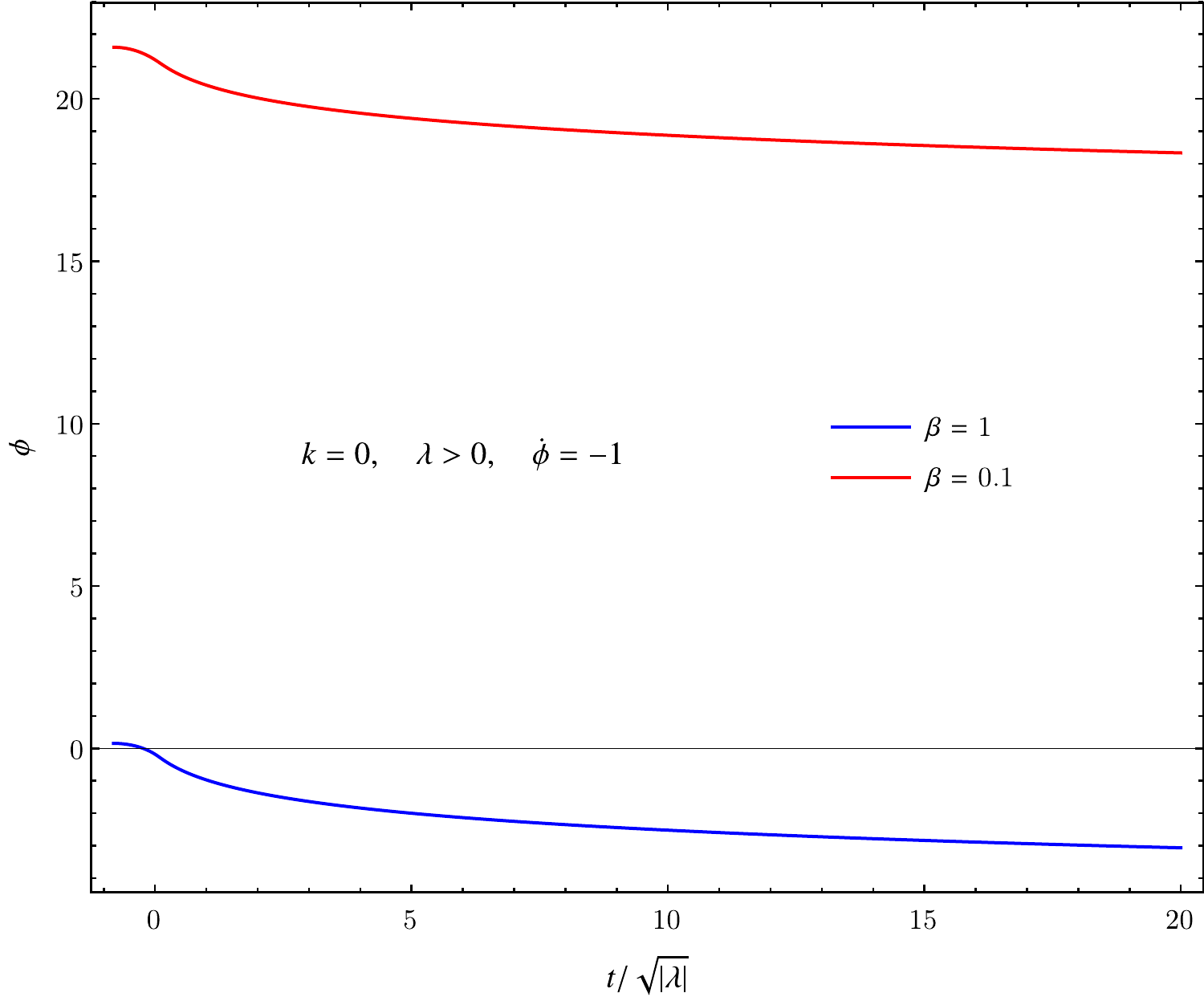}
    \end{subfigure}    
    \caption{The scale factor (left) and the scalar field (right) as a function of rescaled time for the singular solution with $k=0$, $\lambda=1$, first branch.}\label{ephi-aphik=0d=1both}
\end{figure}

As shown in Figure \ref{ephi-aphik=0d=1both}, the scale factor presents an initial singularity from which it grows linearly. Then, the Universe described by this solution asymptotically settles in a GR-like ($a\sim t^{1/3}$) expansion. The scalar field (Figure \ref{ephi-aphik=0d=1both}) is finite at the initial time and its later evolution is also unaffected by the Gauss-Bonnet term. The value of $\beta$ does not influence the qualitative behaviour of the solution. \\ 

\subsubsection*{Negative coupling, $\boldsymbol{\lambda<0}$}
\begin{figure}[!ht]
 \centering
    \includegraphics[width=0.8\textwidth]{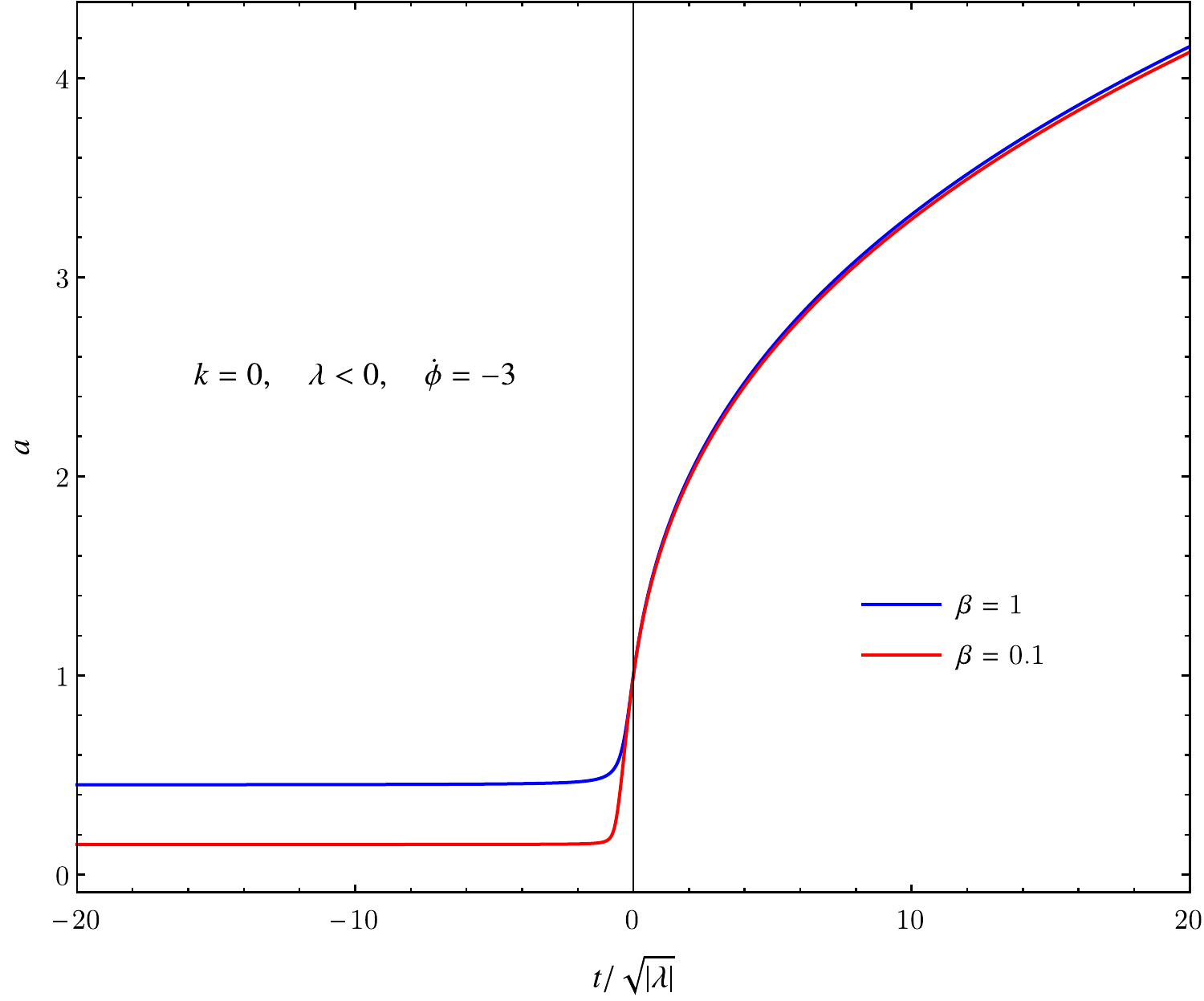}
    \caption{The scale factor as a function of rescaled time for $k=0$, $\lambda=-1$, nonsingular branch.}\label{ephi-ak=0d=-1both2}
\end{figure}

\begin{figure}[!ht]
 \centering
 \begin{subfigure}{.46\textwidth}
    \includegraphics[width=1\textwidth]{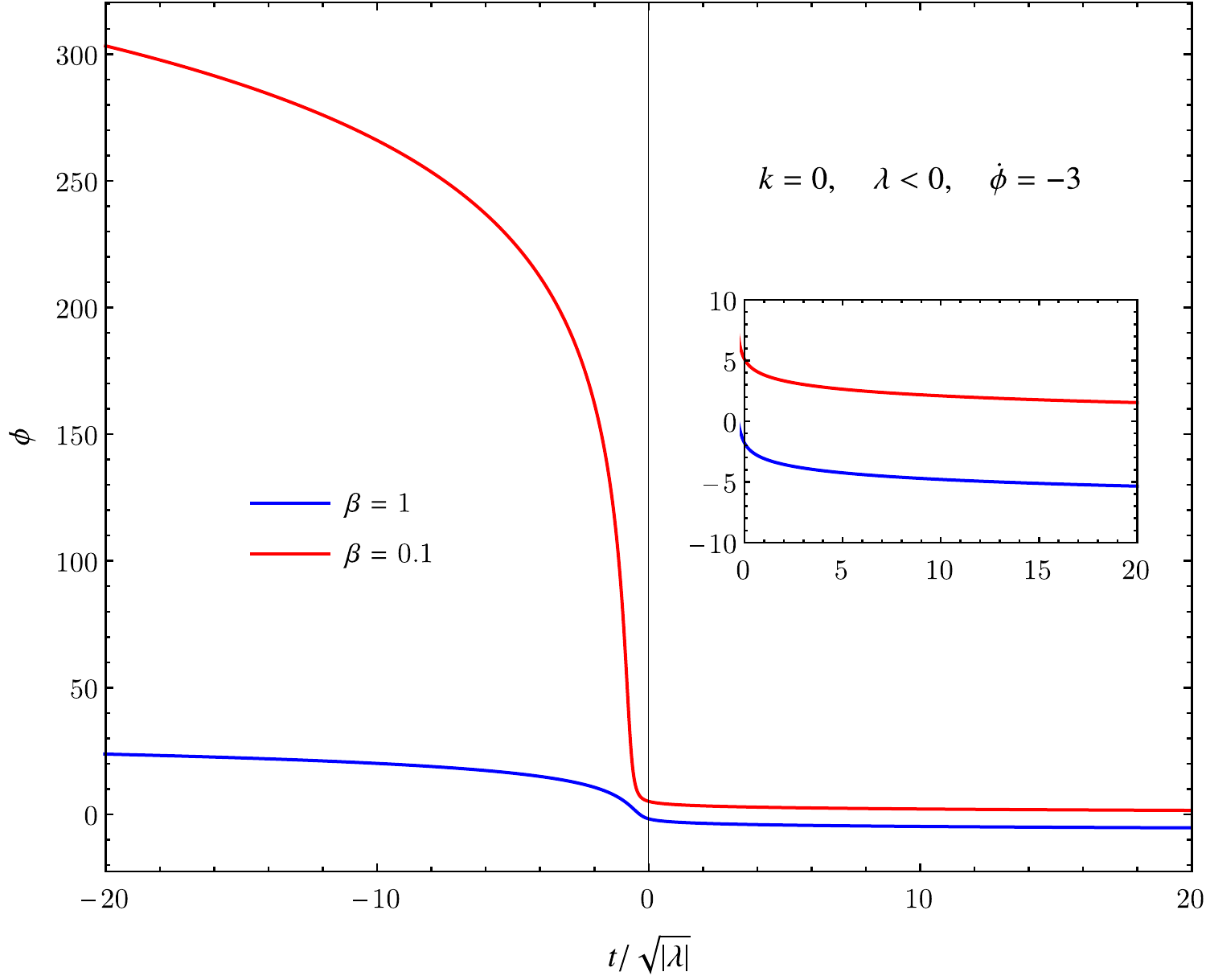}
\end{subfigure}
\begin{subfigure}{.46\textwidth}
    \includegraphics[width=1\textwidth]{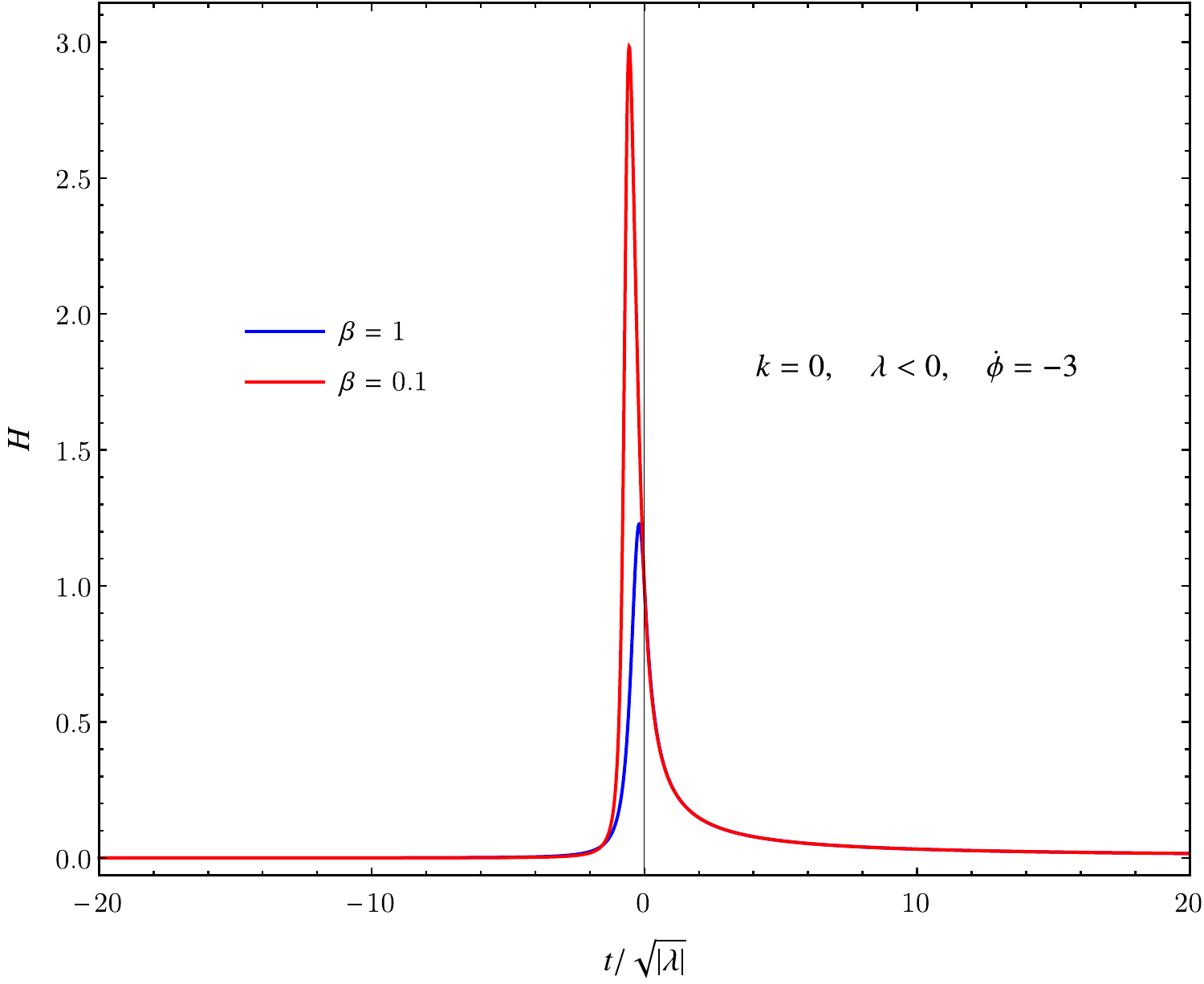}
    \end{subfigure}
    \caption{The scalar field (left) and the Hubble parameter (right) as a function of rescaled time for $k=0$, $\lambda=-1$, second branch.}\label{ephi-Hphik=0d=-1both2}
\end{figure}

In this case, we find a \emph{nonsingular} solution: the scale factor (Figure \ref{ephi-ak=0d=-1both2}) tends to a finite, non zero value in the infinite past, and later switches to a GR-like expansion. The scalar field is finite at every cosmological time, just like the Hubble parameter. Both are displayed in Figure \ref{ephi-Hphik=0d=-1both2}: the scale factor grows logarithmically in the infinite past, while the Hubble parameter tends to zero. At late times the Gauss-Bonnet correction becomes irrelevant and GR solution is approached.\\ 
This is a quite remarkable solution, so we will discuss in detail its stability in Section \ref{ephinonsing}. Here we just add that, varying the value of $\beta$, we find no significant change in the qualitative behaviour of the solution.

\subsection{Positive curvature (k=+1)}
\subsubsection*{Positive coupling, $\boldsymbol{\lambda>0}$}
\begin{figure}[!ht]
 \centering
 \begin{subfigure}{.47\textwidth}
 \centering
    \includegraphics[width=1\textwidth]{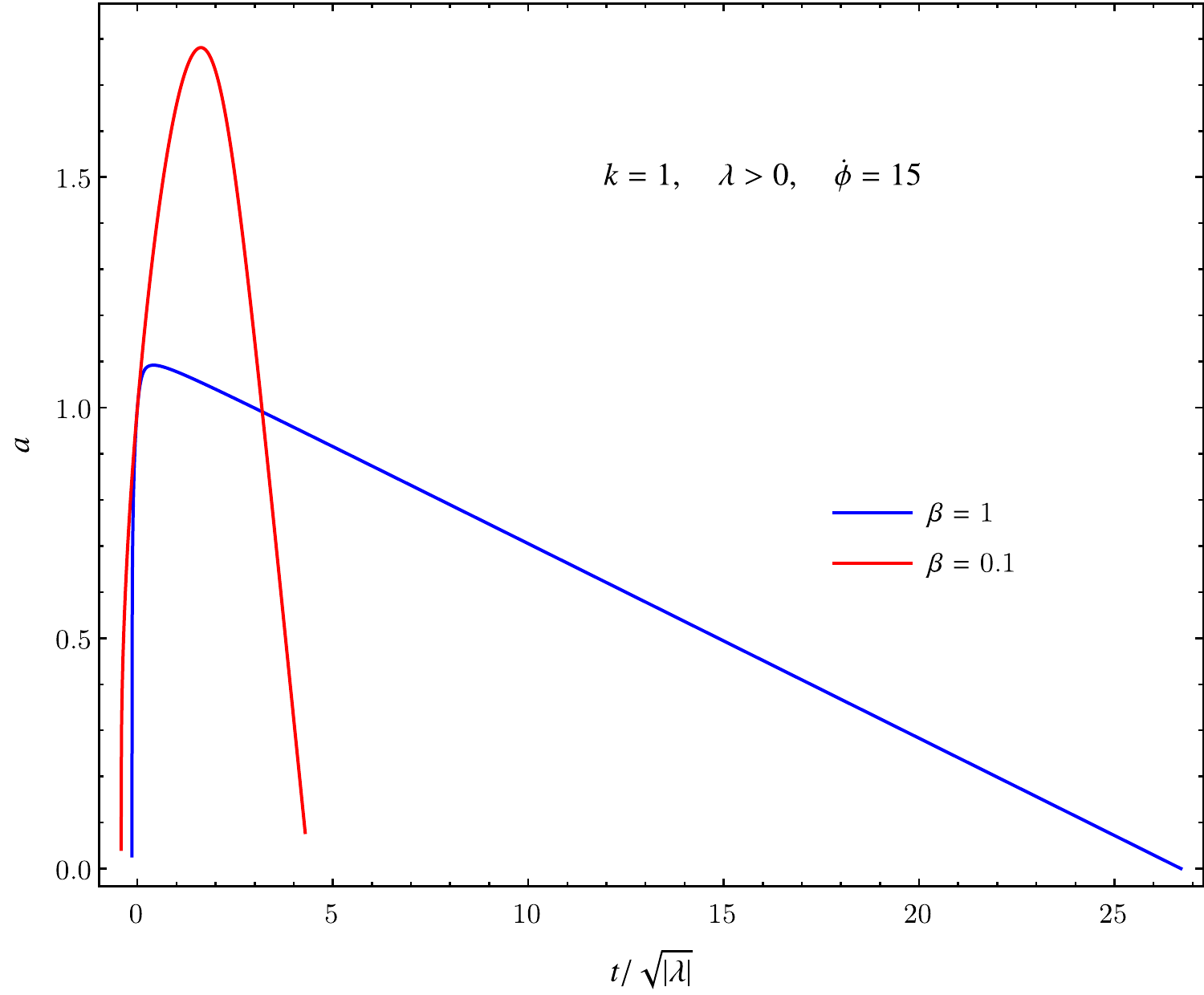}
\end{subfigure}
\begin{subfigure}{.47\textwidth}
\centering
    \includegraphics[width=1\textwidth]{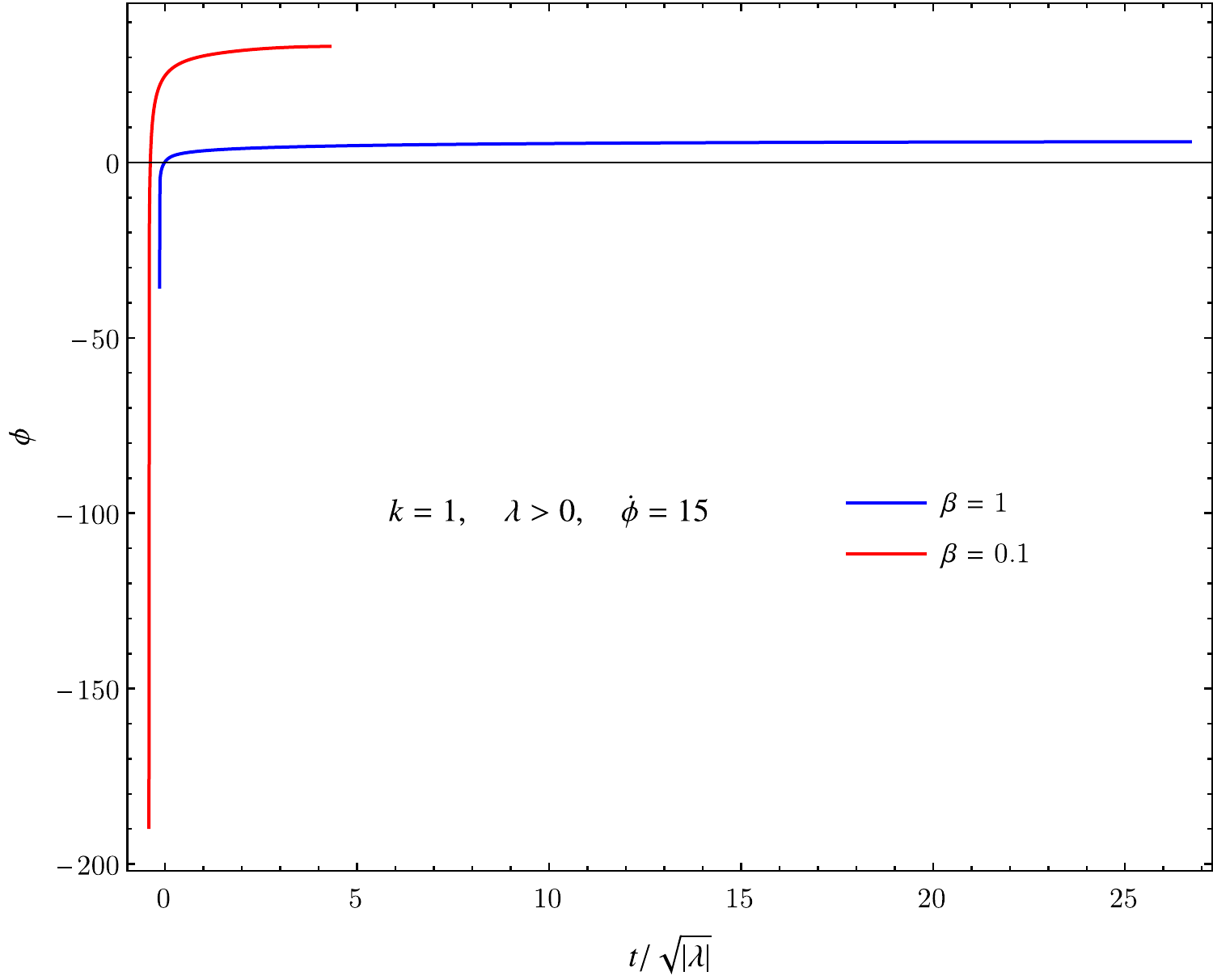}
    \end{subfigure}    
    \caption{The scale factor (left) and the scalar field (right) as a function of rescaled time for a singular solution with $k=1$, $\lambda=1$, first branch.}\label{ephi-aphik=1d=1both}
\end{figure}

Here we need to modify the initial conditions for numerical integration, to prevent the argument of the logarithm in expression \ref{logphi} from becoming negative. With our choice of initial values $ a_0 = 1 $ and $ \dot{a} = 1 $, the conditions are:
\begin{equation}\label{key}
(\dot{\phi}^2-12) >0 \ \text{if} \ \dot{\phi}>0, \quad (\dot{\phi}^2-12)  <0 \ \text{if} \  \dot{\phi}<0.
\end{equation}
The Universe described by the $k=1$, $\lambda>0$ solutions exists for a finite amount of cosmological time, starting and ending with a singularity. The scale factor, shown in Figure \ref{ephi-aphik=1d=1both}, reaches rapidly a maximum and then decreases linearly. The scalar field possesses a finite value both at initial and final time (Figure \ref{ephi-aphik=1d=1both}).\\
The solution doesn't change qualitatively when we tune the value of $\beta$. 
\subsubsection*{Negative coupling, $\boldsymbol{\lambda<0}$}
\begin{figure}[!ht]
 \centering
 \begin{subfigure}{.46\textwidth}
    \includegraphics[width=1\textwidth]{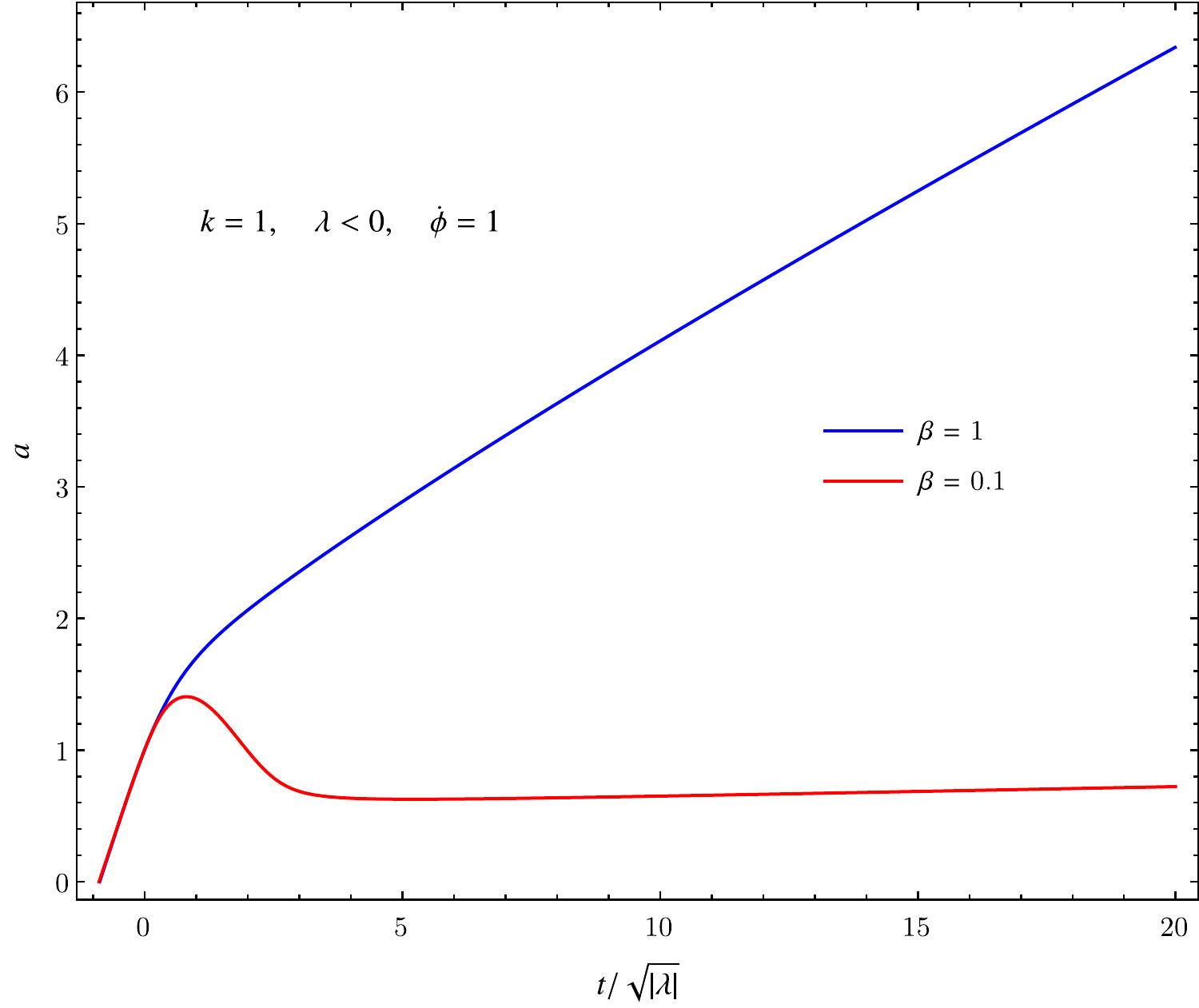}
\end{subfigure}
\begin{subfigure}{.46\textwidth}
    \includegraphics[width=1\textwidth]{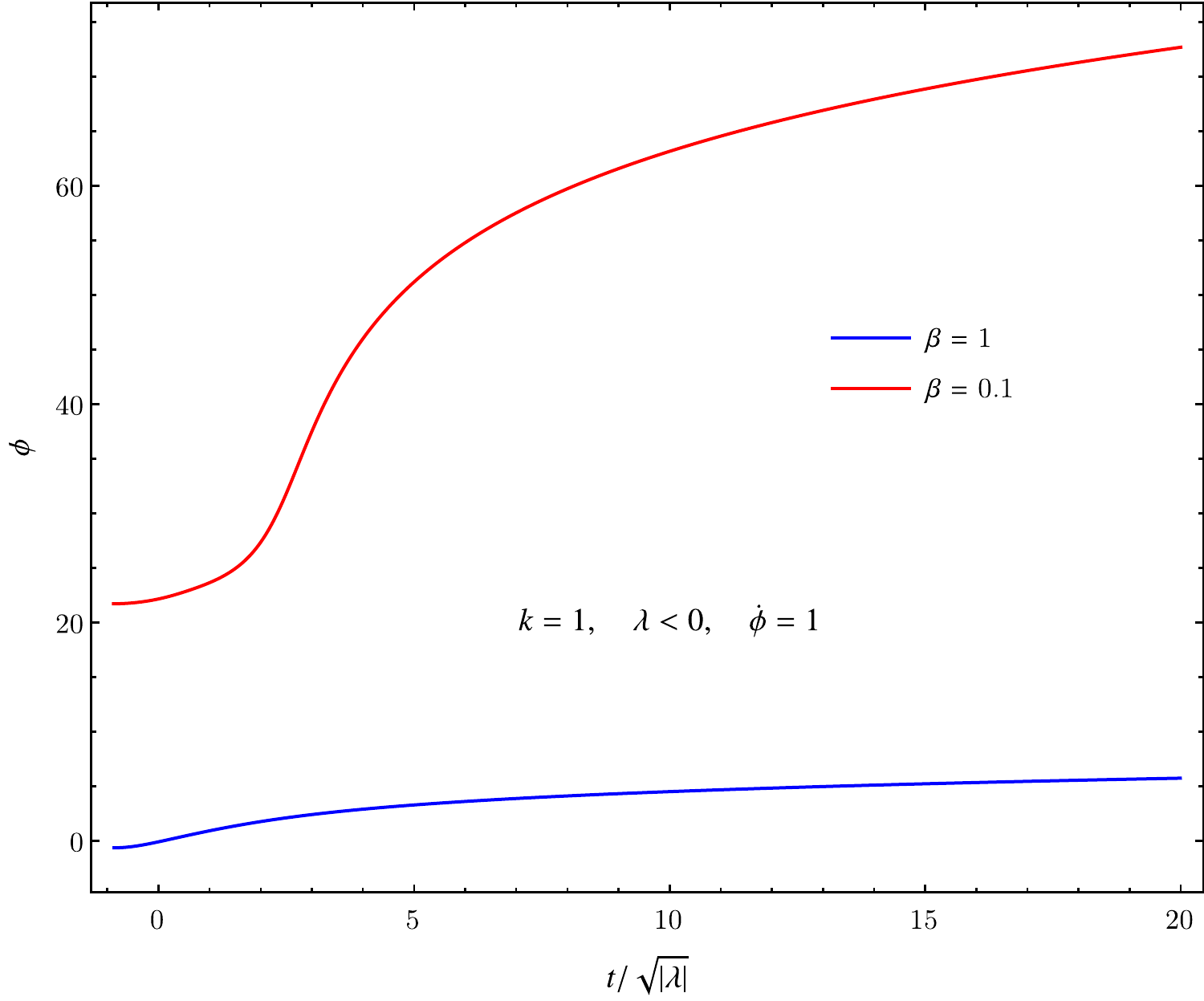}
    \end{subfigure}
   \caption{The scale factor (left) and the scalar field (right) as a function of rescaled time for $k=1$, $\lambda=-1$, first branch.}\label{ephi-phik=0d=-1both2}
\end{figure}

In this case we are not able to find nonsingular solutions: the singular behaviour of the scale factor is shown in Figure \ref{ephi-phik=0d=-1both2}. The Universe, at first, undergoes a fast growth and then expands linearly. The scalar field, displayed in Figure \ref{ephi-phik=0d=-1both2}, is always finite. The Hubble parameter, not shown here, presents an initial singularity and then asymptotically approaches zero.\\
The presence of a local maximum for the scale factor, depends on the value of $\beta$: the maximum is absent when $\beta=1$, while it is present when $\beta$ decreases. After the peak the Universe recovers a linear expansion, but with a smaller rate than in the $\beta=1$ case. The scalar field, on the other hand does not seem to depend qualitatively on $\beta$.\\    

\subsection{Negative curvature (k=-1)}
We are not going to discuss the solutions of the negative curvature case. This is left for a future extension of the present work. 

\clearpage
\section{Stability analysis}
\subsection{Nonsingular solutions}
\label{ephinonsing}
\begin{figure}[!ht]
 \centering
 \begin{subfigure}{.47\textwidth}
 \centering
    \includegraphics[width=1\textwidth]{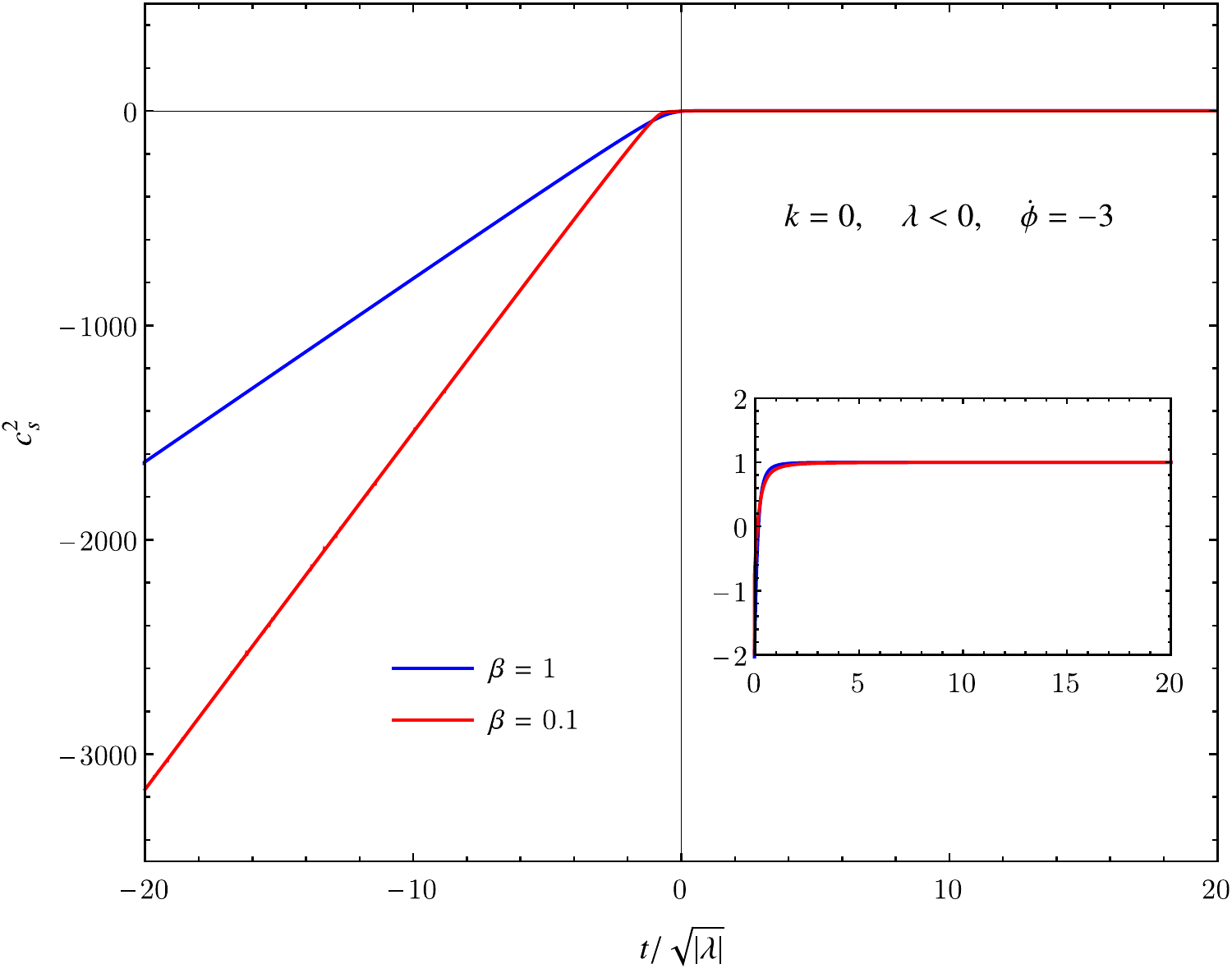}
\end{subfigure}
\begin{subfigure}{.47\textwidth}
\centering
    \includegraphics[width=1\textwidth]{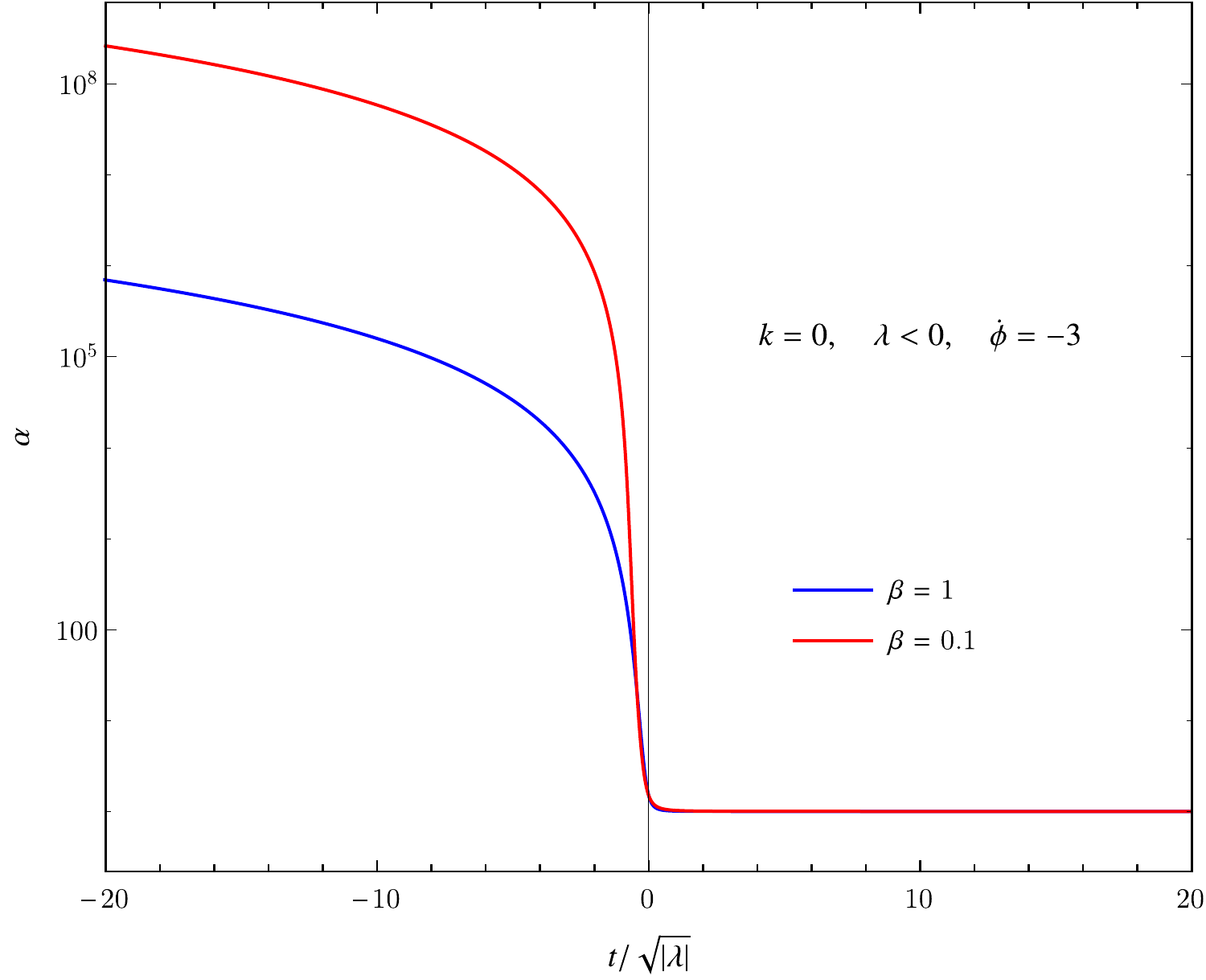}
    \end{subfigure}
    \caption{The squared sound velocity (left) and the ghost factor (right) as a function of rescaled time for the nonsingular solution with $k=0$, $\lambda=-1$.}\label{ephi-csalphak=0d=-1bisboth}
\end{figure}

Whichever coupling function we choose, nonsingular solutions in Gauss-Bonnet theory are doomed to be unstable by the argument outlined in Section \ref{sec:GBstability}. This instability is clearly shown in Figure \ref{ephi-csalphak=0d=-1bisboth} for the flat, nonsingular Universe of the exponential model, described in Section \ref{sec:ephik=0}. No ghost is present, because $\alpha$ is always positive and tends to $+ \infty$ in the infinite past. The propagation of tensor perturbation is sub-luminal throughout the whole solution, but the square sound velocity becomes negative and tends to $- \infty$ in the infinite past. The instability coincides, thus, with the static-Universe phase, when the scale factor is approximately constant. When the Universe exits the static phase and expansion takes place, on the other hand, stability is achieved. In addition, we observe that instability could hardly be confined to short-wavelengths: for every wavelength, however long, there will be a moment in the past where the square sound velocity is sufficiently large to make the mode unstable. Instability could be avoided by setting suitable initial conditions, premising the existence of a natural cut-off for the frequency. However, such initial conditions would be hard to justify, because an arbitrary time in the past, static phase is to be chosen.\\ 
Unstable solutions generally possess also a negative friction term \eqref{frictionterm}, which enhances the instability associated to the negative square sound velocity.

\subsection{Singular solutions}
\label{sec:stabsing}
\begin{figure}[!ht]
 \centering
 \begin{subfigure}{.47\textwidth}
 \centering
    \includegraphics[width=1\textwidth]{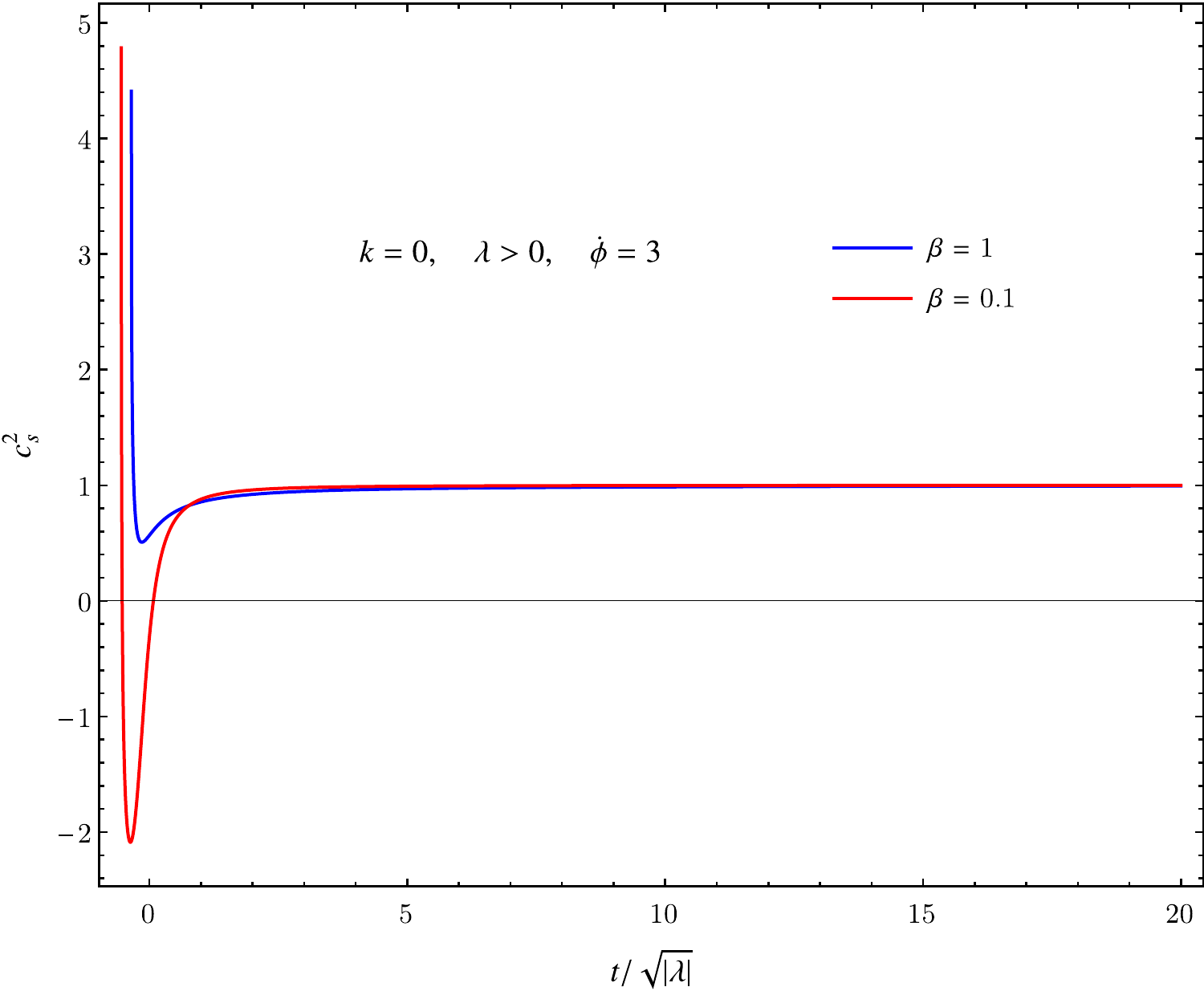}
\end{subfigure}
\begin{subfigure}{.47\textwidth}
\centering
    \includegraphics[width=1\textwidth]{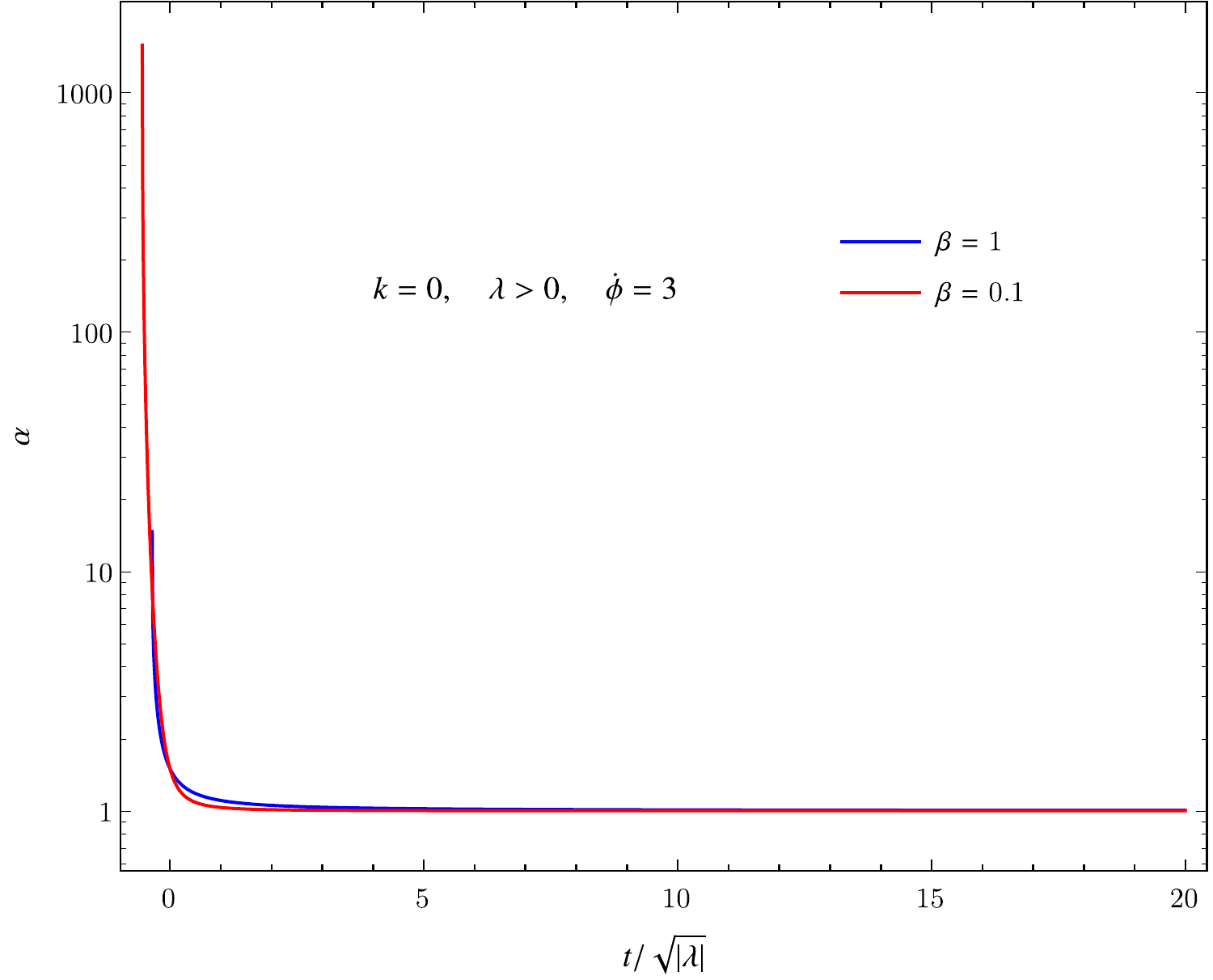}
    \end{subfigure}
    \caption{The squared sound velocity (left) and the ghost factor (right) as a function of rescaled time for a singular solution with $k=0$, $\lambda=-1$, second branch.}\label{ephi-csalphak=0d=1bisboth}
\end{figure}

Singular solutions can be both stable or unstable, depending on the initial conditions and the value of the parameter $\beta$. In Figure \ref{ephi-csalphak=0d=1bisboth}, we display stability obtained by tuning the $\beta$ value, for a singular solution of the exponential model in flat background. \\
However stable, but singular and not inflationary solution is not of particular interest, and if initial conditions were selected as to give sufficient inflationary expansion, the solution would be bound to be unstable by the argument in Section \ref{sec:GBstability}.  
 
\chapter{Gauss-Bonnet cosmology with a massive scalar field}
\label{chap:V}
\section{Action and equations of motion}\label{actioneqV}
Since tensor instability invalidates all inflationary or nonsingular solutions in scalar-Gauss-Bonnet theory, in this Chapter we look for a modification of the theory able to improve stability.\\
We slightly modify the action of the original model introducing a potential $V(\phi)$, depending only on the scalar field and not on the metric. The action under study now is:
\begin{equation}\label{actionV}
\int \! \mathrm{d}^4x \ \sqrt{-g} \left[\dfrac{1}{2}R-\dfrac{1}{2}\partial_{\mu}\phi\partial^{\mu}\phi-V(\phi)+ \dfrac{f(\phi)}{8} R^2_{GB}\right] \ + S_{matter}.
\end{equation}
We study the simple and most natural case, in which the potential is quadratic:
\begin{equation}
V(\phi)=m^2\; \phi^2,
\end{equation}
Throughout this Chapter we will refer to the parameter $m>0$ as the mass, even though this parameter is not what we usually define as the mass of the scalar field. Indeed, the true potential of the scalar field is $V_{\textrm{eff}}=V(\phi)- \frac{f(\phi)}{8} R^2_{GB}$ and its mass should be defined as $m_{\textrm{eff}}^2=\frac{\partial ^2 V_{\textrm{eff}}}{\partial \phi^2}$. However, the parameter m is rightly referred to as the mass, if the mass is defined at asymptotic distance from a curvature source, where space-time becomes flat and $ R^2_{GB} \rightarrow 0 $.\\
In low-energy effective string theory, the potential is absent at the tree level. However the addition of a potential in this context could be justified by non perturbative effects or supersymmetry breaking. SUSY breaking should take place at a scale much smaller that Planck scale, so the scalar field should be very light in our units. Anyway, we will not impose upper limits to the mass on this ground, as we prefer to consider the theory as independent from its string genesis.\\ 
The scalar field equation of the new model can be derived using variational principle on action \eqref{actionV}. We obtain the equation:
\begin{equation}
\dfrac{1}{\sqrt{-g}}\partial_{\mu}\left(\sqrt{-g}\partial^{\mu}\phi\right)=
V'(\phi)-\dfrac{f'(\phi)}{8}R^2_{GB}.\\
\end{equation}
When the potential is quadratic, the equation above becomes
\begin{equation}\label{massivescalareq}
\dfrac{1}{\sqrt{-g}}\partial_{\mu}\left(\sqrt{-g}\partial^{\mu}\phi\right)=
2m^2\phi-\dfrac{f'(\phi)}{8}R^2_{GB}.\\
\end{equation}
Einstein equations are modified, too, by the presence of the potential:
\begin{equation}
G_{\mu\nu}=\partial_{\mu}\phi{\partial}_{\nu}\phi
-\dfrac{1}{2}g_{\mu\nu}\left({\partial} _{\rho}\phi\right)^2-K_{\mu\nu}-g_{\mu\nu}V(\phi),
\end{equation}
where $ K_{\mu\nu} $ is defined by \eqref{Kmunu}. Substituting the explicit form of the potential, Einstein equations become:
\begin{equation}
G_{\mu\nu}=\partial_{\mu}\phi{\partial}_{\nu}\phi
-\dfrac{1}{2}g_{\mu\nu}\left({\partial} _{\rho}\phi\right)^2-K_{\mu\nu}-g_{\mu\nu}m^2\phi^2 \ .
\end{equation}
\section{Cosmological equations}
We focus on a flat, homogeneous and isotropic Universe, using the metric ansatz:
\begin{equation}\label{metricansatz2}
\begin{split}
&ds^2=-dt^2+a(t)^2 \delta_{ij}dx^i dx^j \ .
\end{split}
\end{equation}
The choice of a spatially flat Universe is justified by the fact that it is favoured by current cosmological observations \cite{PLANCK}; moreover, interesting solutions of the previous Chapters appeared in this case, i.e. for $k=0$.\\
When no matter (except for the scalar field) and no cosmological constant is present, equations of motion for the flat FRW Universe read:
\begin{equation}\label{eqfV}
\begin{cases}
\ddot{\phi}+3H\dot{\phi}-3f' H^2\left(H^2+\dot{H}\right)+V'(\phi)=0,\\
3\left(1+\dot{f}H\right)H^2=\dfrac{\dot{\phi}^2}{2}+V(\phi),\\
2\left(1+\dot{f}H\right)\left(H^2+\dot{H}\right)+
\left(1+\ddot{f}\right)H^2=-\dfrac{\dot{\phi}^2}{2}+V(\phi).\\
\end{cases}
\end{equation}
If we define:
\begin{equation}
\Omega_{GB}=\dot{f}H, \quad \quad \Omega_{\phi}=\dfrac{1}{6}
\left( \dfrac{\dot{\phi}^2}{H^2}+2\dfrac{V}{H^2}\right),
\end{equation}
then the constraint equation, namely the second equation in \eqref{eqfV}, can be written as:
\begin{equation}
\Omega_{\phi}+\Omega_{GB}=1.
\end{equation}
By combining the two dimensional parameters of the theory ($[\lambda]=[l]^2$ and $[m]=[l]^{-1}$) we can construct the dimensionless quantity:
\begin{equation}
q=m^2 \lambda.
\end{equation}
One can easily see that, if we rescale time as $t\rightarrow t / \sqrt{\vert \lambda \vert}$, the system of equations depend on the quantity $q$ alone. Then we can fix the value of $\vert \lambda \vert=1$ and study the solutions as functions of only one parameter, $q$ or, equivalently, $m$. \\
When, as in the previous Chapters, initial conditions are arbitrary we still have a big number of free parameters ($\phi_0$, $H_0$, $q$, $A_0$). When initial conditions are partly fixed by the requirement of an inflationary evolution, the problem truly reduces to a two-parameter analysis: $N_0$, the number of e-foldings, and $q$. This is the reason why we will focus on inflationary solutions in our analysis of the model. Moreover, it is physically interesting to study whether inflationary solutions of GB theory become stable in the presence of a mass term.

\section{Strong Energy Conditions and nonsingular solutions}
The Strong Energy conditions for a flat Universe become, with the addition of a potential to the theory:
\begin{equation}
\begin{split}
\rho&+p=2 H\frac{ A}{C},\\
\rho&+3p=\frac{12 H \dot{\phi} \left[6 H^3 \dot{\phi} \left(\dot{\phi}^2 f''+4\right)+6 H^2 V'-12 H V\dot{\phi}-V' \left(2 V+\dot{\phi}^2\right)\right]}{C}.
\end{split}
\end{equation}
where 
\begin{equation*}
\begin{split}
C=&-12 H^2 \left(2 V+\dot{\phi}^2\right)+36 H^4+12 V\dot{\phi}^2+4 V^2+5 \dot{\phi}^4,\\
A=&+12 H^3 \left(\dot{\phi}^4 f''-2 V+3\dot{\phi}^2\right)+12 H^2\dot{\phi} V'+\\
&+H \left(-12 V\dot{\phi}^2+4 V^2+5 \dot{\phi}^4\right)+36 H^5-2 \dot{\phi} V' \left(2 V+\dot{\phi}^2\right).
\end{split}
\end{equation*}
An alternative way to address the problem is to define again the quantity $\Gamma$. Interestingly, by using background equations \eqref{eqfV} we find that it doesn't depend on the potential $V$:
\begin{equation}
\Gamma=\dfrac{p}{\rho}+1=-\dfrac{2}{3}\dfrac{\dot{H}}{H^2}.
\end{equation}
Then, whether the scalar field has a potential or not, violation of Strong Energy conditions requires $\Gamma<2/3$.\\

Although necessary conditions for nonsingular solutions could be easily fulfilled when we add a potential to the basic theory, their appearance is not guaranteed.\\
Alexeyev, Toporensky and Ustiansky, in \cite{ATUNonSing}, claim that nonsingular asymptotics are generally destroyed by the presence of a potential. In the case of a quadratic coupling function $f$, for example, they argue that the potential should be not steeper than $\phi^2$, and that the mass of has an upper limit. However their analysis is based on approximated asymptotics, obtained discarding first order curvature terms. In Section \ref{sec:phi2nonsing} we showed how this approximation can lead to misleading conclusions and can not be trusted \textit{a priori}. Since, as we will later discuss, we do not find nonsingular solutions, our numerical results support their more general statement: \emph{\lq\lq nonsingular solutions present in the string gravity with V = 0 disappear in most cases when the potential is taken into account".}

\section{Tensor perturbations}
The sound velocity of tensor modes retains the same expression \eqref{scond} once the addition of a potential is made. The propagation speed of tensor perturbations depends on the potential only implicitly, i.e. through background solution, which is different when $V=0$ or $V \neq 0$.
Thus, using the background equations \eqref{eqfV}, the conditions for stable and ghost-free perturbations need to be generalized into:
\begin{equation}
\begin{split}
\alpha= H \dot{\phi} f'(\phi)+1>0,\\
\frac{1}{8} \left(3\Gamma -5\right) \alpha +\frac{V(\phi)}{4 H^2}>0 .
\end{split}
\end{equation}
With our choice for the potential, the stability condition becomes:
\begin{equation}
\frac{1}{8} \left(3 \Gamma -5\right) \alpha +\frac{m^2 \phi ^2}{8 H^2}>0 .
\end{equation}
Nonsingular solutions, then, if they existed, would be allowed to be stable, at least for a suitable choice of the parameter $m$. We see that, even if $\Gamma < 5/3$, the stability condition can be fulfilled for sufficiently large values of $m$.\\
The same can be said for inflationary solutions: expressing the stability condition in terms of the slow-roll parameter $\epsilon_H$:
\begin{equation}
\frac{1}{8} \left(2\epsilon_H -5\right) \alpha +\frac{V(\phi)}{4 H^2}>0,
\end{equation} 
we easily see that the requirement for de Sitter-like expansion, i.e. $\epsilon \ll 1$, is compatible with stability.

\section{Slow-roll inflation}
\begin{figure}[!ht]
 \centering
    \includegraphics[width=0.7\textwidth]{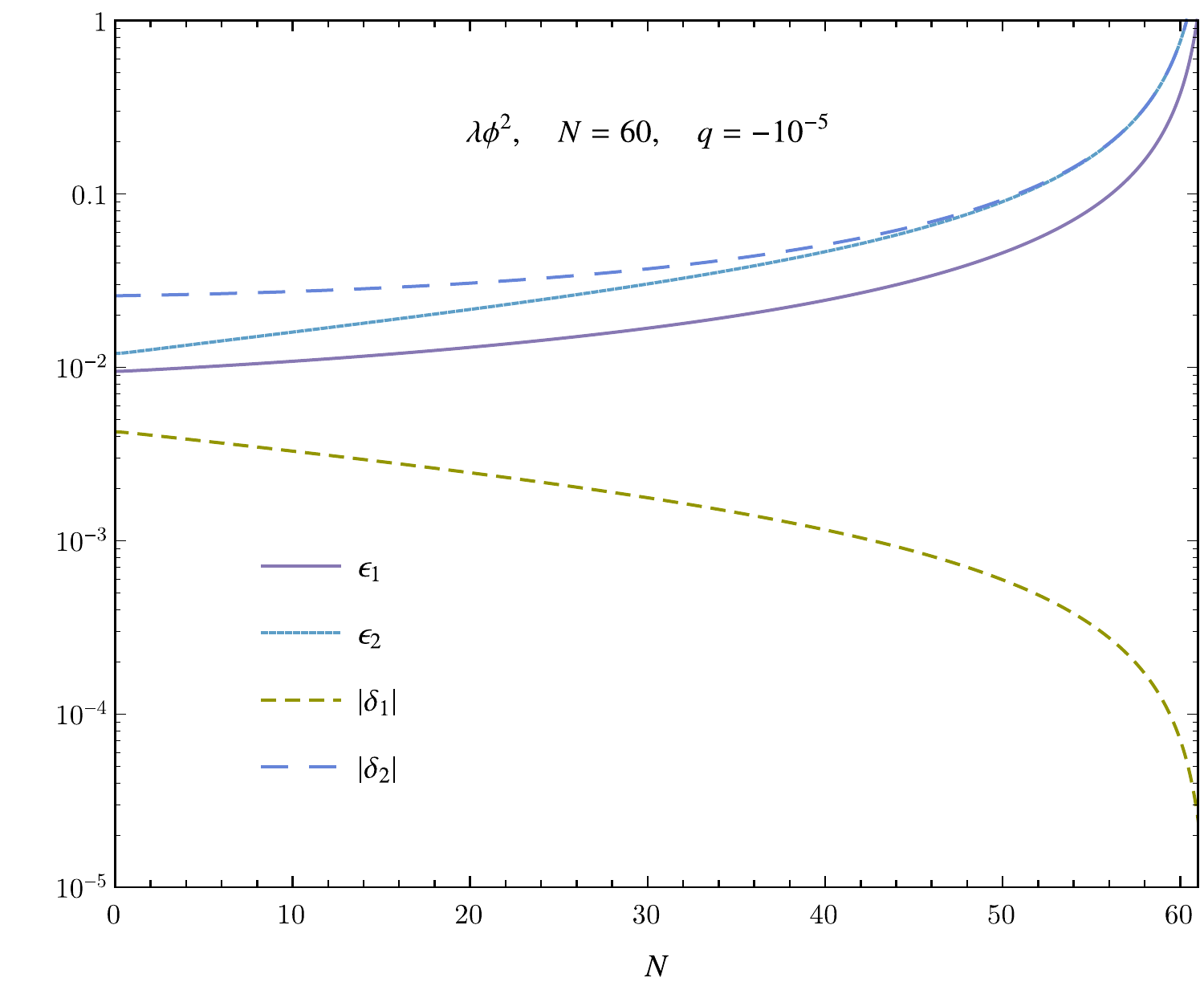}
    \caption{Slow roll parameters as functions of the number of e-foldings in an inflationary solution for the quadratic coupling with $q=-10^{-5}$ and $N_0=60$.}\label{phi2infl-srparamN=60q=-5}
\end{figure}

In this Section we study slow-roll inflation in Gauss-Bonnet gravity. Well-known slow-roll conditions for a minimally coupled inflaton, discussed in Section \ref{sec:slowroll}, are  $\dot{\phi}^2\ll V$ and $\vert \ddot{\phi}\vert \ll 3H\vert \dot{\phi}\vert$, or $\epsilon_H$, $\eta_H \ll 1$. In the presence of a Gauss-Bonnet non minimal coupling, in order obtain sufficient accelerated expansion, we need to impose the additional conditions \cite{GSSlowRoll,KohObsCos}:
\begin{equation}
\vert \dot{f}\vert H^3 \ll V, \quad \quad \vert \ddot{f}\vert \ll \vert \dot{f}\vert H.
\end{equation}
Under the above approximations, the equations of motion for $a(t)$, $\phi(t)$ turn out to be:
\begin{equation}\label{srequations}
\begin{cases}
\begin{split}
&H^2\simeq \dfrac{V}{3}\\
&\dot{\phi}\simeq H^3f'-\dfrac{V'}{3H}
\end{split}
\end{cases}
\end{equation}
We will use these approximated equations to find the initial conditions for inflation. We also define a hierarchy of Hubble flow parameters:
\begin{equation}
\epsilon_1=-\dfrac{\dot{H}}{H^2}, \quad \epsilon_{i+1}=\dfrac{d \ln \vert \epsilon_i \vert}{dN}, \quad i\geq 1.
\end{equation}
Hubble flow parameters can be related to the standard slow roll parameters $\epsilon_H$ and $\eta_H$ defined in Section \ref{sec:slowroll}; for example, $\epsilon_1=\epsilon_H$. To account for the new degrees of freedom introduced by the Gauss-Bonnet coupling, Guo and Schwarz \cite{GSSlowRoll} also define an additional hierarchy of flow parameters:
\begin{equation}
\delta_1=-\dot{f}H, \quad \delta_{i+1}=\dfrac{d \ln \vert \delta_i \vert}{dN}, \quad i\geq 1.
\end{equation} 
Then the slow roll condition in Gauss-Bonnet inflation can be comprehensively expressed as:
\begin{equation}\label{GBSR}
\vert \epsilon_i \vert \ll 1 \quad \textrm{and} \quad \vert \delta_i \vert \ll 1.
\end{equation}
The behaviour, in an example of inflationary solution\footnote{Later in this Section we explain how inflationary solutions are obtained.}, of the first four flow parameters is displayed in Figure \ref{phi2infl-srparamN=60q=-5}. We see that the four parameters are in good agreement with conditions \eqref{GBSR} up until the end of inflation ($N\simeq 60$).

Flow parameters can also be expressed in terms of the potential and the Gauss-Bonnet coupling, through the field equations of the theory:
\begin{equation}
\begin{split}
\epsilon_1&=\frac{Q V'}{2 V},\\
\delta_1&=\frac{1}{3}  Q V f',\\
\epsilon_2&=-Q\left(\frac{Q'}{Q}-\frac{V'}{V}+\frac{V''}{V'}\right),\\
\delta_2 &=-Q \left(\frac{f''}{f'}+\frac{Q'}{Q}+\frac{V'}{V}\right),\\
\end{split}
\end{equation}
where
\begin{equation}
Q=\frac{V'}{V}-\frac{1}{3} V f'.
\end{equation}
The $\epsilon$ parameters defined above are equivalent to the potential slow roll parameters of Eq. \eqref{potentialparam} when $f=0$.\\ 
We are now ready to set the suitable initial conditions for an inflating solution, generalizing to Gauss-Bonnet gravity the conditions of Eq.s \eqref{infcd1}, \eqref{infcd2}. Using a standard procedure \cite{PlanckReheating,GSSlowRoll,KohObsCos}, we denote the values of the inflaton at the beginning and end of inflation as $\phi_i$, $\phi_f$, and require:
\begin{equation}\label{srcd1}
\begin{split}
&\epsilon_1 \vert_{\phi_f}=\dfrac{2}{\phi_f^2}\equiv 1,\\
&N=\ln \left( \dfrac{a_f}{a_i} \right)\equiv N_0 .
\end{split}
\end{equation}
We choose to use the flow parameter $\epsilon_1$ to define the end of the inflationary phase. This choice gives a simple condition and is in good agreement with the behaviour of the flow parameters shown in Figure \ref{phi2infl-srparamN=60q=-5}. However, our procedure could be improved by replacing the first condition in \eqref{srcd1} with: $\textrm{max} \lbrace
\vert \epsilon_i \vert, \vert \delta_i \vert \rbrace \equiv 1$.\\
The number of e-foldings, $N$, can be expressed as a function of the initial value of the inflaton, using slow-roll equations \eqref{srequations}:
\begin{equation}
\begin{split}
N&=\int_{a_i} ^{a_f} \dfrac{da}{a}= \int_{t_i} ^{t_f} H dt = 
\int_{\phi_i} ^{\phi_f} \dfrac{H}{\dot{\phi}} d\phi=
\int_{\phi_i} ^{\phi_f} \dfrac{3 H^2}{3H^4f'-V'}d\phi=\\
&=\int_{\phi_i} ^{\phi_f} \dfrac{V}{V^2f'/3-V'}d\phi.
\end{split}
\end{equation} 
To sum up, initial conditions for an inflationary solution in Gauss-Bonnet theory are:
\begin{equation}\label{icinfl1}
a_0=1, \quad \dot{\phi}_0=0, \quad H_0=+\sqrt{\dfrac{V(\phi_i)}{3}},\\
\end{equation} 
where $\phi_f$ is defined by $\dfrac{2}{\phi_f^2}\equiv 1$, and $\phi_i$ can be extracted from:
\begin{equation}\label{icinfl2}
\int_{\phi_i} ^{\phi_f} \dfrac{V}{V^2f'/3-V'} d\phi\equiv N_0.
\end{equation}
In some cases, the integral above can be solve analytically and, therefore, the value of $\phi_i$ is known exactly. In the other cases $\phi_i$ is computed numerically.\\

The outlined procedure allows to find viable inflationary solutions for both coupling functions, as well as both signs of the coupling parameter $\lambda$. We will not need to distinguish between the two branches of solutions as in the previous Chapters, because inflationary initial conditions will automatically pick up one, according to the values of the parameters.\\

\subsection{Phenomenological predictions}\label{sec:phenompred}
We now study the phenomenological implications of an inflationary solution in Gauss-Bonnet gravity. Studying the effects of a Gauss-Bonnet inflation on the large scale structure of the Universe will allow, through the comparison with experimental data, to constraint the model and its parameters.\\
We focus on scalar and tensor perturbations, deriving their power spectra at linear order in perturbation theory. We follow the steps described in Sections \ref{sec:qflGR}, \ref{sec:psGR} for General Relativity, summarizing  the theoretical results of papers \cite{PerturbInfl,AltriInflPerturb2,AltriInflPerturb,KohObsCos,JHGGBInflation,GSSlowRoll,GSPerturb}.\\
We first recall the definition of the Fourier modes of curvature $\mathcal{R}$ and tensor $h_{ij}$ perturbations:
\begin{equation}
\begin{split}
\mathcal{R}(\tau,\vec{x})&= \dfrac{1}{z_{\mathcal{R}}} \int \; \dfrac{d^3k}{(2\pi)^{3/2}}v_{\mathcal{R}}(\tau,\vec{k})e^{i\vec{k} \cdot \vec{x}},\\
h_{ij}(\tau,\vec{x})&= \dfrac{2}{z_T}  \sum_{\lambda} \int \; \dfrac{d^3k}{(2\pi)^{3/2}}v^{\lambda}_T(\tau,\vec{k}) \epsilon_{\lambda ij} e^{i\vec{k} \cdot \vec{x}},\\
\end{split}
\end{equation}
where $\tau$ is the conformal time, introduced in Eq. \eqref{conft}, $\epsilon_{ij}$ are polarization tensors.
The linearized classical evolution of these Fourier modes can be derived by inserting the perturbed metric into Eq.s \eqref{eqfV} or perturbing action \eqref{actionV} to second order. Curvature modes are found to satisfy:
\begin{equation}\label{inflReq}
v''_{\mathcal{R}}+\left( c^2_{\mathcal{R}}k^2 -\dfrac{z''_{\mathcal{R}}}{z_{\mathcal{R}}} \right)v_{\mathcal{R}}=0,
\end{equation}
where a prime denotes differentiation with respect to conformal time, and $z_{\mathcal{R}}$ and $c_{\mathcal{R}}$ are given by:
\begin{equation}
\begin{split}
z^2_{\mathcal{R}}&=\dfrac{a^2\left( \dot{\phi}^2-3/2 \Delta \dot{f}H^3 \right)}{H^2 \left( 1-1/2 \Delta \right)^2},\\
c^2_{\mathcal{R}}&=1-\dfrac{2\Delta \dot{f} H \dot{H} +1/2 \Delta^2 H^2 (\ddot{f}-\dot{f}H)}{\dot{\phi}^2-3/2 \Delta \dot{f}H^3},
\end{split}
\end{equation}
with $\Delta=\dfrac{-\dot{f}H}{\alpha}$.\\
The Fourier modes of tensor perturbations satisfy a similar equation:
\begin{equation}\label{inflTeq}
v''_{T}+\left( c^2_{T}k^2 -\dfrac{z''_{T}}{z_{T}} \right)v_{T}=0,
\end{equation}
where
\begin{equation}
\begin{split}
z^2_{T}&=a^2 \alpha ,\\
c^2_{T}&= c^2_s= \dfrac{1+\ddot{f}}{\alpha}.
\end{split}
\end{equation}
Both tensor and curvature modes satisfy wave equations with an effective frequency shift $c$ and and effective mass $\frac{z''}{z}$, and are coupled to the scalar field. Both the frequency shift and the effective mass differ from the General Relativity case; see for example Eq. \eqref{DSM}, for scalar modes.\\
In order to solve \eqref{inflReq}, \eqref{inflTeq}, one shall assume that slow-roll parameters are time independent during inflation. This assumption is not completely satisfactory (see Fig. \ref{phi2infl-srparamN=60q=-5}), but is the only one known to provide an analytical solution.\\
The power spectrum of large scale perturbations can be calculated as the Fourier transform of the two-point correlation function of the quantum operators associated to the perturbed variables, in the spirit of inflation. All perturbations of interest cross the horizon soon after $\phi=\phi_i$, so we assume $c k \ll aH$. Using these approximations, one can write down the spectral indices of scalar and tensor perturbations and the tensor-to-scalar ratio, to lowest order in the slow-roll parameters:
\begin{equation}\label{approxparam}
\begin{split}
n_s-1 &= \dfrac{d \ln \mathcal{P}_{\mathcal{R}}}{d \ln k}  \simeq -\frac{2 \epsilon_1 \epsilon_2-\delta_1 \delta_2}{2 \epsilon_1-\delta_1}-2 \epsilon_1 ;\\
n_T &= \dfrac{d \ln \mathcal{P}_T}{d \ln k} \simeq -2 \epsilon_1 ;\\
r&= \dfrac{d \ln \mathcal{P}_T}{d \ln \mathcal{P}_{\mathcal{R}}}  \simeq 8 \left| 2 \epsilon_1-\delta_1 \right| .
\end{split}
\end{equation} 
When slow-roll inflation is implemented in GR, i.e. when $f=0$, the tensor spectral index and the tensor-to-scalar ratio are not independent quantities, but are related by $r=-n_T/2$ (see Section \ref{sec:qflGR}). This so-called \emph{consistency relation} is evidently broken when the Gauss-Bonnet correction is taken into account. We see then that the experimental check of the consistency relation could be an important test for slow-roll inflation in modified gravity theories. Moreover, the loss of degeneracy must be taken into account when analysing CMB data, as one should let $n_T$ vary independently of the tensor-to-scalar ratio. It has been verified that relaxing the consistency relation can lead to slightly different bounds on $n_s$ and $r$ from CMB data \cite{JHGGBInflation}. 

\section{Numerical inflationary solutions}\label{sec:numsolinfl}
Slow-roll inflation with a quadratic potential and an exponential coupling has been studied in \cite{JHGGBInflation}, with the additional parameter $\beta$, mainly comparing predictions on $n_R$ and $r$ with PLANCK data. The case of quadratic potential and coupling is taken into account in \cite{KohObsCos} and, more recently, in \cite{PlanckReheating}.\\
Our numerical results are reviewed in the following sections. A standard constraint on inflationary solutions is $N_0 \gtrsim 50$, required to solve the horizon and flatness problems of standard cosmology.\\
A general result is the absence of nonsingular, inflationary solutions, whether the coupling is quadratic or exponential, positive or negative. Although we have checked this statement numerically only in the explored range of q, we expect such behaviour to be valid in general.

\subsection{Quadratic coupling}
\begin{figure}[!ht]
 \centering
    \includegraphics[width=0.7\textwidth]{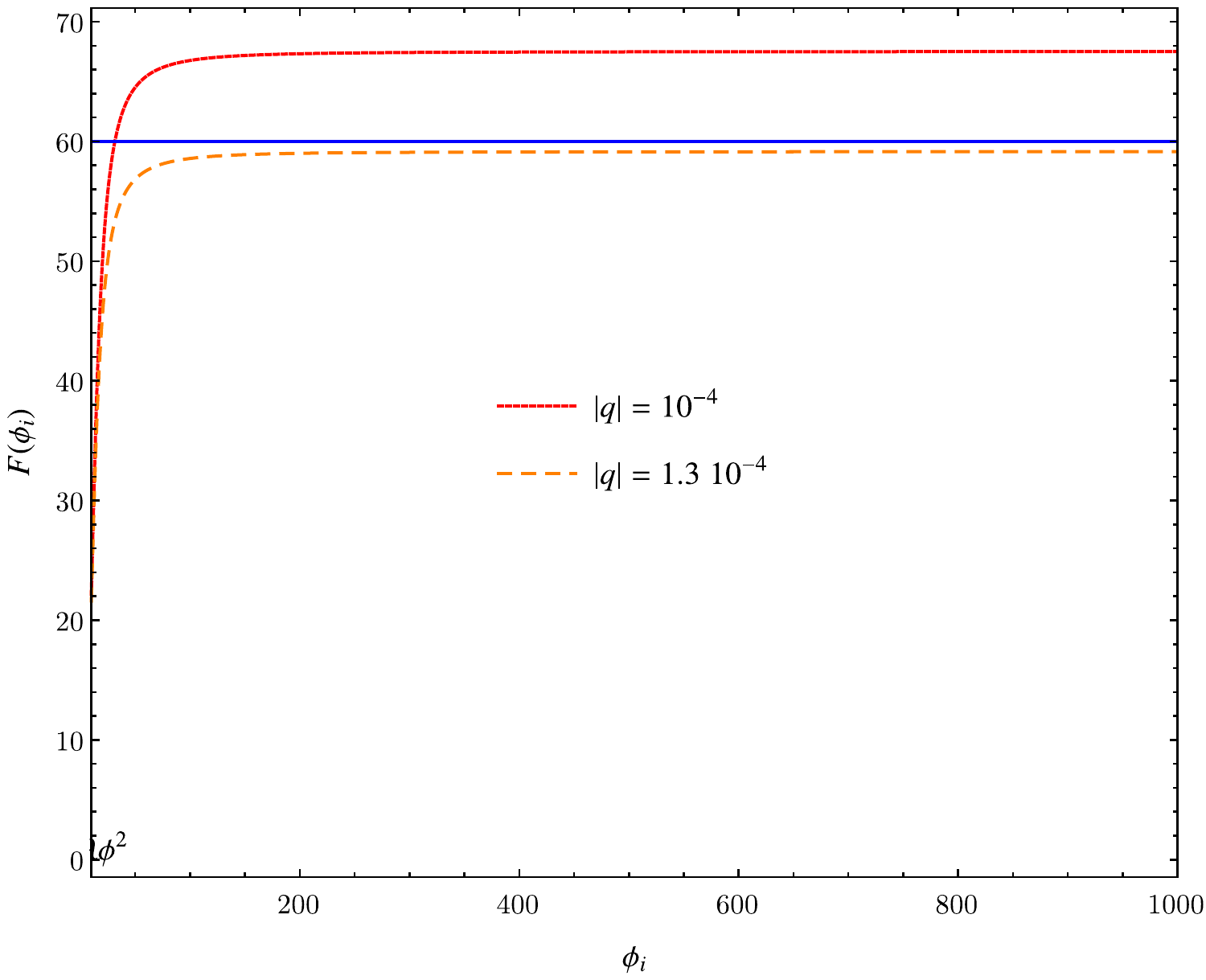}
    \caption{Graphic representation of equation \ref{mN0}. 
    	Intersection with $ F(\phi_i) = N_0 = 60 $ is found for $\vert q \vert=10^{-4}$, while no intersection can be found for $\vert q \vert=1.3$. We also see that, for growing |q|, intersection moves to bigger values of $ \phi_i $, thus justifying our approximation \eqref{limitphi}.}\label{phi2INFL-mmax12N=60}
\end{figure}

With a quadratic coupling, inflationary acceptable solutions ($N_0 \gtrsim 50$) are achievable only if $\vert q \vert \lesssim q_{max}$. To understand why, we explicitly write the slow-roll condition \eqref{icinfl2}. We find $\phi_f= \pm \sqrt{2}$, while $\phi_i$ is implicitly given by:
\begin{equation}\label{mN0}
\begin{split}
F\left( \phi_i \right)&=\frac{\sqrt{\frac{3}{10}} \tan ^{-1}\left(b \; m \phi_i^2\right)}{4 m}-\frac{\sqrt{\frac{3}{10}} \tan ^{-1}(\phi_f^2 b \; m)}{4 m}=\\
&=\frac{\sqrt{\frac{3}{10}} \tan ^{-1}\left(b \; m \phi_i^2\right)}{4 m}-\frac{\sqrt{\frac{3}{10}} \tan ^{-1}(2 b \; m)}{4 m}=N_0,
\end{split}
\end{equation}
with $b=\sqrt{10/3}$ and $\delta=\pm 1$, the sign of $\phi_f$, $\phi_i$ turning out to be irrelevant. Numerically solving the equation above for $m$ approaching its maximum value, we discover that $\phi_i$ grows. Than $m_{max}$ will be reached for $\phi_i \rightarrow \infty$, and we can substitute the asymptotic limit:
\begin{equation}\label{limitphi}
\tan^{-1} \left( b\; m \; \phi_i^2\right)
= \dfrac{\pi}{2} + O\left( \dfrac{1}{\phi_i^2} \right)
\end{equation}
Moreover, expecting $m_{max}$ to be small, we further approximate:
\begin{equation}
\tan^{-1}\left( 2b\; m \right) \simeq 2b \; m + O\left(m^3 \right)
\end{equation} 
Under the approximations above, we can finally analytically solve \eqref{mN0}, finding:
\begin{equation}
m_{max}\simeq \frac{\sqrt{30} \; \pi }{80 \; N_0+40} .
\end{equation} 
For $N_0 = 60$, we get $m_{max} \simeq 3.6 \times 10^{-3}$, while for $N_0=70$ we find $m_{max} \simeq 3.1 \times 10^{-3}$. A numerical study gives a more precise bound: for $N_0=60$, the value we actually choose for our stability analysis, we find $m_{max} \simeq 1 \times 10^{-2}$, while for $N_0=70$ we find $m_{max} \simeq 7 \times 10^{-3}$. If we refer to the parameter $q$, then:
\begin{equation}
\vert q \vert_{max}\simeq 10^{-4} \textrm{  when  } N_0=60, \quad \vert q \vert_{max}\simeq 5 \times 10^{-5} \textrm{  when  } N_0=70 .
\end{equation}
Representative plots are shown in Figure \ref{phi2INFL-mmax12N=60}. Similar constraints are found in \cite{KohObsCos}.\\

\begin{figure}[!ht]
 \centering
 \begin{subfigure}{0.48\textwidth}
    \includegraphics[width=1\textwidth]{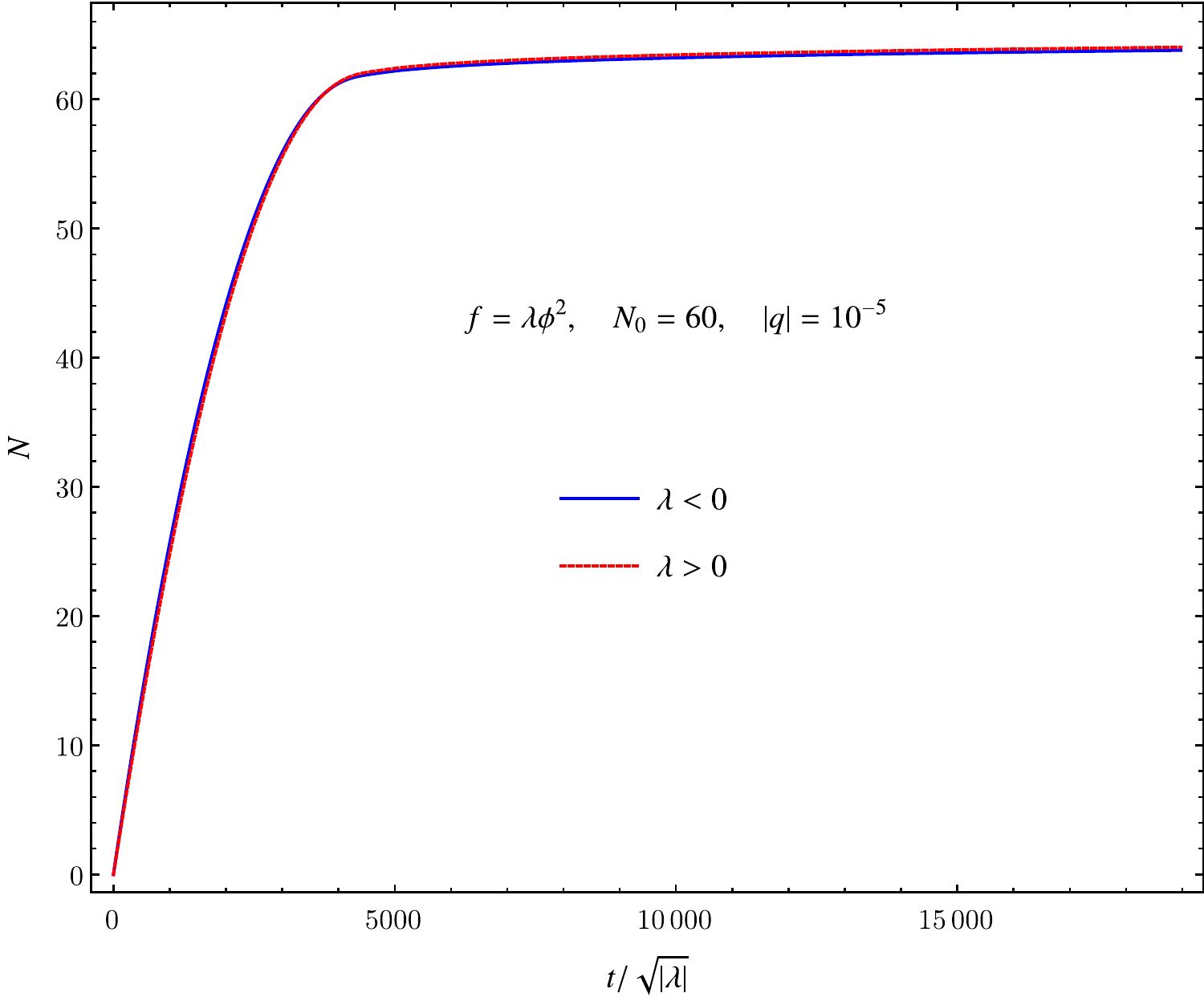}
\end{subfigure}
\begin{subfigure}{0.48\textwidth}
    \includegraphics[width=1\textwidth]{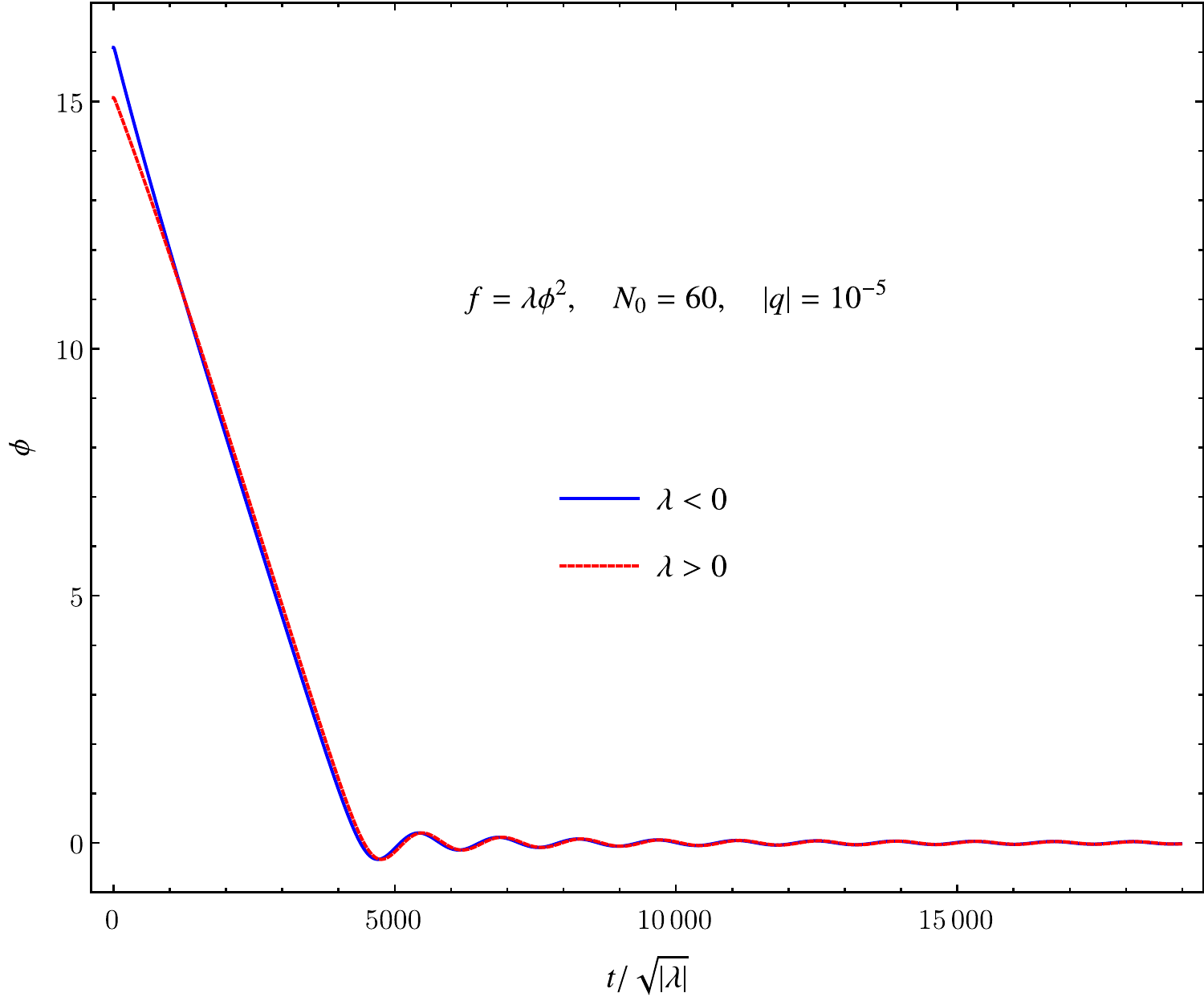}
    \end{subfigure}
    \caption{The number of e-foldings (left) and the scalar field (right) as a function of rescaled time in an inflationary solution with $q=\mp10^{-5}$ and $N_0=60$.}\label{phi2INFL-Nk=0N=60q=-+5}
\end{figure}

In Figure \ref{phi2INFL-Nk=0N=60q=-+5} we show two typical inflationary solutions, for positive and negative coupling, found imposing the initial conditions \eqref{icinfl1} and \eqref{icinfl2}, with $N_0=60$. Our results coincide with the ones recently found in \cite{PlanckReheating}. The scalar field decreases monotonically and then oscillates at frequency $ \omega \simeq \sqrt{2m^2} $ around $\phi=0$. The scale factor reaches the desired number of e-foldings around $t \simeq 20/ m$ and then exits the inflationary phase. 

\subsection{Exponential coupling}
\label{sec:expinfla}
For the exponential coupling, too, suitable inflationary conditions can be found only if $\vert q \vert \lesssim q_{max}$. Unlike the previous case, however, we are not able to prove it analytically, because the indefinite integral in \eqref{icinfl2} is unknown. Numerical, approximated conditions are:
\begin{equation}
\vert q \vert_{max} \simeq 6 \quad (N_0=60), \quad \vert q \vert_{max}\simeq 4 \quad  (N_0=70). 
\end{equation}  

\begin{figure}[!ht]
 \centering
 \begin{subfigure}{0.48\textwidth}
    \includegraphics[width=1\textwidth]{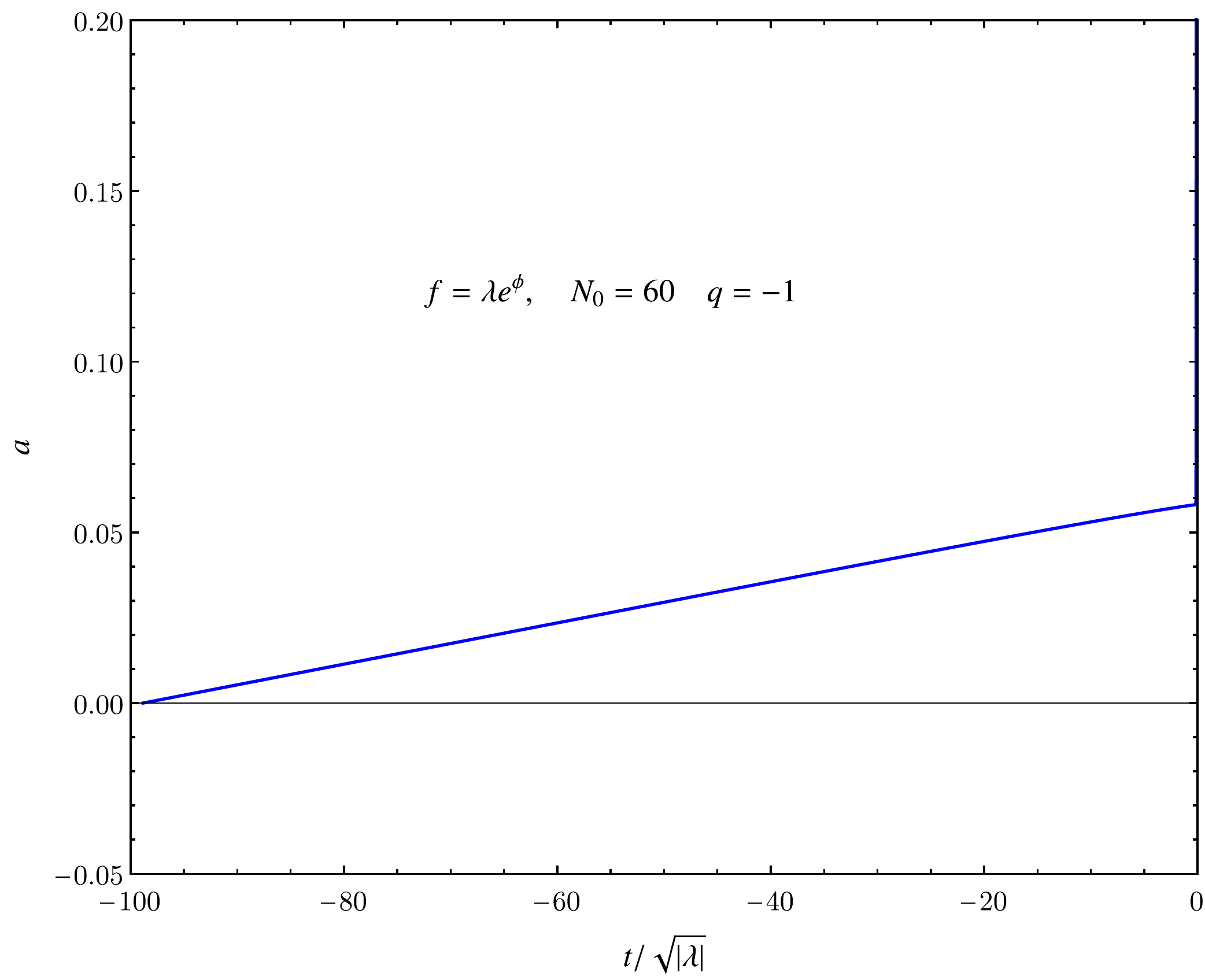}
\end{subfigure}
\begin{subfigure}{0.48\textwidth}
    \includegraphics[width=1\textwidth]{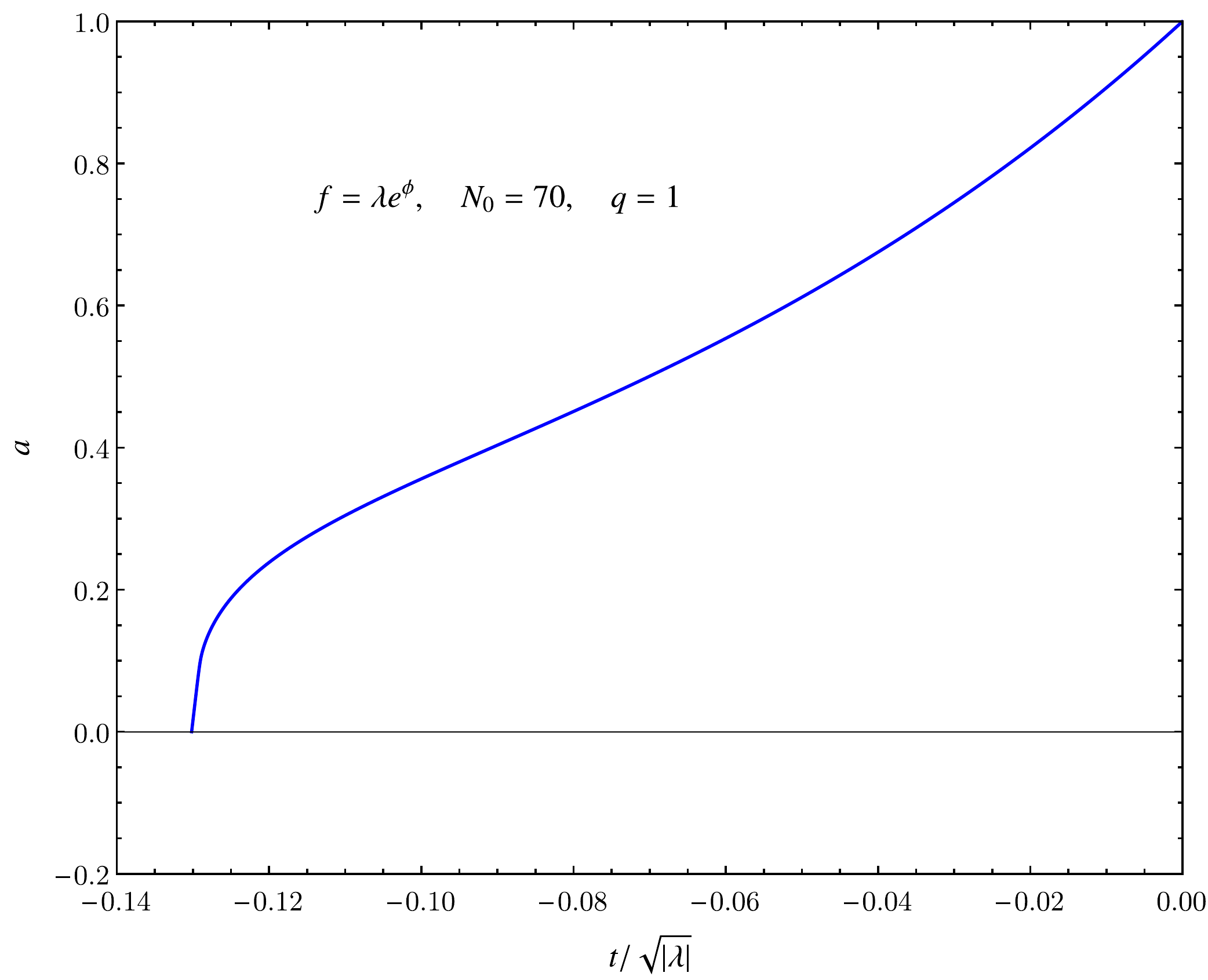}
    \end{subfigure}
    \caption{The scale factor near the singularity, for exponential coupling with $q=-1$ and $N_0=60$ (left) or $q=1$ and $N_0=70$ (right). The plots show the evolution of the scale factor before the inflationary expansion takes place (N<0).}\label{ephiINFL-aN=70q=1}
\end{figure}

Whether the coupling $\lambda$ is positive or negative, inflationary solutions possess an initial singularity, though they approach it quite differently (see the comparison in Figure \ref{ephiINFL-aN=70q=1}). The range $10^{-9} < \vert q \vert < q_{max}$ was explored for both $N_0=60$ and $N_0=70$, but nonsingular solutions could not be found.\\

\begin{figure}[!ht]
 \centering
 \begin{subfigure}{0.48\textwidth}
    \includegraphics[width=1\textwidth]{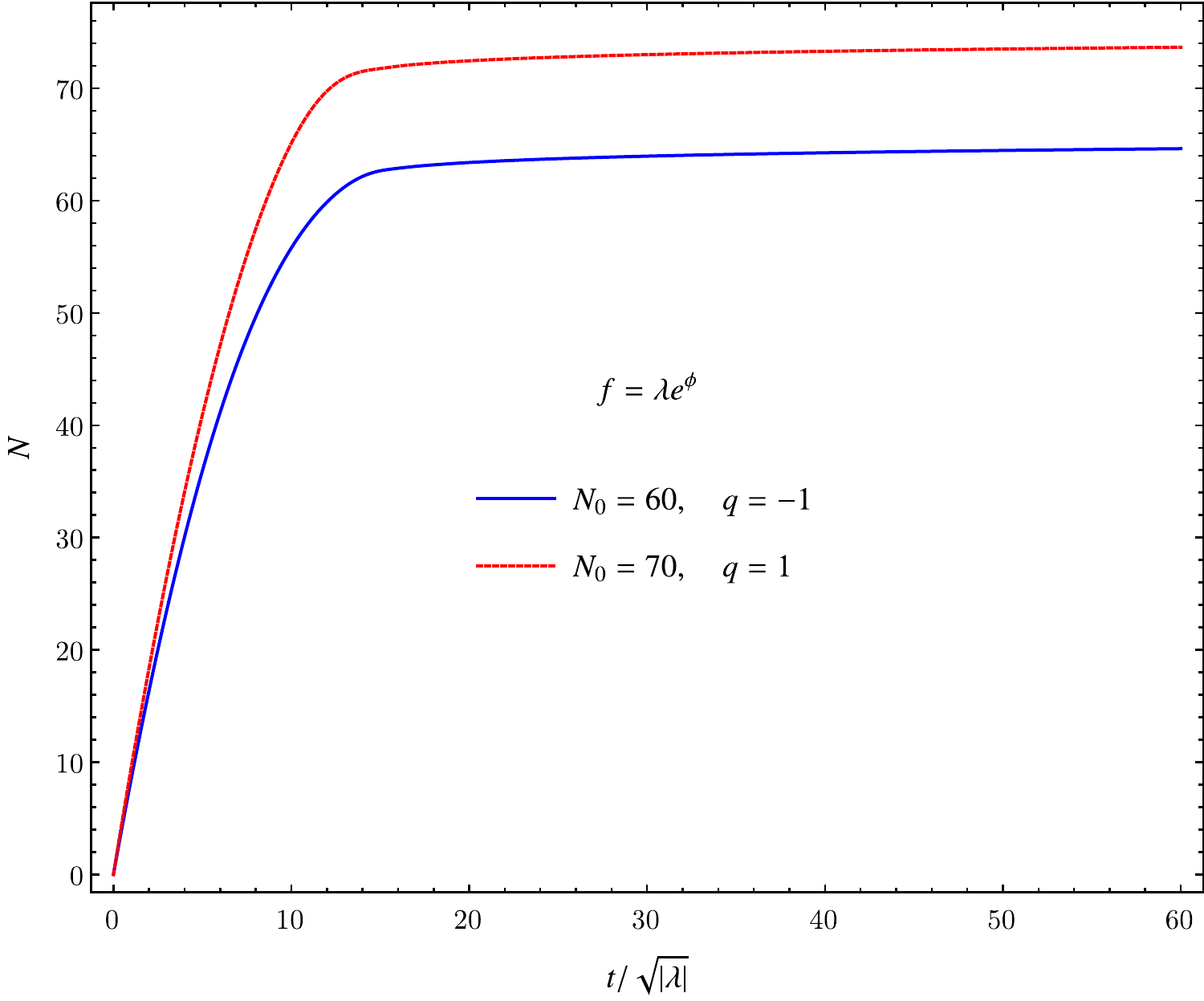}
\end{subfigure}
\begin{subfigure}{0.48\textwidth}
    \includegraphics[width=1\textwidth]{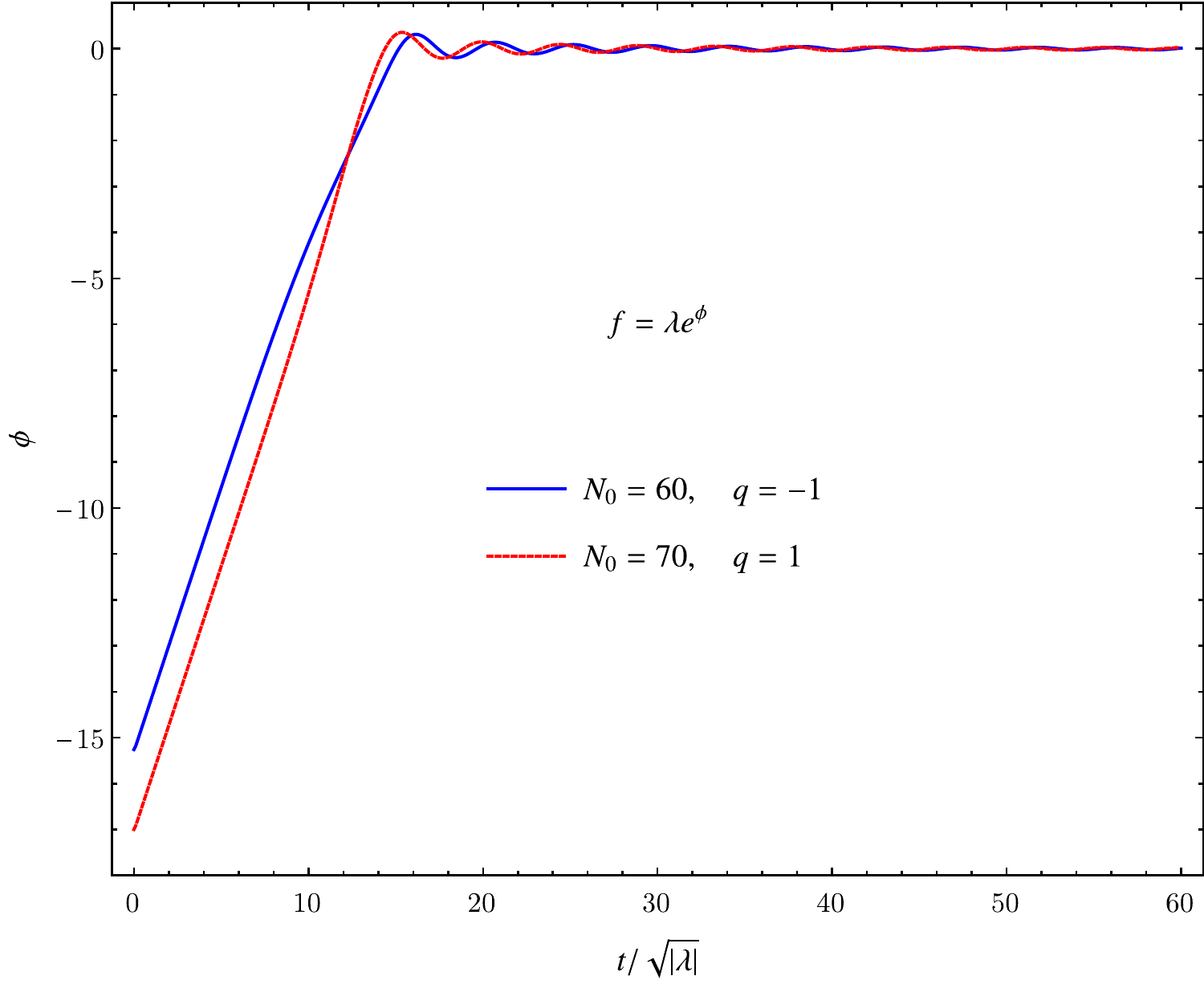}
    \end{subfigure}
    \caption{The number of e-foldings (left) and the scalar field (right) as a function of rescaled time in an inflationary solution for quadratic coupling with $q=\mp1$ and $N_0=60,70$.}\label{ephiINFL-phiN=60q=-1}
\end{figure}

The number of e-foldings and the scalar field through the inflationary phase, $t>0$, are displayed in Figure \ref{ephiINFL-phiN=60q=-1} for different signs of $q$ and different values of the parameter $N_0$. As in the quadratic case, the scalar field exhibits a monotonic
stage, and then shows damped oscillations with frequency determined by the mass, $ \omega\simeq \sqrt{2m^2} $.
 
\clearpage
\section{Stability of inflationary solutions}\label{stabinfla}
We complete our study of Gauss-Bonnet gravity with a massive scalar by analyzing stability of tensor perturbations on inflationary background, in first order perturbation theory. Stability of inflationary solutions in Gauss-Bonnet cosmology with a potential was also studied by Leith and Neupane in \cite{LeiNeuGBcosmologies}, but considering different coupling functions and/or different potentials, and constructing analytical approximated solutions rather than performing an exact numerical analysis, as we did. \\ 
Results are summarized in Table \ref{table:stabInfl} and described in detail in the following Sections. \\
\begin{table}[H]
\centering
\begin{tabular}{|c|c|c|c|}
\hline
Coupling&$N_0$&$\lambda$&Stable\\ \hline
\multirow{4}{*}{$\phi^2$}&\multirow{2}{*}{70}&-1& $q\gtrsim 10^{-5}$\\ 
& &+1&$\nexists \; q$\\ 
&\multirow{2}{*}{60}&-1&$q\gtrsim 10^{-5}$\\ 
& &+1&$\nexists \; q$\\ \hline
\multirow{2}{*}{$e^{\phi}$}&\multirow{2}{*}{60}&-1&$\nexists \; q$\\ 
& &+1&$\forall \; q$\\ \hline
\end{tabular}
\vspace{2mm}
\caption{Numerical stability analysis for inflationary solutions, referred to the range of $q\neq 0$ explored.}
\label{table:stabInfl}
\end{table}
Solutions marked as unstable exhibit instability at early times, \emph{before inflationary expansion takes place}, and are completely stable and non-superluminal during inflation. Such an instability would generally be discarded rather carelessly, claiming it takes place in an epoch where quantum gravity effects dominate. However, Gauss-Bonnet theory should at least partly take into account such effects, if we interpret it as an effective string theory model, and its predictions should be trusted up to a scale very close to the Planck scale. Thus, the issue of pre-inflationary instability requires, in this case, some attention. It is possible that some solutions, here marked as unstable, are in fact stable in the region of time where Gauss-Bonnet theory is valid. This solution could be phenomenologically viable in their region of validity. On the other hand, solutions that become unstable immediately before inflation takes place are, in our opinion, to be rejected. We wish to further address this issue in a future work.\\
All the solutions encountered possess a stage of superluminal propagation at early times. As anticipated in Section \ref{sec:GBstability}, we will not reject solutions on this basis, and we plan to address this issue in the future as well.\\   
 
\begin{figure}[!ht]
 \centering
    \includegraphics[width=0.8\textwidth]{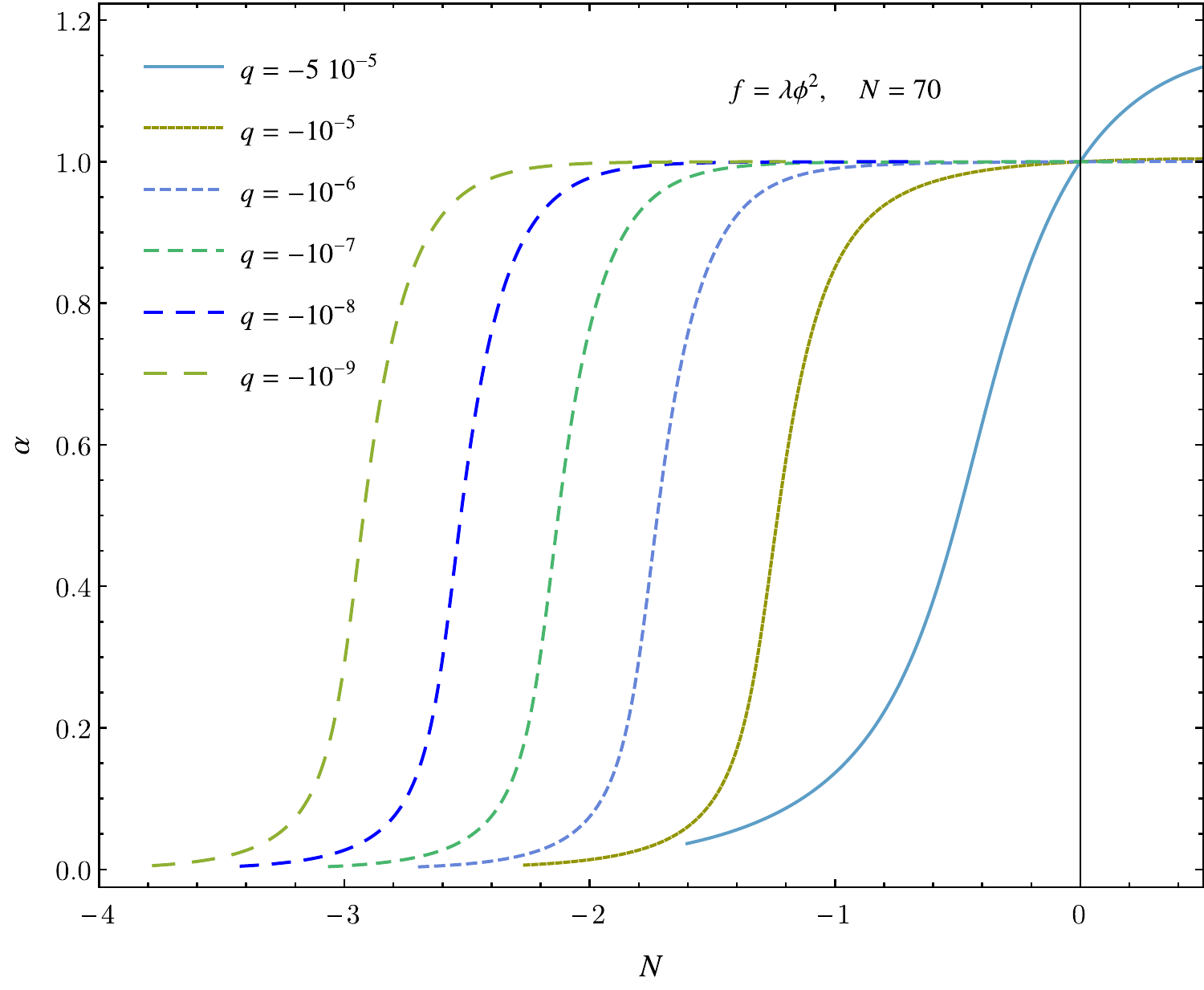}
    \caption{The ghost factor $\alpha$ as a function of the number of e-foldings in inflationary solutions with $q<0$ and $N_0=70$.}\label{phi2INFL-alphaN=70}
\end{figure}

In the following sections we report results for $N_0=60$ mainly, because the number of e-foldings does not affect stability significantly (compare, for example, Figures \ref{phi2INFL-csN=70}). Our analysis only allowed to explore a finite range of $q$s.\\
During our analysis, we always calculate the quantity $\alpha$. However, we find it to be positive for every solution: this means that no ghosts are ever present. We show here one plot only (Figure \ref{phi2INFL-alphaN=70}) as $\alpha$ does not change its behaviour significantly with the coupling or the value of $q$. The other quantity playing a role in the stability of the solution, the friction term, is found to be always positive, whenever the square sound velocity is positive as well.\\

\subsection{Quadratic coupling} 
We study the stability of inflationary scenarios within the range of values, for $N_0=60$ (or $N_0=70$):
\begin{equation}\label{qrange}
10^{-9} < \vert q \vert < 1.2 \times 10^{-4} \quad (5 \times 10^{-5}).
\end{equation}

\begin{figure}[!ht]
 \centering
 \begin{subfigure}{0.48\textwidth}
    \includegraphics[width=1\textwidth]{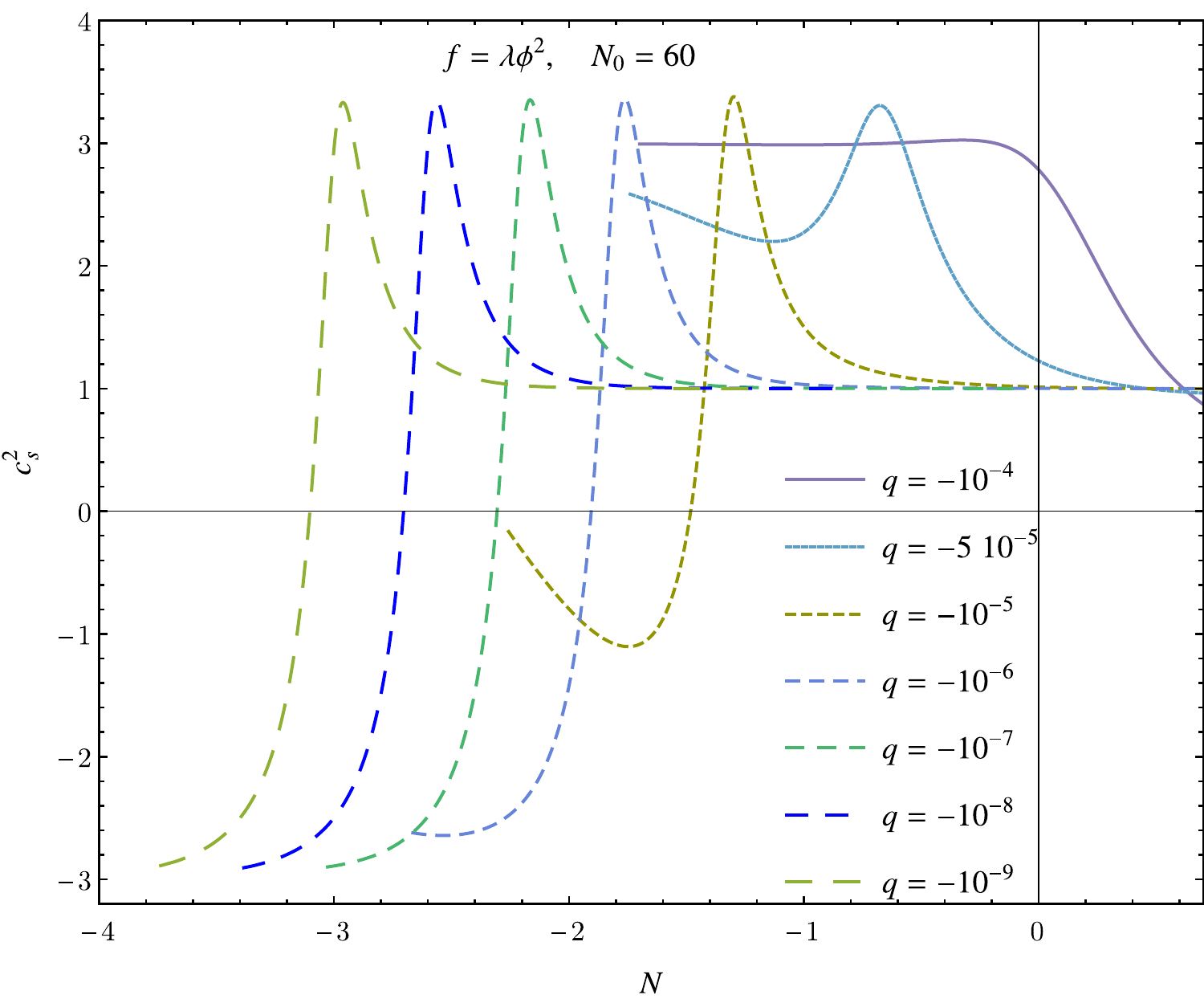}
\end{subfigure}
\begin{subfigure}{0.48\textwidth}
    \includegraphics[width=1\textwidth]{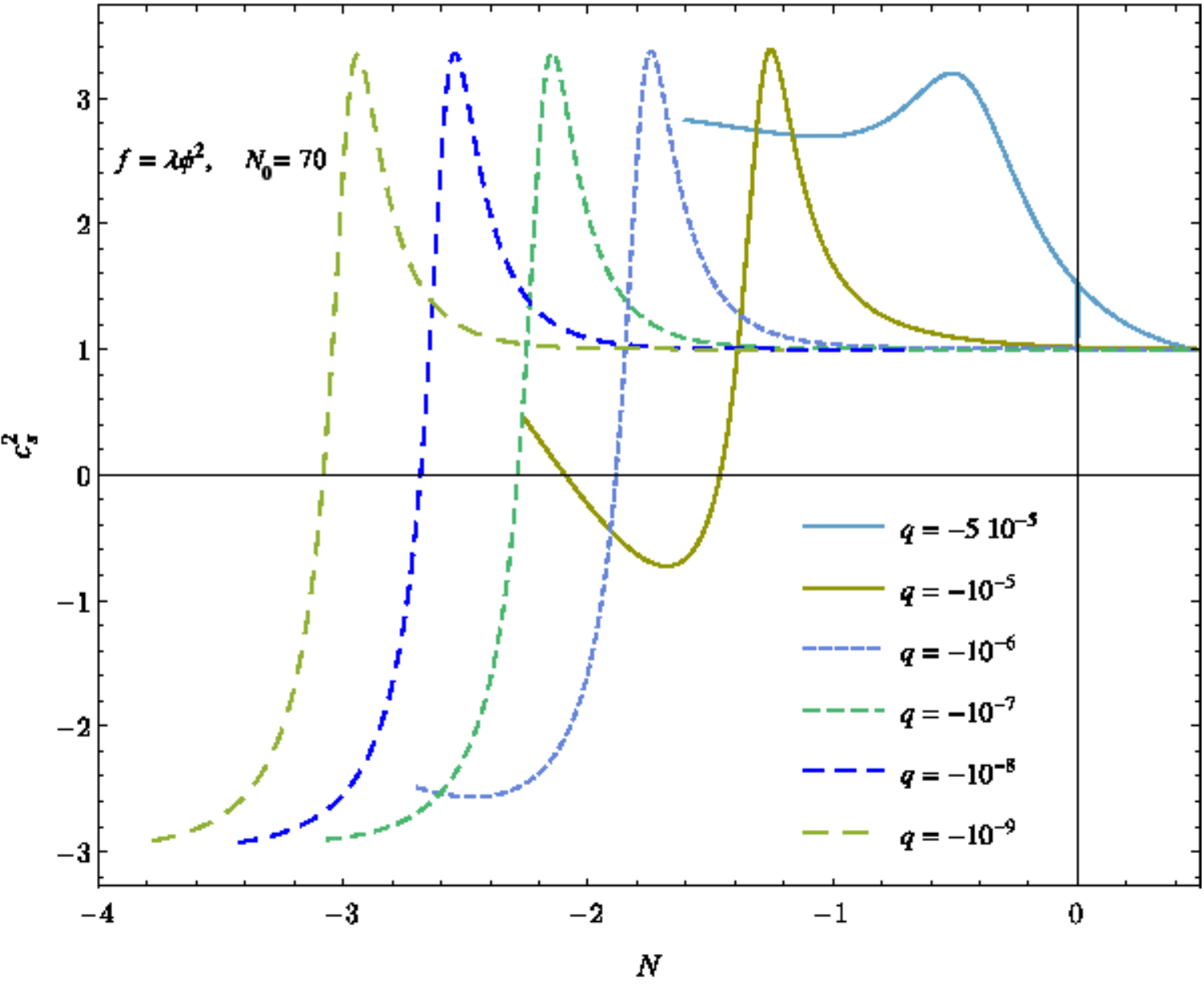}
    \end{subfigure}
    \caption{The squared sound velocity of tensor perturbations as a function of the number of e-foldings in inflationary solutions with $q<0$ and $N_0=60$ (left) or $N_0=70$ (right).}\label{phi2INFL-csN=70}
\end{figure}

\begin{figure}[!ht]
 \centering
    \includegraphics[width=0.8\textwidth]{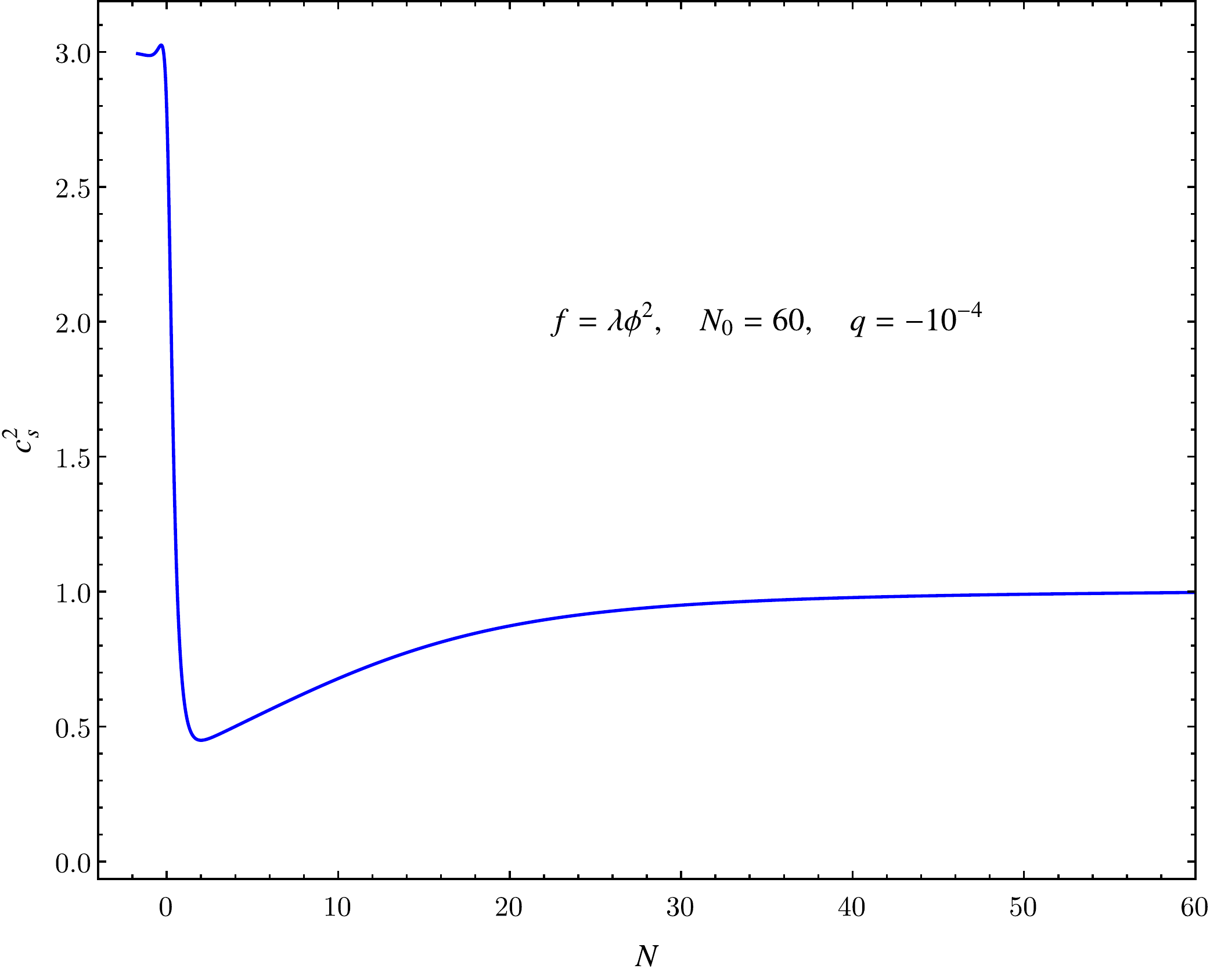}
    \caption{Complete evolution of the squared sound velocity of tensor perturbations as a function of the number of e-foldings in an inflationary solution with $q=-10^{-4}$ and $N_0=60$.}\label{phi2INFL-csN=60q=-4tot}
\end{figure}

\begin{figure}[!ht]
 \centering
    \includegraphics[width=0.8\textwidth]{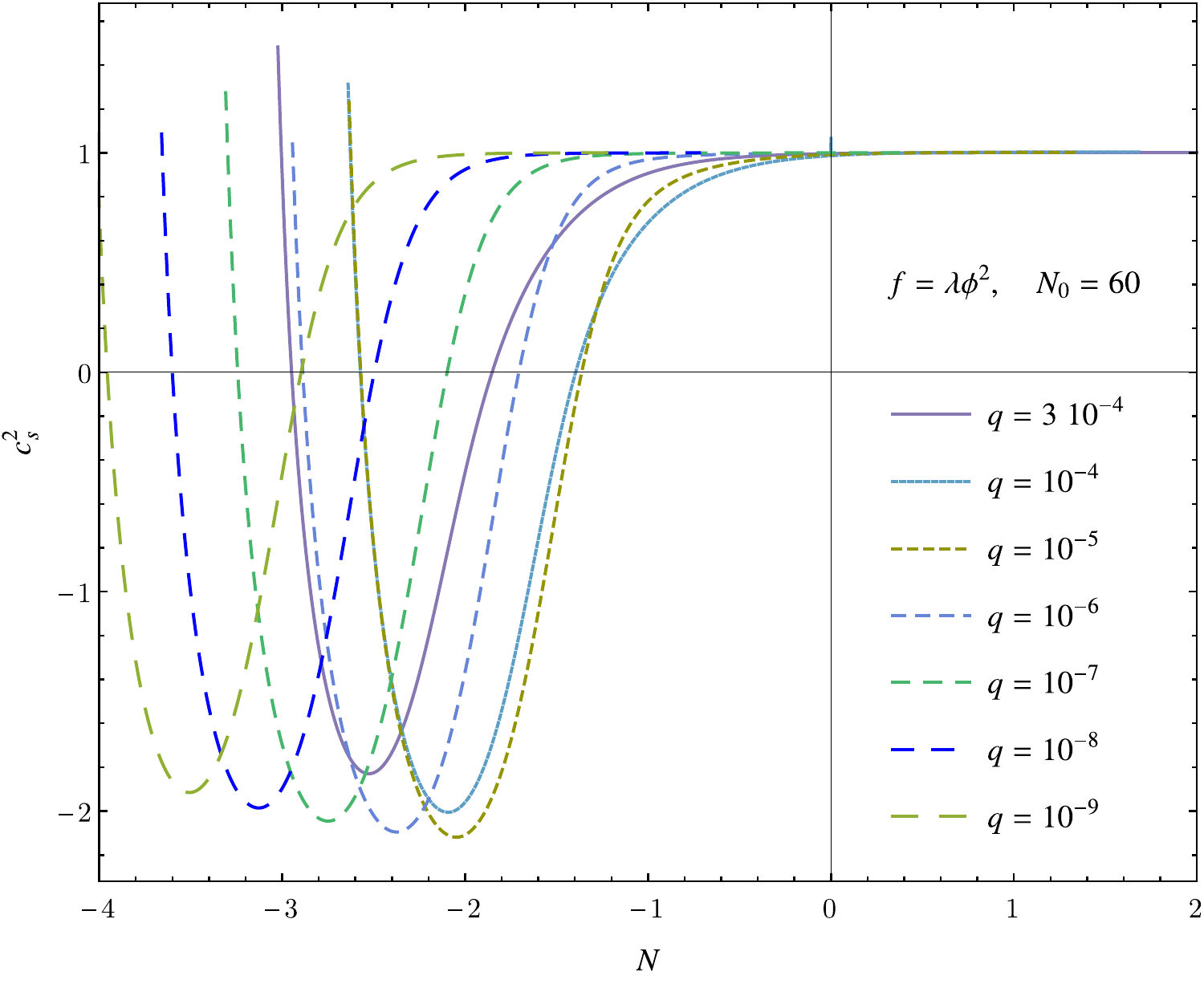}
    \caption{The squared sound velocity of tensor perturbations as a function of the number of e-foldings in inflationary solutions with $q>0$ and $N_0=60$.}\label{phi2INFL-csN=60d=1}
\end{figure}

\begin{figure}[!ht]
 \centering
 \begin{subfigure}{0.48\textwidth}
    \includegraphics[width=1\textwidth]{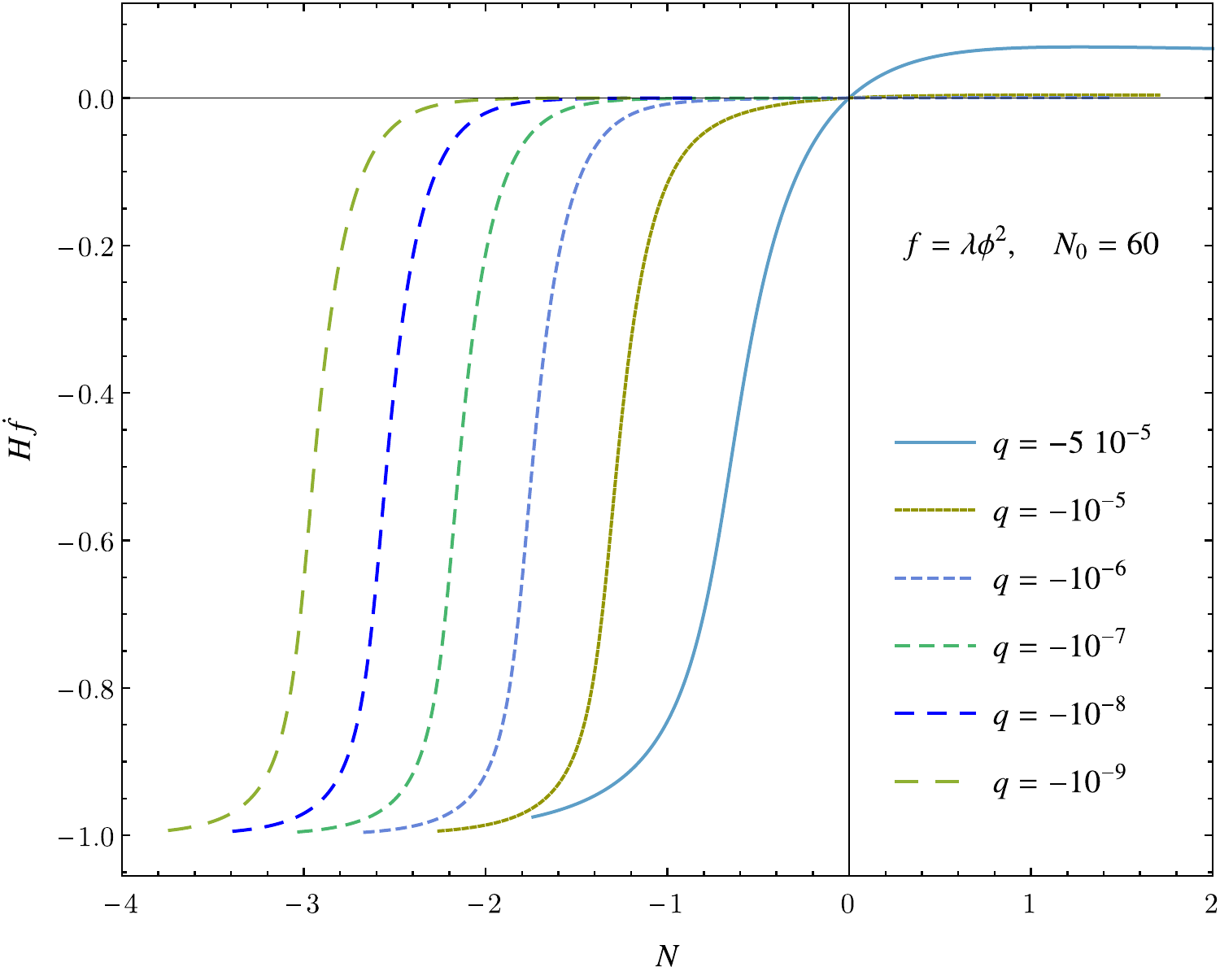}
\end{subfigure}
\begin{subfigure}{0.48\textwidth}
    \includegraphics[width=1\textwidth]{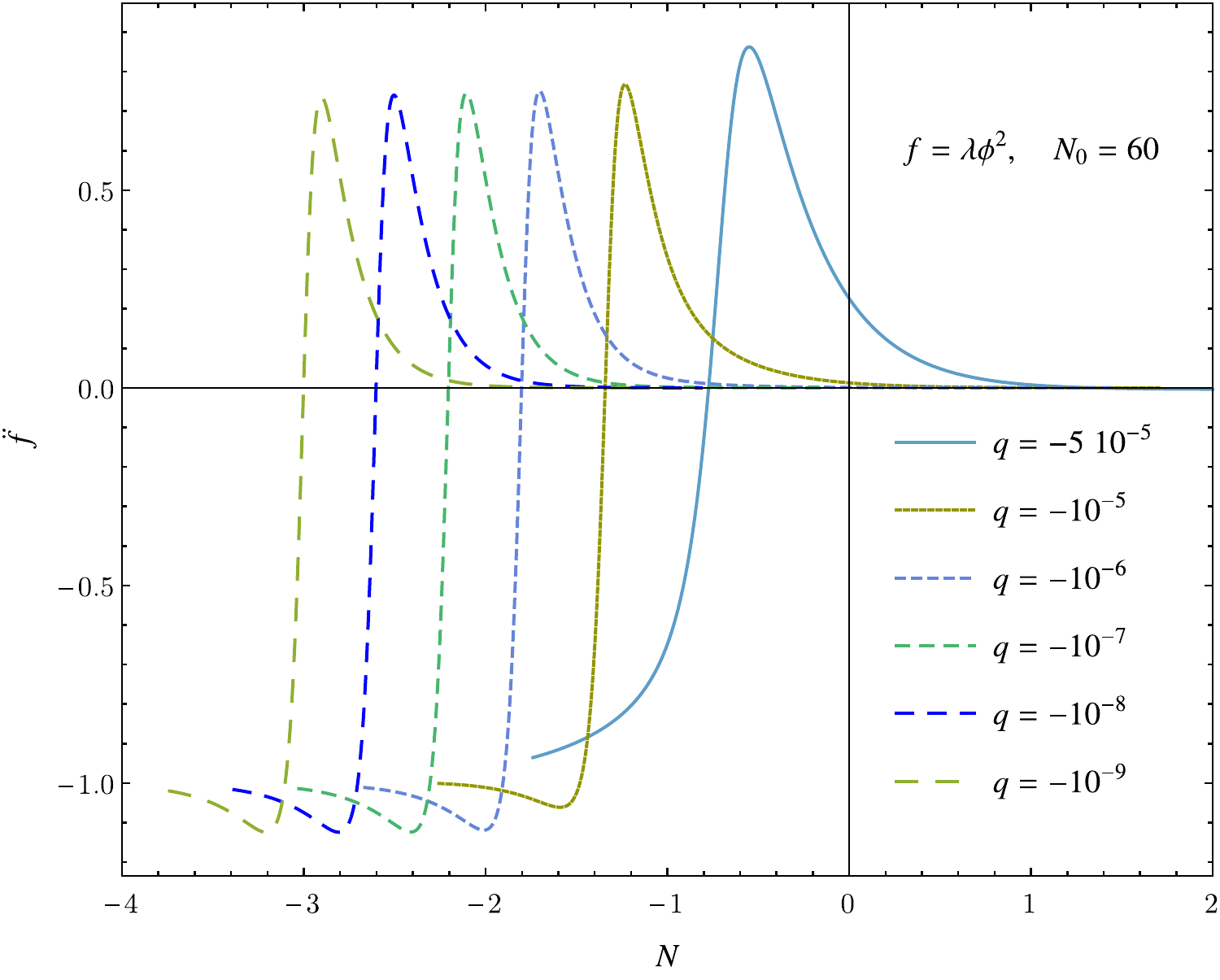}
    \end{subfigure}
    \caption{The $H\dot{f}$ (left) and $\ddot{f}$ (right) terms as functions of the number of e-foldings for inflationary solutions with $q<0$ and $N_0=70$.}\label{phi2INFL-cd2N=60}
\end{figure}

Tensor perturbations of inflationary solutions with both negative quadratic coupling appear to be unstable and superluminal near the singularity (Figure \ref{phi2INFL-csN=70}) as far as $\vert q \vert \lesssim 10^{-5}$. In Figure \ref{phi2INFL-csN=60q=-4tot} we also display the complete evolution of the squared sound velocity for a stable, inflationary solution, showing that it does not become negative during inflation. When the coupling is positive, conversely, we couldn't find stable solutions within the allowed range of $q$ (Figure \ref{phi2INFL-csN=60d=1}).\\
For the negative coupling, we also show the terms associated with the Gauss-Bonnet invariant appearing in equations \ref{eqfV}, \ref{eqf}. Both $H \dot{f}$ and $\ddot{f}$ are order one near the singularity and during the unstable evolution (figure \ref{phi2INFL-cd2N=60}). This means, on the one hand, that we can safely ascribe the instability to the Gauss-Bonnet modification. On the other hand, we don't expect higher order terms arising in effective string-theory to be relevant here: they shouldn't be able to significantly improve inflationary solution's stability.  

\clearpage
\subsection{Exponential coupling}\label{stabexponinfl}
\subsubsection*{Negative coupling, $\boldsymbol{q<0}$}
\begin{figure}[!ht]
 \centering
    \includegraphics[width=0.8\textwidth]{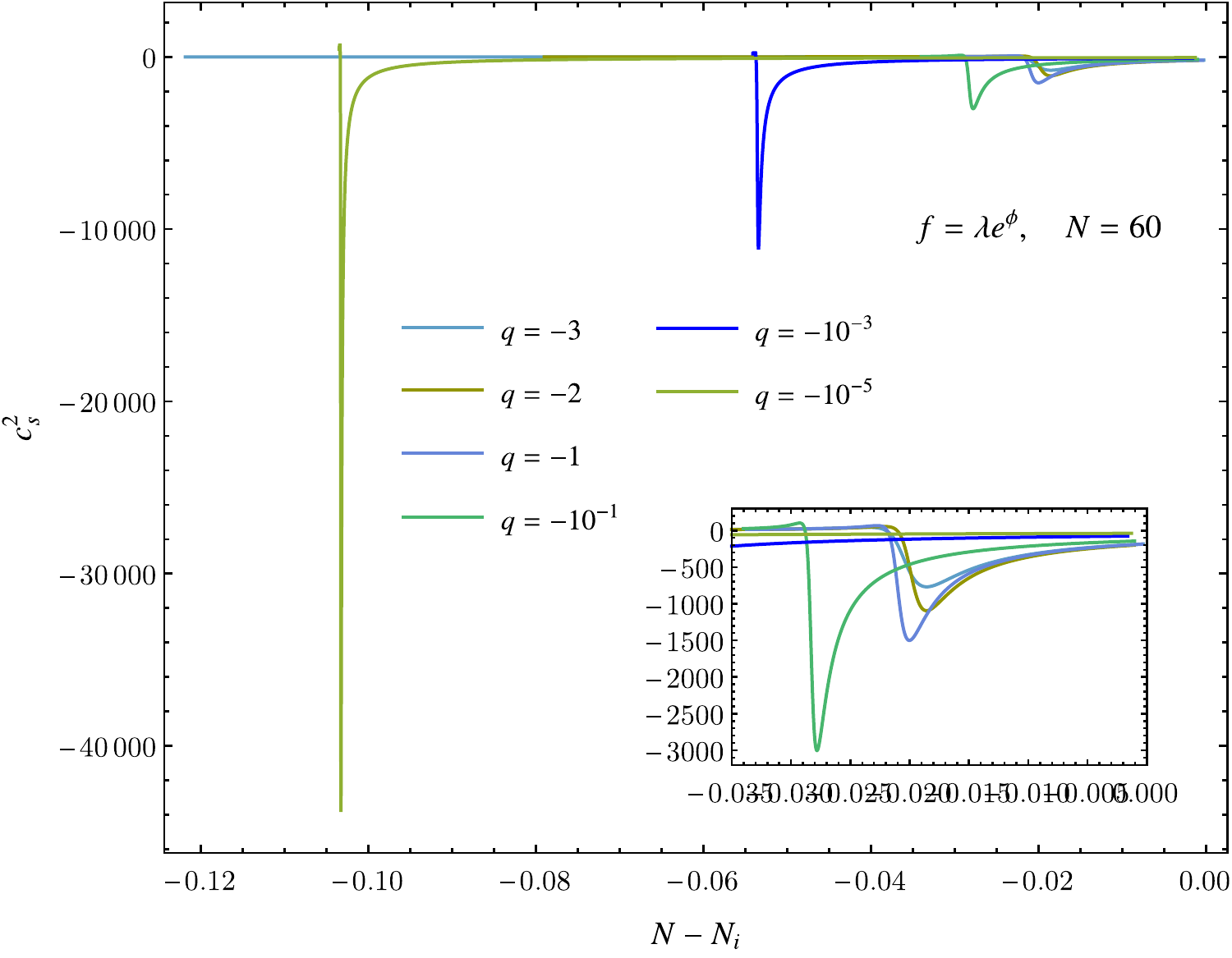}
    \caption{The squared sound velocity of tensor perturbations as a function of number of efolding from singularity avoidance ($N-N_i$) in inflationary, nonsingular solutions with $q<0$ and $N_0=60$.}\label{ephiINFL-csNN=60d=-1}
\end{figure}

Inflationary solutions, found with an exponential coupling when $q<0$, have been described in detail in section \ref{sec:expinfla}. We recognize in them some sort of \lq\lq singularity avoidance" (even though the solution eventually becomes singular), an abrupt change in the scale factor evolution, occurring at different time and e-foldings number ($N=N_i$) according to the value of $q$. This point is peculiar for the solution's stability. In Figure \ref{ephiINFL-csNN=60d=-1} we see that sound velocity of tensor perturbations is always negative at that point. Going back in time, $c_s^2$ reaches a negative minimum and then becomes bigger that one, signalling super-luminal propagation. Even for $q$ approaching its maximum value the solution is unstable. As one could expect, instability always resurfaces near the singularity, although this is not shown in our plots.\\
No stable, inflationary solution is found in the case under study within the range $10^{-7} \leq \vert q \vert < 6$. 

\subsubsection*{Positive coupling, $\boldsymbol{q>0}$}
\begin{figure}[!ht]
 \centering
    \includegraphics[width=0.8\textwidth]{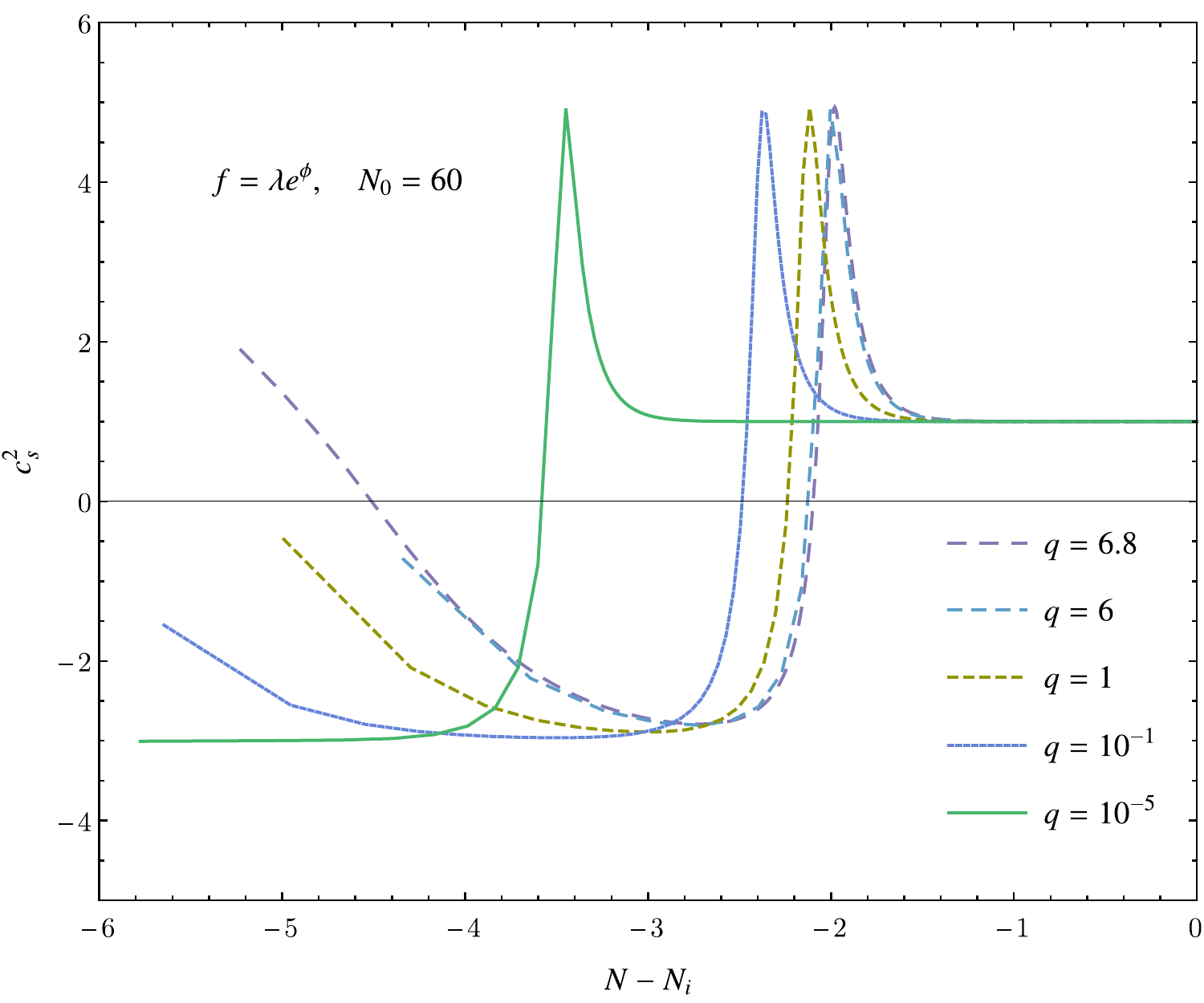}
    \caption{The squared sound velocity of tensor perturbations as a function of number of efolding in inflationary solutions with $q>0$ and $N_0=60$. Inflationary expansion starts at $ N-N_i= 0 $, and the plot terminates approximately at the singularity.}\label{ephiINFL-csNN=60d=1}
\end{figure}

Inflationary solutions with an exponential, positive coupling are found to be unstable and superluminal for every value of q in the considered range, $10^{-5} \leq \vert q \vert < 6.8$. As Figure \ref{ephiINFL-csNN=60d=1} shows, the instability emerges very early in time, i.e. near the singularity; solutions are stable, instead, both during inflation and immediately before it. Solutions with $ N_0 = 70 $, not reported here, behave in a similar way.

\section{Numerical results for primordial perturbations}
In this Section we compare our numerical predictions for the values of $ n_s $ and $ r $ with the constraints based on the full temperature and polarization data of the Planck mission, released in February 2015~\cite{PLANCKinfl}. We use Planck results obtained both from temperature power spectrum (TT), and from temperature power spectrum joined with temperature-polarization cross spectrum (TE), and polarization power spectrum (EE). In both cases the low-multipole polarization power spectrum (LowP) is also taken into account. The Planck mission finds for the scalar spectral index a remarking departure from exact scale invariance ($ n_s < 1 $):
\begin{align}\label{Planckdata}
n_s &= 0.9666 \pm 0.0062 (68\% CL) \quad \text{TT+LowP}, \\
n_s &= 0.9652 \pm 0.0047 (68\% CL) \quad \text{TT,TE,EE,+LowP}.
\end{align}
There is no evidence for inflationary produced tensor modes, since constraints on the tensor-to-scalar ratio are:
\begin{equation}\label{key}
r < 0.106 (95\% \text{CL}) \ \text{TT,+LowP},\quad r < 0.109 (95\% \text{CL}) \  \text{TT,TE,EE,+LowP}.
\end{equation}
In both cases the parameters obtained with TT, TE, EE + LowP and with TT + LowP are consistent with each other.\\
As already stressed in Section \ref{sec:phenompred}, GB inflation predicts the breakdown of the consistency relation between $ r $ and $ n_T $. The constraints provided by the Planck mission, instead, make use of this consistency relation. Planck's constraints on GB inflation could be improved by discarding the consistency relation, although this is
left to a future extension of the present work.\\

Our predictions are based on the numerical inflationary solutions described in Section \ref{sec:numsolinfl} and on the approximated expressions for $ n_s $ and $ r $ \eqref{approxparam}. As always when dealing with inflationary solutions, once $ N_0 $ is fixed, only a limited range of $ q $ is phenomenologically viable (see Section \ref{sec:numsolinfl}). The $ q = 0 $ case is also considered. This is obtained by setting $ \lambda = 0 $, i.e. switching off the Gauss-Bonnet non minimal coupling. We are then left with a simple chaotic inflation model, for which the values of $ n_s $ and $ r $ \eqref{nsrNcmb} are well known.

\subsection{Quadratic coupling}
\begin{figure}[!ht]
 \centering
    \includegraphics[width=0.8\textwidth]{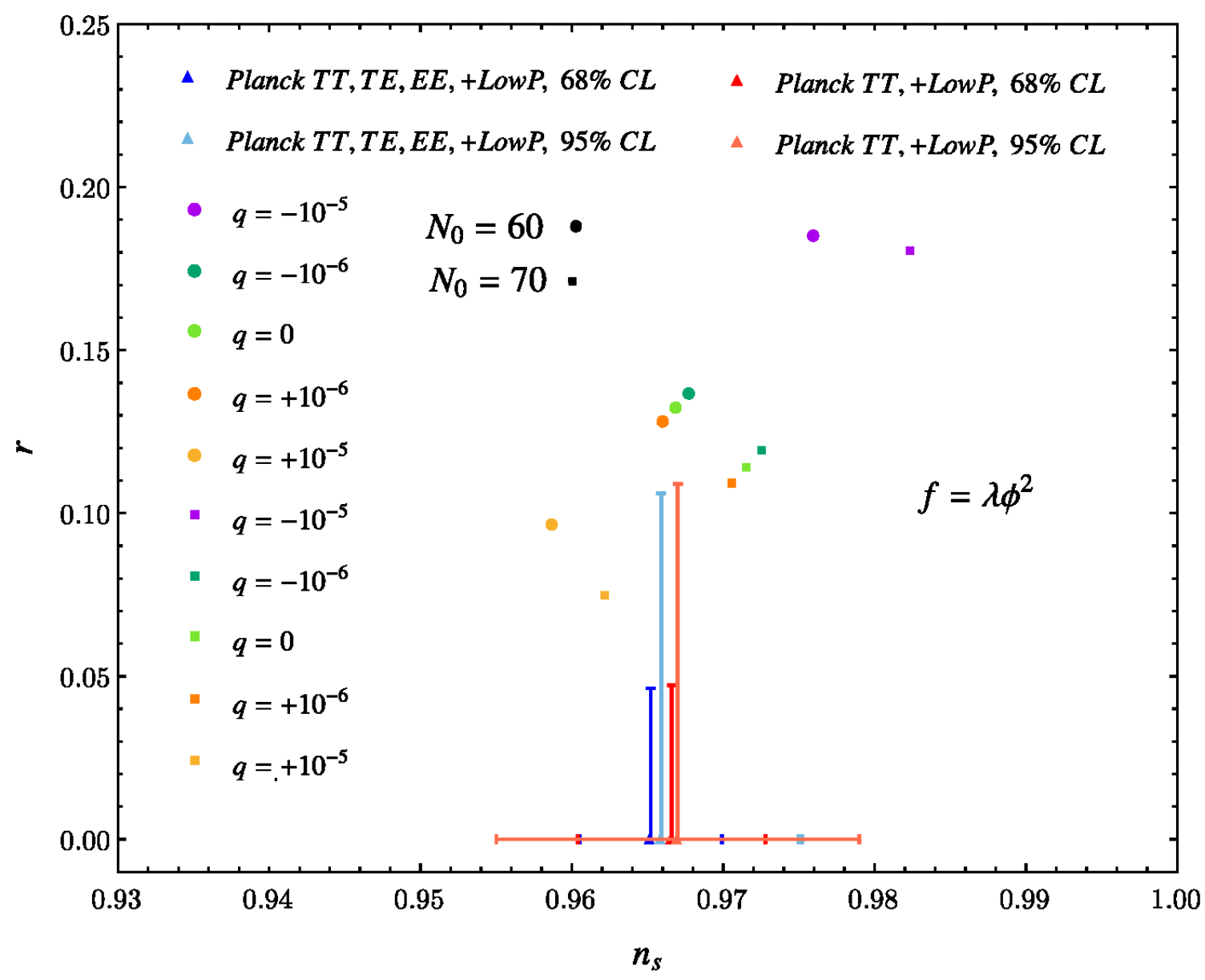}
    \caption{Predicted values of $n_s$ and $r$ in quadratic coupling GB inflation with different values of $q$ and $ N_0 $, together with Planck 68 \% and 95\% CL results.}\label{phi2-nsrN60}
\end{figure}

We first address the quadratic coupling function case, and estimate $ n_s $ and $ r $ for different values and signs of the parameter $ q $, and with $ N_0 = 60 − 70 $. These inflationary parameters were already calculated in \cite{KohObsCos} and compare with PLANCK and BICEP2 2014 data. Their analysis is similar to ours, and results for ns and r are consistent. However, the constraints they refer to are plagued by an incorrect estimation of dust contribution to polarization. Their conclusions cannot be considered valid, in light of the latest cosmological data. More recently, in \cite{PlanckReheating} the model is revisited in light of Planck 2013 constraints.\\
Our results, together with Planck's constraints, are shown in Figure \ref{phi2-nsrN60}. Fixing the number of e-foldings, both ns and r are enhanced when q is negative and suppressed for positive q. When $ N_0 = 60 $, our predictions all lie outside the $ 68\% $ and $ 95\% $ CL regions. Conversely, increasing the number of e-foldings brings $ q > 0 $ predictions inside the $ 95\% $ CL region. We may say that, within the framework of quadratic coupling scalar-GB theory, parameter region $ N_0 \simeq 70 $ and $ q \gtrsim 0 $ is excellently consistent with Planck 2015 constraints. All cosmological solutions associated to the favoured range of q are unstable during the pre-inflationary stage and stable as inflation takes place, as discussed in Section \ref{stabinfla}. However, neither $ N_0 < 70 $ nor $ q < 0 $ are completely ruled out.

\subsection{Exponential coupling}
\begin{figure}[!ht]
	\centering
	\includegraphics[width=0.8\textwidth]{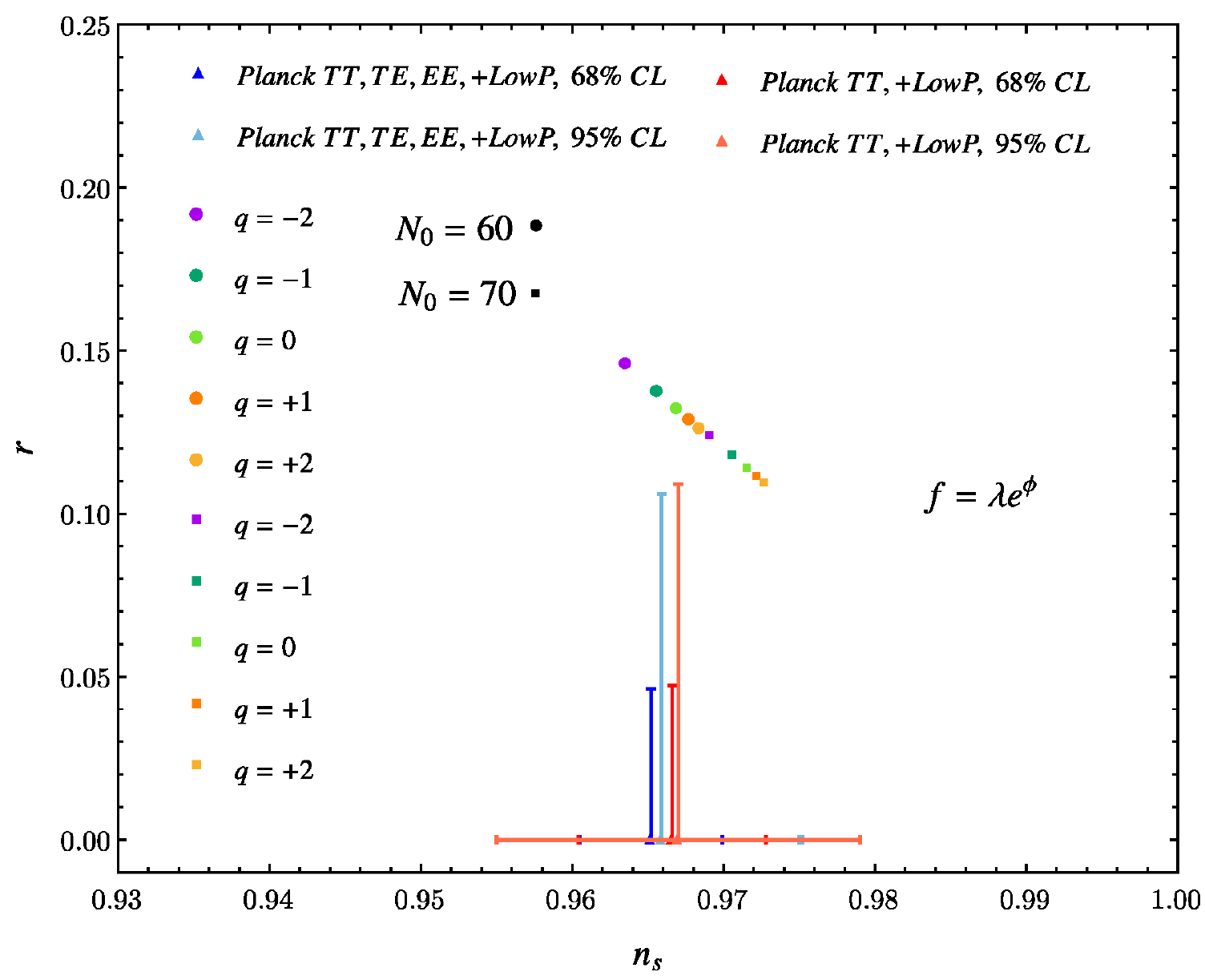}
	\caption{Predicted values of $n_s$ and $r$ in exponential coupling GB inflation with different values of $q$ and $ N_0 $, together with Planck 68 \% and 95\% CL results.}\label{ephiphi-nsrN60}
\end{figure}

Inflationary parameters for the exponential case were first calculated in \cite{JHGGBInflation}. However, their analysis and ours are slightly different, e.g. in the way the parameters to fix and the ones to tune are selected. The referred research also dates back to 2013 and compares GB predictions with Planck+WP2 2013 constraints, while we make use of the latest Planck update.\\
Figure \ref{ephiphi-nsrN60} displays, along with Planck's constraints, our ns and r values for the exponential coupling case. We fix the total number of e-foldings to $ 60 $, and let the coupling q take both signs.\\
When the GB coupling function has an exponential form, we find for $ n_s $ and $ r $ a different dependence upon the sign of $ q $, if compared to the quadratic coupling case. When the coupling becomes more and more negative, the tensor-to-scalar ratio is enhanced, while the scalar spectral index decreases. Vice versa, when the coupling is positive r is suppressed and ns becomes slightly bigger than the reference value associated to $ q = 0 $. Our results all lie outside $ r $'s $ 95\% $ CL region, though $ N_0 = 70 $ results, having a smaller tensor-to-scalar ratio, lie closer to it. We may conclude that 2015 Planck data favour, for exponentially coupled GB gravity, $ N_0 > 60 $ and $ q \gtrsim 0 $. Interestingly, the preferred sign of the coupling ($ q  \gtrsim 0 $) also guarantees that solutions are stable immediately before inflation (see Section \ref{stabexponinfl}).

\chapter{Conclusion}
Throughout this work we studied how scalar-Gauss-Bonnet gravity behaves at early cosmological times. Our aim was to convey a picture of early GB cosmology through new numerical solutions, as well as through a deeper understanding and a critical analysis of new and old findings in the field. By addressing both non singular, bouncing and inflationary solutions in several models, our study wishes to be comprehensive and detailed, at the same time.\\
After re-deriving the well known cosmological solutions of the quadratic coupling model of GB gravity, we addressed the recently proposed pure Gauss-Bonnet approximation \cite{KantiEarlyTime,KantiGBInflation}. We discovered that, despite pure GB being an interesting and handy theory, it cannot be considered as a good approximation of the complete model, at least in the quadratic coupling case. Our comparison between pure and complete GB solutions shows that they do not even share features as fundamental as the initial singularity. We related this to the failed fulfilment of the hypothesis on which the approximation relies. We found that analytical solutions of pure GB equations are not completely satisfying as early time approximations of the complete theory.\\
Wishing to complete the landscape of non singular solutions in GB cosmology, we also studied the previously overlooked case of an exponential coupling. As theoretical arguments suggested, new non singular solutions could be numerically found. As expected, again on theoretical grounds, solutions revealed an instability in the tensor sector, and are therefore to be rejected on phenomenological grounds.\\
Both the quadratic and the exponential GB models can easily account for inflation, once a simple mass term for the scalar field is introduced. Our effort was devoted to the exploration of the inflationary parameter space: we found limits for inflationary solutions and for their early-time stability. The new instabilities found in this thesis pose a new problem, as they could prevent the model from being a successful candidate for inflation: they demand caution and deserve closer examination. Phenomenologically relevant predictions of GB inflation were studied as well, as we estimated the scalar spectral index and tensor-to-scalar ratio. These predictions were commented upon in light of the stability analysis and the most up-to-date Planck constraints.\\
As a final remark, we stress that results obtained in the Planckian region must be treated as only preliminary directions on effects to be later verified by quantum gravity calculations. At the same time, most of our results, such as the nonsingular and/or bouncing solutions, can be regarded as completely classical. First of all, because Gauss-Bonnet gravity is a legitimate classical alternative to GR. Secondly, the time scale and curvature at which the singularity is avoided is determined by the value of the coupling parameter $ \vert \lambda \vert $: a sufficiently large coupling can guarantee a completely classical evolution.\\
We believe that the interesting properties of GB gravity, which clearly emerge in this work, make devoting it further study worth the effort. On the one side, instabilities on spatially curved background deserve a separate treatment, still missing. In addition, the reliability of some of our predictions can be further improved: one could, for instance, take into account the time dependence of inflationary parameters, and compare the running of spectral indexes in GB inflation with the new cosmological constraints.\\
By reading this thesis one can get a taste of how challenging violating the Strong Energy conditions in a healthy manner can be. Although we unsuccessfully looked for stable non singular solutions, allowed for by the scalar potential, hope is not lost that such a result can some day be achieved. Inspiration can be drawn from the recent work of Ijjas and Steinhardt \cite{IjjasSte}, who obtained stable cosmological bounces in Galileon gravity with a reverse-engineering technique. Application of their method to GB cosmology could possibly lead to a similar outcome.

\appendix
\chapter{Numerical methods}
In this Appendix we describe the Mathematica built-in methods for numerical in- tegration \cite{Wolfram}, implemented throughout this thesis to solve GB equations.

\section{Explicit modified Midpoint}
When solving the first order, ordinary differential equation and initial value problem:
\begin{align}\label{problem}
\begin{cases}
y'(t) &= f(t,y(t)),\\
y(t_0) &= y_0,
\end{cases}
\end{align}
each step of the explicit midpoint method is given by:
\begin{equation}\label{explicit}
y_{n+1}=y_{n}+h\, f\left( t_{n}+\frac{h}{2},y_{n}+\frac{h}{2}f(t_n,y_n) \right) +O(h^3).
\end{equation}
Here $ n = 0,1,2... $, while h is the step size ($ t_n = t_{n-1} +h = ... = t_0 +nh $). The global error of the method is $ O(h^2) $ (globally second order method).
In the modified explicit midpoint method, stability is improved by the use of Gragg's smoothing step, which redefines:
\begin{equation}\label{key}
Sy_n = \frac{1}{4} (y_{n-1} + 2y_n + y_{n+1}) . 
\end{equation}
\section{Linearly implicit Euler}
The implicit Euler method approximates the solution of problem \eqref{problem} using:
\begin{equation}\label{key}
y_{n+1} = y_n + hf(t_{n+1}, y_{n+1}) + O(h^2).
\end{equation}

\chapter{Numeric solutions, plots}
\section{Quadratic coupling, $V=0$}
\begin{figure}[!ht]
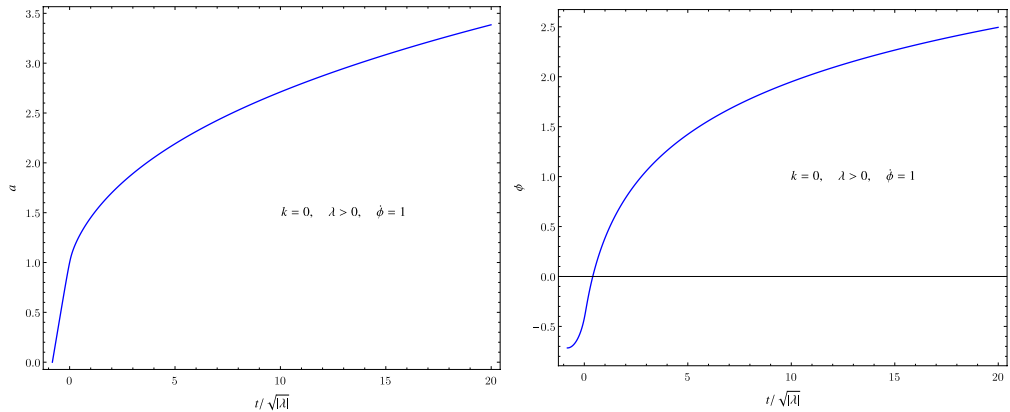

 \centering
 \begin{subfigure}{.47\textwidth}
 \centering
    \includegraphics[width=1\textwidth]{phi2-ak=0d=1}
\end{subfigure}
\begin{subfigure}{.47\textwidth}
\centering
    \includegraphics[width=1\textwidth]{phi2-phik=0d=1}
    \end{subfigure}
    \caption{The scale factor (left) and the scalar field (right) as a function of rescaled time for the singular solution with $k=0$, $\lambda=1$, first branch.}\label{phi2-aphik=0d=1}
\end{figure}

\begin{figure}[!ht]
 \centering
 \begin{subfigure}{.47\textwidth}
 \centering
    \includegraphics[width=1\textwidth]{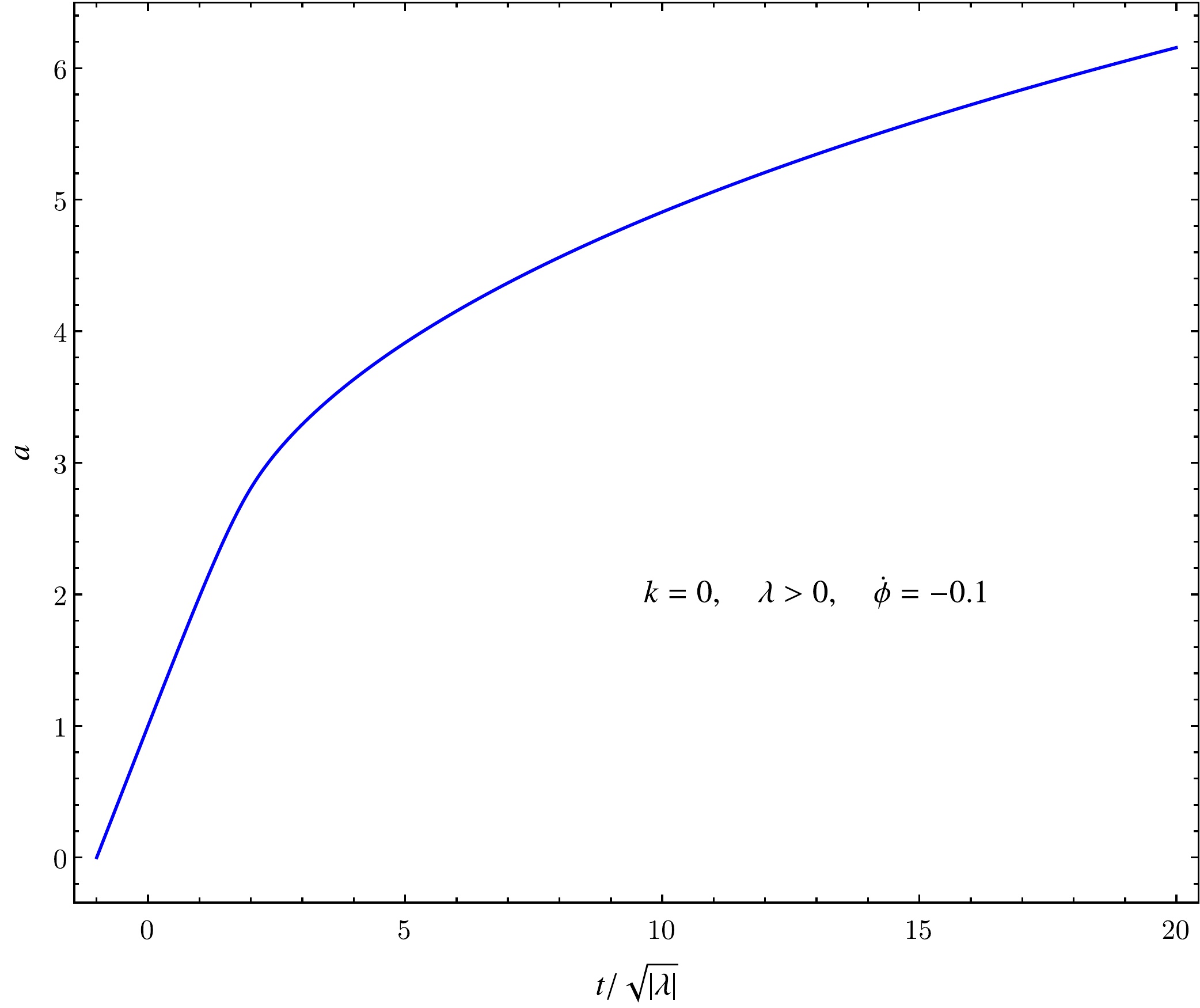}
\end{subfigure}
\begin{subfigure}{.47\textwidth}
\centering
    \includegraphics[width=1\textwidth]{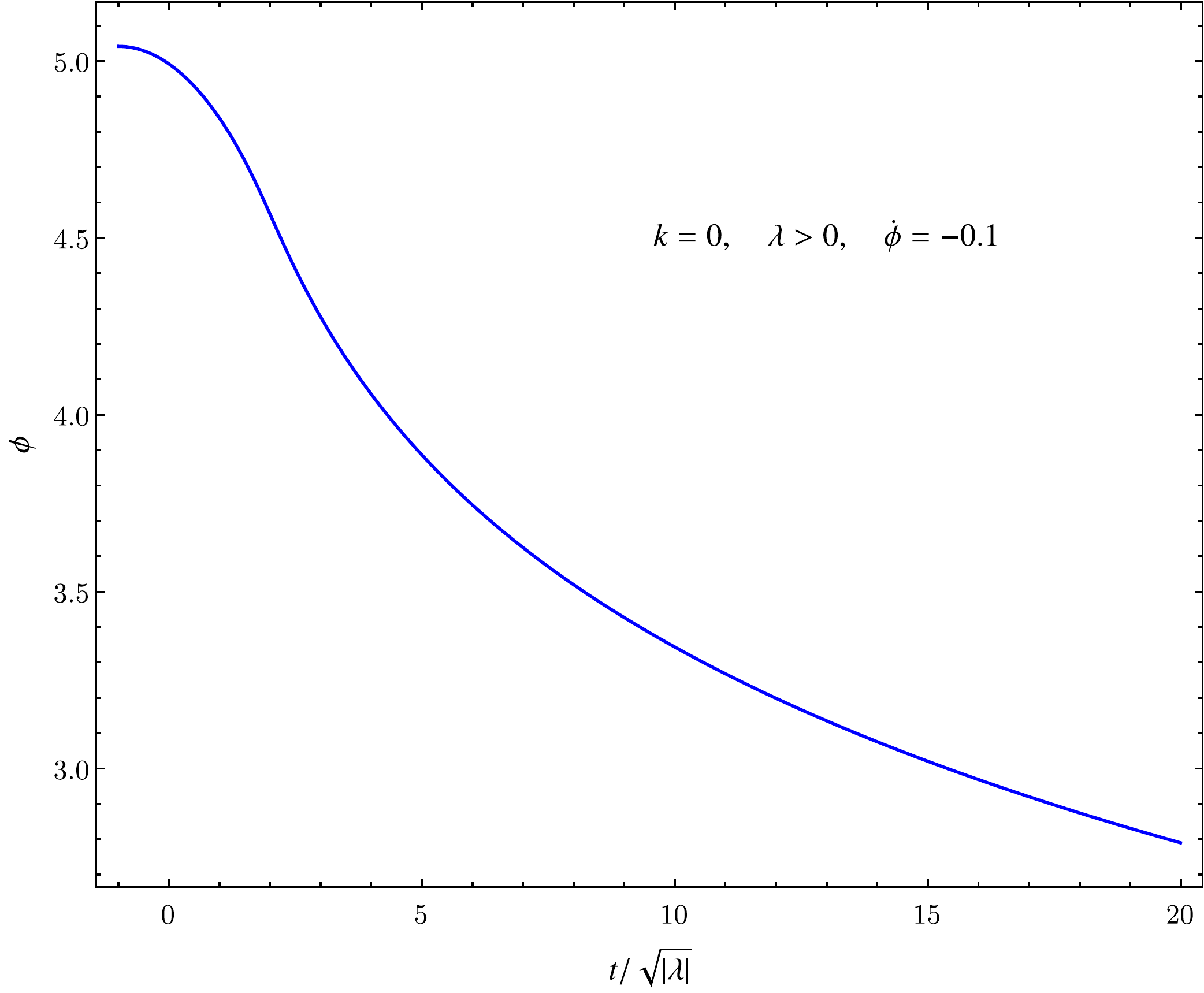}
    \end{subfigure}
    \caption{The scale factor (left) and the scalar field (right) as a function of rescaled time for the singular solution with $k=0$, $\lambda=1$, second branch.}\label{phi2-aphik=0d=1bis}
\end{figure}

\begin{figure}[!ht]
 \centering
 \begin{subfigure}{.47\textwidth}
 \centering
    \includegraphics[width=1\textwidth]{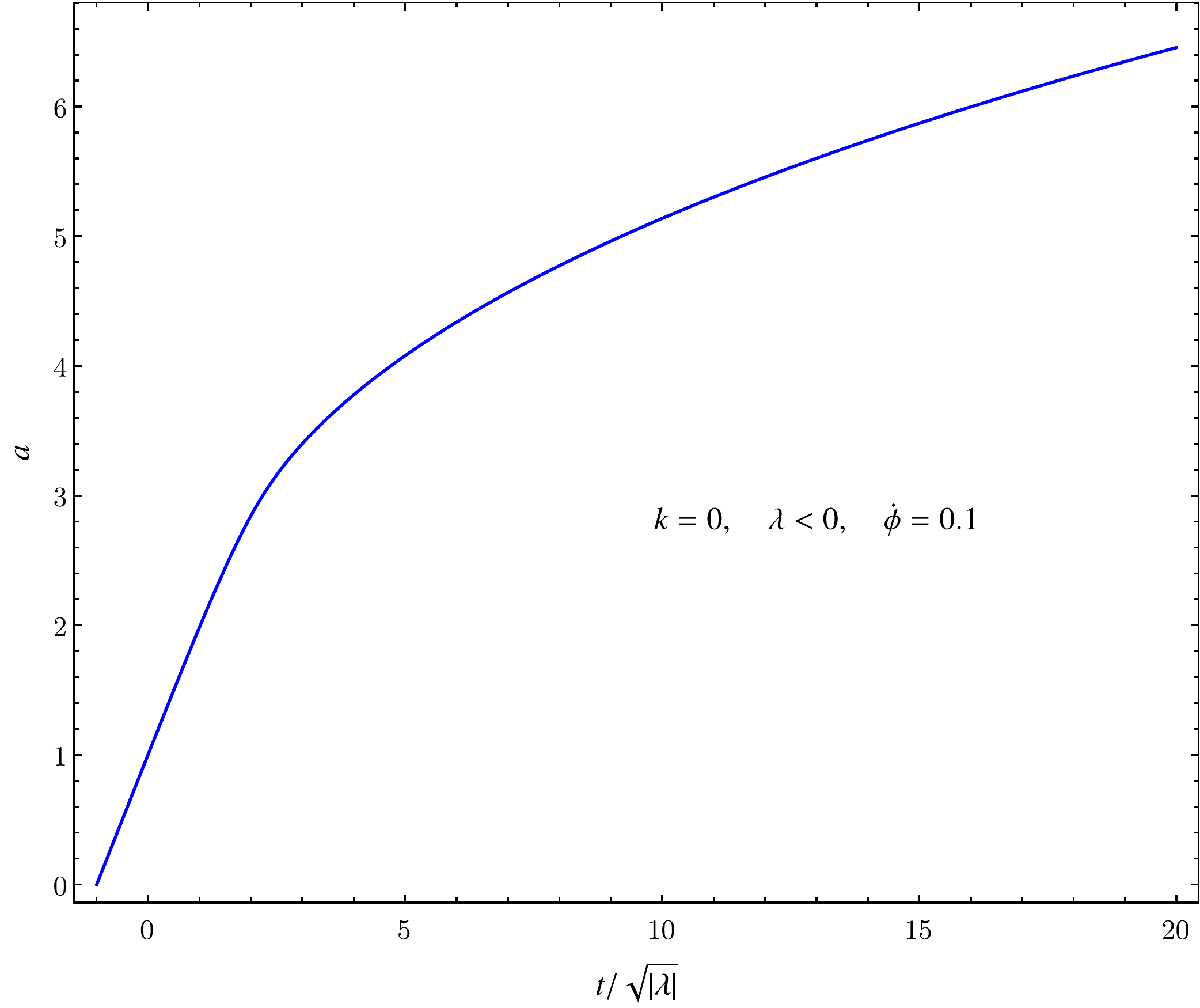}
\end{subfigure}
\begin{subfigure}{.47\textwidth}
\centering
    \includegraphics[width=1\textwidth]{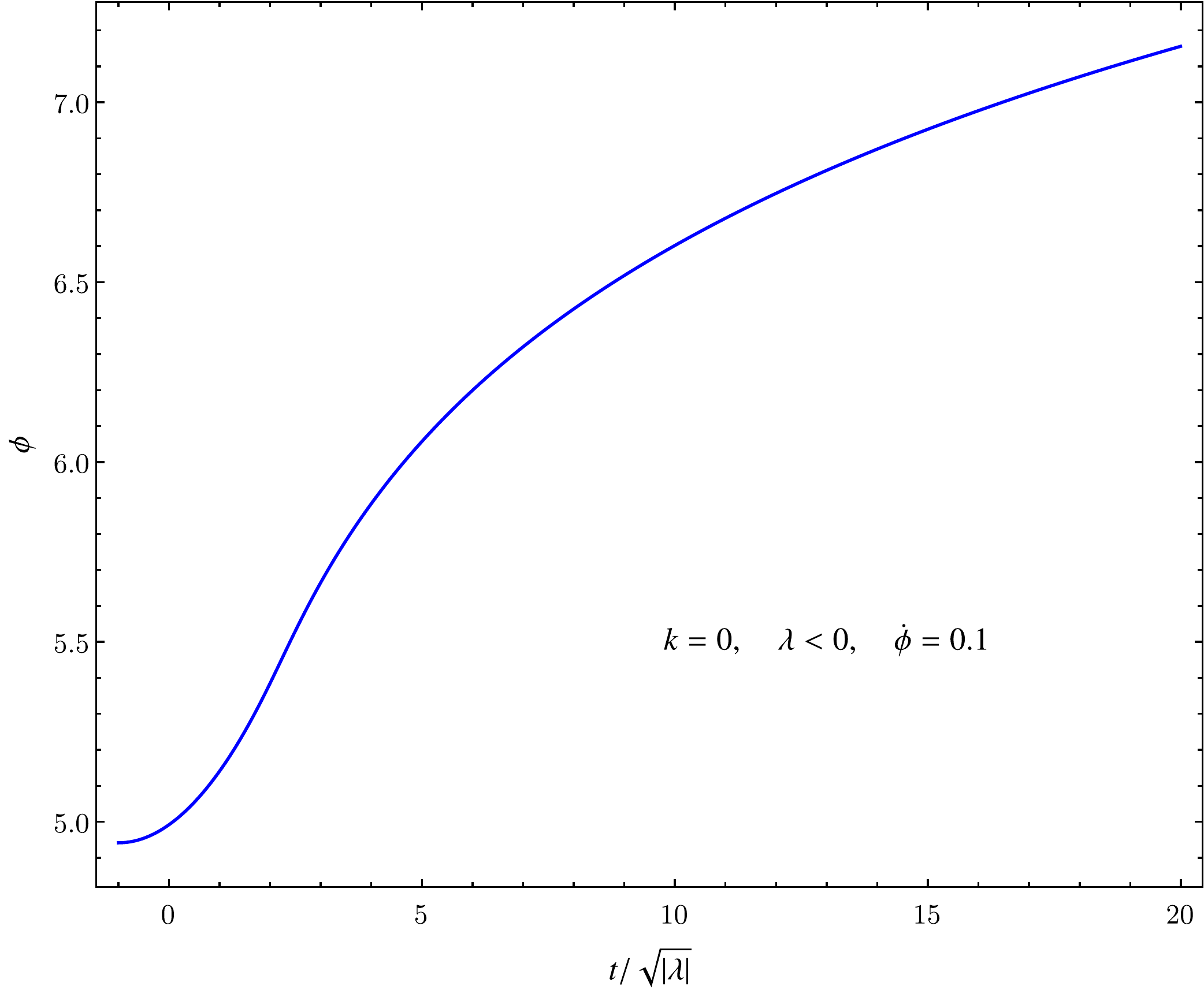}
    \end{subfigure}
    \caption{The scale factor (left) and the scalar field (right) as a function of rescaled time for the solution with $k=0$, $\lambda=-1$, singular branch.}\label{phi2-aphik=0d=-1bis}
\end{figure}

\begin{figure}[!ht]
 \centering
 \begin{subfigure}{.47\textwidth}
 \centering
    \includegraphics[width=1\textwidth]{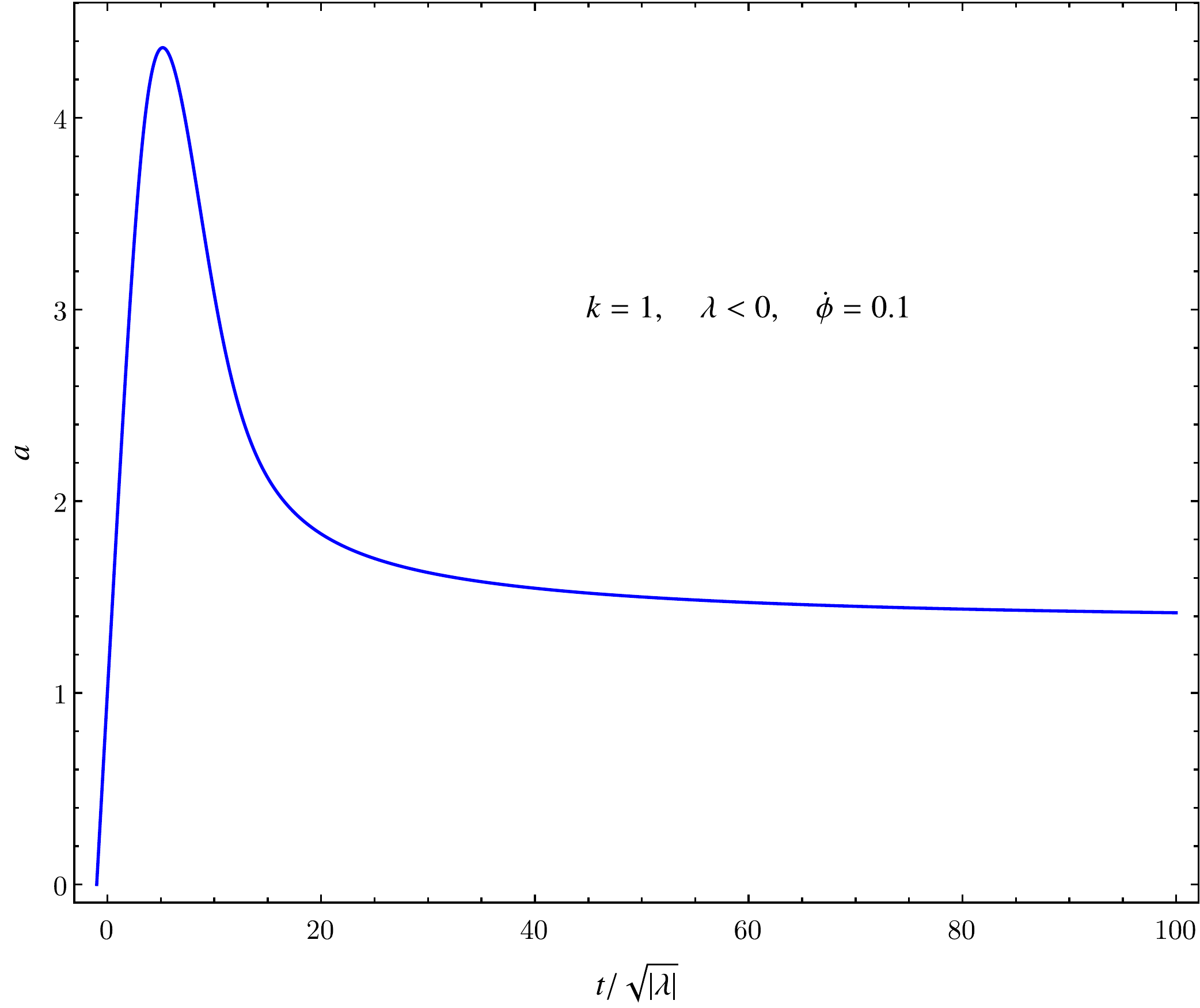}
\end{subfigure}
\begin{subfigure}{.47\textwidth}
\centering
    \includegraphics[width=1\textwidth]{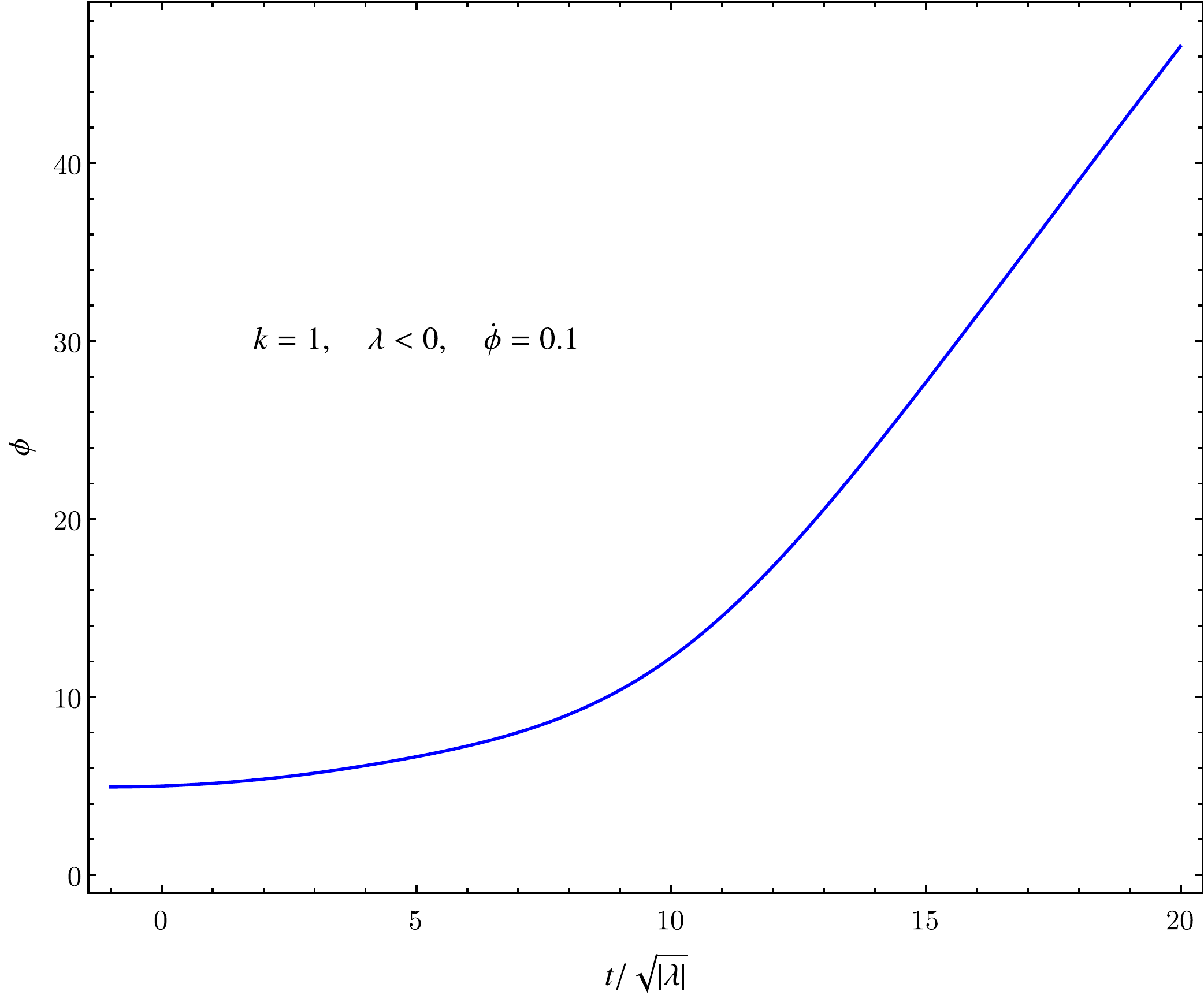}
    \end{subfigure}
    \caption{The scale factor (left) and the scalar field (right) as a function of rescaled time for the singular solution with $k=1$, $\lambda=-1$, second branch.}\label{phi2-aphik=1d=-1bis}
\end{figure}

Here we show some numerical solutions (for the scale factor and the scalar field) of the Guass-Bonnet cosmological equations with a quadratic coupling function, in addition to the more interesting solutions described in Section \ref{sec:NumAnphi2}. In Figures \ref{phi2-aphik=0d=1}, \ref{phi2-aphik=0d=1bis} we display the two branches of singular solutions found with positive coupling and flat space-time. We also display the singular branch of the negative coupling solution with $k=0$, Figure \ref{phi2-aphik=0d=-1bis}. For the positive curvature case, we show (Figure \ref{phi2-aphik=1d=-1bis}) the singular branch of the negative coupling solution.\\ 
 
\clearpage
\section{Exponential coupling, $V=0$}
\begin{figure}[!ht]
 \centering
 \begin{subfigure}{.47\textwidth}
 \centering
    \includegraphics[width=1\textwidth]{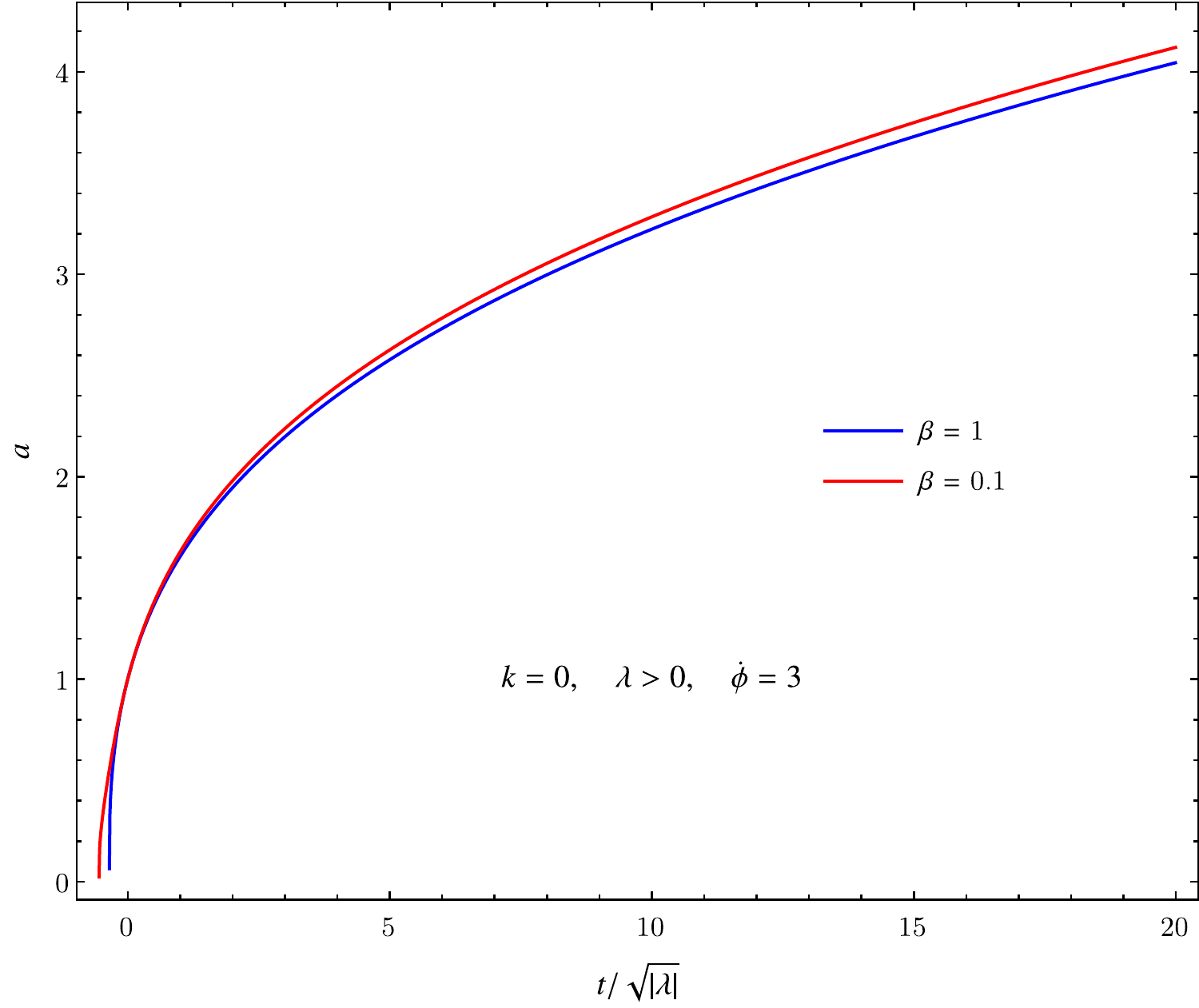}
\end{subfigure}
\begin{subfigure}{.47\textwidth}
\centering
    \includegraphics[width=1\textwidth]{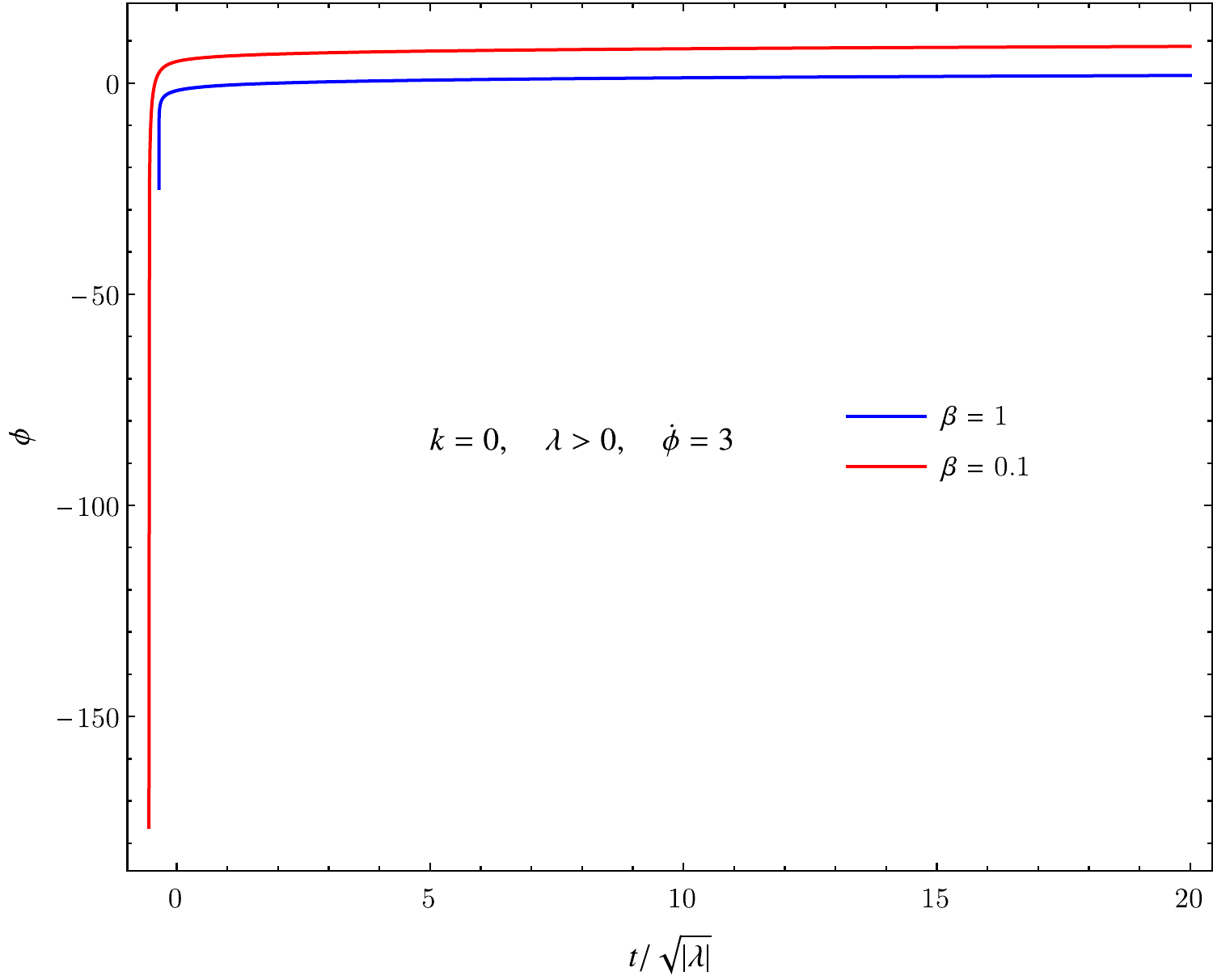}
    \end{subfigure}
    \caption{The scale factor (left) and the scalar field (right) as a function of rescaled time for the singular solution with $k=0$, $\lambda=1$, second branch.}\label{ephi-aphik=0d=1both2}
\end{figure}

\begin{figure}[!ht]
 \centering
 \begin{subfigure}{.47\textwidth}
 \centering
    \includegraphics[width=1\textwidth]{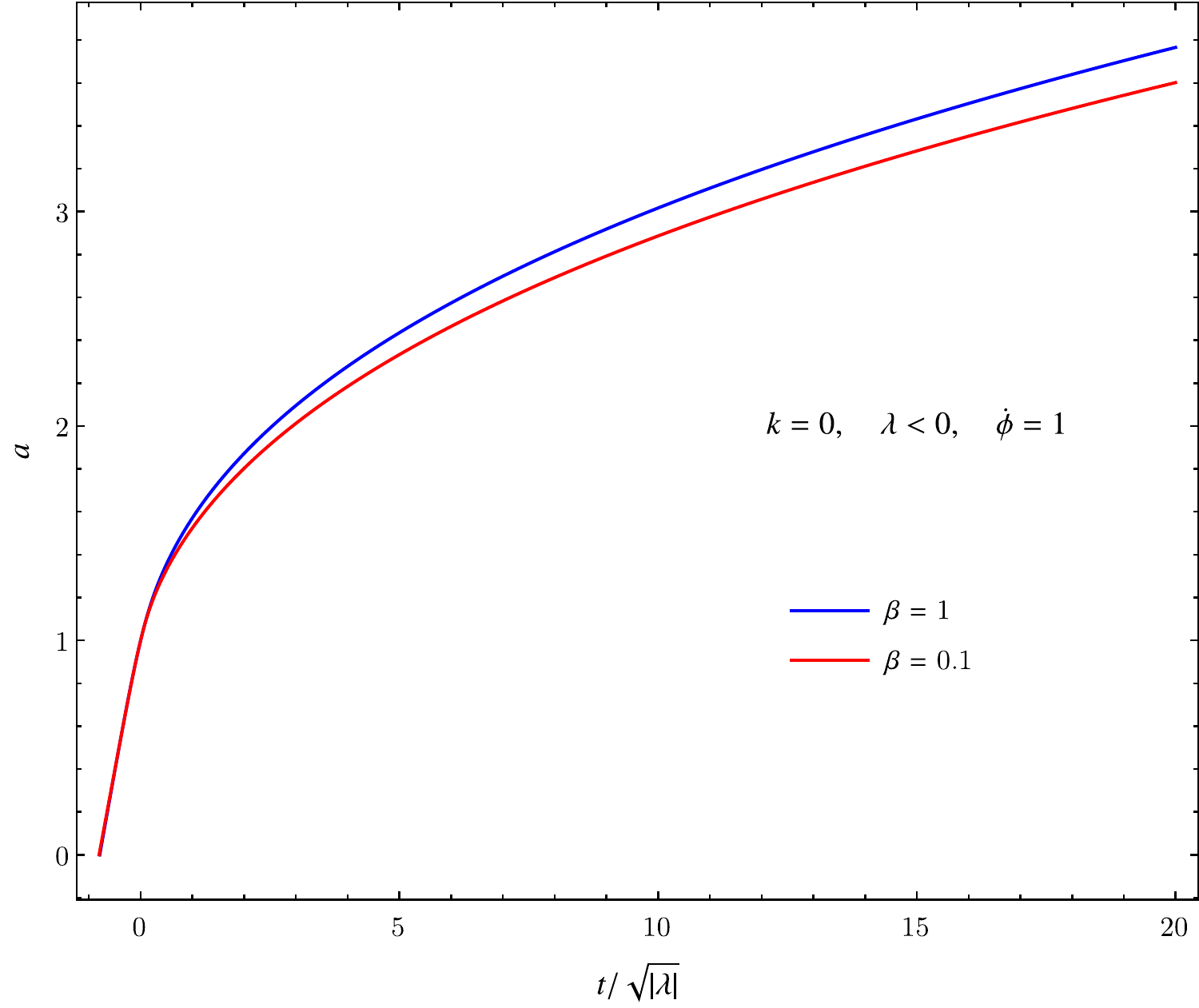}
\end{subfigure}
\begin{subfigure}{.47\textwidth}
\centering
    \includegraphics[width=1\textwidth]{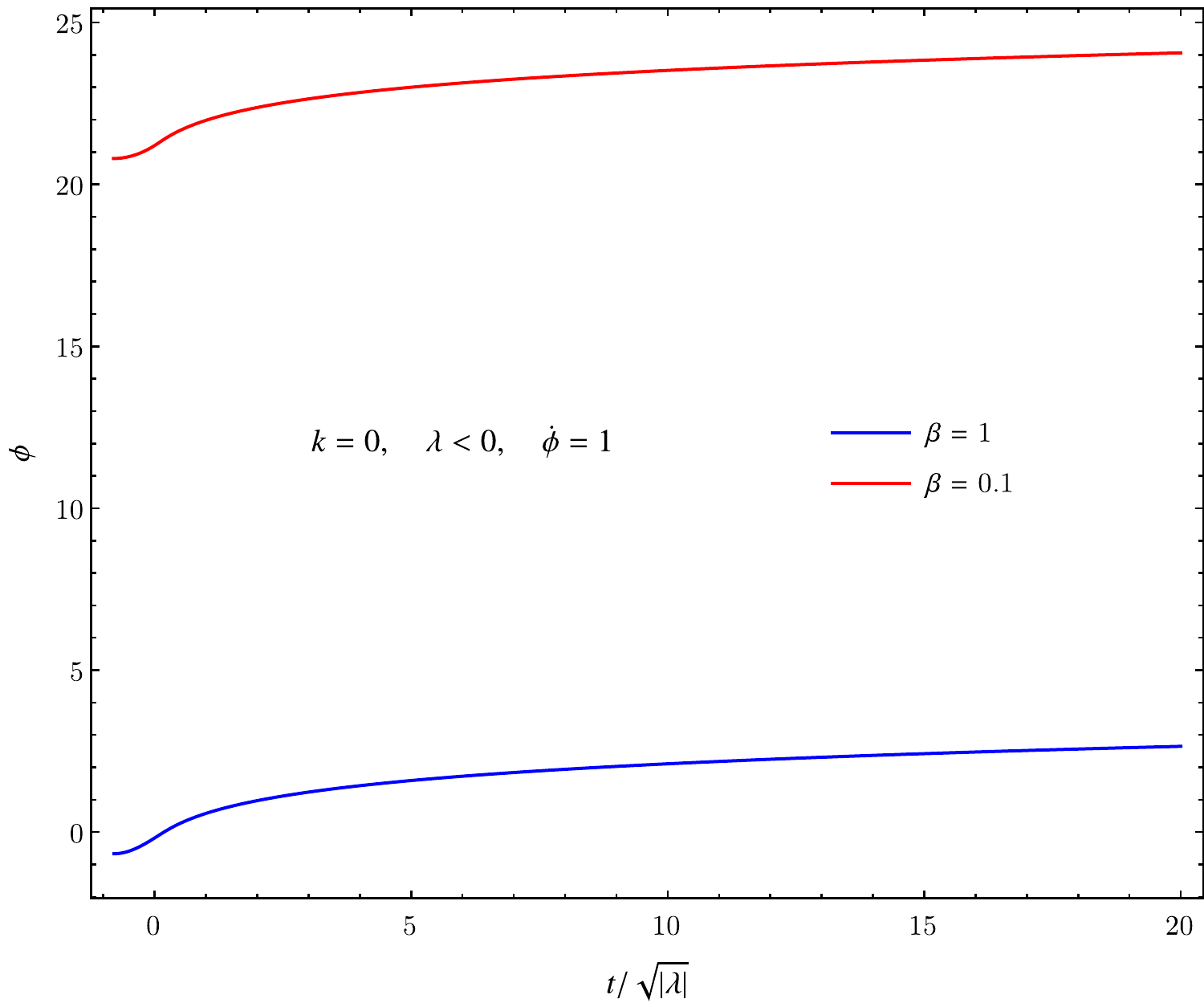}
    \end{subfigure}
    \caption{The scale factor (left) and the scalar field (right) as a function of rescaled time for the singular solution with $k=0$, $\lambda=-1$, first branch.}\label{ephi-aphik=0d=-1both}
\end{figure}

\begin{figure}[!ht]
 \centering
 \begin{subfigure}{.46\textwidth}
    \includegraphics[width=1\textwidth]{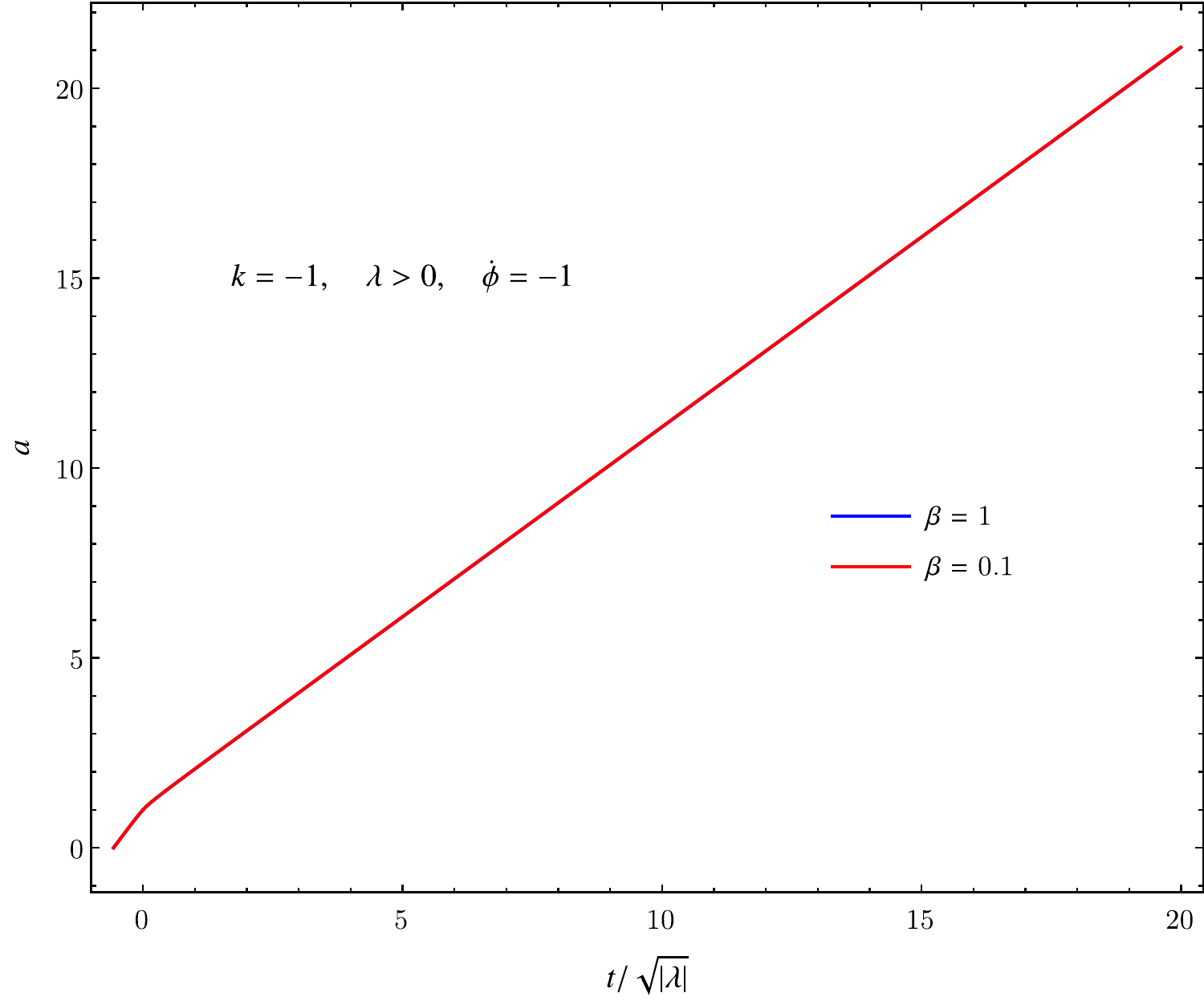}
\end{subfigure}
\begin{subfigure}{.46\textwidth}
    \includegraphics[width=1\textwidth]{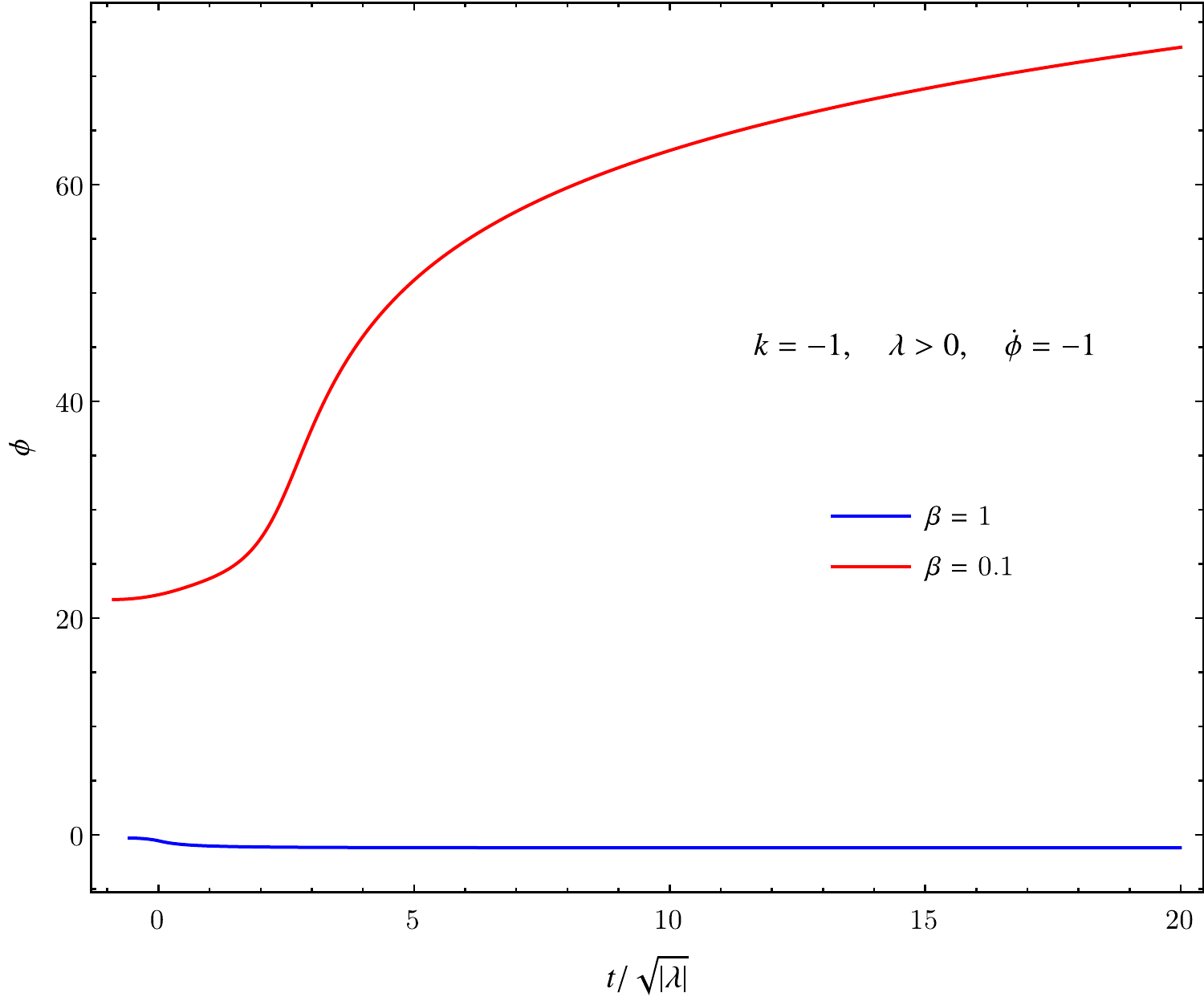}
    \end{subfigure}
    \caption{The scale factor (left) and the scalar field (right) as a function of rescaled time for $k=-1$, $\lambda=1$, first branch.}\label{ephi-aphik=-1d=1bothbi}
\end{figure}

Here we display some numerical solutions of the exponential coupling case, discussed with more detail in Section \ref{sec:numExp}. The singular solutions shown in Figures \ref{ephi-aphik=0d=1both2}, \ref{ephi-aphik=0d=-1both}, are obtain with flat space-time, and with positive and negative coupling respectively. In Figure \ref{ephi-aphik=-1d=1bothbi} we show a positive curvature, negative coupling singular solution.\\

\addcontentsline{toc}{chapter}{Bibliography}
\bibliographystyle{plainnat}

\end{document}